 \documentclass[manuscript]{emulateapj}

 \usepackage{apjfonts}

\slugcomment{The Astrophysical Journal, 785:97, 2014 April 20}

\shorttitle{$UBV$ Color Evolution of Classical Novae}
\shortauthors{Hachisu \& Kato}

\begin{document}

\title{The $UBV$ Color Evolution of Classical Novae. I. Nova-Giant
Sequence in the Color-Color Diagram}


\author{Izumi Hachisu}
\affil{Department of Earth Science and Astronomy, 
College of Arts and Sciences, The University of Tokyo,
Komaba, Meguro-ku, Tokyo 153-8902, Japan} 
\email{hachisu@ea.c.u-tokyo.ac.jp}

\and

\author{Mariko Kato}
\affil{Department of Astronomy, Keio University, 
Hiyoshi, Kouhoku-ku, Yokohama 223-8521, Japan} 
\email{mariko@educ.cc.keio.ac.jp}

%\and
%
%\author{Rosario Gonz\'alez-Riestra} \affil{XMM Science Operation
%Centre, ESAC, P.O. Box 78, 28091 Vallanueva de la Ca\~nada, Madrid,
%Spain}
%\email{Rosario.Gonzalez@sciops.esa.int}

%\author{Angelo Cassatella}
%\affil{INAF, Istituto di Fisica dello Spazio Interplanetario,
%Via del Fosso del Cavaliere 100, 00133 Rome, Italy}
%\email{cassatella@fis.uniroma3.it}

%\and
%
%\author{Taichi Kato}
%\affil{Department of Astronomy, Kyoto University, 
%Sakyo-ku, Kyoto 606-8502, Japan} 
%\email{tkato@kusastro.kyoto-u.ac.jp}

%\altaffiltext{1}{Visiting Astronomer, Cerro Tololo Inter-American Observatory.
%CTIO is operated by AURA, Inc.\ under contract to the National Science
%\altaffiltext{2}{Society of Fellows, Harvard University.}
%\altaffiltext{3}{present address: Center for Astrophysics,
%    60 Garden Street, Cambridge, MA 02138}
%\altaffiltext{4}{Visiting Programmer, Space Telescope Science Institute}
%\altaffiltext{5}{Patron, Alonso's Bar and Grill}


\begin{abstract}
We identified a general course of classical nova outbursts in the $B-V$
versus $U-B$ color-color diagram.  It is reported that novae show
spectra similar to those of A--F supergiants near optical light maximum.
However, they do not follow the supergiant sequence in the color-color
diagram, neither the blackbody nor the main-sequence sequence.
Instead, we found that novae evolve along a new sequence in the pre-maximum 
and near-maximum phases, which we call ``the nova-giant sequence.''
This sequence is parallel to but $\Delta (U-B)\approx -0.2$ mag
bluer than the supergiant sequence.   This is because the mass of a nova
envelope is much ($\sim10^{-4}$ times) less than that of a normal 
supergiant.  After optical maximum, its color quickly evolves back 
blueward along the same nova-giant sequence and reaches the point of 
free-free emission ($B-V=-0.03$, $U-B=-0.97$), which coincides with 
the intersection of the blackbody sequence and the nova-giant sequence,
and remains there for a while.  Then the color evolves leftward 
(blueward in $B-V$ but almost constant in $U-B$), owing mainly to 
the development of strong emission lines.  This is the general course
of nova outbursts in the color-color diagram, which was deduced from
eight well-observed novae in various speed classes.  For a nova with unknown
extinction, we can determine a reliable value of the color excess by 
matching the observed track of the target nova with this general course.
This is a new and convenient method for obtaining the color excesses of
classical novae.  Using this method, we redetermined the color excesses 
of 20 well-observed novae.  The obtained color excesses are 
in reasonable agreement with the previous results, which in turn 
support the idea of our general track of nova outbursts.  
Additionally, we estimated the absolute $V$ magnitudes of about 30
novae using a method for time-stretching nova light curves to analyze
the distance-reddening relations of the novae.
\end{abstract}




\section{Introduction}
\label{introduction}
A classical nova is a thermonuclear runaway event on a mass-accreting
white dwarf (WD) in a binary.  A companion star transfers 
its mass to the WD via Roche-lobe overflow or a wind.
When the mass of the hydrogen-rich envelope of the WD reaches
a critical value, hydrogen at the bottom of the envelope ignites to trigger
a hydrogen shell-flash and the binary becomes a nova.  The photospheric
radius of the WD envelope expands to a giant size and the spectrum of the
nova resembles that of an F-supergiant at or near optical maximum.  After 
optical maximum, the photosphere recedes into progressively deeper
layers as the envelope mass decreases, mainly because of wind mass-loss 
\citep[e.g.][]{hac06kb}. 
The wind mass-loss rate gradually decreases, and optically-thick winds
finally stop before the nova enters a supersoft X-ray phase.
The nova outburst ends when the hydrogen shell burning extinguishes.
This nova envelope evolution was calculated by \citet{kat94h} on the basis
of the optically-thick nova wind theory \citep[e.g.,][]{kat83}. 

%Fig.1 
%\placefigure{light_curve_pu_vul_hr_del_fh_ser_pw_vul_v1668_cyg_v1500_cyg}

%%%\begin{figure}
\begin{figure*}
%%%\epsscale{0.75}
\epsscale{1.0}
%%%\epsscale{1.15}
\plotone{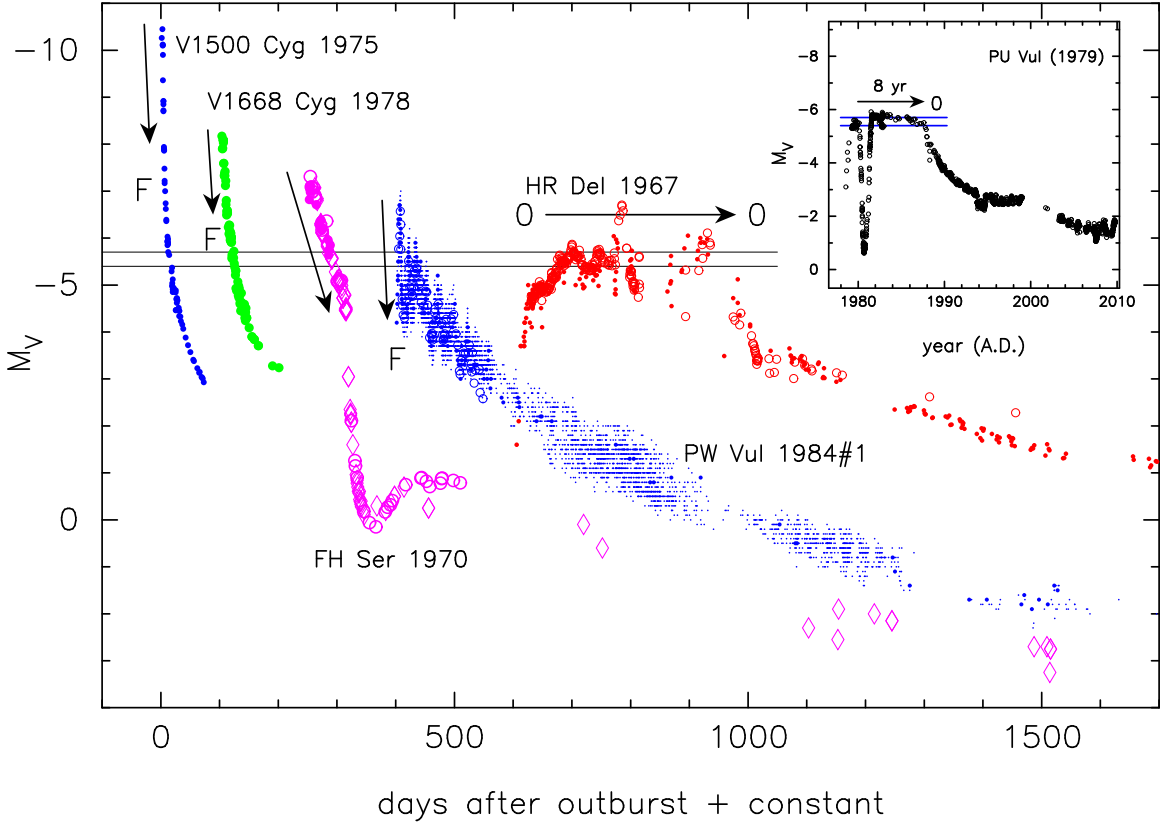}
%\plotone{light_curve_pu_vul_hr_del_fh_ser_pw_vul_v1668_cyg_v1500_cyg.epsi}
%\plotfiddle{evolution1.ps}{5.0cm}{270}{0.4}{0.4}{-170}{220}
\caption{
Optical light curves of six novae with different speed classes, i.e.,
from slow to fast, the symbiotic nova PU~Vul (inset, small black open
circles), very slow nova HR~Del (red symbols),
slow nova PW~Vul (blue symbols), moderately fast nova FH~Ser (magenta
symbols), fast nova V1668~Cyg (green filled squares), and very fast nova
V1500~Cyg (blue filled circles).   
Although their peak brightnesses, shapes, and decline rates are
very different, we will show that their color-color evolution
tracks are very similar.  This indicates that these novae
evolve under common physical conditions.
The sources of optical data are cited
in Sections \ref{pu_vul} (PU~Vul), \ref{hr_del_color} (HR~Del), 
\ref{pw_vul} (PW~Vul), \ref{fh_ser_color} (FH~Ser), 
\ref{v1668_cyg} (V1668~Cyg), and \ref{v1500_cyg} (V1500~Cyg). 
Horizontal lines represent the absolute magnitudes
of $M_V= -5.4$ and $M_V= -5.7$.
The distance modulus of each nova is taken from our results in
Sections \ref{fh_ser_color} --
%%%,  \ref{free_free_color_evolution}, \ref{pu_vul}, and 
\ref{each_slow_novae}.  Black arrows indicate the nova-giant
sequence of each nova, followed by an optically-thick free-free
emission phase (labeled F) or by an optically-thin free-free emission phase
(labeled 0).  See text for more details. 
\label{light_curve_pu_vul_hr_del_fh_ser_pw_vul_v1668_cyg_v1500_cyg}}
\end{figure*}
%%%\end{figure}

Despite their overall similarity, the optical light curves of novae
have a wide variety of timescales and shapes
\citep[e.g.,][see also Figure 
\ref{light_curve_pu_vul_hr_del_fh_ser_pw_vul_v1668_cyg_v1500_cyg}]{pay57,
due81, str10}.  Various empirical time-scaling laws have been
proposed in an attempt to recognize common patterns and unify
the nova light curves \citep[see, e.g.,][for a summary]{hac08kc}.
Recently, \citet{hac06kb} found that, in terms of free-free emission,
the optical and infrared (IR) light curves of several novae follow a 
similar decline law.  Moreover, the time-normalized light curves were 
found to be independent of the WD mass, chemical composition of the ejecta,
and wavelength.  They called this the universal decline law.
Subsequently, \citet{hac07k,hac09ka,hac10k}, \citet{hac08kc}, 
and \citet{kat09hc} applied this universal decline law to a number of
novae ($\gtrsim 30$ novae).  On the basis of this universal decline law,
\citet{hac10k} theoretically explained the maximum magnitude versus
rate of decline (MMRD) law of classical novae.
Therefore, we confidently state that the main part of nova light curves
can be interpreted in terms of free-free emission from nova ejecta 
outside the photosphere.

The evolution of colors is another challenging subject that attracts many
researchers, who have attempted to identify common behavior among various
types of novae.  For example, \citet{due79} noted that the color
evolutions of six novae are remarkably similar in the intrinsic
$(B-V)_0$ versus $(U-B)_0$ color-color diagram,
regardless of their different nova speed classes.
These six novae, however, traced similar but different paths
in the color-color diagram \citep[see Figures 2--7 of][]{due79}.   
\citet{van87} derived the general trends of color evolution
in nova light curves, i.e., $(B-V)_0 = 0.23\pm 0.06$ at maximum
\citep[see also][for an earlier work]{all73} and $(B-V)_0 = -0.02\pm 0.04$
at $t_2$, where $t_m$ ($m=2$ or 3) is the number of days
during which a nova decays by the $m$-th magnitude from its optical
maximum, and $(B-V)_0$ is the intrinsic $B-V$ color of the nova.
These two relations, however, often show large deviations from
the values obtained by other methods.  \citet{mir88} found that the
$B-V$ and $U-B$ color evolution of novae stabilizes soon after 
optical maximum and that this stage showed a general trend of 
$(B-V)_0 = -0.11\pm 0.02$.  He derived the extinctions, $E(B-V)$,
of 23 novae assuming that all of them have the same $(B-V)_0$ color
at the stabilization stage,
i.e., $E(B-V)=(B-V)_{\rm ss} - (B-V)_0 = (B-V)_{\rm ss} + 0.11$,
where $(B-V)_{\rm ss}$ is the observed $B-V$ color at the stabilization stage.
This method looks powerful but sometimes results in a large
difference from the true value.  (We will discuss this in more detail
in Section \ref{discussion}.)

According to Hachisu \& Kato's (2006) universal decline law, the optical
fluxes in the $UBV$ bands could be dominated by free-free emission.
%%%in the stage of Figure \ref{wind_config6ab_premax_no3}(c).
If this is the case, the color is simply estimated to be $(B-V)_0 = 0.13$
and $(U-B)_0 = -0.82$ for the optically-thin free-free
emission ($F_\nu \propto \nu^0$), or to be $(B-V)_0 = -0.03$ 
and $(U-B)_0 = -0.97$ for the optically-thick free-free
emission \citep[$F_\nu \propto \nu^{2/3}$, see, e.g.,][]{wri75},
where $F_\nu$ is the flux at the frequency $\nu$.
The latter value of $(B-V)_0 = -0.03$ is close to $(B-V)_0 = -0.02\pm 0.04$
at $t_2$ derived by \citet{van87}.  However, many novae do not remain
at these pivot points but evolve further blueward.
%%% (see discussion in Section \ref{strong_emission_lines_effect}).

These different trends in nova color evolution may represent different 
sides of the true color evolution, which we do not 
yet fully understand observationally or theoretically. 
The aim of this paper is to find a general path of nova color evolution,
as \citet{due79} tried to do 30 yr ago.  
First, in Section \ref{fh_ser_color} we analyze the moderately-fast nova
FH~Ser in the $(B-V)_0$ - $(U-B)_0$ diagram and find a new sequence
along which novae evolve when the photospheric emission dominates the 
optical spectrum.  We call this new sequence 
``the nova-giant sequence'' after the supergiant sequence.  
Then we examine the color evolution of  slow and fast novae, 
PW~Vul, V1500~Cyg, V1668~Cyg, and V1974~Cyg,
in Section \ref{free_free_color_evolution}.
In Section \ref{pu_vul}, we apply the tracks of color-color evolution 
to PU~Vul and find that the tracks of color-color evolution of the
slow/fast novae are common to this symbiotic nova.
In Section \ref{each_slow_novae}, we examine two very slow novae, 
HR~Del and V723~Cas, and show that the tracks of the color-color evolution
are also common.  Thus, we found that the overall trends in the color-color
evolution are very similar to each other.  
Therefore, we propose a general course of nova
outbursts in the color-color diagram for all speed classes of novae.
In Section \ref{extinction_novae}, we redetermine the reddening 
of novae by fitting the color evolution of a target nova 
with the general course of color evolution in the color-color diagram.
In this way, we propose a new method for estimating the color excess.
Discussion and conclusions follow in Sections
\ref{discussion} and \ref{conclusions}, respectively.
In the Appendix, we estimate the absolute magnitudes of about 30
novae, using the time-stretching method for nova light curves proposed by 
\citet{hac10k} to analyze the distance-reddening relation of each nova.

\begin{figure}
%%\epsscale{0.65}
%%\epsscale{1.0}
\epsscale{1.15}
\plotone{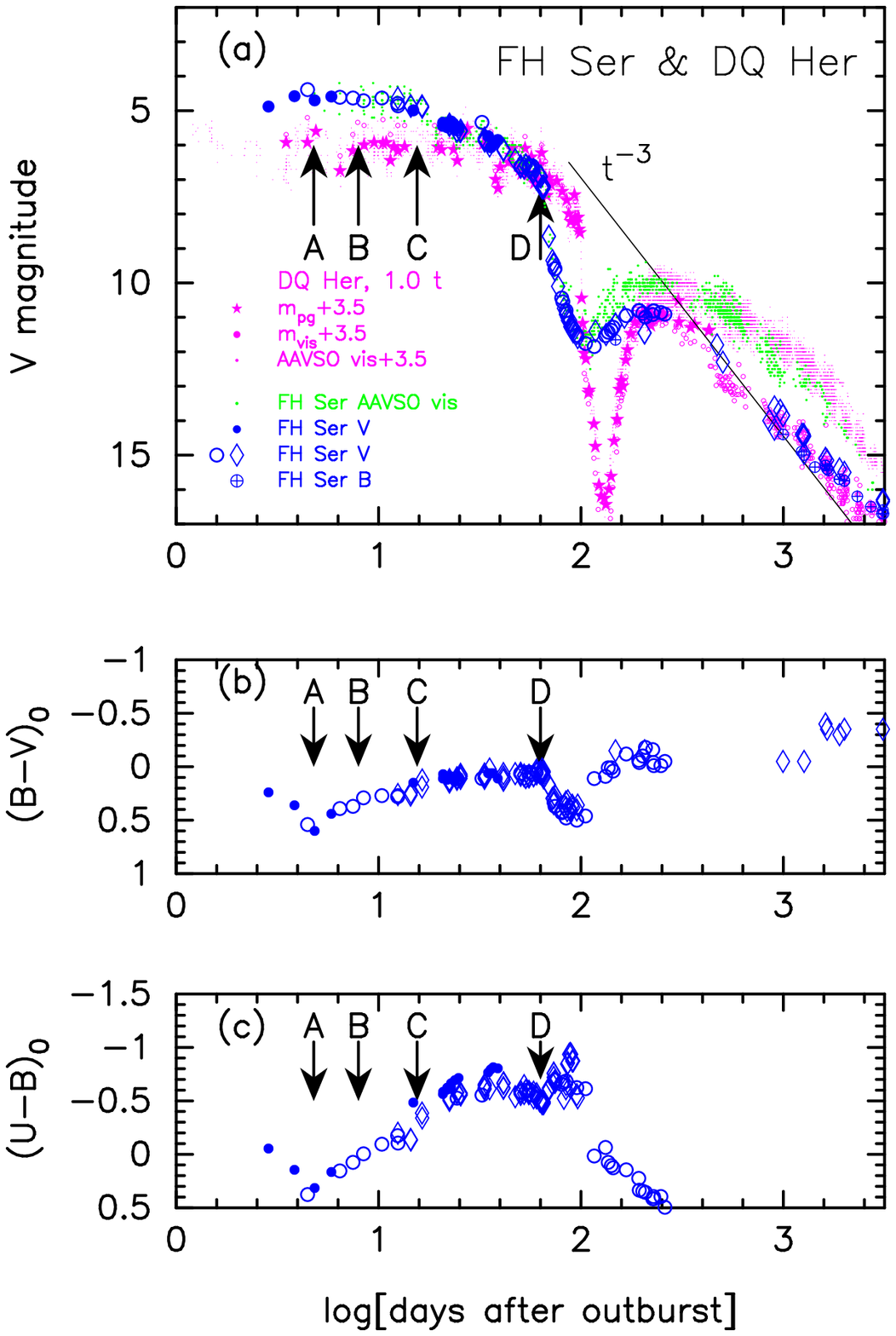}
%\plotone{dq_her_fh_ser_light_v_bv_ub_logscale.epsi}
%\plotfiddle{evolution1.ps}{5.0cm}{270}{0.4}{0.4}{-170}{220}
\caption{
(a) $V$ magnitude, (b) $(B-V)_0$ color, and (c) $(U-B)_0$ color light curves
of FH~Ser.  
The $UBV$ data are taken from \citet[][blue open circles]{bor70}, 
\citet[][blue filled circles]{osa70}, 
and \citet[][blue open diamonds]{bur71}.
The $B-V$ color of Borra \& Anderson (blue open circles) 
is systematically bluer by 0.2 mag than the other data, so
we shifted them down by 0.2 mag.  Four epochs of FH~Ser are specified by 
arrows labeled A, B, C, and D,
where epoch A corresponds to the optical maximum.
We add photographic ($m_{\rm pg}$)
and visual ($m_{\rm vis}$) data for DQ~Her, which are taken from
\citet{gap56}.  The magnitudes of DQ~Her is shifted down by 3.5 mag
to match them with those of FH~Ser in the late $t^{-3}$ law.  
\label{dq_her_fh_ser_light_v_bv_ub_logscale}}
\end{figure}

%Fig.3 
%\placefigure{distance_reddening_fh_ser_only}

\begin{figure}
%%\epsscale{0.75}
\epsscale{1.0}
%%\epsscale{1.15}
\plotone{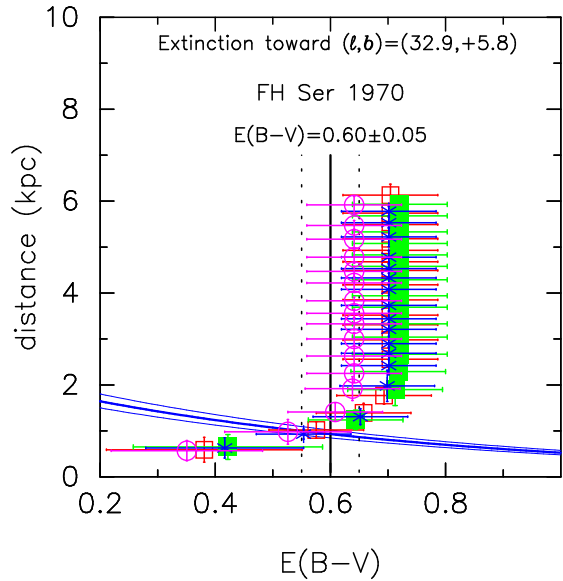}
%\plotone{distance_reddening_fh_ser_only.epsi}
%\plotfiddle{evolution1.ps}{5.0cm}{270}{0.4}{0.4}{-170}{220}
\caption{
Distance-reddening relation toward FH~Ser.
Blue thick solid line with flanking thin solid lines
denote the distance-reddening relation 
calculated from the distance modulus $(m-M)_V=11.7\pm0.2$, i.e., 
Equation (\ref{distance_modulus_fh_ser}) together with
Equation(\ref{distance_modulus_extinction}).
Four sets of data with error bars show distance-reddening relations
in four directions close to FH~Ser:
$(l, b)=(32\fdg75,5\fdg75)$ (red open squares), 
$(33\fdg00,5\fdg75)$ (green filled squares),
$(32\fdg75,6\fdg00)$ (blue asterisks), and 
$(33\fdg00,6\fdg00)$ (magenta open circles);
data are taken from \citet{mar06}.  Vertical black solid line with
flanking dotted lines represent the color excess of $E(B-V)=0.60\pm0.05$.
Two trends, $(m-M)_V=11.7$ (blue solid line) 
and Marshall et al.'s distance-reddening
relations, cross consistently at $E(B-V)\approx0.60$ and $d\approx0.93$~kpc,
which supports our estimate of $E(B-V)=0.60\pm0.05$.
\label{distance_reddening_fh_ser_only}}
\end{figure}

\section{Nova-giant Sequence of FH~Ser}
\label{fh_ser_color}
The first example that we analyze is the moderately-fast nova FH~Ser.
FH~Ser showed a gradual optical decay with $t_2=41$ and $t_3=62$ days
\citep{war95} followed by a sudden drop in brightness about 2.8 mag 
below the optical maximum due to dust shell formation.  
The $V$ light curve is shown in Figure 
\ref{light_curve_pu_vul_hr_del_fh_ser_pw_vul_v1668_cyg_v1500_cyg}
on a linear timescale and the $V$, $(B-V)_0$, and $(U-B)_0$ light
curves of FH~Ser are shown in Figure 
\ref{dq_her_fh_ser_light_v_bv_ub_logscale}
on a logarithmic timescale, where the $UBV$ data are taken from 
\citet[][blue filled circles]{osa70},
\citet[][blue open circles]{bor70}, 
and \citet[][blue open diamonds]{bur71}.
The $B-V$ color of \citet{bor70} is systematically
$\sim0.2$~mag bluer than the others 
whereas their $V$ and $U-B$ data are 
reasonably consistent with those of the other groups.  Therefore,
we shifted Borra \& Andersen's $B-V$ data down by 0.2 mag.

\subsection{Reddening and Distance}
\label{reddening_fh_ser}
\citet{kod70} obtained a value of $E(B-V)=0.6$ for the reddening
toward FH~Ser using the MMRD relation and interstellar reddening relation.
\citet{del97} obtained $E(B-V)=0.82$ from the color at optical maximum 
(i.e., $E(B-V)=(B-V)_{\rm max} - (B-V)_{0, \rm max}= 1.05 - 0.23 = 0.82$),
$E(B-V)=0.61$ from the line ratio of H$\alpha/$H$\beta = 5.7$, and
$E(B-V)=0.5$ from the equivalent width of \ion{Na}{1} $\lambda 5890$.
Then, \citet{del97} adopted an averaged value of $E(B-V)=0.64\pm 0.16$.
Because Kodaira's $E(B-V)=0.60$ and della Valle et al.'s $E(B-V)=0.61$
coincide, we use $E(B-V)=0.60$ in this paper and confirm below 
that this value is reasonable.  

The distance toward FH~Ser was estimated to be $d= 0.92\pm 0.13$~kpc 
by \citet{sla95}, and to be $d=0.95 \pm 0.05$~kpc by \citet{gil07}
both from the expansion parallax method.  We obtained FH~Ser's
distance modulus by comparing its brightness with that of DQ~Her. 
%% As we have already shown in Section \ref{t_pyx_color}, 
The trigonometric parallax distance of DQ~Her
was obtained by \citet{har13} as $d=386^{+33}_{-29}$~pc.
Adopting $A_V=3.1\times E(B-V)=0.31$ \citep{ver87}, we obtain
the distance modulus of DQ~Her as $(m-M)_V=A_V+5\log(d/{\rm10~pc})
=0.31 + 5 \log (386^{+33}_{-29}/ 10)= 8.24\pm0.18$.
Thus, we used $(m-M)_{V,\rm DQ~Her} = 8.2$ for DQ~Her,
Figure \ref{dq_her_fh_ser_light_v_bv_ub_logscale}(a) shows that 
the light curves of FH~Ser and DQ~Her overlap each other in the final
decline phase with a brightness difference of $\Delta V=3.5$.  Because 
the timescales of these two novae are almost the same, we may consider
that the brightness itself should be almost the same.  Then, we obtain
the distance modulus of FH~Ser as
\begin{eqnarray}
(m-M)_{V, \rm FH~Ser} &=& (m-M)_{V,\rm DQ~Her} + \Delta V \cr
&=& 8.2\pm0.2 + 3.5 =11.7\pm0.2.
\label{distance_modulus_fh_ser}
\end{eqnarray}
The distance to FH~Ser is estimated to be $d=0.93$~kpc considering
$E(B-V)=0.60$ and the relation
\begin{equation}
(m-M)_V = 5 \log\left( {{d} \over {10~{\rm pc}}} \right) + 3.1 E(B-V).
\label{distance_modulus_extinction}
\end{equation}
Our distance estimate is perfectly consistent with the other values 
of $d=0.92\pm0.13$ \citep{sla95}
and $d=0.95\pm0.05$~kpc \citep{gil07} mentioned above.

\citet{mar06} recently published a three-dimensional dust extinction map
of our galaxy in the direction of $-100\fdg0 \le l \le 100\fdg0$ 
and $-10\fdg0 \le b \le +10\fdg0$ with grids of $\Delta l=0\fdg25$
and $\Delta b=0\fdg25$, where $(l,b)$ are the galactic coordinates. 
Comparing the set of our $E(B-V)=0.60$ and $d=0.93$~kpc with
the distance-reddening relation toward FH~Ser given by \citet{mar06},
we can examine whether they are reasonable.  The results are shown 
in Figure \ref{distance_reddening_fh_ser_only}, where the galactic
coordinates of FH~Ser are $(l, b)=(32\fdg9090,+5\fdg7860)$
and we plot four relations in directions close to FH~Ser, i.e.,
$(l, b)=(32\fdg75,5\fdg75)$ (red open squares), 
$(33\fdg00,5\fdg75)$ (green filled squares),
$(32\fdg75,6\fdg00)$ (blue asterisks),
and $(33\fdg00,6\fdg00)$ (magenta open circles), with error bars.
The closest one is that of the green filled squares.
The blue solid line of $(m-M)_{V, \rm FH~Ser}=11.7$ crosses 
the trend of these green filled squares at or near $d=0.93$~kpc
and $E(B-V)=0.60$, which is consistent with our adopted values.

%Fig.4 
%\placefigure{color_color_diagram_fh_ser_typical}

\begin{figure}
%%\epsscale{0.75}
%%\epsscale{1.0}
\epsscale{1.15}
\plotone{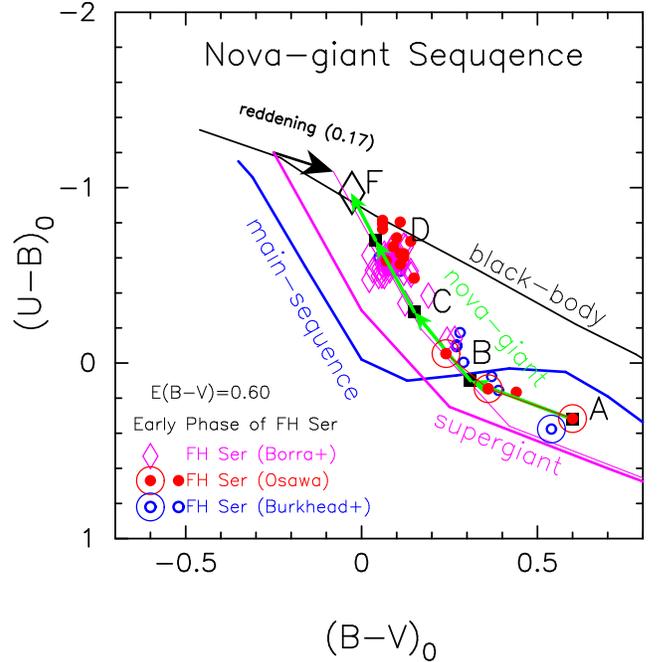}
%\plotone{color_color_diagram_fh_ser_typical.epsi}
%\plotfiddle{evolution1.ps}{5.0cm}{270}{0.4}{0.4}{-170}{220}
\caption{
Dereddened color-color diagram of FH~Ser and the nova-giant sequence.
We show four evolutionary stages of FH~Ser specified by A, B, C, and D
in Figure \ref{dq_her_fh_ser_light_v_bv_ub_logscale}.
Each label is attached to a filled black square.
Open diamond denoted by F indicates landmark for optically-thick 
free-free emission spectra.
We define the nova-giant sequence by a line connecting points
A, B, C, D, and F (green connecting arrows).
Their color data are tabulated in Table \ref{intrinsic_two_color_selected}.
Three other loci are also plotted for the main-sequence (blue solid line),
supergiant (magenta solid line), and blackbody (black solid line) sequences.
The color data of FH~Ser are taken from \citet{osa70} (red filled circles),
\citet{bor70} (magenta open diamonds), and \citet{bur71} (blue open circles). 
Four symbols with a large open circle outside indicate data
in pre-maximum phase.  All other symbols represent data
in post-maximum phase.  Magenta thin solid line represents
supergiant sequence shifted by 0.17 mag in the reddening
direction, as shown at the top of the supergiant sequence by a black arrow.
Most part of the nova-giant sequence overlaps the supergiant sequence
shifted by $\Delta E(B-V)=0.17$ (thin magenta line) except the lower part
(from point A to B) which is about $\Delta (U-B)\approx-0.2$ mag bluer
than the supergiant sequence.
\label{color_color_diagram_fh_ser_typical}}
\end{figure}

%Fig.5 
%\placefigure{wind_config6ab_premax_no3}

\begin{figure*}
%%\begin{figure}
%%\epsscale{0.75}
%%\epsscale{1.0}
\epsscale{1.15}
\plotone{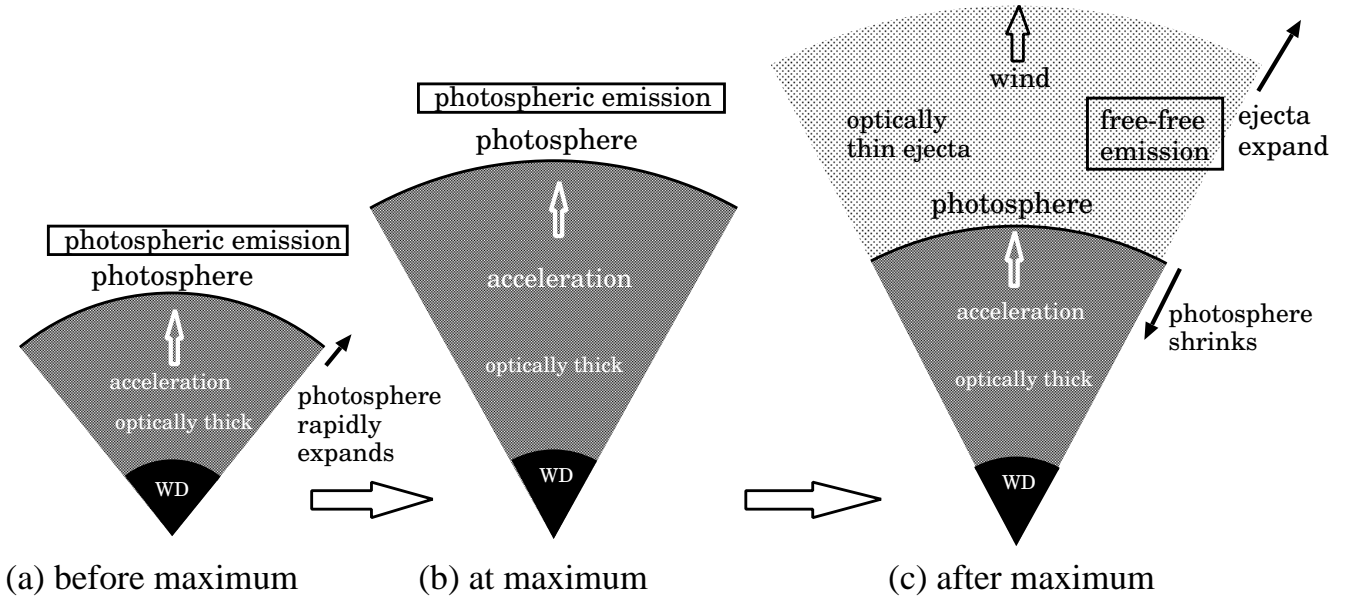}
%\plotone{wind_config6ab_premax_no3.epsi}
%\plotfiddle{evolution1.ps}{5.0cm}{270}{0.4}{0.4}{-170}{220}
\caption{
Schematic illustration of nova envelope evolution for fast and
moderately fast novae.
(a) Surface of expanding envelope moves almost together with
optical photosphere (so-called fireball stage).
(b) Photosphere reaches maximum expansion and then gradually
shrinks, while ejecta are continuously expanding.
At or near optical maximum, the absorption spectra resemble those
of F-supergiants.  
%%We regard this phase as stages 1--3 of PU Vul (see Section \ref{pu_vul}).
(c) After optical maximum, massive, optically-thin ejecta
expand, leaving pseudo-photosphere behind.
More ejecta are continuously supplied by optically-thick winds.
We regard this phase as an optically-thick, free-free emission phase.
The spectral energy distribution (SED) is close to 
$F_\nu\propto \nu^{2/3}$ in the optical (including the $UBV$ bands),
IR, and radio regions.
Here we use the term ``optically-thick wind'' to describe
winds that are accelerated deep inside the photosphere.
On the other hand, the term ``optically-thin wind'' indicates
winds that are accelerated outside the photosphere.
\label{wind_config6ab_premax_no3}}
%%\end{figure}
\end{figure*}

%Table 1
%\placetable{intrinsic_two_color_selected}

\begin{deluxetable}{llll}
\tabletypesize{\scriptsize}
\tablecaption{Intrinsic Two Colors of Selected Novae
\label{intrinsic_two_color_selected}}
\tablewidth{0pt}
\tablehead{
\colhead{Object} & \colhead{Stage} & \colhead{$(B-V)_0$} & 
\colhead{$(U-B)_0$}
} 
\startdata
Free-Free & 0\tablenotemark{a} & +0.13 & $-0.82$ \\
       & F\tablenotemark{b} & $-0.03$ & $-0.97$ \\
FH~Ser & A & $+0.60$ & $+0.32$ \\
       & B & $+0.31$ & $+0.10$ \\
       & C & $+0.15$ & $-0.29$ \\
       & D & $+0.04$ & $-0.70$ \\ 
       & F & $-0.03$ & $-0.97$ \\
PU~Vul & 1 & $+0.15$ & $-0.34$ \\
       & B & $+0.31$ & $+0.10$ \\
       & 2 & $+0.45$ & $+0.21$ \\
       & B & $+0.31$ & $+0.10$ \\
       & 3($=$C) & $+0.15$ & $-0.29$ \\
       & 4 & $+0.14$ & $-0.91$ \\
       & 5 & $-0.20$ & $-0.91$ \\
%
%#1  0.15 -0.342
%#2  0.45 +0.208
%#3  0.15 -0.292
%#4  0.14 -0.912
%#5 -0.20 -0.912
%
%
PW~Vul & D & $+0.04$ & $-0.70$ \\ 
       & C & $+0.15$ & $-0.29$ \\
       & 4 & $+0.14$ & $-0.91$ \\
       & 4' & $-0.08$ & $-1.10$ \\
       & 5' & $-0.60$ & $-0.70$ \\
V1500~Cyg & D & $+0.04$ & $-0.70$ \\ 
       & C & $+0.15$ & $-0.29$ \\
       & 4' & $-0.08$ & $-1.10$ \\
       & 5' & $-0.60$ & $-0.70$ \\
V1668~Cyg & D & $+0.04$ & $-0.70$ \\ 
       & C & $+0.15$ & $-0.29$ \\
       & 4'' & $-0.05$ & $-0.91$ \\
       & 5'' & $-0.50$ & $-0.91$ \\
V1974~Cyg & D & $+0.04$ & $-0.70$ \\ 
       & 4'' & $-0.05$ & $-0.91$ \\
       & 5'' & $-0.50$ & $-0.91$ 
\enddata
\tablenotetext{a}{Optically-thin ($F_\nu \propto \nu^0$).} 
\tablenotetext{b}{Optically-thick ($F_\nu \propto \nu^{2/3}$).} 
\end{deluxetable}

%0.60 0.32
%0.31 0.10
%0.15 -0.292
%0.04 -0.70

\subsection{Nova-giant Sequence in the Color-Color Diagram}
\label{nova_giant_sequence_fh_ser}
The observed colors are dereddened by
\begin{equation}
(U-B)_0 = (U-B) - 0.64 E(B-V),
\label{dereddening_eq_ub}
\end{equation}
\begin{equation}
(B-V)_0 = (B-V) - E(B-V),
\label{dereddening_eq_bv}
\end{equation}
where the factor of $0.64$ is taken from \citet{rie85}.  The dereddened
colors of FH~Ser are 
plotted in Figure \ref{color_color_diagram_fh_ser_typical}.
Here we only depict the data
before the dust blackout started about 70 days after the outburst.
We also plot three known sequences, the blackbody, supergiant,
and main-sequence sequences, the data for which are taken from \citet{all73}.
We also added a point, i.e., optically-thick
free-free emission spectra (open diamond denoted by F) for 
$F_\nu \propto \nu^{2/3}$ \citep{wri75}.

We have frequently seen in the literature statements that the spectra of
novae near maximum are similar to those of A--F type supergiants.
For FH~Ser, \citet{kod70} wrote ``Nova Serpentis 1970 showed a spectrum
similar to F-type near its light maximum.''
However, the track of FH~Ser in the color-color diagram does not follow
the supergiant sequence, as clearly shown in Figure
\ref{color_color_diagram_fh_ser_typical}.  The shape of the FH~Ser track
is very similar to that of the supergiant sequence, 
but it is about $\Delta (U-B)\approx -0.2$ mag bluer than the 
supergiant sequence.  Therefore, we are forced to define a new sequence 
based on the data for FH~Ser, which is designated by points A, B, C, D, and F
from redder to bluer.  In what follows, we will see that many novae evolve
along this sequence when their photospheric spectra are similar to those of 
A--F type supergiants.  Therefore, we call this track 
``the nova-giant sequence'' after the supergiant sequence.
The color data of points A--D and F are 
listed in Table \ref{intrinsic_two_color_selected}.

In the pre-maximum phase, FH~Ser descends along the nova-giant sequence
as shown in Figure \ref{color_color_diagram_fh_ser_typical}
by the symbols enclosed within a larger open circle.
After optical maximum, it returns along the same nova-giant sequence.

\citet{geh88} divided physical development of nova ejecta 
into a few distinct stages, i.e., fireball expansion, 
optically-thin gas expansion, and dust formation. 
Figure \ref{wind_config6ab_premax_no3} illustrates
the expanding nova ejecta before dust formation occurs. 
Figure \ref{wind_config6ab_premax_no3}(a) shows the so-called fireball stage,
in which the optical photosphere moves almost together with the surface 
of the expanding envelope.  We directly observe the photospheric emission.  
This photospheric emission resembles those of A--F type supergiants.  
In Figure \ref{wind_config6ab_premax_no3}(b), the photosphere reaches
maximum expansion and detaches from the top of the ejecta.  Then the
photosphere recedes into progressively deeper layers while the ejecta
expand continuously.  Figure \ref{wind_config6ab_premax_no3}(c)
corresponds to the stage after optical maximum, in which
massive, optically-thin ejecta expand leaving the photosphere behind.
More ejecta are continuously supplied by optically-thick winds.
Here we use the term ``optically-thick wind'' to describe winds that are
accelerated deep inside the photosphere, that is, in the optically-thick
region.  On the other hand, ``optically-thin wind'' indicates winds that
are accelerated outside the photosphere, that is, in the optically-thin
region.

We will show later that all novae follow the nova-giant sequence
near the maximum, regardless of their speed classes and, further, 
that faster novae tend to have shorter excursions along the nova-giant
sequence mainly because their envelopes are less massive.  
Slower novae tend to have longer redward journeys because of their more
massive envelope masses.  Among our examples, FH~Ser reached the reddest
part of the nova-giant sequence, up to point A
in Figure \ref{color_color_diagram_fh_ser_typical}.
Novae return along the nova-giant sequence after optical maximum and
approach point F, where free-free emission dominates the spectrum,
as shown schematically in Figure \ref{wind_config6ab_premax_no3}(c).
In FH~Ser, however, the color-color track was strongly affected
by the dust blackout at point D just before it approached point F
(see Figure \ref{dq_her_fh_ser_light_v_bv_ub_logscale}).   

The nova-giant sequence is about $\Delta (U-B)\approx -0.2$ mag
bluer than the supergiant sequence.
This is because nova spectra resemble those of A--F supergiants
near maximum regardless of the nova speed classes and we directly observe
the photospheric emission of novae. 
However, the envelope masses of novae are much ($\sim10^{-4}$) smaller
than those of normal supergiants.  Thus, the Balmer jump could be shallower
($U$ could be brighter) in novae than in normal supergiants,
as shown in Section \ref{pu_vul}.
This is why the position of nova-giant sequence is parallel to 
but $\Delta (U-B)\approx -0.2$ mag bluer than the supergiant sequence,
although nova spectra resemble those of A--F supergiants. 

It may seem that the nova-giant sequence is simply the supergiant sequence
and it falsely appears to differ because we happened
to underestimate the extinction by an amount of $\Delta E(B-V)\sim0.2$.
We shifted the supergiant sequence by $\Delta E(B-V)=0.17$ in the direction
of reddening in Figure \ref{color_color_diagram_fh_ser_typical} (thin 
magenta solid line).
The bluer part of the shifted track almost coincides with the 
bluer part of the nova-giant sequence, but
the very reddest part of the nova-giant sequence deviates from
that of the shifted supergiant sequence.
Therefore, these two sequences are intrinsically different. 
This $\Delta (U-B)\sim -0.2$ mag difference in the redder part 
was already discussed by \citet{bel89} for PU~Vul,
as will be introduced in Section \ref{nova_giant_sequence_123}.

%Fig.6 
%\placefigure{pw_vul_v1500_cyg_v_bv_ub_color_curve}

\begin{figure}
%%\epsscale{0.75}
%%\epsscale{1.0}
\epsscale{1.15}
\plotone{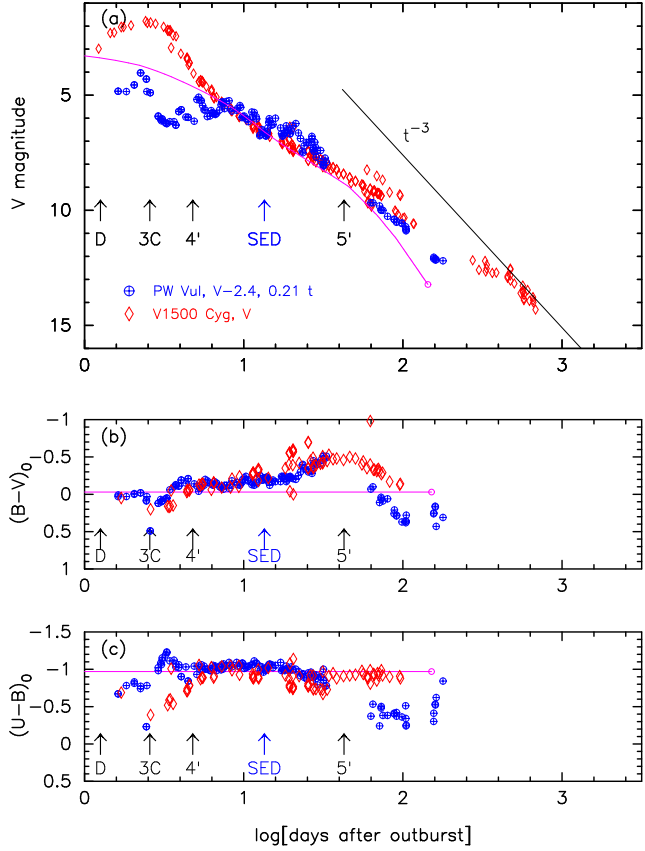}
%\plotone{pw_vul_v1500_cyg_v_bv_ub_color_curve.epsi}
%\plotfiddle{evolution1.ps}{5.0cm}{270}{0.4}{0.4}{-170}{220}
\caption{
Same as Figure \ref{dq_her_fh_ser_light_v_bv_ub_logscale}, but for
PW~Vul (blue open circles with plus sign inside) and 
V1500~Cyg (red open diamonds).
We squeezed the light curves of PW~Vul by 0.21 in the time direction
(horizontal shift of $\Delta \log t=-0.68$)
and shifted its magnitudes by $-2.4$ against those of V1500~Cyg,
as indicated in the figure.  We adjusted the time at $V$ maximum
of PW~Vul according to that of V1500~Cyg.  Magenta solid lines denote
optical and color light curves of free-free emission. 
Several epochs are specified by labels D, C ($=$3), 4', and 5' which
are defined in Figures \ref{color_color_diagram_fh_ser_typical} and
\ref{color_color_diagram_pw_vul_v1500_cyg_v1668_cyg_1974_cyg}.
Their positions in the color-color diagram are tabulated in
Table \ref{intrinsic_two_color_selected}.
  We also indicate the epoch of PW~Vul
64 days after the outburst by ``SED'', at which the spectrum
of Figure \ref{sed_pw_vul_iue_ubvrijhk_free_free_bb} was secured.
\label{pw_vul_v1500_cyg_v_bv_ub_color_curve}}
\end{figure}

%Fig.7
%\placefigure{v1668_cyg_v1500_cyg_v1974_cyg_v_bv_ub_color_curve_logscale}

\begin{figure}
%%\epsscale{0.75}
%%\epsscale{1.0}
\epsscale{1.15}
\plotone{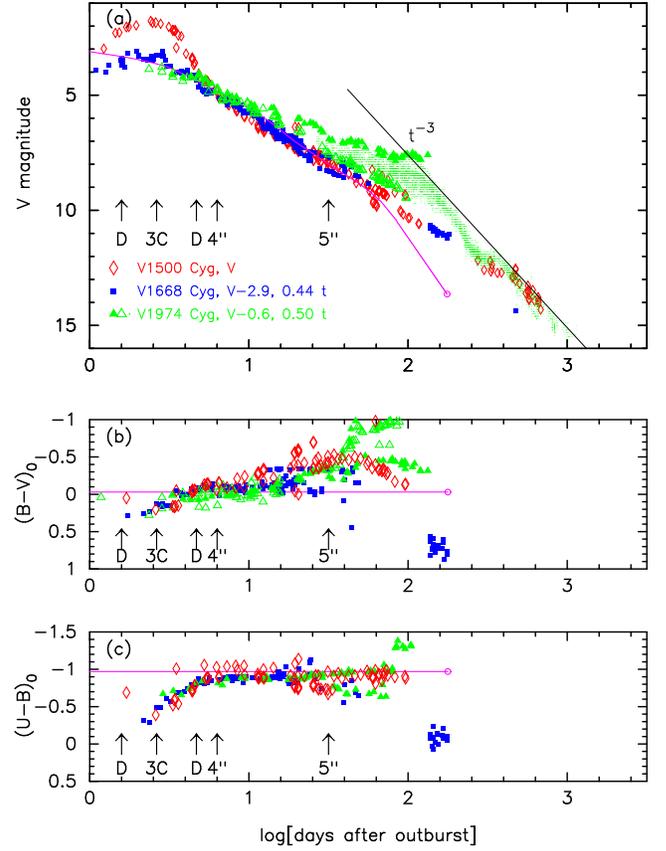}
%\plotone{v1668_cyg_v1500_cyg_v1974_cyg_v_bv_ub_color_curve_logscale.epsi}
%\plotfiddle{evolution1.ps}{5.0cm}{270}{0.4}{0.4}{-170}{220}
\caption{
Same as Figure \ref{pw_vul_v1500_cyg_v_bv_ub_color_curve}, but for
V1668~Cyg (blue filled squares), V1974~Cyg (green triangles and dots),
and V1500~Cyg (red open diamonds).
We squeezed the light curves of V1668~Cyg and V1974~Cyg by 0.54 and 0.63
and shifted their magnitudes by $-2.6$ and $-0.4$, respectively,
against those of V1500~Cyg, as indicated in the figure.  
We adjusted the time at $V$ maximum
of V1668~Cyg and V1974~Cyg according to that of V1500~Cyg.
Several epochs are specified by labels D, C ($=$3), D, 4'', and 5'',
which are defined in Figures \ref{color_color_diagram_fh_ser_typical} and 
\ref{color_color_diagram_pw_vul_v1500_cyg_v1668_cyg_1974_cyg}.
Their positions in the color-color diagram are tabulated in
Table \ref{intrinsic_two_color_selected}.
\label{v1668_cyg_v1500_cyg_v1974_cyg_v_bv_ub_color_curve_logscale}}
\end{figure}

%Fig.8
%\placefigure{color_color_diagram_pw_vul_v1500_cyg_v1668_cyg_1974_cyg}

\begin{figure*}
%%\begin{figure}
\epsscale{0.85}
%%\epsscale{1.15}
\plotone{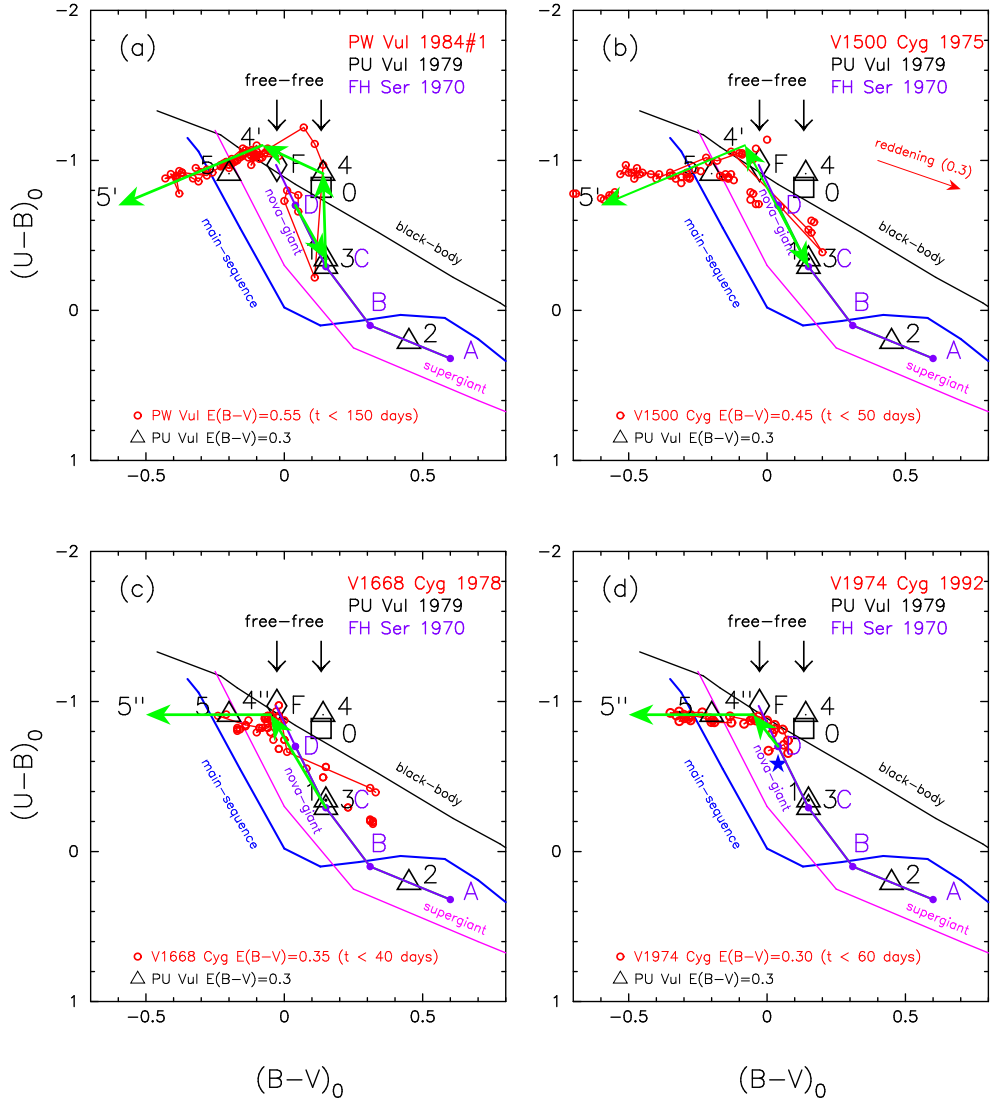}
%\plotone{color_color_diagram_pw_vul_v1500_cyg_v1668_cyg_1974_cyg.epsi}
%\plotfiddle{evolution1.ps}{5.0cm}{270}{0.4}{0.4}{-170}{220}
\caption{
Color-color evolution of four classical novae 
in the intrinsic $(B-V)_0$ vs. $(U-B)_0$ diagram.
(a) PW~Vul 1984\#1, (b) V1500~Cyg 1975, (c) V1668~Cyg 1978,
and (d) V1974~Cyg 1992.  Color data for each nova are denoted by 
red open circles.  Purple solid line indicates the nova-giant sequence
of FH~Ser, and green arrows indicate the path of each nova.  
Blue filled star symbol in (d) denotes
a datum in the pre-maximum phase of V1974~Cyg.
Other various symbols and lines have the same meanings
as in Figure \ref{color_color_diagram_fh_ser_typical}.
We add 10 points, 0, 1, 2, 3, 4, 5, 4', 5', 4'', and 5'',
and their data are listed in Table \ref{intrinsic_two_color_selected}.  
See text for more details.
\label{color_color_diagram_pw_vul_v1500_cyg_v1668_cyg_1974_cyg}}
%%\end{figure}
\end{figure*}

%Fig.9
%\placefigure{sed_pw_vul_iue_ubvrijhk_free_free_bb}

\begin{figure}
%%\begin{figure*}
%%\epsscale{0.75}
%%\epsscale{1.0}
\epsscale{1.15}
\plotone{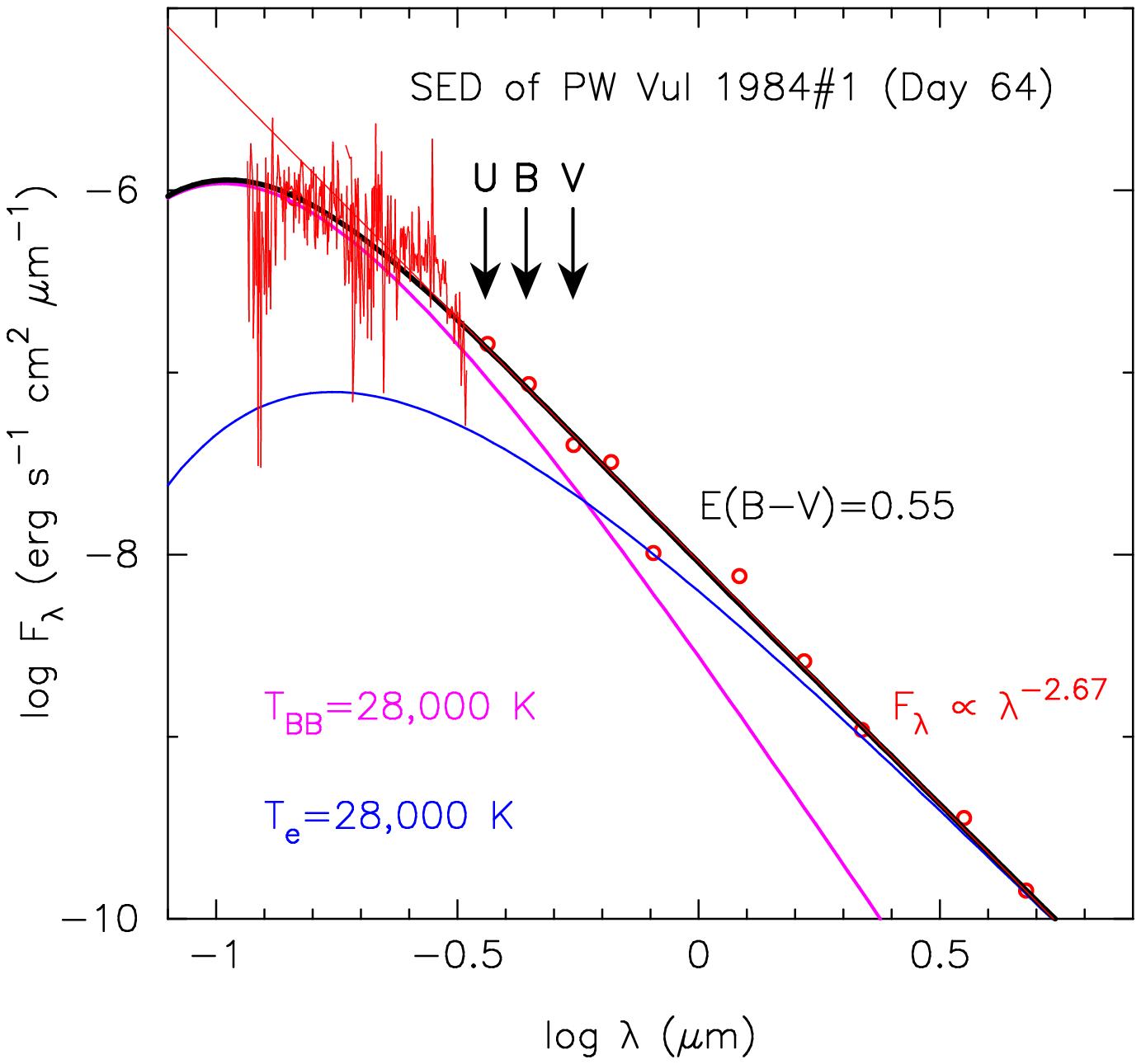}
%\plotone{sed_pw_vul_iue_ubvrijhk_free_free_bb.epsi}
%\plotfiddle{evolution1.ps}{5.0cm}{270}{0.4}{0.4}{-170}{220}
\caption{
Dereddened spectrum of PW~Vul 1984\#1 64 days after the outburst
(September 30).  Red solid line: 
{\it IUE} spectra, SWP24088 and LWP04458, are taken from the INES data
archive server.  Open red
circles: $UBVRI$ data from \citet{rob95}
and $JHKLM$ data from \citet{geh88}.
All data are dereddened with $E(B-V)=0.55$.
Global features of spectrum can be fitted with combination
(thick black solid line) of blackbody with temperature 
of $T_{\rm BB}=28,000$~K (thick magenta line) and optically-thick, 
free-free emission with electron
temperature of $T_{\rm e}=28,000$~K (thick blue line). 
We also add a red thin solid straight line
of $F_\lambda \propto \lambda^{-2.67}$ (corresponding to point F in Figure
\ref{color_color_diagram_pw_vul_v1500_cyg_v1668_cyg_1974_cyg})
as the limiting case of optically-thick free-free emission for
$T_{\rm e}=\infty$.
\label{sed_pw_vul_iue_ubvrijhk_free_free_bb}}
%%\end{figure*}
\end{figure}

\section{Free-free Emission Phase of Slow and Fast Novae}
\label{free_free_color_evolution}
In this section, we carefully study the color-color evolution of four 
well-observed classical novae, PW~Vul, V1500~Cyg, V1668~Cyg, and V1974~Cyg,
all of which show the optically-thick, free-free emission dominated phase. 
Their $V$, $(B-V)_0$, and $(U-B)_0$ light curves are plotted in Figures
\ref{pw_vul_v1500_cyg_v_bv_ub_color_curve} and 
\ref{v1668_cyg_v1500_cyg_v1974_cyg_v_bv_ub_color_curve_logscale}
on a logarithmic timescale.

\subsection{PW~Vul}
\label{pw_vul}
     PW~Vul 1984\#1 was discovered by Wakuda on UT 1984 July 27.7 
\citep{kos84}, about a week before the optical maximum
of $m_{V, {\rm max}} = 6.3$ on UT August 4.1.  
It shows a wavy structure on a smoothly-decaying light curve with
$t_2=83$ and $t_3=147$ days \citep{war95}.  Because no estimate of
the outburst day was found, we assume it to be JD~2445908.0
(UT 1984 July 26.5).  The light curve of PW~Vul is plotted in Figure 
\ref{light_curve_pu_vul_hr_del_fh_ser_pw_vul_v1668_cyg_v1500_cyg}
on a linear timescale and in Figure
\ref{pw_vul_v1500_cyg_v_bv_ub_color_curve} on a logarithmic timescale.  
PW~Vul was observed well with {\it International Ultraviolet Explore}
({\it IUE}) during the UV bright phase \citep[e.g.,][]{and91}
and also covered well by IR observation \citep[e.g.,][]{eva90, geh88}.

     The reddening toward PW~Vul was estimated to be
$E(B-V)=0.58 \pm 0.06$ from the \ion{He}{2} $\lambda1640/\lambda4686$
ratio, and $E(B-V)=0.55 \pm 0.1$ from the interstellar absorption feature
at 2200\AA\ both by \citet{and91},
$E(B-V)=0.60 \pm 0.06$ according to \citet{sai91} from the \ion{He}{2}
$\lambda1640/\lambda4686$ ratio, $E(B-V)=0.45 \pm 0.1$ according to 
\citet{due84} from the galactic extinction in the direction of the nova.
For the last galactic absorption, we examined the galactic dust absorption
map in the NASA/IPAC Infrared Science 
Archive\footnote{http://irsa.ipac.caltech.edu/applications/DUST/},
which is calculated on the basis of recent data from \citet{schl11} and
gives $E(B-V)=0.43 \pm 0.02$ in the direction of PW~Vul.
Thus, we adopted the arithmetic mean of these four values,
i.e., $E(B-V)=0.55 \pm 0.1$, in this paper.  

The distance toward PW~Vul was 
estimated to be $d=1.8 \pm 0.05$~kpc by \citet{dow00} on the basis of the
expansion parallax method.  Then the distance modulus in $V$ 
becomes $(m-M)_V = 5 \log (1800/10) + 3.1\times0.55= 13.0$.
The maximum magnitude is calculated to be
$M_{V, \rm max}= m_{V, \rm max} - (m-M)_V= 6.3-13.0= -6.7$. 

     Figure \ref{color_color_diagram_pw_vul_v1500_cyg_v1668_cyg_1974_cyg}(a)
shows the intrinsic color-color diagram of PW~Vul,
where the $UBV$ data of \citet{rob95}
are dereddened with $E(B-V)=0.55$ together with
Equations (\ref{dereddening_eq_ub}) and (\ref{dereddening_eq_bv}).
The nova started its color-color evolution at point D (near point F)
on the nova-giant sequence and descended to near point C (or point 3),
then jumped up to point 4.  The nova further went up slightly
and then moved downward and to the left.  Then, PW~Vul stayed at point 4'
for a while, as shown in Figure \ref{pw_vul_v1500_cyg_v_bv_ub_color_curve}
and moved gradually to point 5'.  Here points 1, 2, 3, 4, and 5 will be
introduced in Section \ref{pu_vul} as a template of the symbiotic nova
PU~Vul (``PU~Vul template''), the data for which are tabulated in Table
\ref{intrinsic_two_color_selected}.
We specify a template for the color-color evolution of PW~Vul
by a green arrow from point D to point C ($=$3) and then by three
green arrows from point 3 to point 4, from point 4 to point 4', and
from point 4' to point 5' (``PW~Vul template'').  
The data for the PW~Vul template are also tabulated
in Table \ref{intrinsic_two_color_selected}.

\subsection{Free-Free Emission Phase}
\label{free_free_dominated}
     In this way, the evolutionary path of PW~Vul first follows 
the nova-giant sequence from point D to point C.  Then it departs
from the nova-giant sequence and jumps up to points 4, and 4', and then
to point 5' in the color-color diagram.
This jump to point 4' is mainly due to free-free emission.

We can interpret this transition from point C to 4' as follows:
in the early expanding phase, from point D to C, we observed the photospheric
emission (A--F type supergiants), as schematically illustrated in Figures 
\ref{wind_config6ab_premax_no3}(a) and (b).
The nova-giant sequence is much redder than that of the blackbody sequence
in the $(U-B)_0$ color because of a large contribution from the Balmer jump.
When the photosphere shrinks leaving the ejecta behind, or 
winds begin to blow, the configuration of the envelope changes from that 
in Figure \ref{wind_config6ab_premax_no3}(b) to that in
Figure \ref{wind_config6ab_premax_no3}(c).
Then the Balmer jump absorption becomes shallower and is eventually
filled with emission lines.  The resultant $(U-B)_0$ color becomes
bluer to approach that of a blackbody (or free-free emission).
This type of spectral change was well documented by
\citet{bel89} for PU~Vul (see their Figure 6),
which will be mentioned later in Section \ref{nova_giant_sequence_123}.

To understand this property, we analyze the 
spectral energy distribution (SED) of PW~Vul 64 days after
the outburst, that is, when PW~Vul stayed between point 4' and
point 5', as shown in Figure \ref{pw_vul_v1500_cyg_v_bv_ub_color_curve}
(denoted by ``SED'').
We assume that the spectrum is approximated by a summation of
the blackbody emission at the photospheric temperature
$T_{\rm ph}=T_{\rm BB}$ and optically-thick free-free emission at
the electron temperature of $T_{\rm e}$, i.e., 
\begin{equation}
F_{\nu} = f_1 B_{\nu}(T_{\rm ph}) + f_2 S_{\nu}(T_{\rm e}),
\label{wind_spectrum_combination}
\end{equation}
where $B_{\nu}(T_{\rm ph})$ is the Planckian at 
$T_{\rm ph}$, and $S_{\nu}(T_{\rm e})$ is
the free-free spectrum at $T_{\rm e}$, and
$f_1$ and $f_2$ are numeric constants \citep[e.g.,][]{nis08}.
Nishimaki et al. adopted this formulation to analyze the spectra of 
Wolf-Rayet (WR) stars.
The optically-thick free-free spectrum of $S_{\nu}(T_{\rm e})$
was obtained by \citet{wri75} as
\begin{equation}
S_{\nu}(T_{\rm e}) = B_{\nu}(T_{\rm e}) K^{2/3}_{\nu}(T_{\rm e}),
\label{free_free_thick_eq1}
\end{equation}
where the linear free-free absorption coefficient $K_{\nu}(T_{\rm e})$
is given by
\begin{equation}
K_{\nu}(T_{\rm e}) = 3.7 \times 10^8 \left[ 1- 
\exp\left(-{{h \nu} \over {k T_{\rm e}}}\right)
\right] Z^2 g_{\nu}(T_{\rm e}) T^{-1/2}_{\rm e} \nu^{-3},
\end{equation}
in cgs units \citep{wri75}, and $g_\nu(T_{\rm e})$ is the Gaunt factor.
The Gaunt factor generally depends weakly on the frequency
and temperature, and we simply assume that it is unity in this paper.
Thus, we have four fitting parameters in this expression,
i.e., $f_1$, $f_2$, $T_{\rm ph}$, and $T_{\rm e}$. 
In the radio and IR regions of the spectrum where $h\nu \ll k T$,
Equation (\ref{free_free_thick_eq1}) can be expressed in a simple form as
\begin{equation}
S_\nu \propto \nu^{2/3}.
\label{free_free_thick_eq2_simple}
\end{equation}

Figures \ref{sed_pw_vul_iue_ubvrijhk_free_free_bb} shows the broadband
spectrum of PW~Vul, a combination of {\it IUE} spectra from Short 
Wavelength Prime (SWP)24088 and Long Wavelength Prime (LWP)04458 taken
from the INES archive data 
sever\footnote{http://sdc.cab.inta-csic.es/ines/index2.html},  
and the $UBVRI$ data from \citet{rob95}
and the $JHKLM$ data from \citet{geh88}.
All the data were dereddened with $E(B-V)=0.55$.
Assuming that $T_{\rm ph} = T_{\rm e}$, we varied $T_{\rm ph}$
in 500~K steps and obtained a temperature of 
$T_{\rm ph} = T_{\rm e} = 28,000$~K.
The global features of the SED is approximated well by a combination
(black solid line) of the blackbody (thick magenta line)
and free-free emission (thick blue line) spectra.  
The slope of the combined spectrum at the $UBV$ bands is very close to
that of optically-thick free-free emission 
($F_\lambda \propto \lambda^{-8/3}\approx \lambda^{-2.67}$),
where $F_\lambda$ is the energy flux at the wavelength $\lambda$.
This explains why the position of stage 4' is close to both the
blackbody sequence and optically-thick free-free emission, i.e.,
point F in Figure 
\ref{color_color_diagram_pw_vul_v1500_cyg_v1668_cyg_1974_cyg}(a).

PW~Vul stayed at point 4' (near point F) for a while during the free-free
emission phase, and moved gradually from point 4' to point 5', as shown
in Figures \ref{pw_vul_v1500_cyg_v_bv_ub_color_curve} and
\ref{color_color_diagram_pw_vul_v1500_cyg_v1668_cyg_1974_cyg}(a).
In the color-color diagrams of Figure 
\ref{color_color_diagram_pw_vul_v1500_cyg_v1668_cyg_1974_cyg},
the other three novae also do not stay only at or near point F
but evolve blueward from point 4' to 5' 
(or 4'' to 5'') while maintaining an almost constant $(U-B)_0$.
From point 4' to 5',
optically-thick free-free emission dominates the spectrum in 
the optical and IR region, as shown in Figure
\ref{sed_pw_vul_iue_ubvrijhk_free_free_bb}.
Then, the optical light curve almost follows the universal decline law
\citep{hac06kb} and its $(B-V)_0$ and $(U-B)_0$ colors should be
$-0.03$ and $-0.97$, respectively.  We added a free-free emission light
curve and $(B-V)_0$ and $(U-B)_0$ colors
in Figure \ref{pw_vul_v1500_cyg_v_bv_ub_color_curve}
(magenta solid lines).  These light and color curves roughly reproduce
the trends of the observed data except for the gradual deviation 
of the $(B-V)_0$ color after point 4' (or 4'').  This supports the fact that
free-free emission dominates the spectrum during the phase between
4' and 5' (or 4'' and 5'').     

%Fig.10
%\placefigure{skopal_emission_effect}

\begin{figure}
%%\begin{figure*}
%%\epsscale{0.75}
\epsscale{1.15}
\plotone{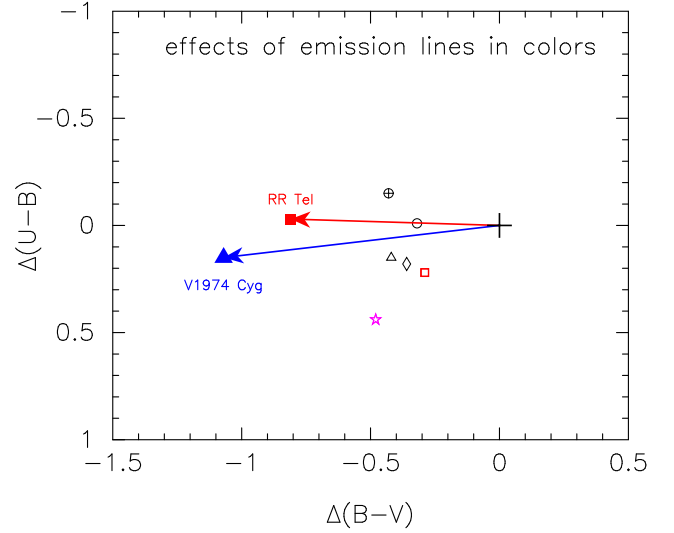}
%\plotone{skopal_emission_effect}
%\plotfiddle{evolution1.ps}{5.0cm}{270}{0.4}{0.4}{-170}{220}
\caption{
Effects of emission lines in the color-color diagram for
seven stars.  Here, $\Delta (B-V)$ and $\Delta (U-B)$ are the
excesses due to emission lines.  Compiling the data in
\citet{sko07}, we plot AG~Peg (red open square), Z~And (open diamond),
AR~Pav (open triangle), AX~Per in eclipse (open circle),
AX~Per out of eclipse (open circle with a plus sign), 
RR~Tel (red filled square), V1016~Cyg (magenta open star symbol),
and V1974~Cyg (blue filled triangle).  RR~Tel and V1974~Cyg
show large and almost horizontal blueward excursions 
from the origin (black plus sign) because of the effects of emission lines.
See text for more details.
\label{skopal_emission_effect}}
%\end{figure*}
\end{figure}

\subsection{Effects of Strong Emission Lines}
\label{emission_lines_effect}
Figure \ref{skopal_emission_effect} shows the excesses of 
$B-V$ and $U-B$ colors due to emission lines for several classical
and symbiotic novae and symbiotic stars.
Here, the excesses due to emission lines in the $U$, $B$, and $V$
magnitudes are defined as $\Delta U$, $\Delta B$, and $\Delta V$,
respectively, and the data for them are taken from \citet{sko07}.
In the classical nova V1974~Cyg (blue filled triangle)
and symbiotic nova RR~Tel (red filled square), 
$\Delta U$ and $\Delta B$ are much
brighter than $\Delta V$, but $\Delta U$
is as large as $\Delta B$, so the resultant $\Delta (U-B)$ color
changes very little even if $\Delta(B-V)$ takes a long blueward journey.
Unfortunately, PW~Vul is not included in Skopal's analysis.  Instead,
we show the blueward excursions of the fast nova V1974~Cyg about 210
days after light maximum.  The length of the blueward excursion is
significant; i.e., $\Delta (B-V)\approx -1.1$.
Thus, we conclude that the gradual blueward trip from point 4' to 5' 
(or from point 4'' to 5'') is mainly due to the growth of strong 
emission lines, especially in the $U$ and $B$ bands.  

In this work, we stop following the color evolution when the $V$ magnitude
drops by 3 mag from the maximum because strong emission lines make 
an increasingly large contribution to the $(U-B)_0$ and $(B-V)_0$ colors,
and their effects cloud the general evolution of colors.

\subsection{V1500~Cyg}
\label{v1500_cyg}
V1500~Cyg is an extremely fast nova, exhibiting 
probably the fastest and largest eruption among novae.
The light curve is shown in Figure 
\ref{light_curve_pu_vul_hr_del_fh_ser_pw_vul_v1668_cyg_v1500_cyg}
on a linear timescale and the $V$ light curve and $(B-V)_0$ and
$(U-B)_0$ color curves are plotted in Figures
\ref{pw_vul_v1500_cyg_v_bv_ub_color_curve} 
and \ref{v1668_cyg_v1500_cyg_v1974_cyg_v_bv_ub_color_curve_logscale}
on a logarithmic timescale.  
It rose to the maximum, $m_V= 1.85$, on 1975 August 31 from a
pre-outburst brightness of $m_V > 21$ \citep{you76}.
The distance of $1.2 \pm 0.2$~kpc \citep{lan88} and
interstellar extinction of $E(B-V)= 0.5 \pm 0.05$ \citep{fer77}
suggest a peak absolute luminosity of $M_V = -10.0 \pm 0.3$,
which is about 4 mag brighter than the Eddington luminosity.
An extensive summary of the observational results and modelings
can be found in the review by \citet{fer86}.

     \citet{gal76} obtained the magnitudes of the three optical
bands ($V$, $R$, and $I$) and the eight IR bands (1.2, 1.6, 2.2, 3.6, 4.8,
8.5, 10.6, and $12.5~\mu$m) during the 50 days following the discovery.
They estimated the outburst day to be UT August 28.9 from 
the angular expansion of the photosphere.
They concluded that the SED was approximately
that of a blackbody during the first 3 days, whereas it is close to
$F_\nu =$~constant after the 4th day.  These $F_\nu \propto \nu^0$
(constant) spectra resemble those usually ascribed
to optically-thin, free-free emission.

     On the basis of the IR photometry from 1 to $20~\mu$m,
\citet{enn77} also concluded that the nova's spectrum changed from
blackbody to bremsstrahlung emission at day $\sim 4-5$,
that is, from Rayleigh-Jeans, $F_\nu \propto \nu^2$,
to thermal bremsstrahlung, $F_\nu \propto \nu^0$.
Thus we regard that the nova enters a free-free emission phase
about 5 days after the outburst.
They also obtained an outburst date of JD~2442653.0$\pm 0.5$
from an analysis of the photospheric expansion similar to that of
\citet{gal76}.  Thus, we adopted the outburst day of V1500
Cyg as $t_{\rm OB}=$JD~2442653.0 (defined as $t=0$).

     The distance to V1500~Cyg has been discussed by many
authors.  \citet{you76} estimated the distance to be $1.4 \pm 0.1$~kpc
for $E(B-V)= 0.45$ \citep{tom76} from the reddening-distance law toward
the nova.  A firm upper limit to the apparent distance modulus was
obtained as $(m-M)_V \le 12.5$ by \citet{and76} from
the galactic rotational velocities of interstellar H and K absorption
lines.  The nebular expansion parallax method is a different way
to estimate the distance.  \citet{bec80} first imaged an expanding
nebular ($0\farcs 25$~yr$^{-1}$) of V1500~Cyg and estimated a distance
of 1350 pc assuming the expansion velocity of 
$v_{\rm exp} = 1600$~km~s$^{-1}$.
However, \citet{wad91} resolved an expanding nebula and
obtained a much lower expansion rate of $0\farcs 16$~yr$^{-1}$; they
estimated the distance to be 1.56~kpc, with a much smaller expansion 
velocity of $v_{\rm exp} = 1180$~km~s$^{-1}$
observed by \citet{coh85}.  \citet{sla95} obtained a similar expanding
angular velocity of the nebula ($0\farcs16$~yr$^{-1}$)
and obtained a distance of 1550 pc
assuming $v_{\rm exp} = 1180$~km~s$^{-1}$.
Here, we adopted a distance of $d=1.5$~kpc
and a reddening of $E(B-V)=0.45$.
Thus, the distance modulus is $(m-M)_V= 12.3$, and the maximum brightness
is $M_{V,\rm max}= m_{V,\rm max} - (m-M)_V = 1.85 - 12.3 = -10.45$.

     The dereddened color-color diagram is plotted in Figure
\ref{color_color_diagram_pw_vul_v1500_cyg_v1668_cyg_1974_cyg}(b),
where the $UBV$ magnitudes are taken from \citet{pfa76},
\citet{ark76}, \citet{due77}, and \citet{con80}.
%
%W. Pfau, 1976, IBVS, No.1106,
%V. P. Arkhipova & G. V. Zaitseva, 1976, Sov. Astron. Lett. vol. 2, 35-36
%M. E. Contadakis, 1980, IBVS, No.1818
%
%JD          V       B-V     U-B
%2442654.7  2.04    0.50     -0.40
%2442655.6  1.85    0.65     -0.10
%2442656.6  2.40    0.60     -0.25
%
%The nova started from a position close to point F, then followed down
%along the line F -- 1 and then came back to point 4'.  After that, the nova
%went blueward to point 5'.  
In the figure we connect only the data set of \citet{due77} with a red, 
thin, solid line to show the development of the colors.
V1500~Cyg follows the nova-giant sequence in the pre-maximum stage,
i.e., starting from point D and reaching point C ($=$3) at optical
light maximum.  It did not reach point B, point 2, or point A.  
In the post-maximum phase, it returns from point C ($=$3)
and reaches point 4' (near point F) along the nova-giant sequence.

As introduced in the previous Sections \ref{pw_vul} and 
\ref{free_free_dominated}, optically-thick, free-free emission
dominates the spectrum in the optical and IR region from point 4' to 5'
(Figure \ref{sed_pw_vul_iue_ubvrijhk_free_free_bb}).
When the continuum of free-free emission dominates the spectrum,
the optical light curve follows the universal decline law
\citep{hac06kb}, and its $(B-V)_0$ and $(U-B)_0$ colors remain
at point F, i.e., $(B-V)_0=-0.03$ and $(U-B)_0=-0.97$
(magenta solid lines in Figures \ref{pw_vul_v1500_cyg_v_bv_ub_color_curve}
and \ref{v1668_cyg_v1500_cyg_v1974_cyg_v_bv_ub_color_curve_logscale}).
When strong emission lines contribute to the $UBV$ broadband,
the light curves deviate from these values.
We specify a template (``V1500~Cyg template'') consisting of the four 
points in Figure 
\ref{color_color_diagram_pw_vul_v1500_cyg_v1668_cyg_1974_cyg}(b)
to illustrate the track of color-color evolution, from point D to
C ($=$3), 4', and then to 5'. 
We tabulate each position of the V1500~Cyg template in Table 
\ref{intrinsic_two_color_selected}.

%Fig.11
%\placefigure{all_mass_os_and_v1668_cyg_x35z02c10o20_new}

\begin{figure*}
%%\begin{figure}
%%\epsscale{0.75}
\epsscale{1.0}
%%\epsscale{1.15}
\plotone{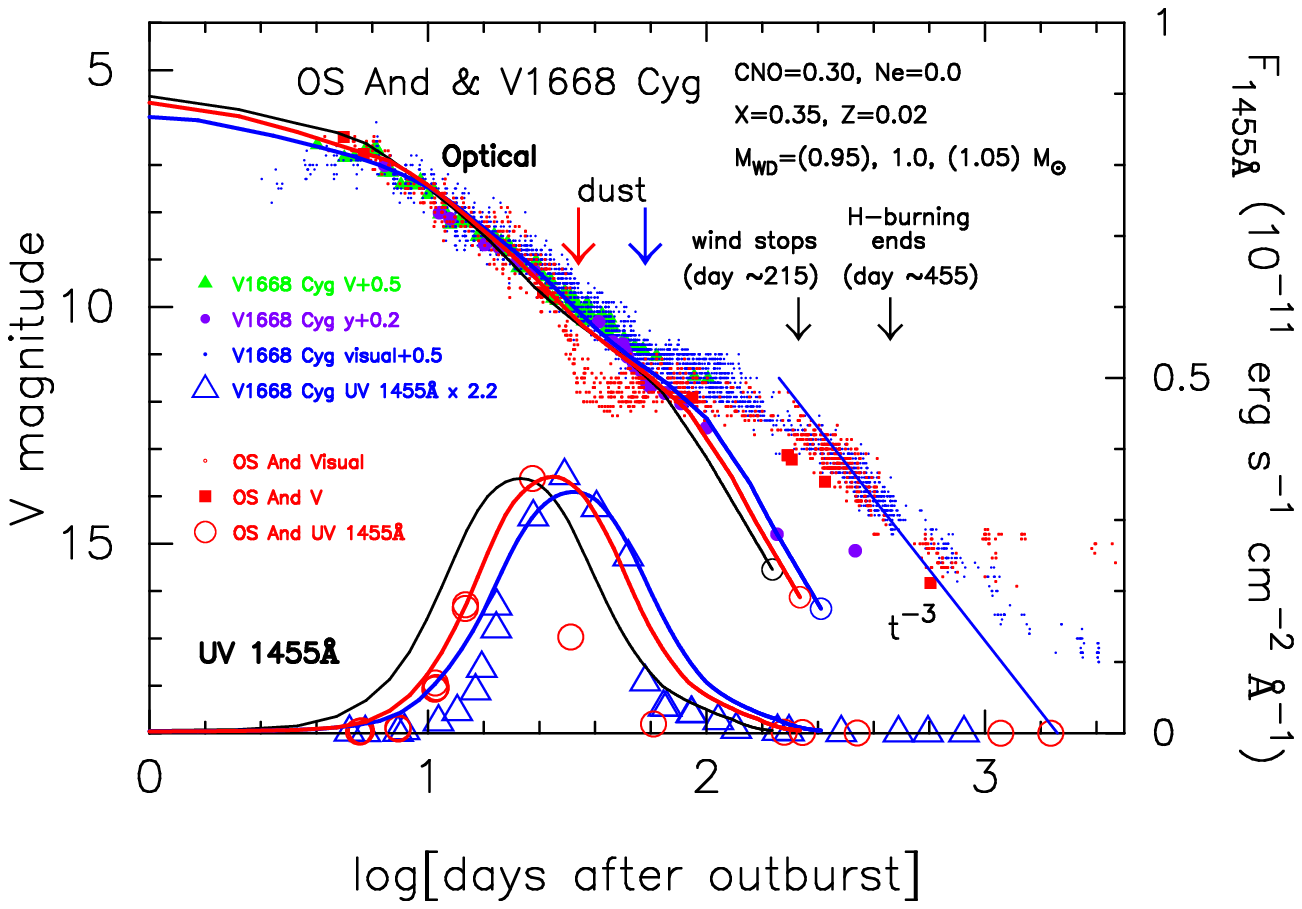}
%\plotone{all_mass_os_and_v1668_cyg_x35z02c10o20_new.epsi}
%\plotfiddle{evolution1.ps}{5.0cm}{270}{0.4}{0.4}{-170}{220}
\caption{
Optical/UV 1455\AA~ light curves of OS~And 1986 and V1668~Cyg 1978.
The UV~1455\AA~ data of OS~And are taken from \citet{cas02}. 
The optical/UV 1455\AA~  data for V1668~Cyg are the same as those
in Figure 16 of \citet{hac06kb}. 
The optical $V$ data (red filled squares) of OS~And were taken from 
\citet{kik88}, \citet{ohm87}, and IAU Circular Nos. 4282, 4293, 
4306, 4342, and 4452.  Visual data (red small dots) of OS~And are from 
the AAVSO archive.  We also add our model free-free emission light curves 
and UV~1455\AA~  model fluxes for both OS~And and V1668~Cyg.
The best fit model of OS~And is a $1.0~M_\sun$ WD
among three WDs with 0.95 (blue solid line), 1.0 (red solid), 
and 1.05 (black solid) $M_\sun$ and 
an assumed chemical composition of $X=0.35$,
$Y=0.33$, $Z=0.02$, and $X_{\rm C+O}=0.30$.
On the other hand, the best fit model of V1668~Cyg is a $0.95~M_\sun$ WD. 
The sudden drops in the UV flux of OS~And on day $\sim 30$ 
and of V1668 Cyg on day $\sim 60$ are caused by the formation of 
an optically-thick/thin dust shell (denoted by arrows with ``dust'').  
\label{all_mass_os_and_v1668_cyg_x35z02c10o20_new}}
\end{figure*}
%%\end{figure}

%Fig.12
%\placefigure{v1668_cyg_absorption}

\begin{figure}
%%\epsscale{0.75}
%%\epsscale{1.0}
\epsscale{1.15}
\plotone{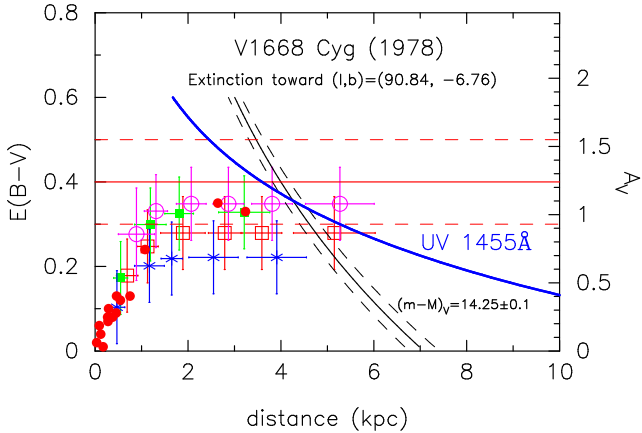}
%\plotone{v1668_cyg_absorption.epsi}
%\plotfiddle{evolution1.ps}{5.0cm}{270}{0.4}{0.4}{-170}{220}
\caption{
Distance-reddening relation toward V1668~Cyg.
Blue thick solid line denotes the distance-reddening relation
for UV~1455\AA\  flux fitting.
Red horizontal solid line flanked by red dashed lines 
show the reddening of $E(B-V)=0.4\pm0.1$ determined from 
the 2200\AA\  feature \citep{sti81}.   Black solid line flanked
by dashed lines corresponds to a distance modulus of
$(m-M)_V=14.25\pm0.1$.  Red filled circles represent distance-reddening
relation data from \citet{slo79}.  
%Large red filled circles indicate
%a distance-reddening relation toward V1668~Cyg obtained by \citet{slo79}.
Four sets of data with error bars show distance-reddening relations
in four directions close to V1668~Cyg:
$(l, b)=(90\fdg75,-6\fdg75)$ (red open squares), 
$(91\fdg00, -6\fdg75)$ (green filled squares),
$(90\fdg75,  -7\fdg00)$ (blue asterisks), 
and $(91\fdg00,  -7\fdg00)$ (magenta open circles); 
data were taken from \citet{mar06}.  
These trends/lines cross at $d\approx 4.3$~kpc and
$E(B-V)\approx0.35$.
\label{v1668_cyg_absorption}}
\end{figure}

\subsection{V1668~Cyg}
\label{v1668_cyg}
     V1668~Cyg was discovered on UT 1978 September 10.24 \citep{mor78},
two days before its optical maximum of $m_{V, {\rm max}} = 6.04$.
The light curve of V1668~Cyg is plotted in Figure 
\ref{light_curve_pu_vul_hr_del_fh_ser_pw_vul_v1668_cyg_v1500_cyg}
on a linear timescale and in Figures 
\ref{v1668_cyg_v1500_cyg_v1974_cyg_v_bv_ub_color_curve_logscale} and
\ref{all_mass_os_and_v1668_cyg_x35z02c10o20_new} on a logarithmic timescale.  
The distance-reddening law in the direction of V1668~Cyg
was obtained by \citet{slo79}, as shown in Figure \ref{v1668_cyg_absorption}
(large red filled circles), although the number of stars is small
and the data are scattered for $d > 1$~kpc.
They also obtained a reddening of
$E(B-V)= 0.38$ from the interstellar feature of \ion{K}{1} (7699 \AA)
and thus a distance of $d = 3.3$ kpc.
\citet{due80} criticized Slovak \& Vogt's work and proposed
a distance of $d= 2.3$ kpc from their newly obtained
distance-reddening law and $E(B-V)= 0.35$, using the same
stars as those in \citet{slo79}.  These distance-reddening laws, however,
rely on only three stars beyond 1~kpc \citep{slo79}, so
we cannot judge these two different estimates.
Assuming that the optical maximum is the Eddington luminosity,
\citet{sti81} estimated the distance to be $d= 2.2$ kpc, together
with their $E(B-V)= 0.40$ from the 2200 \AA\   feature.  However,
we have no evidence that the maximum luminosity of V1668~Cyg is just 
the Eddington limit.

The distance to the nova can also be estimated from the MMRD relation.
\citet{kla80} obtained $M_{V, {\rm max}} = -8.0 \pm 0.2$ from
Schmidt-Kaler's \citep{sch57} relation together with $t_3 = 24.3$ days
\citep{mal79}.  This gives a distance modulus of
$(m-M)_V=14.04$ and a distance of $d = 3.7$~kpc, together with
$m_{V, {\rm max}} = 6.04$ and $A_V = 3.1 E(B-V)= 3.1 \times 0.40=1.24$.

Using the model light curves of free-free emission, \citet{hac10k}
analyzed the multiwavelength light curves of V1668~Cyg and V1974~Cyg.
They calibrated the absolute magnitudes of the model light curves
for various WD masses and tabulated them in their Tables 2 and 3.
If we use these calibrated model light curves, the absolute magnitudes
of nova light curves are determined by fitting.  In the following,
we show an example of this fitting.  We plot the light curves of V1668~Cyg
(and OS~And, see Section \ref{os_and_color})
in Figure \ref{all_mass_os_and_v1668_cyg_x35z02c10o20_new}
with the free-free emission model light curves taken from \citet{hac10k}.  
This figure also shows the UV~1455\AA\  light curves corresponding to
each optical light curve model.  This narrowband flux represents
well the evolution of the photospheric temperature of novae 
\citep[e.g.,][]{cas02, hac06kb, hac08kc}.  
In the figure, we shift down the $V$ and visual magnitudes by 0.5 mag 
and the $y$ magnitudes of V1668~Cyg by 0.2 mag so that they
overlap with the $V$ light curve of OS~And.
The UV~1455\AA\  fluxes of V1668~Cyg are also scaled up by a factor of 2.2.
The $0.95~M_\sun$ WD (blue solid line) model shows the best fit with
both the observed {\it IUE} fluxes and optical $V$ light curve of V1668~Cyg
when we increase the WD mass in $0.05~M_\sun$ steps.  Here, we assumed 
a chemical composition of $X=0.35$, $Y=0.33$, $Z=0.02$, $X_{\rm C+O}=0.30$.
From the optical light curve fitting, we obtained the distance modulus of
V1668~Cyg as
\begin{eqnarray}
(m-M)_V&=&m_{\rm w}-M_{\rm w}\cr
&=&(16.35-0.5)-(+1.6)=14.25,
\label{v1668_cyg_distance_modulus}
\end{eqnarray}
where $m_{\rm w}= 16.35$ is read directly from the end point
of the free-free emission model light curve (large, open circle at
the bottom of the line) in
Figure \ref{all_mass_os_and_v1668_cyg_x35z02c10o20_new}
and $M_{\rm w}$ is taken from Table 2 of \citet{hac10k} as
$M_{\rm w}= +1.6$  for the $M_{\rm WD}=0.95~M_\sun$ model.
Thus, we obtain a distance modulus of $(m-M)_V=14.25\pm0.1$, where
$\pm0.1$ is the possible fitting error.

\citet{hac10k} also proposed another distance-reddening relation
calculated from the UV 1455\AA\   flux fitting as
\begin{equation}
2.5\log \left( F_{\lambda}^{\rm obs}/F_{\lambda}^{\rm mod}\right)
= 8.3 E(B-V) + 5\log \left( {{d} \over {10~{\rm kpc}}} \right),
\label{v1668_cyg_uv1455_distance_modulus}
\end{equation}
where, as shown in Figure \ref{all_mass_os_and_v1668_cyg_x35z02c10o20_new},
$F_{\lambda}^{\rm mod}$ is the model flux (blue solid line) at the
distance of 10~kpc without absorption,
$F_{\lambda}^{\rm obs}$ is the observed flux (blue large open triangles),
and the absorption
is calculated from $A_\lambda=8.3 E(B-V)$ for $\lambda=1455$\AA\  
\citep{sea79}.

Figure \ref{v1668_cyg_absorption} shows various distance-reddening
relations toward V1668~Cyg.  
Here, we omit the MMRD relation mentioned above
because it is not very reliable.
The first is the distance reddening relation given by Equation
(\ref{distance_modulus_extinction}) together with
the distance modulus of $(m-M)_V=14.25\pm0.1$ calculated above
from the calibrated model light curve.
The second is the distance-reddening relation given by
Equation (\ref{v1668_cyg_uv1455_distance_modulus}) for fitting UV~1455\AA.
The third is an estimated value of $E(B-V)=0.4\pm0.1$ from
the 2200 \AA\  feature \citep{sti81}.
The fourth is a relation given by \citet{mar06},
where the galactic coordinates of V1668~Cyg are
$(l, b)=(90\fdg8373,-6\fdg7598)$.  %%%% 090.8373 -06.7598 
The four sets of data with error bars show the distance-reddening relations
in four directions close to V1668~Cyg:
$(l, b)=(90\fdg75,-6\fdg75)$ (red open squares), 
$(91\fdg00, -6\fdg75)$ (green filled squares),
$(90\fdg75,  -7\fdg00)$ (blue asterisks), 
and $(91\fdg00,  -7\fdg00)$ (magenta open circles). 
These trends/lines cross at $d\approx 4.3$~kpc and 
$E(B-V)\approx 0.35$.  We adopted $d=4.3$~kpc and 
$E(B-V)=0.35$ in this paper.  Then, the maximum magnitude is 
$M_{V, \rm max}= m_{V,\rm max} - (m-M)_V= 6.04 - 14.25\approx -8.2$.

     Figure \ref{color_color_diagram_pw_vul_v1500_cyg_v1668_cyg_1974_cyg}(c)
shows the dereddened color-color diagram of V1668~Cyg, where
the $UBV$ data are taken from \citet{der78}, \citet{due80}, 
and \citet{kol80}.  It started from somewhere on or near the nova-giant
sequence (probably point D), although we did not find data on the 
pre-maximum (rising) phase of V1668~Cyg.  Then, it reached near 
point C ($=$3).  The nova quickly evolved to point 4'' (near point F)
along the nova-giant sequence and stayed at point 4'' for a while.
Optically-thick free-free emission dominates the spectrum in 
the optical and IR region from point 4'' to 5'',
as denoted by magenta solid lines in Figure 
\ref{v1668_cyg_v1500_cyg_v1974_cyg_v_bv_ub_color_curve_logscale}.
Then the nova moved horizontally (blueward in $(B-V)_0$ toward point 5''
but was almost constant in $(U-B)_0$).  In the figure we only connect
the points observed by \citet{der78} to follow
the early color evolution.  We do not connect other data observed
by \citet{due80} and \citet{kol80} to avoid a confusing presentation
of many line-connections.
Thus, we specify a template for V1668~Cyg (``V1668~Cyg template'')
by three points, that is, (probably from point D to point C), 
from point C to point 4'', and then from point 4'' to 5''. 
These points are tabulated in Table \ref{intrinsic_two_color_selected}.

%Fig.13
%\placefigure{mass_v_uv_x_v1974_cyg_x55z02o10ne03_real_scale}

\begin{figure*}
%%\begin{figure}
%%\epsscale{0.75}
\epsscale{1.0}
%%\epsscale{1.15}
\plotone{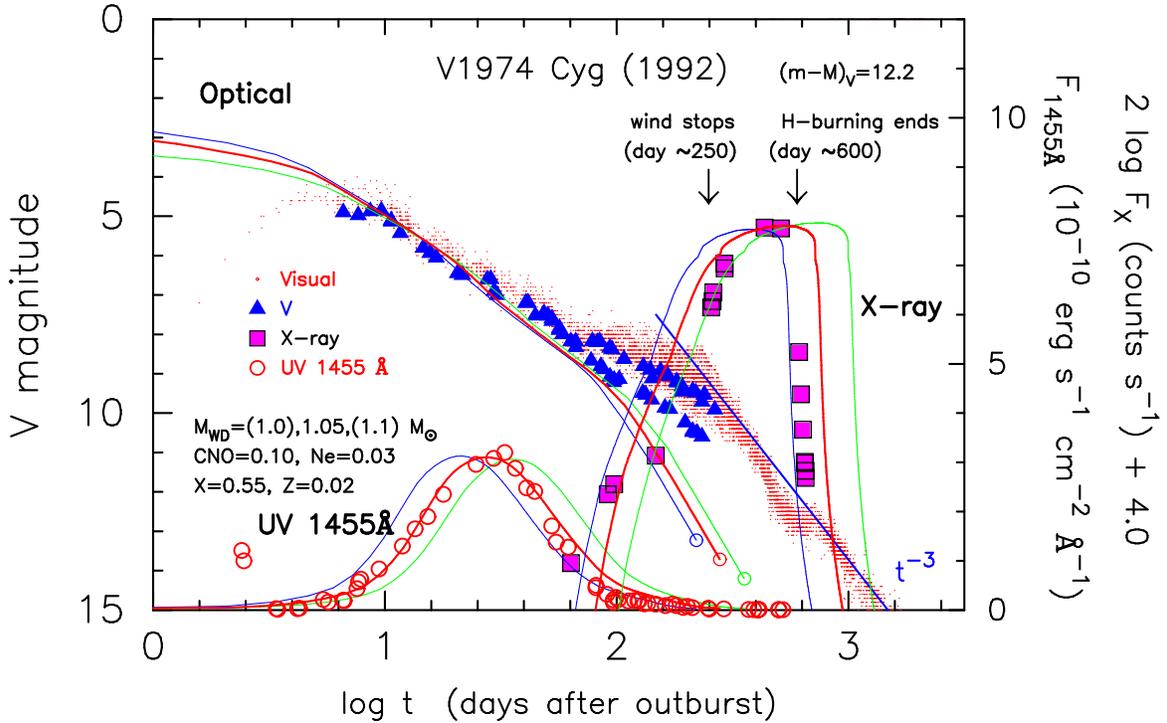}
%\plotone{mass_v_uv_x_v1974_cyg_x55z02o10ne03_real_scale.epsi}
%\plotfiddle{evolution1.ps}{5.0cm}{270}{0.4}{0.4}{-170}{220}
\caption{
Optical/UV 1455\AA\ /X-ray light curves of V1974~Cyg.
The optical/UV 1455\AA\ /X-ray data for V1974~Cyg are the same as those
in Figure 19 of \citet{hac06kb}.  The best fit model is a $1.05~M_\sun$ WD
among three WDs with 1.0 (green solid line), 1.05 (red thick solid line), 
and 1.1 (blue solid line) $M_\sun$ WDs and an assumed chemical
composition of $X=0.55$, $Y=0.30$, $Z=0.02$, $X_{\rm C+O}=0.10$,
and $X_{\rm Ne}=0.03$.
\label{mass_v_uv_x_v1974_cyg_x55z02o10ne03_real_scale}}
\end{figure*}
%%%\end{figure}

%Fig.14
%\placefigure{v1974_cyg_reddening_distance_marshall2006}

\begin{figure}
%%\epsscale{0.75}
%%\epsscale{1.0}
\epsscale{1.15}
\plotone{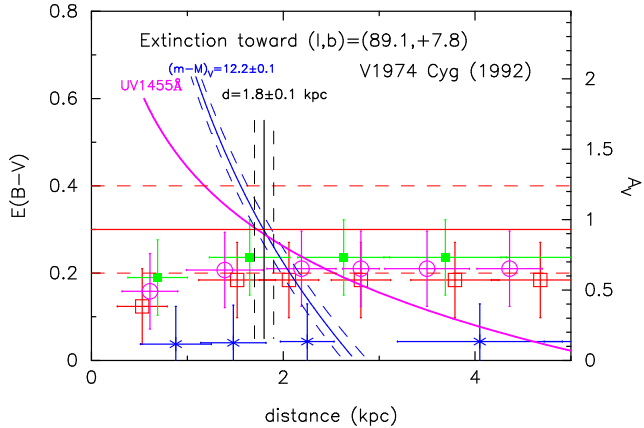}
%\plotone{v1974_cyg_reddening_distance_marshall2006}
%\plotfiddle{evolution1.ps}{5.0cm}{270}{0.4}{0.4}{-170}{220}
\caption{
Same as Figure \ref{v1668_cyg_absorption}, but for V1974~Cyg.
Magenta thick solid line denotes the distance-reddening relation
for UV~1455\AA\  flux \citep[see Figure 19 of][]{hac06kb}.
Red horizontal solid line flanked by red dashed lines 
shows reddening of $E(B-V)=0.3\pm0.1$.
Blue solid line flanked by dashed lines 
corresponds to $(m-M)_V=12.2\pm0.1$.
Black vertical solid line flanked by dashed lines 
corresponds to the distance $d=1.8\pm0.1$~kpc.
Four sets of data with error bars show distance-reddening relations
in four directions close to V1974~Cyg:
$(l, b)=(89\fdg00,7\fdg75)$ (red open squares), 
$(89\fdg25, 7\fdg75)$ (green filled squares),
$(89\fdg00,  8\fdg00)$ (blue asterisks), 
and $(89\fdg25,  8\fdg00)$ (magenta open circles); 
data are taken from \citet{mar06}.  
These trends/lines cross at $d\approx 1.8$~kpc and
$E(B-V)\approx0.30$.
\label{v1974_cyg_reddening_distance_marshall2006}}
\end{figure}

\subsection{V1974~Cyg}
\label{v1974_cyg}
     V1974~Cyg was discovered at $m_V \sim 6.8$ on UT 1992 February 19.07
\citep[JD 2448671.57;][]{col92} on the way to its optical maximum
of $m_{V, {\rm max}} \approx 4.2$ around February 22 (JD 2448674.5).
This is the first nova ever observed in all the wavelengths
from gamma-ray to radio, and was especially well-observed by an
X-ray satellite {\it ROSAT} and a UV satellite {\it IUE}.
The light curve and color curves of V1974~Cyg are plotted in Figure 
\ref{v1668_cyg_v1500_cyg_v1974_cyg_v_bv_ub_color_curve_logscale}
and the UV~1455\AA\  and X-ray light curves are shown in Figure
\ref{mass_v_uv_x_v1974_cyg_x55z02o10ne03_real_scale}
both on a logarithmic timescale.  {\it ROSAT} first detected 
the beginning and end of supersoft X-ray emission from classical 
novae \citep[e.g.,][]{kra96, bal98}.

To determine the reddening toward V1974~Cyg, we plot
the distance-reddening relation in Figure 
\ref{v1974_cyg_reddening_distance_marshall2006}.
The distance to V1974~Cyg was carefully determined by 
\citet{cho97a} to be $d=1.77\pm 0.11$~kpc using an expansion
parallax method.  Here we adopted Chochol et al.'s distance value of 
$d=1.8\pm0.1$~kpc (black vertical solid line flanked by thin, dashed lines).
The galactic coordinates of V1974~Cyg are
$(l, b)=(89\fdg1338, 7\fdg8193)$.  %%%%089.1338 +07.8193  
The four sets of data points with error bars show the distance-reddening
relations in four directions close to V1974~Cyg:
$(l, b)=(89\fdg00,7\fdg75)$ (red open squares), 
$(89\fdg25, 7\fdg75)$ (green filled squares),
$(89\fdg00,  8\fdg00)$ (blue asterisks), 
and $(89\fdg25,  8\fdg00)$ (magenta open circles); 
the data are taken from \citet{mar06}.  
The closest one is that denoted by green filled squares, which
crosses the $d=1.8$~kpc line at $E(B-V)=0.25\pm0.05$. 
The reddening was also estimated by many researchers.
\citet{cho97a} presented their mean value of $E(B-V)=0.26\pm0.03$
from various estimates based mainly on the MMRD relations.
\citet{aus96} obtained the reddening toward V1974~Cyg mainly on the basis
of the UV and optical line ratios for days 200 through 500, i.e.,
$E(B-V)=0.36\pm 0.04$.  The NASA/IPAC galactic dust absorption map
gives $E(B-V)=0.35 \pm 0.01$ in the direction toward V1974~Cyg.
Here we adopted the mean value of these four estimates, i.e.,
$E(B-V)=0.3 \pm 0.1$ and plot this in Figure
\ref{v1974_cyg_reddening_distance_marshall2006} (horizontal,
red, solid line flanked by thin, dashed lines). 

The distance-reddening relation is also obtained from
the UV~1455\AA\  flux fitting in Figure
\ref{mass_v_uv_x_v1974_cyg_x55z02o10ne03_real_scale}.
Comparing the model flux of $1.05~M_\sun$ WD and the observed ones,
we obtain the relation in Equation (\ref{v1668_cyg_uv1455_distance_modulus})
for V1974~Cyg, which is plotted by a magenta thick solid line in Figure
\ref{v1974_cyg_reddening_distance_marshall2006}.
We also plot the distance-reddening relation, i.e., Equation 
(\ref{distance_modulus_extinction}), using the
distance modulus of $(m-M)_V=12.2\pm0.1$ from the free-free model 
light curve fitting, i.e.,
$(m-M)_V=m_w - M_w = 13.8 - (+1.6)=12.2$ in Figure
\ref{mass_v_uv_x_v1974_cyg_x55z02o10ne03_real_scale}.
These trends/lines consistently cross at $d\approx 1.8$~kpc and
$E(B-V)\approx0.30$.  Therefore, we confidently adopt
$d=1.8$~kpc and $E(B-V)=0.30$ for V1974~Cyg.

Using $E(B-V)=0.30$, we dereddened the color-color evolution of V1974~Cyg
in Figure \ref{color_color_diagram_pw_vul_v1500_cyg_v1668_cyg_1974_cyg}(d),
where the $UBV$ data are taken from IAU Circulars and \citet{cho93}.
It started from a position near point D in the color-color diagram 
before an iron curtain developed in the UV wavelength region.
This point is highlighted by a blue filled star symbol because it is only 
the data in the pre-maximum phase \citep{kos92}.  The blue star symbol 
is located on the nova-giant sequence.  This starting point
is very similar to that of PW~Vul and corresponds to the so-called
fireball stage.  Then, the nova ascended to point 4'' (near point F) 
along the nova-giant sequence.  After the nova reached point 4'', 
it stayed at or near point 4'' for 10--20 days.  Then it
gradually moved horizontally leftward, as shown in
Figures \ref{v1668_cyg_v1500_cyg_v1974_cyg_v_bv_ub_color_curve_logscale}
and \ref{color_color_diagram_pw_vul_v1500_cyg_v1668_cyg_1974_cyg}(d).
This track is almost the same as that of V1668~Cyg.  We specify it
by points D, 4'', and 5'' (``V1974~Cyg template'').
These positions are tabulated in Table \ref{intrinsic_two_color_selected}.

Note that V1974~Cyg did not make a long journey 
toward point C ($=$3).   It did not reach point C whereas
the other three novae, PW~Vul, V1500~Cyg, and V1668~Cyg, did.
This behavior is very consistent with the development of the
UV1455\AA\  continuum flux (see Figure 
\ref{mass_v_uv_x_v1974_cyg_x55z02o10ne03_real_scale}).
In V1974~Cyg the UV flux was relatively high
a few days before optical maximum and then almost vanished 
near optical maximum, followed by a quick rise just after the 
optical maximum.  On the other hand, V1668~Cyg maintained almost zero
UV1455\AA\  flux for a relatively long time (10--20~days)
before and after optical maximum, followed by a quick rise with
a timescale similar to that of V1974~Cyg \citep{cas02, kat07h}, 
although $t_3$ is shorter in V1668~Cyg (26 days) than in V1974~Cyg (42 days).
This fact simply means that the photospheric temperature at the
optical maximum in V1974~Cyg is much higher than that of 
V1668~Cyg.  This is why the excursion toward
point C is much shorter in V1974~Cyg than in V1668~Cyg.
From the viewpoint of nova theory, this suggests
a much lower envelope mass in V1974~Cyg than in V1668~Cyg.

In this way, we found that the five well-observed novae follow 
a similar path in the color-color diagram.  Our new finding of
the nova-giant sequence is a characteristic property near and around
optical maximum that is common among these novae.

%Fig.15
%\placefigure{v723_cas_pu_vul_light_simple}

%%\begin{figure}
\begin{figure*}
%%%\epsscale{0.75}
\epsscale{1.0}
%%%\epsscale{1.05}
%%%\epsscale{1.15}
\plotone{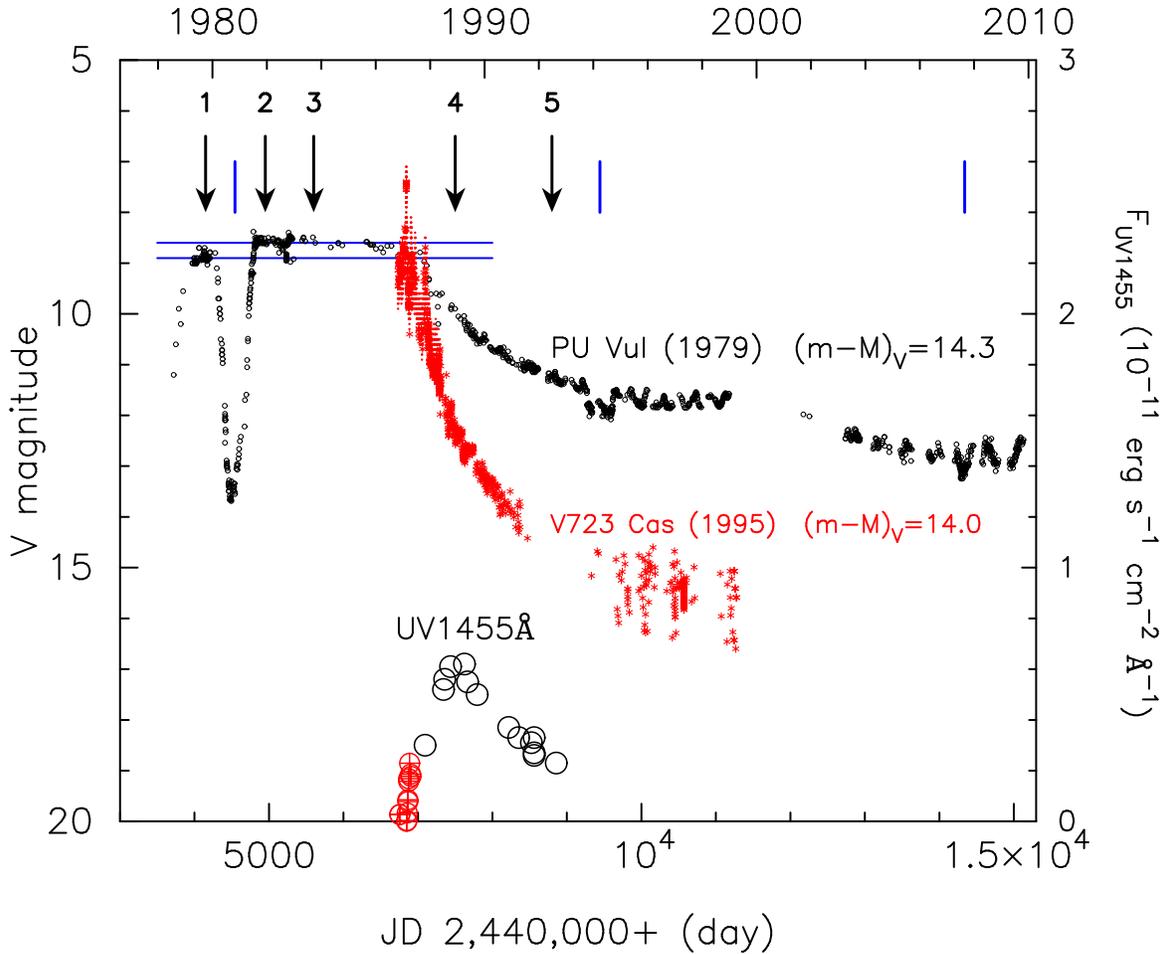}
%\plotone{v723_cas_pu_vul_light_simple.epsi}
%\plotfiddle{evolution1.ps}{5.0cm}{270}{0.4}{0.4}{-170}{220}
\caption{
Optical and UV 1455 \AA\  light curves of PU~Vul (black symbols)
and V723~Cas (red symbols).  
Small open black circles: $V$ magnitudes of PU~Vul.
Large open black circles: {\it IUE} UV~1455\AA\  flux of PU~Vul.
All data for PU~Vul are those in \citet{kat11hcg, kat12mh}.
Three blue vertical lines indicate the central times of eclipses
in 1980, 1994, and 2007.
Five downward arrows denote five stages of color evolution of PU~Vul
as defined in Section \ref{pu_vul}.
Two horizontal solid lines denote
absolute magnitudes of $M_V=-5.4$ and $-5.7$
in the flat peak of PU~Vul in 1979 and 1981-1983, respectively. 
Red asterisks and dots show $V$ and visual magnitudes of V723~Cas,
which are taken from \citet{cho97b, cho98} 
and the AAVSO archive, respectively.  
Large open red circles with a plus sign inside: 
{\it IUE} UV~1455\AA\  flux of V723~Cas, the values of 
which are multiplied by ten.
Light curves of V723~Cas are shifted by about 8 yr leftward 
to match the start of the UV 1455 \AA\  outburst of these two novae;
$V$ and visual light curves of V723~Cas are shifted by 0.3 mag
upward to match the absolute magnitudes of these two novae.  
\label{v723_cas_pu_vul_light_simple}}
\end{figure*}
%%\end{figure}

%Fig.16
%\placefigure{color_color_diagram_pu_vul_points_premax}

\begin{figure}
%%\begin{figure*}
%%\epsscale{0.75}
\epsscale{1.15}
\plotone{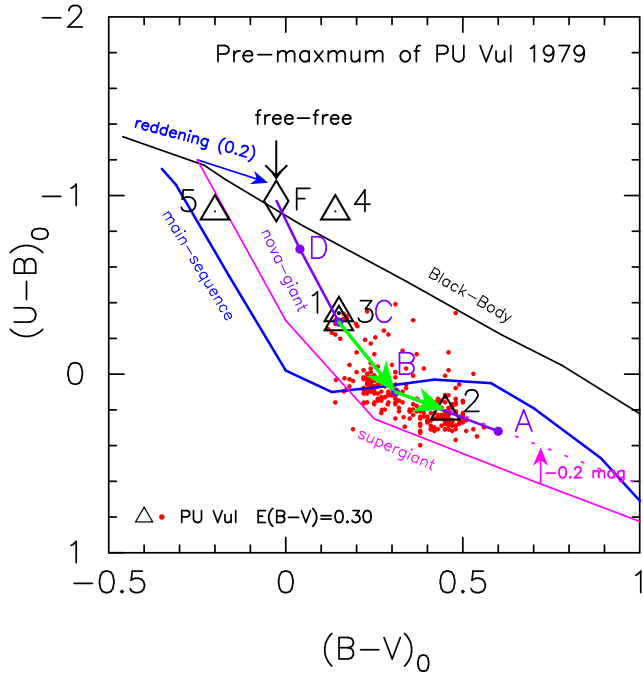}
%\plotone{color_color_diagram_pu_vul_points_premax.epsi}
%\plotfiddle{evolution1.ps}{5.0cm}{270}{0.4}{0.4}{-170}{220}
\caption{
Same as Figure \ref{color_color_diagram_fh_ser_typical}, but for
PU~Vul (denoted by red filled circles) in the pre-maximum phase.
The data are taken from \citet{shu12}.
Five evolutionary epochs of PU~Vul in Figure \ref{v723_cas_pu_vul_light_simple}
are indicated by open triangles with numbers.
A landmark of optically-thick, free-free emission spectra is
added, i.e., point F denoted by an open diamond.
The track of PU~Vul (green arrows) almost coincides with the nova-giant
sequence (from point 1 to 2 through point B),
which is located at about $\Delta (U-B)\approx -0.2$ mag 
bluer than the supergiant sequence.
The reddening direction is shown at the top of 
the supergiant sequence by a blue arrow, the length of which
corresponds to $E(B-V)=0.2$.  See text for more details.
\label{color_color_diagram_pu_vul_points_premax}}
%%\end{figure*}
\end{figure}

%Fig.17
%\placefigure{color_color_diagram_pu_vul_points_postmax}

\begin{figure}
%%\begin{figure*}
%%\epsscale{0.75}
\epsscale{1.15}
\plotone{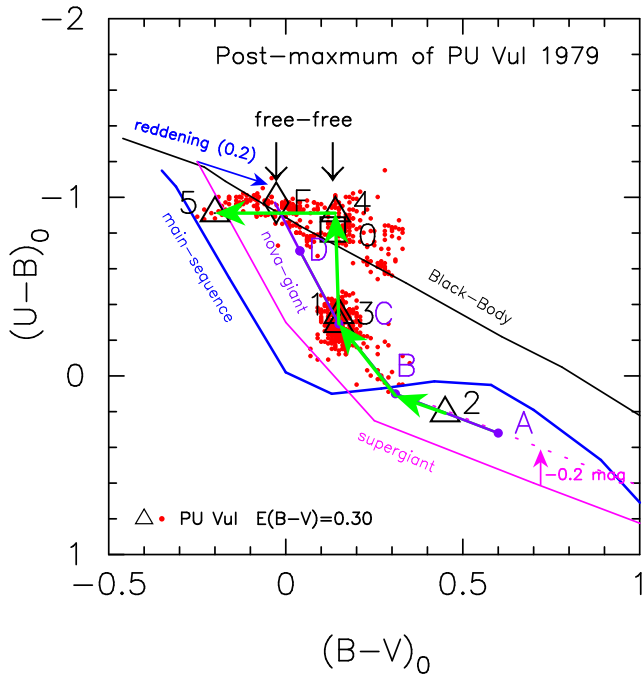}
%\plotone{color_color_diagram_pu_vul_points_postmax.epsi}
%\plotfiddle{evolution1.ps}{5.0cm}{270}{0.4}{0.4}{-170}{220}
\caption{
Same as Figure \ref{color_color_diagram_pu_vul_points_premax}, but
in the post-maximum phase (from points 2, B, 3, 4, and 5).
Two landmarks of free-free emission spectra are added:
optically-thin (open square denoted by 0)
and optically-thick (open diamond denoted by F).
\label{color_color_diagram_pu_vul_points_postmax}}
%%\end{figure*}
\end{figure}

%Fig.18
%\placefigure{sed_pu_vul_iue_ubv_free_free_bb}

\begin{figure}
%%\begin{figure*}
%%\epsscale{0.75}
\epsscale{1.15}
\plotone{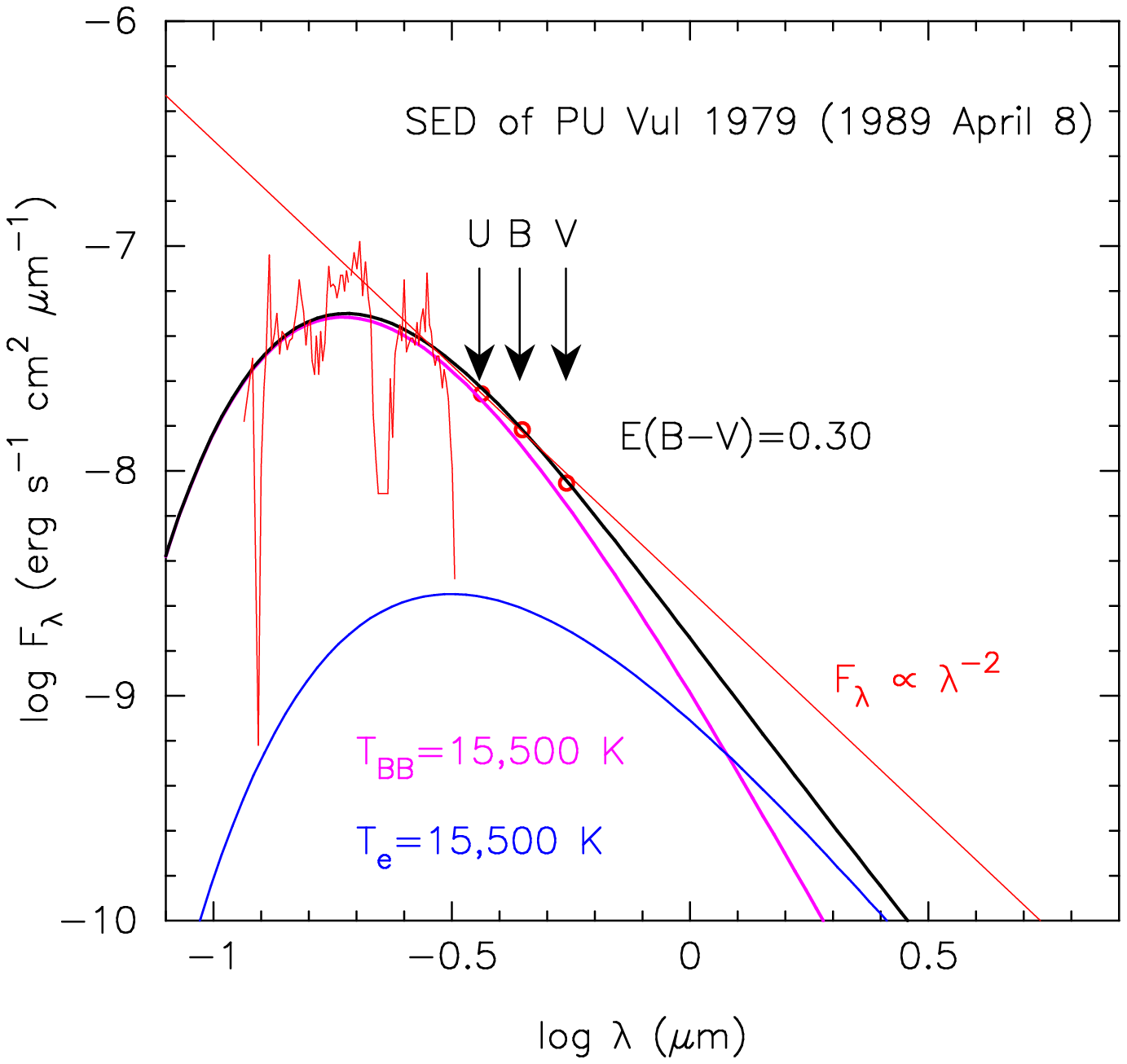}
%\plotone{sed_pu_vul_iue_ubv_free_free_bb.epsi}
%\plotfiddle{evolution1.ps}{5.0cm}{270}{0.4}{0.4}{-170}{220}
\caption{
Same as Figure \ref{sed_pw_vul_iue_ubvrijhk_free_free_bb}, but for
PU~Vul about 11 yr after the outburst.  Open red
circles: $U$, $B$, and $V$ band fluxes \citep[taken from][]{kol95}, 
which are denoted by three black downward arrows.
Red solid line: UV fluxes ({\it IUE} SWP35966 and LWP15323).
All fluxes are dereddened with $E(B-V)=0.30$.
Global features of spectrum can be fitted with combination
(black solid line)
of blackbody with temperature $T_{\rm BB}=15,500$~K (magenta solid line)
and optically-thick, free-free emission with electron
temperature $T_{\rm e}=15,500$~K (blue solid line). 
We also plot an optically-thin, free-free emission spectrum
of $F_\nu \propto \nu^0$, that is, $F_\lambda \propto \lambda^{-2}$
(red thin straight solid line).  In the $UBV$ bands,
the spectral energy distribution (SED) is close enough to
that of optically-thin, free-free emission of 
$F_\lambda \propto \lambda^{-2}$.
\label{sed_pu_vul_iue_ubv_free_free_bb}}
%%\end{figure*}
\end{figure}

\section{Color Evolution of Symbiotic Nova PU~Vul}
\label{pu_vul}
Our next example is the symbiotic nova PU~Vul, because 
there are many $UBV$ color data \citep[e.g.,][]{shu12} during
the long lasting outburst.  It is an eclipsing binary with an orbital
period of $\sim 4900$ days (13.4 yr) \citep{kol95, nus96, gar96, shu12}.
It exhibited an outburst in 1978 \citep[see e.g.][]{bel89},
and its very slow evolution provides us with dense optical
spectroscopic/photometric data as well as
{\it IUE/Hubble Space Telescope} ({\it HST}) UV observations. 
The $V$ light curve of PU~Vul is plotted 
in Figure \ref{v723_cas_pu_vul_light_simple},
together with the light curve of the very slow nova V723~Cas,
which we discuss later in Section \ref{each_slow_novae}. 
PU~Vul consists of a mass-accreting WD and
a mass-donating, semi-regularly pulsating M6 red giant (RG).
\citet{kat11hcg, kat12mh} obtained $\sim0.6~M_\sun$ for the WD and 
$\sim0.8~M_\sun$ for the RG.  

\citet{kat12mh} carefully analyzed the available data for PU~Vul
and obtained a reddening of $E(B-V)=0.30$ with a possible systematic
error of 0.05 and a distance of $d=4.7$~kpc from four combinations
of three independent methods.  The NASA/IPAC galactic dust absorption map
gives $E(B-V)=0.29 \pm 0.01$ in the direction of PU~Vul, whose galactic
coordinates are $(l, b)=(62\fdg5753, -8\fdg5317)$; this is perfectly
consistent with Kato et al.'s reddening estimate.
%%% Schlafly \& Finkbeiner 2011 (ApJ 737, 103).
Therefore, we adopted $E(B-V)=0.30$ and $d=4.7$~kpc in this paper.
Thus, the distance modulus in the $V$ band is
$(m-M)_V=A_V+5\log(d/{\rm10~pc})=3.1\times0.3+5\log(470)=14.3$.

%%% 062.5753 -08.5317 

To follow the color evolution, we chose five epochs
that represent typical evolutionary stages of PU~Vul, as indicated
by downward arrows in Figure \ref{v723_cas_pu_vul_light_simple}.
These stages were chosen to avoid the three total eclipses in 1980, 1994,
and 2007, which are indicated by three, blue, vertical lines in Figure
\ref{v723_cas_pu_vul_light_simple}, because the hot component (the WD) is 
totally occulted by the cool component (the RG) during the eclipses.
We briefly summarize the features of each stage in the $V$ magnitude
as follows:\\
stage 1: Pre-maximum phase, 0.3 mag below the maximum. \\
stage 2: Optical maximum. \\
stage 3: Post-maximum flat peak before the nebular phase. \\
stage 4: Onset of the nebular phase, $\sim 1.5$ mag
decayed from the maximum. \\
stage 5: Mid nebular phase, $\sim 2.5$ mag decayed from the maximum. \\

For each stage, we took a mean value of the magnitudes in Figure 3 of 
\citet{shu12}, which shows many color data with a small scatter around
the mean value.  Using the 1979--1993 data of \citet{shu12},
we show the color-color evolution of PU~Vul in 
Figure \ref{color_color_diagram_pu_vul_points_premax} for the
pre-maximum phase (from stage 1 to 2) and in Figure 
\ref{color_color_diagram_pu_vul_points_postmax}
for the post-maximum phase (from stage 2 to stages 3, 4, and 5).

\subsection{Nova-giant Sequence: Stages 1--3}
\label{nova_giant_sequence_123}
During the flat peak (stages 1 -- 3), PU~Vul evolved from point 1 
to 2 in Figure \ref{color_color_diagram_pu_vul_points_premax}
and from point 2 to 3 in Figure
\ref{color_color_diagram_pu_vul_points_postmax}.
\citet{kan91a} reported pure absorption spectra
for F-type supergiants, which we attribute to the photospheric emission
of PU~Vul.  The paths from points 1 to 2 through point B 
and from points 2 to 3 through point B are
parallel to, but $\Delta (U-B) \approx -0.2$ mag bluer than,
the supergiant sequence, as shown in Figures
\ref{color_color_diagram_pu_vul_points_premax} and
\ref{color_color_diagram_pu_vul_points_postmax}.
This part of the track coincides perfectly with the nova-giant sequence 
defined by the FH~Ser data in Section \ref{fh_ser_color}. 
This bluer position can be seen in the spectra of PU~Vul obtained 
by \citet[][p. 123]{bel89}, who reported, ``There is a good 
agreement of the energy distribution of PU~Vul and normal supergiants 
in the spectrum region from 3000\AA\  to 7000\AA\ .  
In the region from 3200\AA\  to 3800\AA\ , the UV excess is clearly seen.
The amount of this excess in 1983 agrees well with the photometric estimate
$\Delta (U-B)=-0.2$ mag.''  (See the spectrum in Figure 6 
of \citet{bel89} for more details.)   We will show below
in Sections \ref{each_slow_novae} and \ref{extinction_novae}
that this nova-giant sequence is common among many novae.

\subsection{Wind phase: stage 4}
The optical spectrum of PU~Vul changed to that of a Wolf-Rayet (WR) star
in 1987 \citep{iij89}, indicating a transition to the nebular phase
\citep[see also][]{bel89, tom91}.  \citet{vog92} and \citet{sio93} also
reported that {\it IUE} spectra changed to that of a WR type wind. 
These changes are consistent with Kato et al.'s (2012) UV light curve
analysis that optically-thin winds with a mass-loss rate of several
times $10^{-7} M_\sun$~yr$^{-1}$ started in 1987.
At this epoch, PU~Vul quickly ascends in the color-color diagram almost
vertically to point 4 (at stage 4) above the blackbody sequence. 
Point 4 is very close to point 0, which is denoted by an open square
in Figure \ref{color_color_diagram_pu_vul_points_postmax}, corresponding
to the position of optically-thin free-free emission ($F_\nu \propto \nu^0$),
i.e., $(B-V)_0=+0.13$ and $(U-B)_0=-0.82$.

We can interpret this transition from stage 3 to 4 as follows:
in the early expanding phase, PU~Vul moved along the nova-giant sequence
from stage 1 to 2 and then returned to 3.  We observed the photospheric
emission (pure absorption feature of F-supergiants) during stages 1 -- 3, 
as schematically illustrated in Figure \ref{wind_config6ab_premax_no3}(b).
The nova-giant sequence is much redder than the blackbody sequence in the 
$(U-B)_0$ color because of the large contribution from the Balmer jump.
When winds began to blow, the configuration of the envelope changed from 
that in Figure \ref{wind_config6ab_premax_no3}(b) to that in
Figure \ref{wind_config6ab_premax_no3}(c).  Then, the Balmer jump became
shallower \citep[see the spectrum in Figure 6 of][]{bel89}
and was filled with emission lines.  Thus, the spectrum is approaching
that of a blackbody or free-free emission.  The resultant $(U-B)_0$ color
becomes bluer to approach that of a blackbody or free-free emission.

To understand this property, we analyze the nova spectra
from UT 1989 April 8, in the nebular phase (stage 4).  
We assume that the spectrum is a summation of the blackbody emission
at the photospheric temperature $T_{\rm ph}=T_{\rm BB}$ and optically
thick free-free emission at the electron temperature of $T_{\rm e}$,
i.e., Equation (\ref{wind_spectrum_combination}).
Figures \ref{sed_pu_vul_iue_ubv_free_free_bb} shows the broad band spectrum
of PU~Vul, a combination of {\it IUE} spectra of SWP35966 and LWP15323 
taken from the INES archive data sever
and optical $UBV$ fluxes observed on UT 1989 April 8 \citep{kol95}.
Assuming that $T_{\rm ph} = T_{\rm e}$, we changed $T_{\rm ph}$
in 500~K steps and obtained a temperature of 
$T_{\rm ph} = T_{\rm e} = 15500$~K.  This is roughly consistent with the
photospheric temperature of $T_{\rm ph}\sim 20000$~K
theoretically calculated by \citet{kat12mh}.
As shown in Figure \ref{sed_pu_vul_iue_ubv_free_free_bb}, 
we need a free-free emission component to fit the spectrum although
its contribution is rather small at the $UBV$ bands.
The slope of the resultant spectrum at the $UBV$ bands is very close to
that of optically-thin free-free emission ($F_\lambda \propto \lambda^{-2}$).
This explains why the position of stage 4 is close to both the
blackbody sequence and optically-thin, free-free emission, i.e.,
point 0 in Figure \ref{color_color_diagram_pu_vul_points_postmax}.

\subsection{Effects of Strong Emission Lines: Stage 5}
\label{strong_emission_lines_effect}
In the color-color diagram of Figure 
\ref{color_color_diagram_pu_vul_points_postmax},
PU~Vul further evolves blueward from point 4 to 5 (stage 4 to 5)
while maintaining an almost constant $(U-B)_0$.
This blueward change is mainly due to the growth of strong emission lines,
especially in the $U$ and $B$ bands, as already explained in Section
\ref{emission_lines_effect}.  

Figure \ref{skopal_emission_effect} shows the effect
of strong emission lines, i.e., $\Delta(B-V)$ and $\Delta(U-B)$,  
in the color-color diagram.  Among the seven stars
in Figure \ref{skopal_emission_effect}, six are symbiotic stars
and three are symbiotic novae.
Symbiotic novae can be divided into two groups according to
their spectral evolution. The first group exhibits a long
(several years) ``supergiant phase,'' in which they resemble 
an A--F supergiant while the star undergoes a nova outburst.
In the second group, a nebular phase begins almost immediately
after optical maximum, and a ``supergiant phase,'' if there is one,
has a very short duration \citep{mue94}.
The first group includes PU~Vul, RR~Tel, AG~Peg, and RT~Ser. 
The second group includes V1016 Cyg, HBV 475, and HM Sge.
Because PU~Vul is not included in \citet{sko07}, we highlight
RR~Tel in Figure \ref{skopal_emission_effect}
because it belongs to the same group as PU~Vul and
its spectral evolution is similar to that of PU~Vul.
It is very clear that the blueward excursion in the color-color diagram
is due to emission lines, because the position moves almost
horizontally by $\Delta (B-V)\approx -0.8$.

Thus, we specify the color-color evolution of PU~Vul by points 1, 2,
3, 4, and 5 (``PU~Vul template'').
These points are tabulated in Table \ref{intrinsic_two_color_selected}.
Here we stop following the color evolution when the $V$ magnitude drops by 
about 3 mag from the maximum because strong emission lines make 
increasingly large contributions to the colors, and their effects
cloud the overall color evolution.

It may be surprising that the color-color evolution of PU~Vul follows 
almost the same tracks as those of slow/moderately fast/fast novae.
This is because the physics of emission (Figure 
\ref{wind_config6ab_premax_no3})
is common among these novae regardless of the speed class or
light curve shape.  In a classical nova, the evolution is very fast,
and it quickly passes stages (a) and (b) in Figure 
\ref{wind_config6ab_premax_no3} and enters the optically
thick wind phase (c).  On the other hand, in PU~Vul, the optically-thick
winds were not accelerated and it slowly evolved from stages (a) 
to (b) and then entered the optically-thin wind phase.
In the next section, we further show that very slow novae also follow
this common track in color-color evolution.

%Fig.19
%\placefigure{light_curve_v723_cas_hr_del_v5558_sgr}

\begin{figure}
%%%\begin{figure*}
%%\epsscale{0.75}
\epsscale{1.15}
\plotone{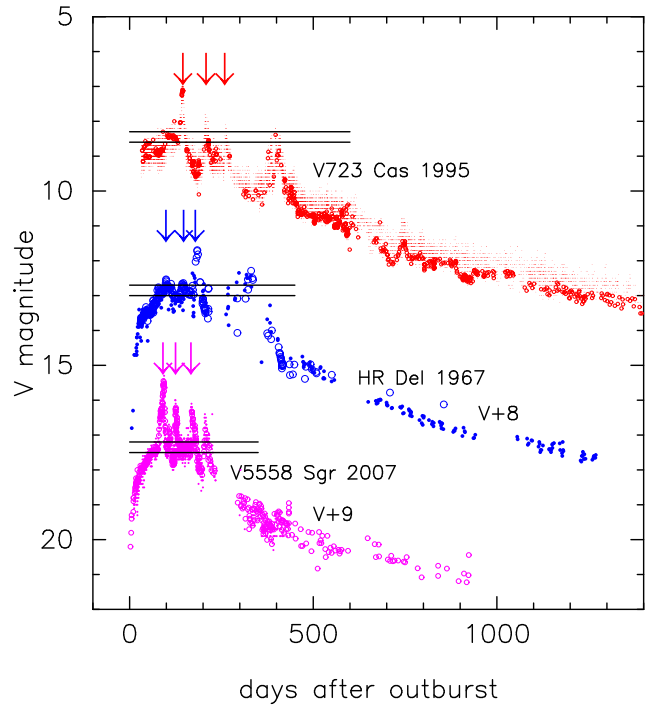}
%\plotone{light_curve_v723_cas_hr_del_v5558_sgr.epsi}
%\plotfiddle{evolution1.ps}{5.0cm}{270}{0.4}{0.4}{-170}{220}
\caption{
Optical light curves of three very slow novae, V723~Cas (red),
HR~Del (blue), and V5558~Sgr (magenta), from top to bottom.
V723~Cas: small red dots are visual magnitudes taken from the AAVSO
archive; red open circles are $V$ magnitudes taken from \citet{cho97b}.
HR~Del: blue small filled circles are photographic magnitudes
($m_{\rm pg}$) taken from \citet{ter68}; blue large open circles
are $V$ magnitudes taken from \citet{mol69}, \citet{man70},
\citet{bar70}, \citet{ond68}, and \citet{dre77}.
V5558~Sgr: magenta dots are visual magnitudes taken from the AAVSO
archive; magenta open circles are $V$ magnitudes taken from
the AAVSO and Variable Star Observers League in Japan
(VSOLJ) archives.  Two horizontal solid lines denote
absolute magnitudes of $M_V=-5.4$ and $-5.7$, which correspond
to the flat peaks of PU~Vul in 1979 and 1981--1983, respectively. 
First three flaring peaks are denoted by downward arrows for each nova.
\label{light_curve_v723_cas_hr_del_v5558_sgr}}
\end{figure}
%%%\end{figure*}

\section{Color Evolution of Very Slow Novae, HR~Del, V723~Cas, and V5558~Sgr}
\label{each_slow_novae}
In this section, we examine the color evolutions of three very slow novae,
HR~Del, V723~Cas, and V5558~Sgr.  They have similar 
light curves (see Figure \ref{light_curve_v723_cas_hr_del_v5558_sgr}) and
spectral evolutions \citep[e.g.,][]{fri92, fri02, eva03, mun07c}.

\subsection{The Nature of Outbursts: Transition from Static to Wind Evolution}
\label{transition_from_static_to_wind}
HR~Del, V723~Cas, and V5558~Sgr show multiple peaks in their light curves,
as shown in Figure \ref{light_curve_v723_cas_hr_del_v5558_sgr}, whereas
PU~Vul shows a smooth light curve with a flat maximum phase except for
the eclipses.  \citet{kat09h} proposed the concept of multiple solutions
of novae as an explanation for such multipeak structures in the optical light
curves.  They pointed out that there are two different types of nova
evolution; one is evolution with optically-thick winds
for $M_{\rm WD} \gtrsim 0.5~M_\sun$ \citep[e.g.,][]{kat94h} and the other
is that without them for $M_{\rm WD} \lesssim 0.7~M_\sun$.  Both types
can be realized in slow novae in a certain range of WD masses, 
$0.5~M_\sun \lesssim M_{\rm WD} \lesssim 0.7~M_\sun$.
For example, the evolution of GQ Mus 1983 is explained by a sequence
of optically-thick wind solutions on a $\sim 0.7~M_\odot$ WD \citep{hac08kc},
whereas the evolution of PU Vul 1979 is described by a sequence of
hydrostatic solutions on a $\sim 0.6~ M_\sun$ WD \citep{kat11hcg}.
These two evolutions show a remarkable difference in the optical light curves.
In a wind-type nova, massive optically-thick winds carry
away a large part of the envelope, producing a quick decay of the
light curve.  Thus, the optical light curve of the nova shows a sharp peak.
On the other hand, in low mass WDs no optically-thick winds are accelerated,
so the nova evolves very slowly and retains an extended photosphere with
low temperatures for a long time, which produces a long lasting flat optical
peak before the magnitude slowly decays as in PU~Vul.

\citet{kat11h} explained theoretically that the transition from static
evolution to wind evolution could occur during an outburst.
In such a case, the nova shows a flat optical peak with no indication
of strong mass loss in the early phase of the outburst, like PU~Vul,
followed by a quick decay phase, as in normal novae with strong
optically-thick winds.  They further suggested that such a transition
from a static to a wind structure is accompanied by behavior such as
oscillations in the brightness, because the internal structures of
the static/wind solutions differ greatly, and a relaxation process
in the transition should induce some oscillatory features.  
They suggested that the light curve behaviors of HR~Del, V723~Cas, 
and V5558~Sgr in Figure \ref{light_curve_v723_cas_hr_del_v5558_sgr}
correspond to this type of transition. 

\citet{kat11h} proposed the following transition mechanism:
The static evolution itself is stable, 
as seen in the 8-yr flat peak of PU~Vul (Figures 
\ref{light_curve_pu_vul_hr_del_fh_ser_pw_vul_v1668_cyg_v1500_cyg} 
and \ref{v723_cas_pu_vul_light_simple}).
In close binaries, however, the envelope structures are
affected because of the effect of a companion star's gravity, which may
trigger the transition.  For PU~Vul ($P_{\rm orb}=4900$ days),
however, the transition was not triggered because its companion star is
far outside the photosphere of the bloated WD envelope. 
For HR~Del ($P_{\rm orb}=0.214$ days), V723~Cas ($P_{\rm orb}=0.692$ days),
and V5558~Sgr ($P_{\rm orb}=$ unknown), their outbursts started
as static evolution, but their transitions were triggered a few hundred 
days after the outburst, owing to the effect of the companion located
deep in the nova envelope.  If this is the case,
these three novae should exhibit not only similar light curve evolution 
but also similar color evolution.
In this section we first examine the distances and absorptions of these
three novae to obtain the dereddened colors and then discuss
the color evolution of these three very slow novae.

\subsection{Absolute magnitude at pre-maximum halt in slow novae}
\label{pre-maximum_halts}
First, we examine the absolute magnitude of these very slow novae.
As already explained in Section \ref{transition_from_static_to_wind},
\citet{kat11h} modeled the pre-maximum phase of these
very slow novae with a static evolution followed by 
the transition from a static to a wind structure.
They predicted that this transition occurs in a narrow range of WD masses,
$0.5 ~M_\sun \lesssim M_{\rm WD} \lesssim 0.7 ~M_\sun$.
Thus, the brightness at the pre-maximum phase is similar to that of
PU~Vul ($\sim0.6~M_\sun$), i.e.,
$M_V= -5.4$ in stage 1 and $M_V=-5.7$ in stage 2.
We apply these two absolute magnitudes of PU~Vul  to 
the $V$ light curves of the three novae.
Because these light curves show oscillatory behavior,
it is not easy to define the stages corresponding to stages 1 and 2
in PU~Vul.  However, we finally found reasonable fittings, as shown
in Figure \ref{light_curve_v723_cas_hr_del_v5558_sgr}.  Thus,
we obtained a distance modulus of
$(m-M)_V=10.4$, 14.0, and 13.9 for HR~Del, V723~Cas, and V5558~Sgr,
respectively.

\begin{figure}
%%\epsscale{0.75}
%%\epsscale{1.0}
\epsscale{1.15}
\plotone{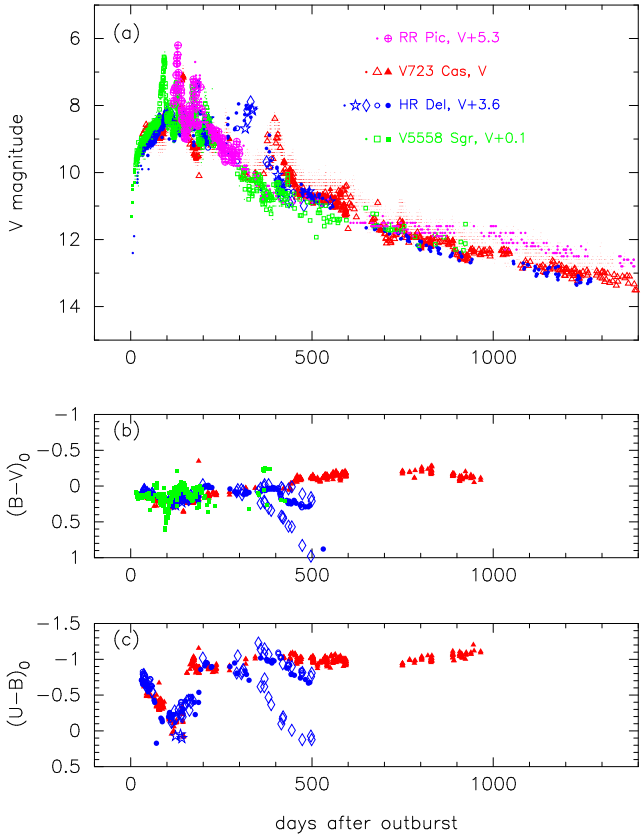}
%\plotone{v723_cas_hr_del_v5558_sgr_rr_pic_ub_bv_color_light_curve_revised.epsi}
%\plotfiddle{evolution1.ps}{5.0cm}{270}{0.4}{0.4}{-170}{220}
\caption{
Light curves in (a) $V$ band, and color evolution in (b) $(B-V)_0$, 
and (c) $(U-B)_0$ for RR~Pic (magenta open circles with a plus sign), 
HR~Del (blue symbols), V723~Cas (red triangles), and V5558~Sgr 
(green squares).  To overlap them for as long as possible,
we shifted the $V$ magnitudes of RR~Pic, HR~Del, and V5558~Sgr downward
by $5.3$, $3.6$, and $0.1$ mag, respectively, and arranged the origin
of the time of each nova against those of V723~Cas.
The visual and photometric data for RR~Pic are taken from the AAVSO archive
and \citet{daw26}, respectively.
\label{v723_cas_hr_del_v5558_sgr_rr_pic_ub_bv_color_light_curve_revised}}
\end{figure}

\begin{figure}
%%\epsscale{0.75}
%%\epsscale{1.0}
\epsscale{1.15}
\plotone{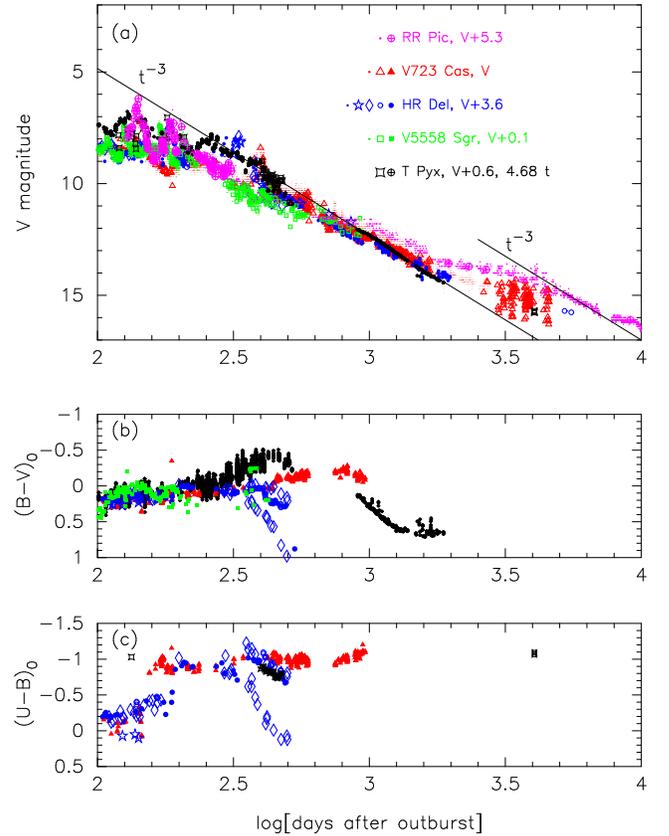}
%\plotone{v723_cas_hr_del_v5558_sgr_ub_bv_color_light_curve_revised_logscale.epsi}
%\plotfiddle{evolution1.ps}{5.0cm}{270}{0.4}{0.4}{-170}{220}
\caption{
Same as Figure 
\ref{v723_cas_hr_del_v5558_sgr_rr_pic_ub_bv_color_light_curve_revised},
but on a logarithmic timescale.  Data for T~Pyx (black symbols)
are added for comparison (see Section \ref{t_pyx_color}).  
The data for T~Pyx are taken from the AAVSO archive.
We stretched the light curve of T~Pyx by 4.68 
and shifted the magnitudes down by 0.6 mag against that of V723~Cas.
The late phase $V$ light curves of these novae follow a $t^{-3}$ law
for a uniform expansion of constant nebular mass. 
\label{v723_cas_hr_del_v5558_sgr_ub_bv_color_light_curve_revised_logscale}}
\end{figure}

We confirmed and calibrated the distance moduli of these three novae
obtained above by comparing their light curves with that of RR~Pic,
because the distance to RR~Pic was recently obtained using trigonometric 
parallax, i.e., $d=521^{+54}_{-45}$~pc \citep{har13}.  
The distance modulus of RR~Pic is calculated to
be $(m-M)_V=5\log 521^{+54}_{-45}/10 + 0.13 = 8.7\pm0.2$,
where we used $A_V=0.13$ after \citet{har13}.
The $V$ light curve of RR~Pic is plotted in Figure 
\ref{v723_cas_hr_del_v5558_sgr_rr_pic_ub_bv_color_light_curve_revised}
on a linear timescale and in Figure 
\ref{v723_cas_hr_del_v5558_sgr_ub_bv_color_light_curve_revised_logscale}
on a logarithmic timescale with those of HR~Del, V723~Cas, and V5558~Sgr.
We shifted the $V$ light curves of RR~Pic, V5558~Sgr, and HR~Del
vertically so that they overlap that of V723~Cas.
The light curve of RR~Pic is very similar to those of these three novae,
so RR~Pic belongs to the same type of novae as HR~Del, V723~Cas, and 
V5558~Sgr, although the very early (rising) phase of RR~Pic was not observed.
Because these four novae have almost the same timescale of decline, 
we simply assumed that their brightnesses are all the same.  
The difference in $V$ magnitude compared to V723~Cas is $-5.3$ for RR~Pic,
$-3.6$ for HR~Del, and $-0.1$ for V5558~Sgr.  Therefore, the difference
$\Delta V$ from RR~Pic is calculated as $\Delta V= -3.6+5.3$ for HR~Del,
$\Delta V= -0.0+5.3$ for V723~Cas, and $\Delta V= -0.1+5.3$ for V5558~Sgr. 
The distance moduli of these three novae are $(m-M)_{V,\rm HR~Del}=10.4$, 
$(m-M)_{V,\rm V723~Cas}=14.0$, and $(m-M)_{V,\rm V5558~Sgr}=13.9$.  
Thus we have
\begin{eqnarray}
(m-M)_{V, \rm RR~Pic} &=& 8.7\pm0.2\cr
&=&(m-M)_{V,\rm HR~Del} - \Delta V \cr
&=& 10.4 -(-3.6+5.3) = 8.7\cr
&=&(m-M)_{V,\rm V723~Cas} - \Delta V \cr
&=& 14.0 -(-0.0+5.3) = 8.7\cr
&=&(m-M)_{V,\rm V5558~Sgr} - \Delta V \cr
&=& 13.9 -(-0.1+5.3) = 8.7.
\label{rr_pic_distance_mod}
\end{eqnarray}
These values are very consistent with each other.

%Fig.20
%\placefigure{v723_cas_hr_del_v5558_sgr_rr_pic_ub_bv_color_light_curve_revised}

%Fig.21
%\placefigure{v723_cas_hr_del_v5558_sgr_ub_bv_color_light_curve_revised_logscale}

%Fig.22
%\placefigure{bv_color_hr_del_v723_cas_v5558_sgr_premax}

\begin{figure}
%%\begin{figure*}
%%\epsscale{0.75}
\epsscale{1.15}
\plotone{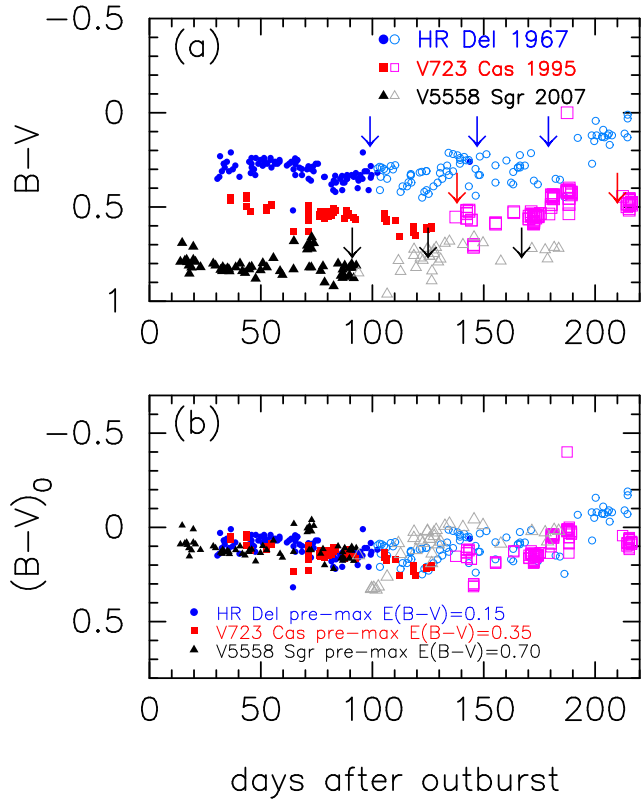}
%\plotone{bv_color_hr_del_v723_cas_v5558_sgr_premax.epsi}
%\plotfiddle{evolution1.ps}{5.0cm}{270}{0.4}{0.4}{-170}{220}
\caption{
(a) Observed $B-V$ colors of three novae, HR~Del, V723~Cas, and
V5558~Sgr.  Downward arrows indicate first three flaring peaks
of novae, which are the same as those in Figure
\ref{light_curve_v723_cas_hr_del_v5558_sgr}.
We use filled and open symbols to denote data before and after
the first flaring peak, respectively.
The data for HR~Del are taken from \citet{oco68},
\citet{man70}, \citet{bar70}, and \citet{ond68}.
The data for V723~Cas are taken from \citet{cho97b}.
The data for V5558~Sgr are taken from the AAVSO and VSOLJ archives. 
(b) Intrinsic $(B-V)_0$ colors of the three novae.  We estimated
the interstellar extinctions by assuming that the colors of these three novae
are the same in the pre-maximum phase.  
Here we fix the extinction of HR~Del to be
$E(B-V)=0.15$.  The obtained extinctions are $E(B-V)=0.35$ for V723~Cas
and $E(B-V)=0.70$ for V5558~Sgr.  
\label{bv_color_hr_del_v723_cas_v5558_sgr_premax}}
%%\end{figure*}
\end{figure}

%Fig.23
%\placefigure{sed_v723_cas_dereddening_free_free_bb}

\begin{figure}
%%\epsscale{0.75}
\epsscale{1.15}
\plotone{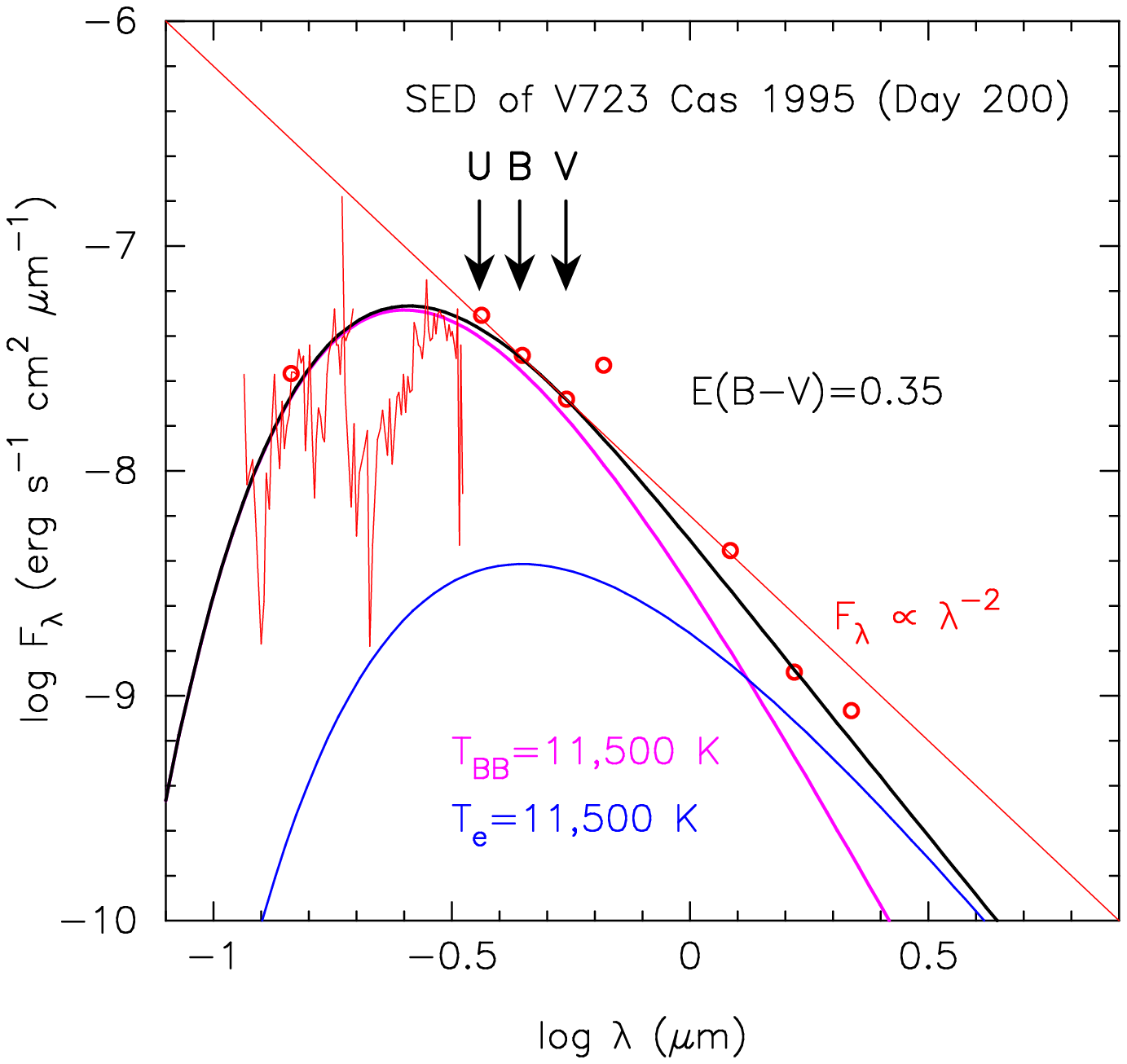}
%\plotone{sed_v723_cas_dereddening_free_free_bb.epsi}
%\plotfiddle{evolution1.ps}{5.0cm}{270}{0.4}{0.4}{-170}{220}
\caption{
Same as Figure \ref{sed_pu_vul_iue_ubv_free_free_bb}, but
for V723~Cas about 200 days after the outburst.
Observed $UBVR$ fluxes are taken from \citet{cho97b} and $JHK$ fluxes
are taken from \citet{kam99}.  {\it IUE} UV spectra (SWP56799
and LWP31977) are taken from the INES archive data sever.
We also plot the UV1455\AA\  flux.
All fluxes denoted by red open circles and red thin solid lines
are dereddened with $E(B-V)=0.35$.
Global features of spectrum can be fitted with combination
(black solid line) of blackbody with temperature
$T_{\rm BB}=15,500$~K (magenta solid line)
and optically-thick, free-free emission with electron
temperature $T_{\rm e}=15,500$~K (blue solid line). 
In the region of the $UBV$ bands, the SED is close to
that of optically-thin, free-free emission of 
$F_\lambda \propto \lambda^{-2}$ (red thin solid line).  
\label{sed_v723_cas_dereddening_free_free_bb}}
\end{figure}

\subsection{HR~Del 1967}
\label{hr_del_color}
The very slow nova HR~Del was discovered by Alcock \citep{can67}
at $m_v= 5.0$ on UT 1967 July  8.9 (JD 2439680.4).
The light curve of HR~Del is plotted in Figures 
\ref{light_curve_v723_cas_hr_del_v5558_sgr},
\ref{v723_cas_hr_del_v5558_sgr_rr_pic_ub_bv_color_light_curve_revised}, and
\ref{v723_cas_hr_del_v5558_sgr_ub_bv_color_light_curve_revised_logscale}.
It reached 4.7 mag at optical maximum. 
Since the outburst day is not known, we adopted UT 1967 June 8.5  
as the outburst day, i.e., $t_{\rm OB}=$JD~2439653.0,
from the figure of \citet{rob68}.  
\citet{ver87} obtained $E(B-V)=0.15\pm 0.03$ for the extinction
toward HR~Del.  The NASA/IPAC galactic dust absorption map
gives $E(B-V)=0.11 \pm 0.006$ in the direction of HR~Del,
whose galactic coordinates are $(l, b)=(63\fdg4304, -13\fdg9721)$; 
this is roughly consistent with Verbunt's value.  \citet{dow00} obtained
a distance of $d=0.76\pm 0.13$~kpc to HR~Del from the nebular
expansion parallax.  More recently, \citet{har03} obtained a new
value of the distance, $d=0.97\pm 0.07$~kpc, also from the expansion
parallax method using {\it HST} imaging.  Other, older, estimates are
all between the above two estimates, i.e., $d=0.940\pm 0.155$~kpc from
various expansion parallax methods 
\citep{mal75, koh81, due81, sol83, coh83, sla94, sla95} 
or $d=0.835\pm 0.092$~kpc from other techniques \citep{dre77}.
If we adopted the new distance estimate of $d=0.97$~kpc \citep{har03}
and the extinction of $E(B-V)=0.15$ \citep{ver87},
the distance modulus is $(m-M)_V=5\log 970/10 + 3.1\times 0.15
=10.4$, which is perfectly consistent with the value in 
Equation (\ref{rr_pic_distance_mod}).
Therefore, we used $E(B-V)=0.15$ and $d=0.97$~kpc for HR~Del.

%%% 063.4304 -13.9721 

\subsection{V723~Cas 1995}
\label{v723_cas}
V723 Cas is also a very slow nova; it was discovered at mag 9.2
on UT 1995 August 24.57 (JD~2449954.07).  \citet{mun96} proposed 
UT July 20.5 as the outburst day, i.e., $t_{\rm OB}=$JD~2449919.0,
so we adopted this day in this paper.
The interstellar extinction and distance toward V723~Cas were
estimated by many authors.  \citet{cho97b} obtained $(m-M)_V= 13.8$, 
$E(B-V)=0.57$, and $d=2.38$~kpc from various MMRD relations
as well as the $V_{15}$ and $B_{15}$ brightnesses (absolute $M_V$ and 
$M_B$ 15 days after the maximum, respectively) of novae. 
\citet{iij98} gave $(m-M)_V= 13.2$, $A_V=0.88$ ($E(B-V)=A_V/3.1=0.28$),
and $d=3$~kpc from the interstellar absorption, but \citet{iij06} revised 
the value to $(m-M)_V= 14.0$ for $E(B-V)=0.57\pm0.05$ and $d=2.8$~kpc. 
\citet{eva03} obtained $(m-M)_V= 14.9$, $A_V=1.9$ ($E(B-V)=A_V/3.1=0.61$),
and $d=4$~kpc using the MMRD relation and the Eddington limit of
a $0.67~M_\sun$ WD ($M_V= -6.59$), together with the apparent flat peak
magnitude of $m_V=8.25$.  However, their assumed absolute magnitude
of $M_V= -6.59$ is too bright for a nova on a low-mass WD 
\citep[see, e.g., Equations (3) and (4) of][]{hac04k}.  
As a result, their obtained value of $(m-M)_V= 14.9$ is too large.
If we adopt $M_V= -5.3$ \citep{hac04k} for a low-mass WD and 
$m_V=8.25$ above, the distance modulus becomes 
$(m-M)_V= 13.6$, which is much smaller than $(m-M)_V= 14.9$ 
and roughly consistent with the other values mentioned above.

\citet{nes08} criticized the application of the MMRD relations to very
irregular light curves. Instead, they assumed that the absolute magnitude
of V723~Cas is the same as that of HR~Del, because they are similar
types of novae.  They obtained $(m-M)_V= 13.7$, $E(B-V)=0.5\pm 0.1$,
and $d=2.7^{+0.4}_{-0.3}$~kpc. 
\citet{hac04k} obtained the distance modulus of $(m-M)_V= 13.9$ 
for a $0.59~M_\sun$ WD with a solar composition envelope,
by comparing their blackbody light curves with observations.  
Hachisu \& Kato could not specify $E(B-V)$ but proposed $d=2.5$~kpc 
for $E(B-V)=0.6$ and $d=4.0$~kpc for $E(B-V)=0.3$.
Thus, the resultant distance modulus to V723~Cas was not very scattered
and fell in a relatively narrow range of $(m-M)_V= 13.6$--14.0.

The interstellar extinction toward V723 Cas was also estimated
by many authors, but their values are quite scattered.
In increasing order, \citet{rud02} obtained 
$E(B-V) = 0.20\pm 0.12$ in 1999 August and $0.25\pm 0.1$ in 2000 July
from the Paschen and Brackett lines.  \citet{iij98} obtained
$E(B-V) = 0.29$ from the reddening of field stars near the location
of V723~Cas.  \citet{mun96} obtained $E(B-V) = 0.45$ from interstellar
\ion{Na}{1}~D double lines.  \citet{nes08} estimated the extinction
to be  $E(B-V) = 0.5\pm 0.1$ from various values in the literature
and their $N_{\rm H}$ value from X-ray spectrum model fits.  
\citet{cho97b} obtained $E(B-V) = 0.57$ from the intrinsic colors
at maximum and at two magnitude below maximum.  
\citet{gon96} gave $E(B-V) = 0.60$ from the 2200\AA\  dust absorption feature. 
\citet{eva03} derived $E(B-V) = 0.78\pm 0.15$ from the IR \ion{H}{1} 
recombination lines.

The galactic extinction is $E(B-V)= 0.39$ in the direction of V723~Cas 
from the dust map of \citet{sch98},
where the galactic coordinates of V723~Cas are 
$(l, b)=(124\fdg9606, -8\fdg8068)$.
The HEASARC $N_{\rm H}$ tool yields $N_{\rm H}= 2.1 \times 10^{21}$
for LAB \citep[Leiden/Argentine/Bonn;][]{kal05} 
and $2.4 \times 10^{21}$~cm$^{-2}$  for DL \citep{dic90},
respectively.  If we use the relation of 
$E(B-V)= N_{\rm H}/5.8\times 10^{21}$~cm$^{-2}$ \citep{boh78}
(or the recent relation of 
$E(B-V)= N_{\rm H}/8.3\times 10^{21}$~cm$^{-2}$ \citep{lis14}),
the extinction is $E(B-V)=0.4$ (or $E(B-V)=0.29$).
The recent NASA/IPAC dust map gives $E(B-V)=0.34\pm0.01$.
Therefore, the extinction could be around $E(B-V)\sim 0.35$
if the circumstellar absorption is negligible.

\citet{gon96} obtained a large value of $E(B-V)=0.6$ on the basis of
the strength of the 2175 \AA\ absorption feature in the {\it IUE} 
spectra taken in 1996 January, approximately 150 days after the outburst.
This value must be taken as an upper limit, because it was obtained from
spectra taken while the nova was still in an optically-thick state, 
and may have been affected by an additional circumstellar absorption.
To confirm this effect, R. Gonz\'alez-Riestra (2012, private communication)
examined {\it IUE} spectra taken later in the outburst, and obtained
color excess values of 0.34, 0.33, and 0.25 for days 160, 172,
and 176 after the outburst, respectively, all with an uncertainty of 
$\pm0.05$.  These values are much smaller than that of 0.60 obtained
150 days after the outburst.  Therefore, she concluded
that the reddening toward V723~Cas is $E(B-V)=0.30\pm0.05$.

The reddening can be estimated by another method based on the similarity
of the three novae, HR~Del, V723~Cas, and V5558~Sgr.
Figure \ref{bv_color_hr_del_v723_cas_v5558_sgr_premax}(a) shows 
the $B-V$ color evolution of these three novae during the first
200 days, that is, in the flat pre- and post-maximum phase. 
The positions of flaring pulses are indicated by downward arrows.
Figure \ref{bv_color_hr_del_v723_cas_v5558_sgr_premax}(b)
shows HR~Del dereddened with $E(B-V)= 0.15$ (see Section \ref{hr_del_color}).
We also plot the data for V723~Cas and V5558~Sgr shifted upward by 0.35 
and 0.70, respectively, to match their data with those of HR~Del.
In our fitting process, we changed the $E(B-V)$ value by steps of 0.05,
so the possible systematic error is 0.05.  If the intrinsic colors
$(B-V)_0$ of these three very slow novae are the same in the pre-maximum
phase, $E(B-V)$ is about $0.35$ for V723~Cas, which is consistent with
the new estimates by Gonz\'alez-Riestra and the value in the dust map on
the NASA/IPAC web site.

Recently, \citet{lyk09} obtained a distance of
$d=3.85^{+0.23}_{-0.21}$~kpc to V723~Cas from the expansion parallax
method.  If we adopt this distance and reddening
$E(B-V)=0.35$, we obtain the distance modulus
$(m-M)_V=5\log 3850/10 + 3.1\times 0.35 =14.0$, which is
perfectly consistent with Equation (\ref{rr_pic_distance_mod}). 
Therefore, we use $d=3.85$~kpc and $E(B-V)=0.35$ for V723~Cas.

Figure \ref{sed_v723_cas_dereddening_free_free_bb} shows the dereddened
spectrum of V723~Cas about 200 days after the outburst.
The broadband spectrum can be reproduced by a combination of a blackbody
($T_{\rm BB}=11,500$~K) and free-free ($T_{\rm e}=11,500$~K) emission,
as introduced in the previous section.  The $UBV$ magnitudes are closely
fitted with only a blackbody of $T_{\rm BB}=11,500$~K and, at the same time,
are fitted with the spectrum of optically-thin free-free emission
($F_\nu\propto \nu^0$ or $F_\lambda\propto \lambda^{-2}$ represented
by a red thin solid line).   Therefore, its position is close to 
point 0 in the color-color diagram.  We discuss the color-color evolution
of V723~Cas in Section \ref{resemblance_to_puvul} below.

%Fig.24
%\placefigure{v5558_sgr_distance_reddening_observation}

\begin{figure}
%%\epsscale{0.75}
\epsscale{1.15}
\plotone{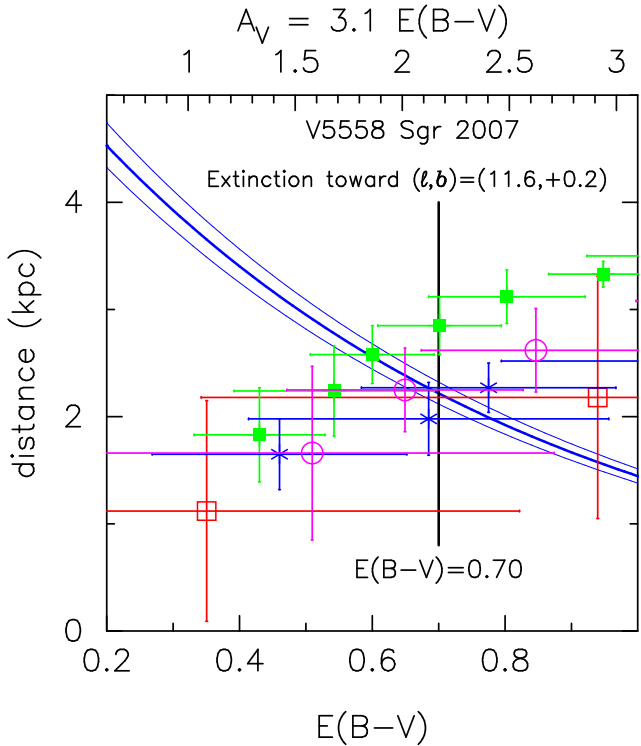}
%\plotone{v5558_sgr_distance_reddening_observation.epsi}
%\plotfiddle{evolution1.ps}{5.0cm}{270}{0.4}{0.4}{-170}{220}
\caption{
Distance-reddening relation toward V5558~Sgr.
Blue thick solid line flanked by blue thin solid lines
denotes the distance-reddening relation
of $(m-M)_V=13.9\pm0.1$.
Four sets of data with error bars show distance-reddening relations
in four directions close to V5558~Sgr:
$(l, b)=(11\fdg5,+0\fdg0)$ (red open squares), 
$(11\fdg75,+0\fdg0)$ (green filled squares),
$(11\fdg5,+0\fdg25)$ (blue asterisks),
and $(11\fdg75,+0\fdg25)$ (magenta open circles); data are
taken from \citet{mar06}.  Vertical black solid line
represents color excess $E(B-V)=0.70$.
\label{v5558_sgr_distance_reddening_observation}}
\end{figure}

\subsection{V5558~Sgr 2007}
\label{v5558_sgr}
V5558~Sgr was discovered by Sakurai \citep{nak07a} at mag 10.3
on UT 2007 April 14.777.  Sakurai also reported that nothing is visible
on an image taken on UT April 9.8 (limiting mag 11.4). 
The star was also detected by Haseda \citep{yam07b} at mag 11.2
on UT April 11.792.  Because the outburst day is not known,
we adopted UT 2007 April 8.5 as the outburst day, i.e., 
$t_{\rm OB}=$JD~2454199.0.  The optical light curve and color evolution
are shown in    
Figures \ref{light_curve_v723_cas_hr_del_v5558_sgr},
\ref{v723_cas_hr_del_v5558_sgr_rr_pic_ub_bv_color_light_curve_revised},
\ref{v723_cas_hr_del_v5558_sgr_ub_bv_color_light_curve_revised_logscale}, and 
\ref{bv_color_hr_del_v723_cas_v5558_sgr_premax}.

We found in the literature two different estimates for the reddening:
one $E(B-V)=0.36$ was obtained by \citet{mun07c} from 
the \ion{Na}{1}~D lines, and the other $E(B-V)=0.8$ was obtained by 
\citet{rud07b} from the \ion{O}{1} lines.  Our estimate in the previous
subsection, $E(B-V)=0.70$, is consistent with Rudy et al.'s value
but much larger than Munari et al.'s.
Because these two values are very different, we further examine $E(B-V)$
from a different point of view.  In the previous subsections, we showed 
the similarity of the absolute magnitudes of HR~Del, V723~Cas,
and PU~Vul in the pre-maximum phase (see Figure
\ref{light_curve_v723_cas_hr_del_v5558_sgr}).
Thus, we derived $(m-M)_{V,\rm V5558~Sgr}=13.9$, assuming that
V5558~Sgr has the same absolute magnitude in the pre-maximum phase.
This distance modulus was confirmed and calibrated using the similarity 
of the light curves to that of RR~Pic (see Equation 
(\ref{rr_pic_distance_mod})).   Using this distance modulus in the $V$ band, 
we derived a distance-reddening relation, i.e.,
$(m-M)_V = 13.9 \pm 0.1$, together 
with Equation (\ref{distance_modulus_extinction}).
Figure \ref{v5558_sgr_distance_reddening_observation} shows
this distance-reddening relation as a blue thick solid line flanked 
with two blue thin solid lines, which correspond to $\pm 0.1$ mag error.

This figure also shows the distance-reddening relations for four
directions near V5558~Sgr,
whose galactic coordinates are $(l, b)=(11\fdg6107,+0\fdg2067)$.
The closest distance-reddening relation, which is denoted by blue asterisks,
crosses the blue thick line at or near
the vertical black solid straight line of $E(B-V)=0.70$. 
Thus, we use $E(B-V)=0.70$ and $d=2.2$~kpc in this paper.

The distance to V5558~Sgr was also estimated by \citet{pog10}.
Using various MMRD relations, she obtained a distance modulus of
$(m-M)_V=12.4$--12.8, which is much smaller than our value of $(m-M)_V=13.9$.
This suggests that the MMRD relations are problematic in calculating
the absolute magnitude of very slow novae such as V5558~Sgr.

%Fig.25
%\placefigure{color_color_diagram_hr_del_v723_cas_premax}

\begin{figure}
%%\begin{figure*}
%%\epsscale{0.75}
\epsscale{1.15}
\plotone{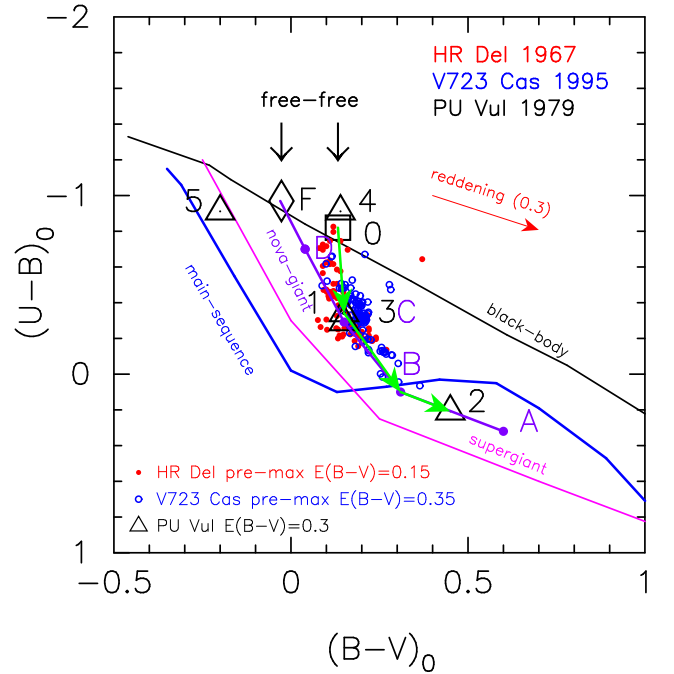}
%\plotone{color_color_diagram_hr_del_v723_cas_premax.epsi}
%\plotfiddle{evolution1.ps}{5.0cm}{270}{0.4}{0.4}{-170}{220}
\caption{
Same as Figure \ref{color_color_diagram_pu_vul_points_premax},
but for the pre-maximum phases of HR~Del (red filled circles)
and V723~Cas (blue open circles).  
Large open square denotes stage 0 for the $F_\nu\propto\nu^0$ 
optically-thin, free-free emission.  
Green arrows indicate evolutionary paths of HR~Del and V723~Cas.
It seems that both HR~Del and V723~Cas started at stage 0 and then
moved to 1, B, and 2.  We indicated the direction of reddening
by a red arrow, the length of which corresponds to $E(B-V)=0.3$.
\label{color_color_diagram_hr_del_v723_cas_premax}}
%%\end{figure*}
\end{figure}

%Fig.26
%\placefigure{color_color_diagram_hr_del_v723_cas_postmax}

\begin{figure}
%%\begin{figure*}
%%\epsscale{0.75}
%%\epsscale{1.0}
\epsscale{1.15}
\plotone{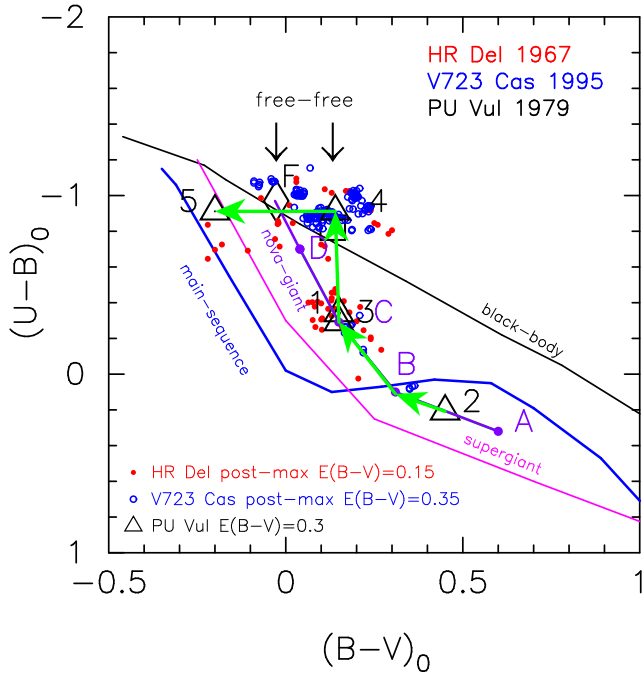}
%\plotone{color_color_diagram_hr_del_v723_cas_postmax.epsi}
%\plotfiddle{evolution1.ps}{5.0cm}{270}{0.4}{0.4}{-170}{220}
\caption{
Same as Figure \ref{color_color_diagram_hr_del_v723_cas_premax},
but for the post-maximum phases of HR~Del (red filled circles)
and V723~Cas (blue open circles).
\label{color_color_diagram_hr_del_v723_cas_postmax}}
%%\end{figure*}
\end{figure}

%Fig.27
%\placefigure{nova_pre_post_max_config}

\begin{figure*}
%%\begin{figure}
%%\epsscale{0.75}
%%\epsscale{1.0}
\epsscale{1.15}
\plotone{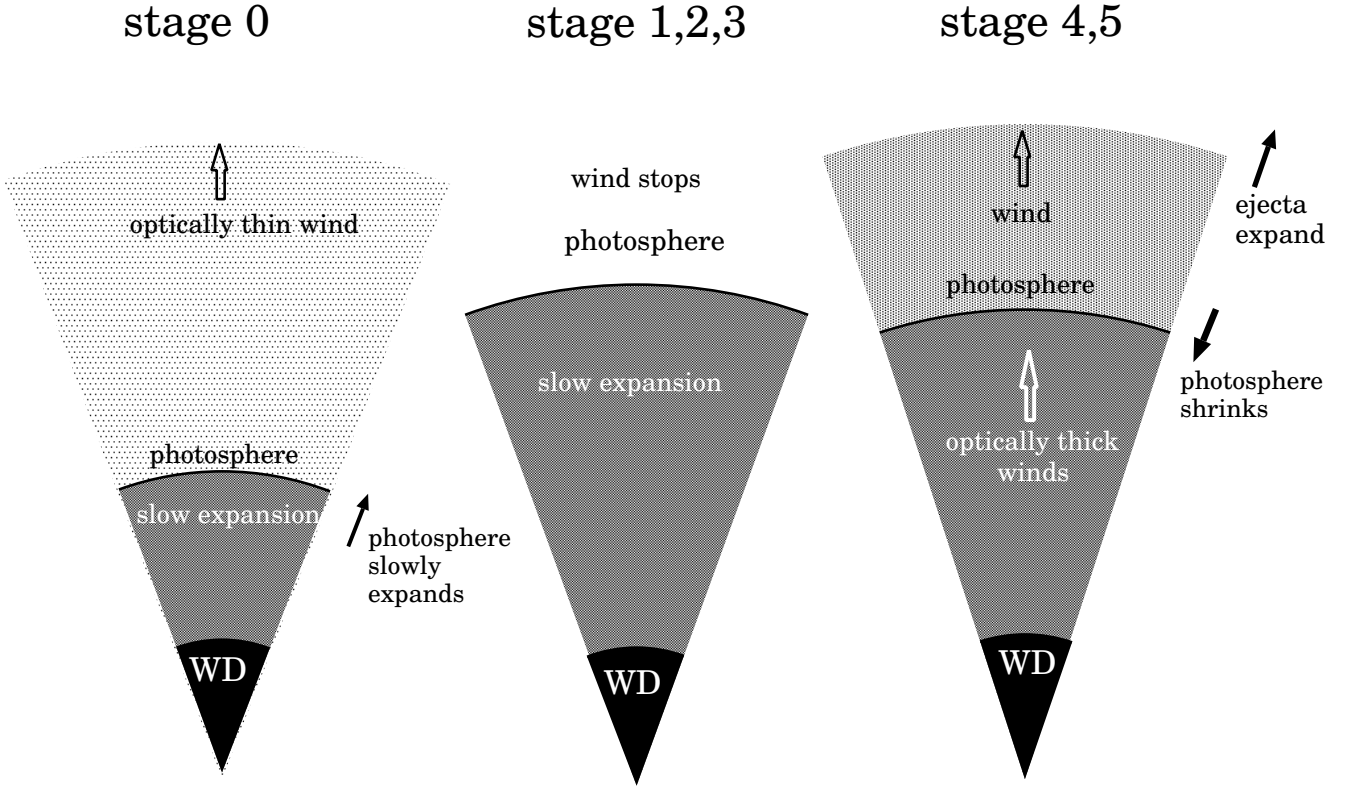}
%\plotone{nova_pre_post_max_config.epsi}
%\plotone{nova_pre_post_max_config.eps}
%\plotfiddle{evolution1.ps}{5.0cm}{270}{0.4}{0.4}{-170}{220}
\caption{
Schematic illustration of envelope structures in very slow novae
such as HR~Del, V723~Cas, and V5558~Sgr.
(a) Optically-thin winds blow in the very early (rising) phase,
which corresponds to stage 0 in Figure 
\ref{color_color_diagram_hr_del_v723_cas_premax}.
The photospheric temperature is high enough to blow
optically-thin winds in the rising phase of the novae.
Here the term ``optically-thin wind'' means that winds are accelerated
outside the photosphere, i.e., in the optically-thin region.
The WD envelope is close to a static configuration, so 
the wind velocity is much higher than that of the photospheric expansion.
(b) Emission lines gradually weaken with time, and absorption
components develop.  The optically-thin winds virtually stop.
Near optical maximum, the nova spectra resemble the
absorption spectra of F-type supergiants.  
We regard this phase as stage 1, 2, or 3.
(c) After the transition from a static to a wind configuration, the nova
entered the optically-thick wind phase.  Here the term ``optically-thick
wind'' means that winds are accelerated deep inside the photosphere,
i.e., in the optically-thick region.
The position in the color-color diagram returns
again to the wind phase (stage 4 near stage 0). 
Final excursion from stage 4 to 5 is mainly due to the development
of strong emission lines.
\label{nova_pre_post_max_config}}
%%\end{figure}
\end{figure*}

\subsection{Color-color Evolution of Very Slow Novae HR~Del and V723~Cas}
\label{resemblance_to_puvul}
Many authors have discussed the resemblance between
HR~Del and V723~Cas in their light curves and spectral evolution
\citep[e.g.,][]{fri02, eva03}.  We noted another resemblance in 
the color evolution of HR~Del and V723~Cas in the pre-maximum phase. 
Figure \ref{color_color_diagram_hr_del_v723_cas_premax} depicts 
the color-color evolution of V723~Cas (blue open circles) and HR~Del
(red filled circles) in the pre-maximum phase.  These novae follow
almost the same track as PU~Vul (large, black, open triangles).
Unfortunately, no $U$-band data are available for V5558~Sgr,
but we expect that this nova also shows a similar evolution
in the color-color diagram.

The only difference with respect to PU~Vul is that HR~Del and V723~Cas
started their very early evolutions from point 0, which represents
the position of optically-thin, free-free emission.
This is because they had optically-thin
winds in the very early phase of their outbursts.
\citet{iij98} reported prominent emission lines
of \ion{H}{1} and \ion{Fe}{2} with a P-Cygni profile
in an early stage of the pre-maximum phase of V723~Cas.
We regard this early pre-maximum phase with winds as stage 0
(see Figure \ref{color_color_diagram_hr_del_v723_cas_premax}).
Similar winds in the very early pre-maximum phase were
also reported for HR~Del \citep{raf78} and V5558~Sgr \citep{tan11}. 
\citet{iij98} observed that the emission lines of V723~Cas
gradually weakened with time as absorption components developed.
Then, the pure absorption spectra of F-type supergiants appeared near
optical maximum.  Similar spectral features were also observed in HR~Del
\citep{raf78} and V5558~Sgr \citep{tan11}.
We regard this late, pre-maximum phase with virtually
no winds as stage 1, point B, and stage 2 (Figure 
\ref{color_color_diagram_hr_del_v723_cas_premax}).  
This is consistent with our interpretation that the pre-maximum evolution
of V723~Cas is represented essentially by a static evolution 
with virtually no wind mass loss, like that of in PU~Vul. 

Figure \ref{color_color_diagram_hr_del_v723_cas_postmax} shows
the color-color diagram in the post-maximum phase of V723~Cas and HR~Del.  
In \citet{kat11h}, the nova entered the optically
thick wind phase after the transition from a static to a wind
configuration, i.e., after the relaxation oscillation.
V723~Cas moved from stages 3 to 4 (near stage 0)
after optical maximum (and the relaxation oscillation).
The position of stage 4 is close to point 0
(optically-thin, free-free emission), because the wind mass-loss rate
is relatively small and the free-free emission makes
a small contribution to the $UBV$ band region, as shown in
Figures \ref{sed_pu_vul_iue_ubv_free_free_bb}
and \ref{sed_v723_cas_dereddening_free_free_bb}.
The blueward excursion in $(B-V)_0$ from stages 4 to 5
is mainly due to the effects of strong emission lines,
as already discussed in the previous section.
In this way, we found that V723~Cas evolved from stage 0
to stages 1, 2, 3, 4, and 5.

The evolution of HR~Del is also plotted in Figures
\ref{color_color_diagram_hr_del_v723_cas_premax}
and \ref{color_color_diagram_hr_del_v723_cas_postmax}.
It also started at stage 0 and moved to 1 and then went 
to a midway point with respect to 2; it then returned to 3, 
jumped up to 4 and finally moved to 5.
Even in the post-maximum phase, the color evolutions of HR~Del,
V723~Cas, and PU~Vul almost overlap each other
(Figure \ref{color_color_diagram_hr_del_v723_cas_postmax}).
In PU~Vul, the phase from 3 to 4 corresponds to the epoch
in which optically-thin winds began to blow.
In HR~Del and V723~Cas, this phase corresponds to
the phase after the transition from a static to a wind configuration.
In much less massive WDs ($\sim0.6~M_\sun$), optically-thick winds
are relatively weak so the wind mass-loss rate is not as large as
in fast novae, as already described 
in Section \ref{free_free_color_evolution}.
As a result, these very slow novae evolve to point 4, which is very
close to point 0, which corresponds to optically-thin, free-free emission.  
This is why the evolutionary paths resemble that of PU~Vul.

Figure \ref{nova_pre_post_max_config} shows a schematic illustration 
of the evolution of these very slow novae.  In the very early stage of
the pre-maximum phase, optically-thin winds are accelerated
because the photospheric temperature is high.  In very slow novae,
the rising phase lasts long enough to be spectroscopically observed.
This stage corresponds to stage 0 on the color-color diagram (e.g.,
Figure \ref{color_color_diagram_hr_del_v723_cas_premax}) and in 
Figure \ref{nova_pre_post_max_config}(a).
The position of point 0 is defined by the spectrum of optically-thin 
free-free emission ($F_\nu\propto\nu^0$) but it happens to be close 
to the blackbody sequence.  The spectra of slow novae
at this early stage are close to that of a blackbody 
in the $UBV$ wavelength region 
because prominent emission lines of \ion{H}{1} and \ion{Fe}{2} dominate 
and fill the Balmer jump so the resultant $U-B$ color is close
to that of a blackbody. 
The emission lines gradually weakened with time, and absorption
components developed as the photospheric temperature decreased.
The optically-thin winds virtually stop.  The absorption spectra
of F-type supergiants appeared at this optical 
maximum stage.  The nova enters stages 1, 2, and then 3
in Figures \ref{color_color_diagram_hr_del_v723_cas_premax},
\ref{color_color_diagram_hr_del_v723_cas_postmax}, and
in Figure \ref{nova_pre_post_max_config}(b).
During these stages, we directly observe the photospheric emission of
an F-type supergiant. 
Therefore, the nova evolves along the nova-giant sequence.
After the transition from a static to a wind configuration, the nova
entered the wind phase, in which optically-thick winds are accelerated
and the mass-loss rate is much higher than in the early rising phase.
Then the position in the color-color diagram returns 
again to the wind phase at point 4 slightly above point 0.
This stage is illustrated in Figure \ref{nova_pre_post_max_config}(c),
and the spectrum of V723~Cas is plotted in Figure
\ref{sed_v723_cas_dereddening_free_free_bb}.
The spectrum in the $UBV$ region is very close to that of optically-thin
free-free emission, i.e., $F_\nu\propto\nu^0$ 
or $F_\lambda\propto\lambda^{-2}$ (thin red line).
The final excursion from stages 4 to 5 is mainly due to the development
of strong emission lines in the nebula (winds), as discussed 
in the previous sections.

In this way, these very slow novae follow essentially the same path
on the color-color diagram as that deduced from FH~Ser and other
traditional novae.  The small difference comes from the difference 
in the wind mass-loss rates, that is, optically-thin/thick winds
(Figure \ref{nova_pre_post_max_config}).
Thus, the color-color evolutions are almost the same among these novae,
because the physics involved is the same.

%Fig.28
%\placefigure{color_color_diagram_four_typical_example}

\begin{figure*}
%%\begin{figure}
\epsscale{1.0}
%%\epsscale{1.15}
\plotone{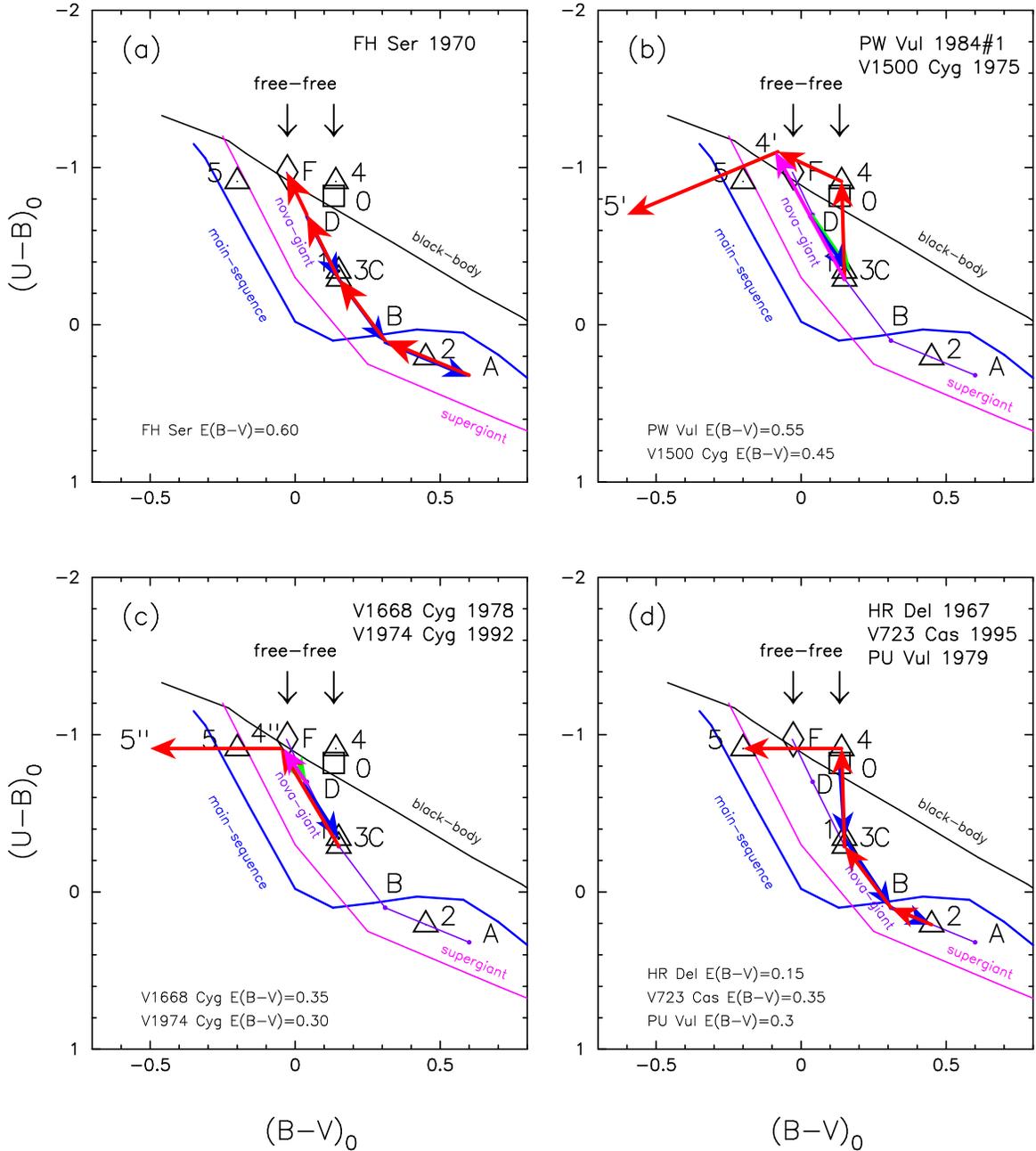}
%\plotone{color_color_diagram_four_typical_example.epsi}
%\plotfiddle{evolution1.ps}{5.0cm}{270}{0.4}{0.4}{-170}{220}
\caption{
Typical templates for novae in the intrinsic $(B-V)_0$ vs. $(U-B)_0$
color-color diagram. (a) FH~Ser, 
(b) PW~Vul (blue/red) and V1500~Cyg (green/magenta),
(c) V1668~Cyg (blue/red) and V1974~Cyg (green/magenta),
(d) PU~Vul, HR~Del, and V723~Cas.
Various symbols and lines have the same meanings
as in Figures \ref{color_color_diagram_fh_ser_typical}
and \ref{color_color_diagram_pu_vul_points_postmax}.
Blue/green arrows show rising phase of novae; red/magenta
arrows indicate declining phase.  See text for more details.
\label{color_color_diagram_four_typical_example}}
%%\end{figure}
\end{figure*}

\section{Estimates of Extinction toward Various Novae}
\label{extinction_novae}

\subsection{General Course of $UBV$ Color-Color Evolution}
\label{summary_color_evolution}
\begin{figure*}
\epsscale{1.0}
%%\epsscale{1.15}
\plotone{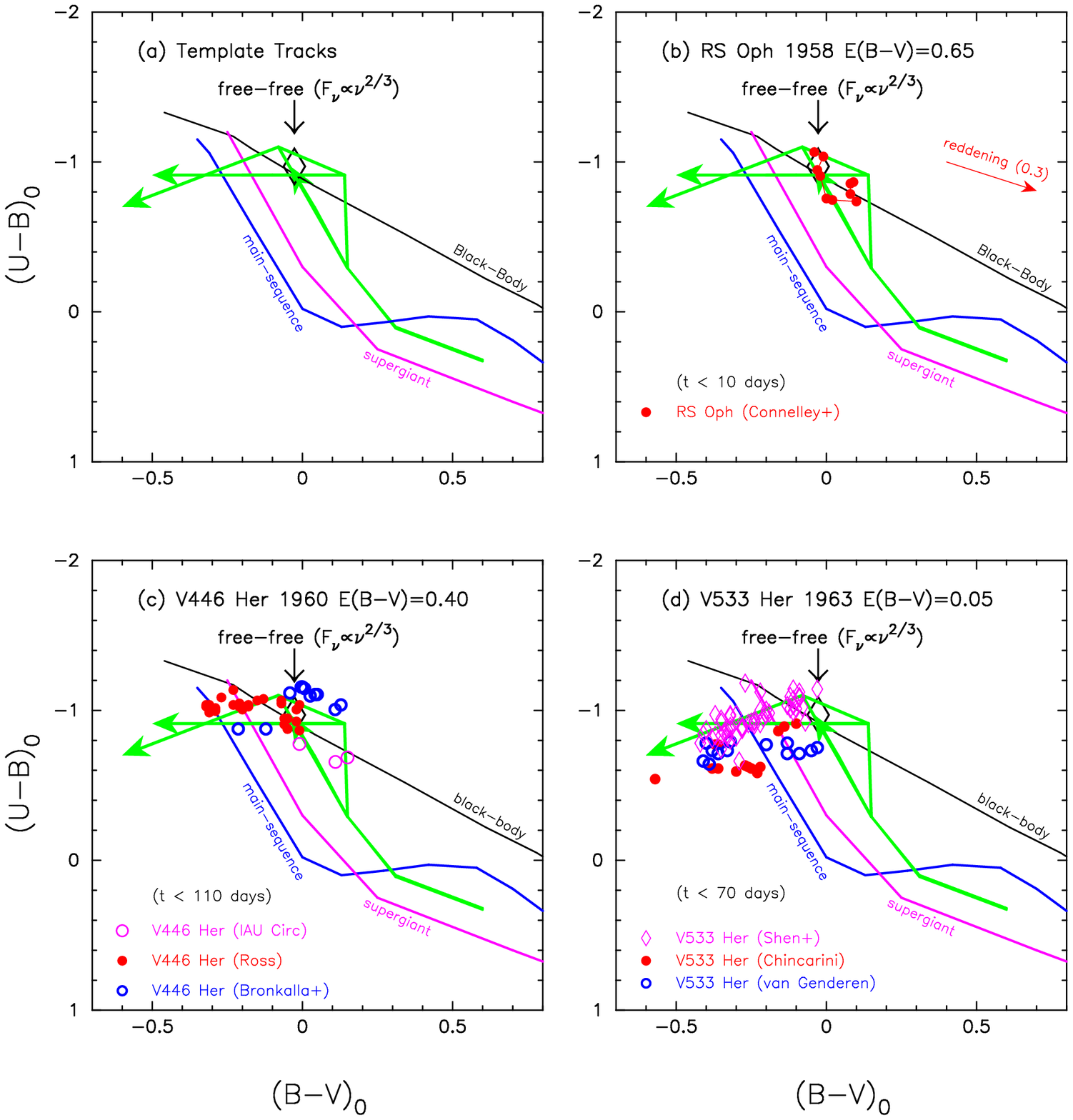}
%\plotone{color_color_diagram_templ_rs_oph_v446_her_v533_her.epsi}
%\plotfiddle{evolution1.ps}{5.0cm}{270}{0.4}{0.4}{-170}{220}
\caption{
Same as Figure
\ref{color_color_diagram_pw_vul_v1500_cyg_v1668_cyg_1974_cyg}, but
for (a) typical templates of tracks, (b) RS~Oph 1958, (c) V446~Her 1960,
and (d) V533~Her 1963.
Green lines with an arrow indicate several general tracks of nova
color-color evolution, as shown in Figure
\ref{color_color_diagram_four_typical_example}. 
We obtained the following color excesses for these three novae:
(b) $E(B-V)= 0.65$ for RS~Oph, (c) $E(B-V)= 0.40$ for V446~Her, and
(d) $E(B-V)= 0.05$ for V533~Her.
The $UBV$ data are taken from the literature as noted in each figure. 
\label{color_color_diagram_templ_rs_oph_v446_her_v533_her}}
\end{figure*}

We studied the color-color evolution of novae in the $(B-V)_0$ 
- $(U-B)_0$ diagram and showed that well-observed novae follow very similar
evolutionary courses in the intrinsic color-color diagram.
There are slight differences among them, and we summarize these differences
in Figure \ref{color_color_diagram_four_typical_example}.
Here, we specified four templates, (1) the moderately fast nova FH~Ser,
a prototype of the nova-giant sequence,
(2) the slow nova PW~Vul and the very fast nova V1500~Cyg, 
(3) the fast novae V1668~Cyg and V1974~Cyg, and (4) the symbiotic
nova PU~Vul and the very slow novae V723~Cas and HR~Del.
The tracks are characterized by several specified
points (A, B, C, D, F, 0, 1, ..., 5, 4', 5', 4'', and 5''; see Table
\ref{intrinsic_two_color_selected} for their values).
Figure \ref{color_color_diagram_templ_rs_oph_v446_her_v533_her}(a) shows
the templates of these eight well-observed novae.
It is again remarkable that these different types of novae follow
similar tracks in the color-color diagram.  Because these eight 
well-observed novae follow the templates in Figure
\ref{color_color_diagram_templ_rs_oph_v446_her_v533_her}(a),
we expect that many other novae also follow the same path.
In other words, if all the novae follow this template,
we can determine the color excess of a target nova  
by directly comparing its track in the color-color diagram with
our general track in Figure
\ref{color_color_diagram_templ_rs_oph_v446_her_v533_her}(a).
This is a new method for obtaining the color excesses of classical novae.

\begin{figure*}
\epsscale{1.0}
%%%\epsscale{1.15}
\plotone{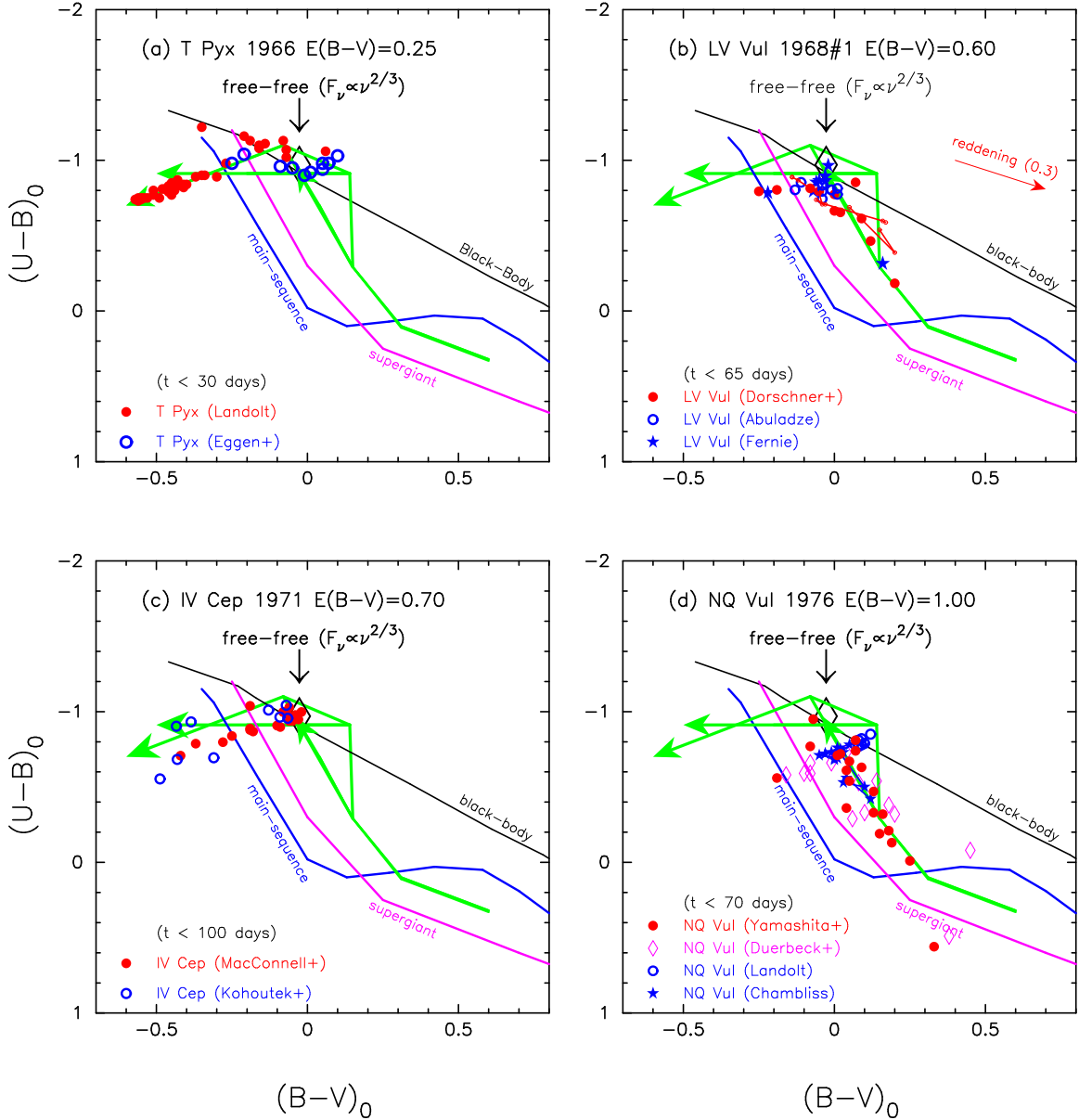}
%\plotone{color_color_diagram_t_pyx_lv_vul_iv_cep_nq_vul.epsi}
%\plotfiddle{evolution1.ps}{5.0cm}{270}{0.4}{0.4}{-170}{220}
\caption{
Same as Figure
\ref{color_color_diagram_templ_rs_oph_v446_her_v533_her}, but
%%%\ref{color_color_diagram_pw_vul_v1500_cyg_v1668_cyg_1974_cyg}, but
for (a) T~Pyx 1966, (b) LV~Vul 1968\#1, where we add loci of V1500~Cyg 
(thin red solid lines) only for the very early phase, 
(c) IV~Cep 1971, and (d) NQ~Vul 1976.  
\label{color_color_diagram_t_pyx_lv_vul_iv_cep_nq_vul}}
\end{figure*}

\begin{figure*}
%%\begin{figure}
%%\epsscale{0.75}
\epsscale{1.0}
%%\epsscale{1.15}
\plotone{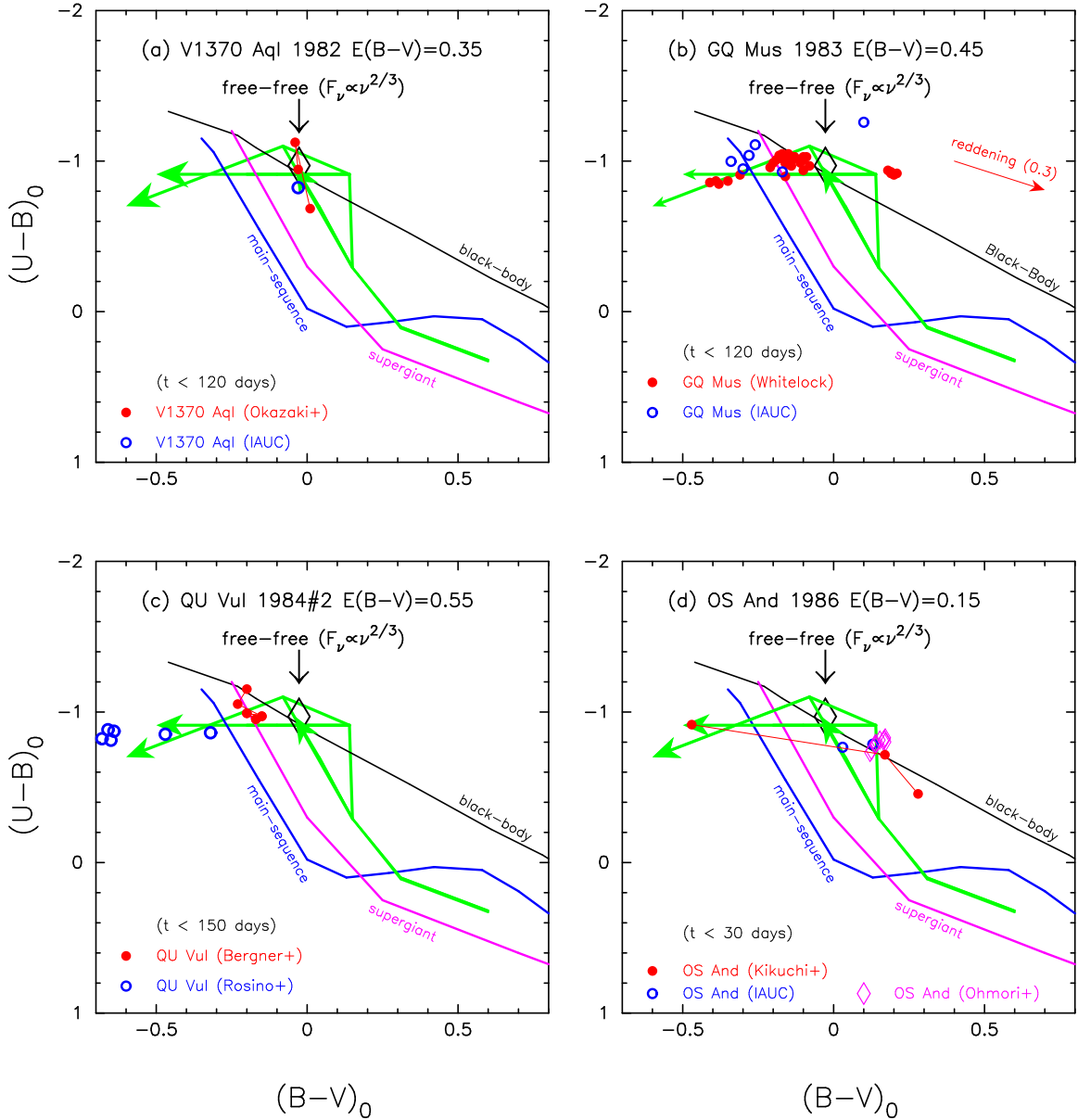}
%\plotone{color_color_diagram_v1370_aql_gq_mus_qu_vul_os_and.epsi}
%-->\plotone{color_color_diagram_v1370_aql_gq_mus_qu_vul_os_and.epsi}
%\plotfiddle{evolution1.ps}{5.0cm}{270}{0.4}{0.4}{-170}{220}
\caption{
Same as Figure
\ref{color_color_diagram_templ_rs_oph_v446_her_v533_her}, but
for (a) V1370~Aql 1982,  (b) GQ~Mus 1983, (c) QU~Vul 1984\#2,
and (d) OS~And 1986.
\label{color_color_diagram_v1370_aql_gq_mus_qu_vul_os_and}}
\end{figure*}

\begin{figure*}
%%\begin{figure}
%\epsscale{0.75}
\epsscale{1.0}
%%\epsscale{1.15}
\plotone{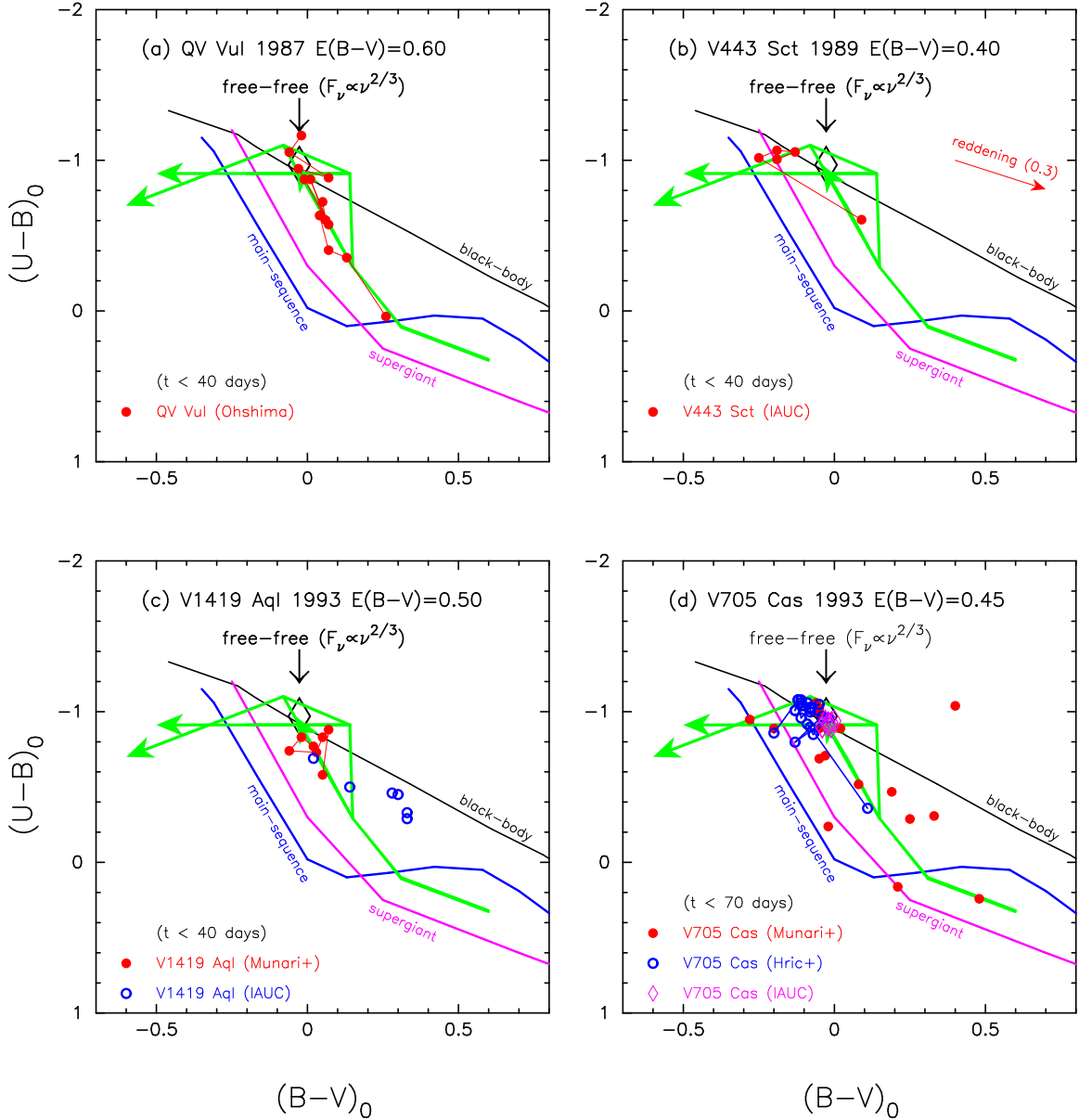}
%\plotone{color_color_diagram_qv_vul_v443_sct_v1419_aql_v705_cas.epsi}
%\plotfiddle{evolution1.ps}{5.0cm}{270}{0.4}{0.4}{-170}{220}
\caption{
Same as Figure
\ref{color_color_diagram_templ_rs_oph_v446_her_v533_her}, but
for (a) QV~Vul 1987,  (b) V443~Sct 1989, (c) V1419~Aql 1993, 
and (d) V705~Cas 1993.
\label{color_color_diagram_qv_vul_v443_sct_v1419_aql_v705_cas}}
%%\end{figure}
\end{figure*}

\begin{figure*}
%%\epsscale{0.75}
\epsscale{1.0}
%%\epsscale{1.15}
\plotone{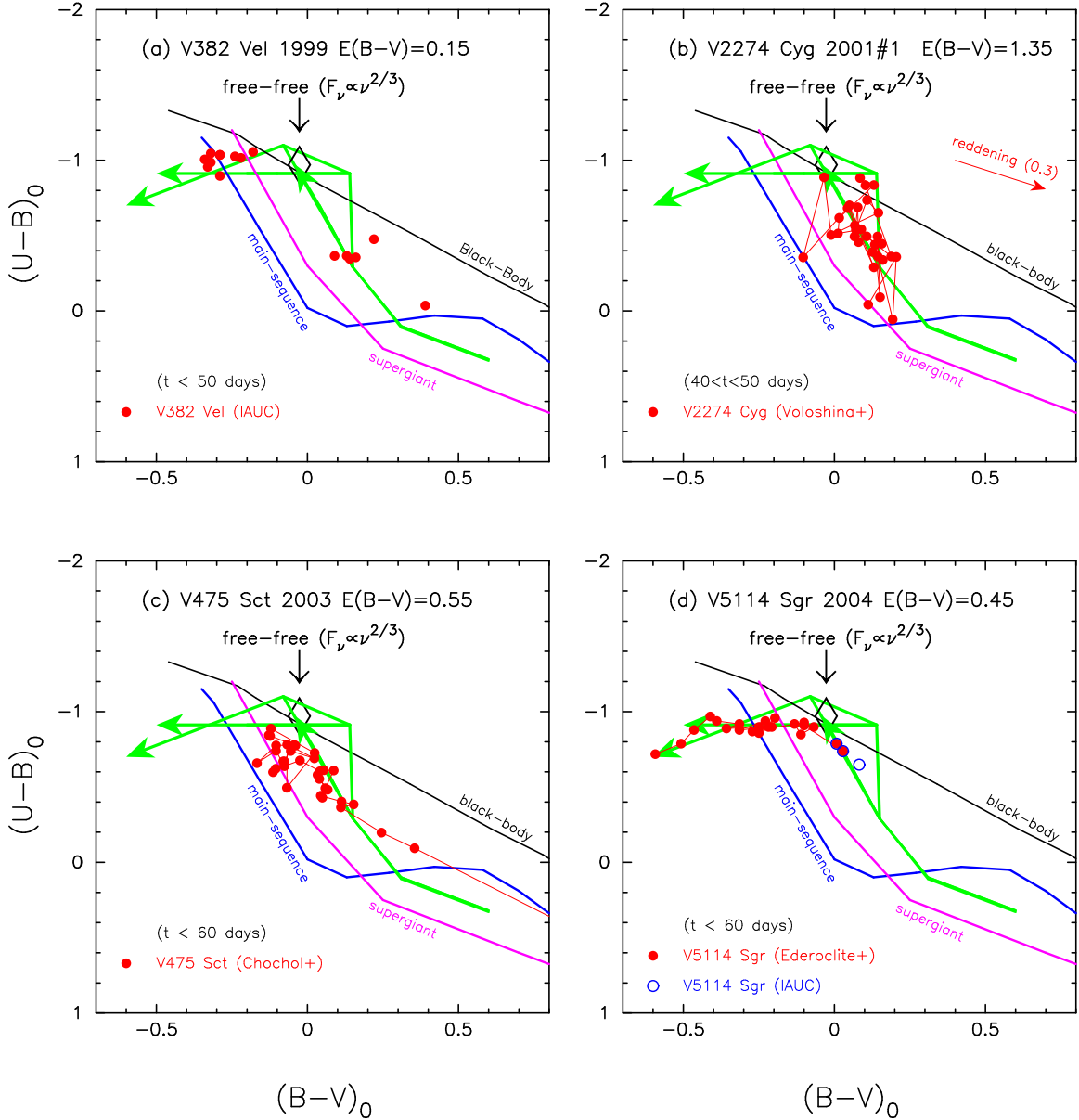}
%\plotone{color_color_diagram_v382_vel_v2274_cyg_v475_sct_v5114_sgr.epsi}
%\plotfiddle{evolution1.ps}{5.0cm}{270}{0.4}{0.4}{-170}{220}
\caption{
Same as Figure
\ref{color_color_diagram_templ_rs_oph_v446_her_v533_her}, but
for (a) V382~Vel 1999,  (b) V2274~Cyg 2001\#1, (c) V475~Sct 2003,
and (d) V5114~Sgr 2004.
\label{color_color_diagram_v382_vel_v2274_cyg_v475_sct_v5114_sgr}}
\end{figure*}

\begin{figure*}
%%\epsscale{0.75}
%%\epsscale{0.8}
\epsscale{1.0}
%%\epsscale{1.15}
\plotone{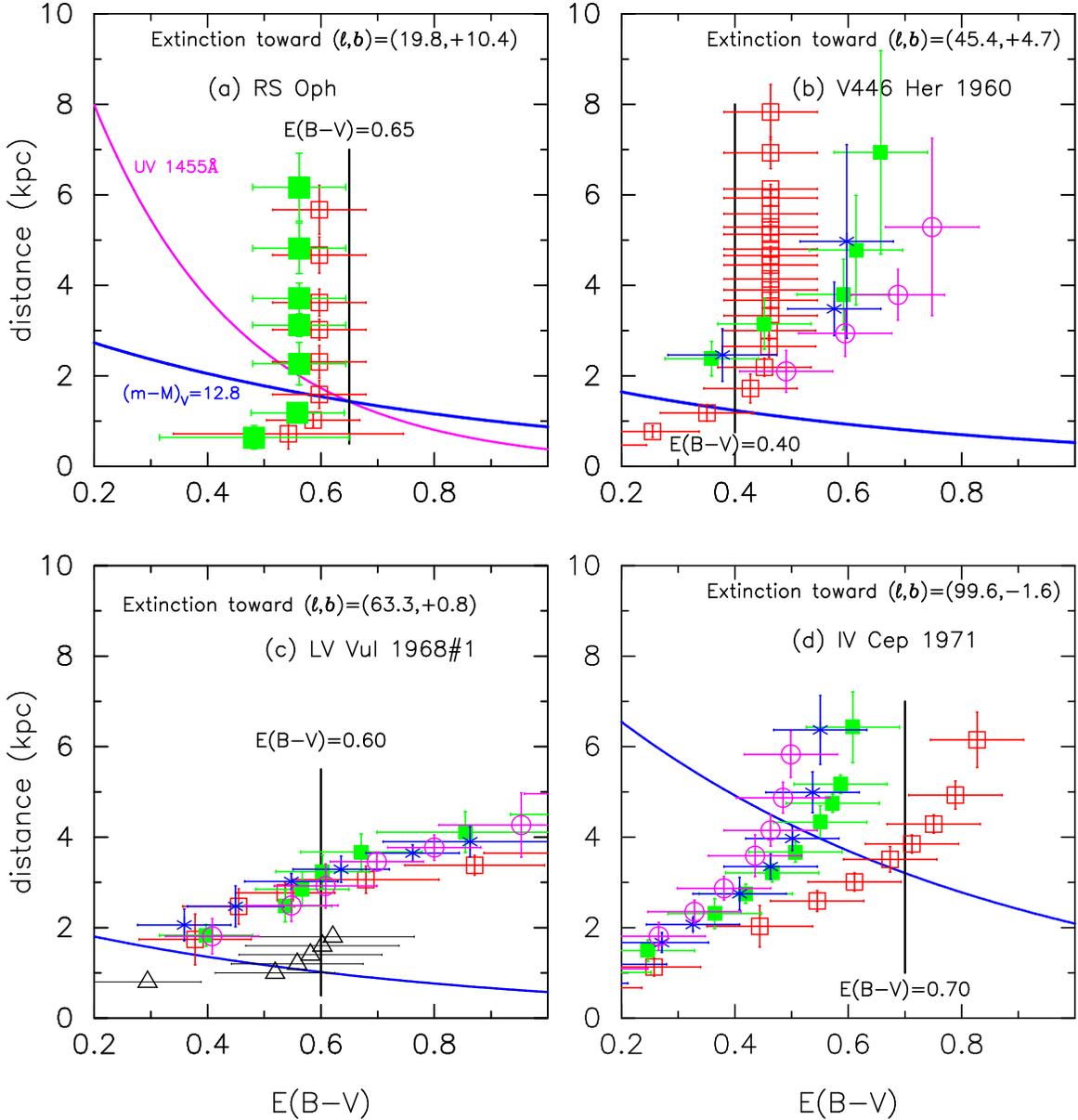}
%\plotone{distance_reddening_rs_oph_v446_her_lv_vul_iv_cep.epsi}
%\plotfiddle{evolution1.ps}{5.0cm}{270}{0.4}{0.4}{-170}{220}
\caption{
Same as Figure \ref{distance_reddening_fh_ser_only}, but for (a) RS~Oph,
(b) V446~Her, (c) LV~Vul, and (d)IV~Cep.
Blue solid lines indicate relations (a) $(m-M)_V=
-5 + 3.1 E(B-V) + 5 \log d = 12.8$, (b) $(m-M)_V=11.7$,(c) $(m-M)_V=11.9$,
where black open triangles indicate the galactic extinction model
of \citet{hak97}, and (d) $(m-M)_V=14.7$.  Magenta solid line in (a) 
denotes the distance-reddening relation of UV~1455\AA\  fitting 
calculated in the Appendix (Figure 
\ref{rs_oph_v446_her_v1500_cyg_v1668_cyg_v382_vel_v_color_logscale}).
\label{distance_reddening_rs_oph_v446_her_lv_vul_iv_cep}}
\end{figure*}

\begin{figure*}
%%\begin{figure}
%%\epsscale{0.75}
\epsscale{1.0}
%%\epsscale{1.15}
\plotone{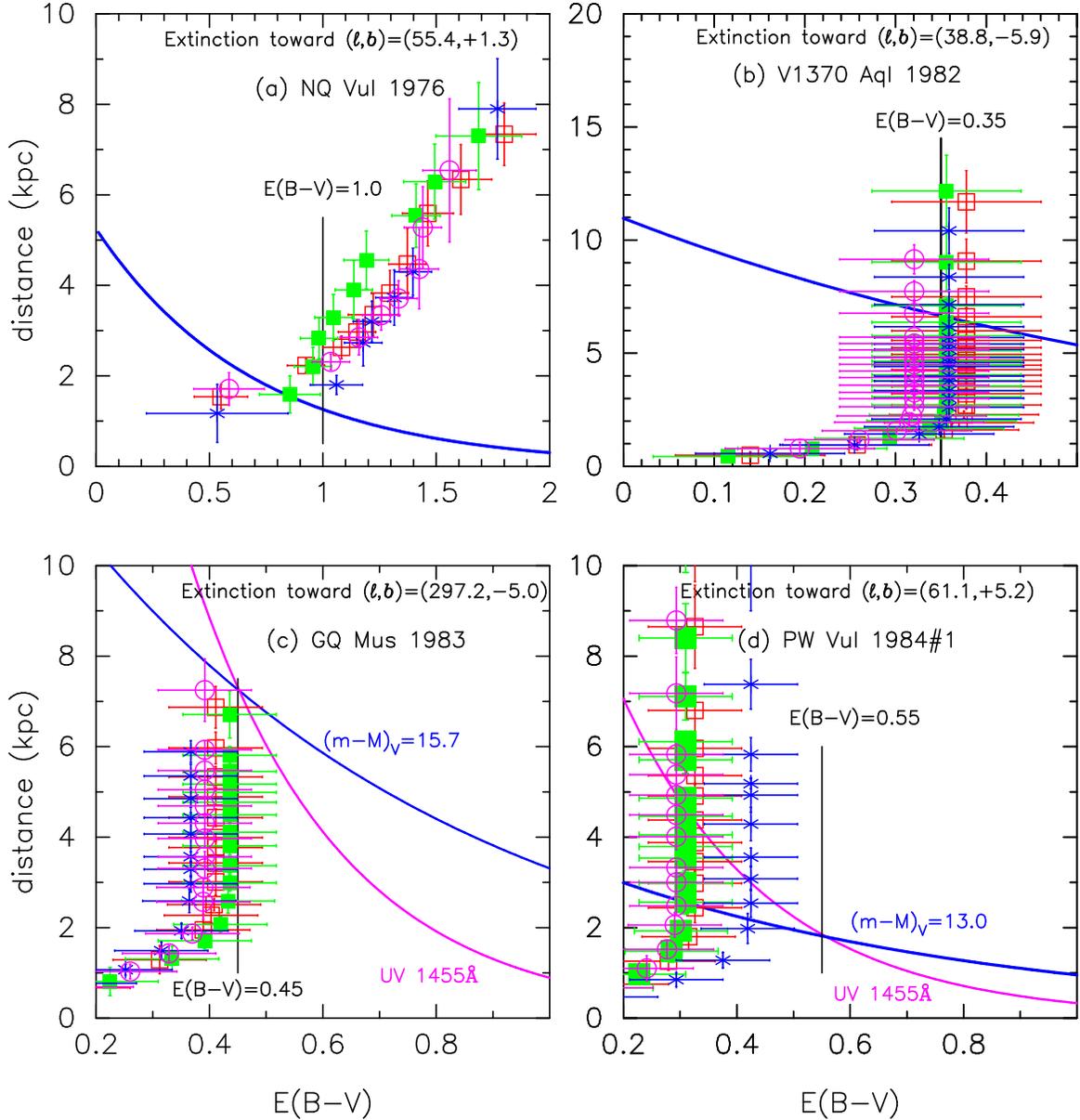}
%\plotone{distance_reddening_nq_vul_v1370_aql_gq_mus_pw_vul.epsi}
%\plotfiddle{evolution1.ps}{5.0cm}{270}{0.4}{0.4}{-170}{220}
\caption{
Same as Figure 
\ref{distance_reddening_rs_oph_v446_her_lv_vul_iv_cep},
but for (a) NQ~Vul, (b) V1370~Aql, (c) GQ~Mus, and (d) PW~Vul.
Blue solid lines indicate relations (a) $(m-M)_V=13.6$,
(b) $(m-M)_V=15.2$, (c) $(m-M)_V=15.7$, and (d) $(m-M)_V=13.0$. 
Magenta solid lines in (c) and (d) 
denote distance-reddening relations of UV~1455\AA\  fitting 
for GQ~Mus and PW~Vul, respectively, calculated in the Appendix (Figure 
\ref{qu_vul_pw_vul_gq_mus_v_bv_ub_color_logscale}).
\label{distance_reddening_nq_vul_v1370_aql_gq_mus_pw_vul}}
\end{figure*}

\begin{figure*}
%%\epsscale{0.75}
\epsscale{1.0}
%%\epsscale{1.15}
\plotone{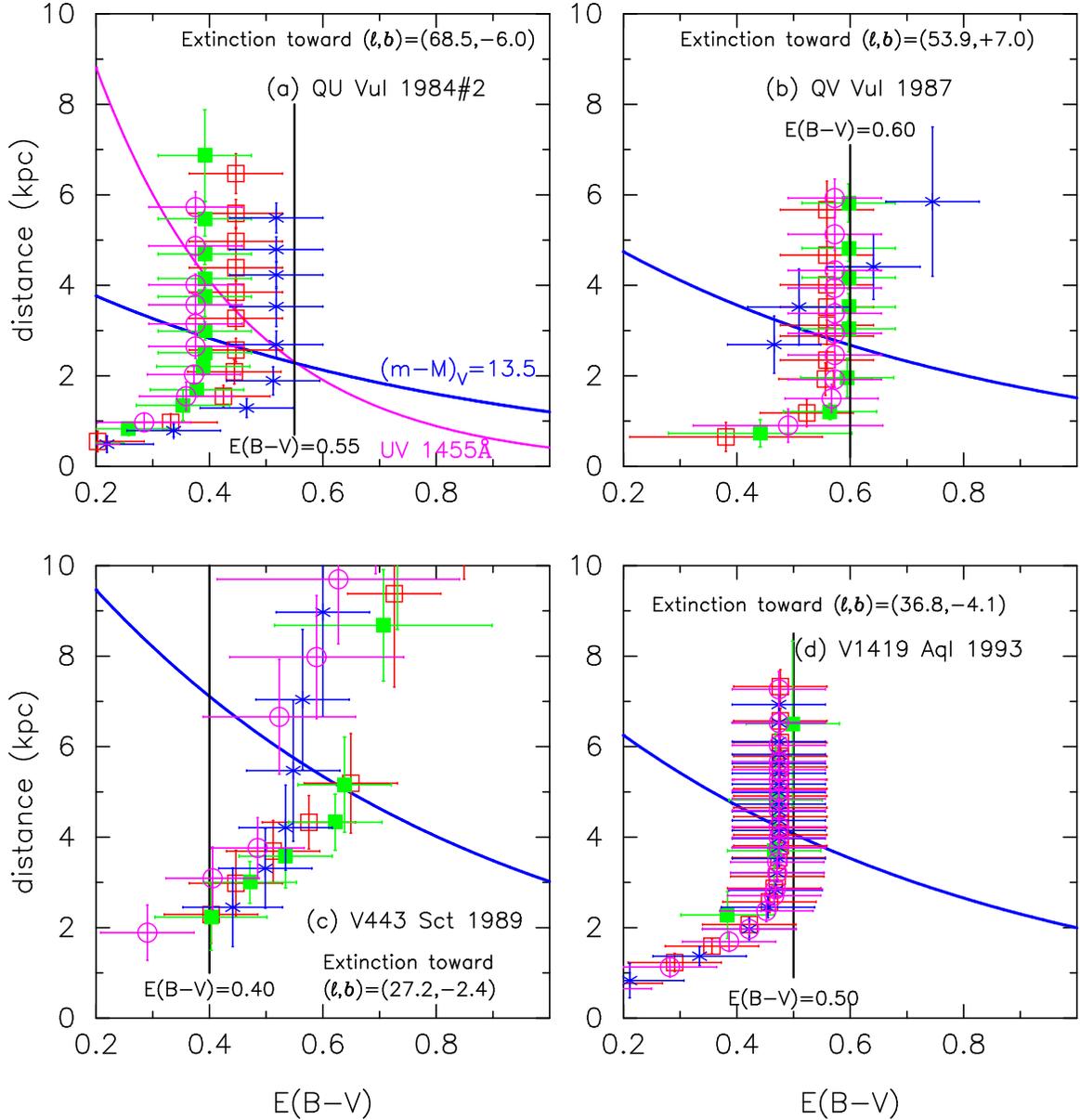}
%\plotone{distance_reddening_qu_vul_qv_vul_v443_sct_v1419_aql.epsi}
%\plotfiddle{evolution1.ps}{5.0cm}{270}{0.4}{0.4}{-170}{220}
\caption{
Same as Figure \ref{distance_reddening_rs_oph_v446_her_lv_vul_iv_cep}, 
but for (a) QU~Vul, (b) QV~Vul, (c) V443~Sct, and (d) V1419~Aql.
Blue solid lines indicate relations 
(a) $(m-M)_V=13.5$, (b) $(m-M)_V=14.0$, (c) $(m-M)_V=15.5$, 
and (d) $(m-M)_V=14.6$.  Magenta solid line in (a) 
denotes distance-reddening relation of UV~1455\AA\  fitting 
for QU~Vul calculated in the Appendix (Figure 
\ref{qu_vul_pw_vul_gq_mus_v_bv_ub_color_logscale}).
\label{distance_reddening_qu_vul_qv_vul_v443_sct_v1419_aql}}
\end{figure*}

\begin{figure*}
%%\epsscale{0.75}
\epsscale{1.0}
%%\epsscale{1.15}
\plotone{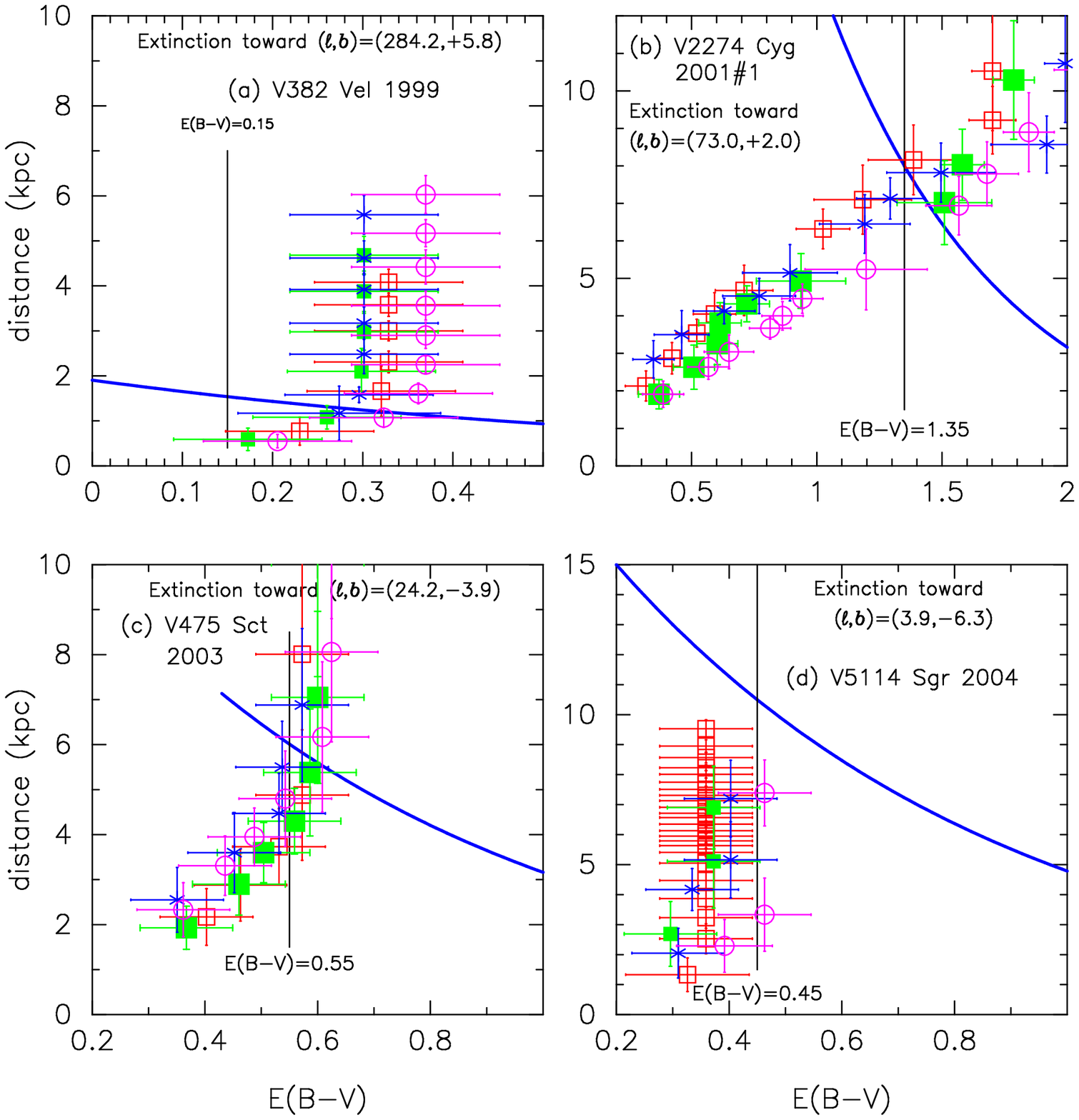}
%\plotone{distance_reddening_v382_vel_v2274_cyg_v475_sct_v5114_sgr.epsi}
%\plotfiddle{evolution1.ps}{5.0cm}{270}{0.4}{0.4}{-170}{220}
\caption{
Same as Figure \ref{distance_reddening_rs_oph_v446_her_lv_vul_iv_cep}, 
but for (a) V382~Vel, (b) V2274~Cyg, (c) V475~Sct, and (d) V5114~Sgr.
Blue solid lines indicate distance-reddening relations
(a) $(m-M)_V=11.4$, (b) $(m-M)_V=18.7$, (c) $(m-M)_V=15.6$,
and (d) $(m-M)_V=16.5$. 
\label{distance_reddening_v382_vel_v2274_cyg_v475_sct_v5114_sgr}}
\end{figure*}

In this section, we determine the color excesses of various novae using
our general color-evolution tracks found in this work.  We collected 
as many novae from the literature as possible that have sufficient data 
points (usually more than 10).  We assume that all the novae follow 
the color-color evolution tracks in Figure
\ref{color_color_diagram_templ_rs_oph_v446_her_v533_her}(a).
(Compare specifically one of the tracks in Figure
\ref{color_color_diagram_four_typical_example}.) 
To obtain $E(B-V)$, we changed $E(B-V)$ by steps of 0.05 to fit the observed 
track of a target nova with our general tracks 
in Figure \ref{color_color_diagram_four_typical_example}.
Thus, the possible error of the resultant $E(B-V)$ is $\pm 0.05$.
In what follows, we will show our target 19 novae
in the order of discovery.  Their color-color diagrams are summarized
in Figures
\ref{color_color_diagram_templ_rs_oph_v446_her_v533_her} ---
\ref{color_color_diagram_v382_vel_v2274_cyg_v475_sct_v5114_sgr}
and the distance-reddening relations toward the target novae 
are shown in Figures
\ref{distance_reddening_rs_oph_v446_her_lv_vul_iv_cep} ---
\ref{distance_reddening_v382_vel_v2274_cyg_v475_sct_v5114_sgr}.

%Table 2

\begin{deluxetable}{llllll}
\tabletypesize{\scriptsize}
\tablecaption{Extinctions and Distances of Selected Novae
\label{extinction_from_two_color_diagram}}
\tablewidth{0pt}
\tablehead{
\colhead{Object} & \colhead{Year} & \colhead{Present\tablenotemark{a}} 
& \colhead{V\&Y(1987)\tablenotemark{b}} & 
\colhead{M(1988)\tablenotemark{c}} & \colhead{Present} \\
  &  & $E(B-V)$ &  $E(B-V)$ &  $E(B-V)$ & $(m-M)_V$
} 
\startdata
OS~And & 1986 & 0.15 & $(0.21/0.27)$\tablenotemark{d} & (0.34)\tablenotemark{e} & 14.7 \\
V1370~Aql & 1982 & 0.35 & \nodata & \nodata & 15.2 \\
V1419~Aql & 1993 & 0.50 & $(0.5/0.6)$ & (0.62) & 14.6 \\
%%%V1493~Aql & 1999\#1 & 1.10 & \nodata & \nodata & 17.7 \\
%%%V1494~Aql & 1999\#2 & 0.50 & \nodata & \nodata & 13.1 \\
V705~Cas & 1993 & 0.45 & $(0.33/0.36)$ & (0.43) & 13.4 \\
V723~Cas & 1995 & 0.35 & $(0.48/-)$ & (0.57) & 14.0 \\
%%%V1065~Cen & 2007 & 0.50 & $(0.29/0.45)$ & \nodata & 15.5 \\
IV~Cep & 1971 & 0.70 & 0.66 & 0.75 & 14.7 \\
V1500~Cyg & 1975 & 0.45 & 0.37 & 0.47 & 12.3 \\
V1668~Cyg & 1978 & 0.35 & 0.31 & 0.36 & 14.25 \\
V1974~Cyg & 1992 & 0.30 & $(0.17/0.27)$ & (0.35) & 12.2 \\
V2274~Cyg & 2001\#1 & 1.35 & \nodata & \nodata & 18.7 \\
%%%V2275~Cyg & 2001\#2 & 1.05 & $(0.9/1.0)$ & \nodata & 16.3 \\
%%%V2362~Cyg & 2006 & 0.60 & $(-/0.56)$ & \nodata & 15.9 \\
%%%V2467~Cyg & 2007 & 1.40 & $(-/1.16)$ & \nodata & 16.2 \\
%%%V2468~Cyg & 2008 & 0.80 & \nodata & \nodata & 15.6 \\
%%%V2491~Cyg & 2008 & 0.23 & $(0.23/0.22)$ & \nodata & 16.5 \\
HR~Del & 1967 & 0.15 & $(0.24/-)$ & (0.24) & 10.4 \\
V446~Her & 1960 & 0.40 & 0.27 & 0.46 & 11.7 \\
V533~Her & 1963 & 0.05 & 0.0 & 0.05 & 11.1 \\
GQ~Mus & 1983 & 0.45 & $(-/0.37)$ & 0.42 & 15.7 \\
RS~Oph  & 1958 & 0.65 & (0.67/0.65) & (0.75) & 12.8 \\
%%%V2615~Oph  & 2007 & 0.95 & (0.89/0.91) & \nodata & 16.3 \\
T~Pyx  & 1966 & 0.25 & (0.16/0.20) & (0.33) & 13.8 \\
%%%V1280~Sco & 2007\#1 & 0.35 & \nodata & \nodata & 11.0 \\
V443~Sct & 1989 & 0.40 & $(-/0.25)$ & \nodata & 15.5 \\
V475~Sct & 2003 & 0.55 & (0.71/0.47) & (0.72) & 15.6 \\
%%%V496~Sct & 2009 & 0.55 & $(0.57/-)$ & \nodata & 13.6 \\
FH~Ser & 1970 & 0.60 & 0.61 & 0.79 & 11.7 \\
V5114~Sgr & 2004 & 0.45 & $(0.43/0.40)$ & (0.55) & 16.5 \\
V5558~Sgr & 2007 & 0.70 & $(0.63/-)$ & (0.83) & 13.9 \\
V382~Vel & 1999 & 0.15 & $(0.08/-)$ & \nodata & 11.4 \\
LV~Vul & 1968\#1 & 0.60 & 0.58 & 0.70 & 11.9 \\
NQ~Vul & 1976 & 1.00 & 1.0 & 1.2 & 13.6 \\
PU~Vul & 1979 & 0.30 & (0.52/0.29) & (0.55) & 14.3 \\
PW~Vul & 1984\#1 & 0.55 & 0.39 & 0.51 & 13.0 \\
QU~Vul & 1984\#2 & 0.55 & $(-/0.42)$ & \nodata & 13.5 \\
QV~Vul & 1987 & 0.60 & $(1.0/0.70)$ & (0.75) & 14.0 
%%%V458~Vul & 2007\#1 & 0.65 & $(-/0.55)$ & \nodata & 15.6 
\enddata
\tablenotetext{a}{Estimated from color-color diagram fit.} 
\tablenotetext{b}{Taken from \citet{van87}.} 
\tablenotetext{c}{Taken from \citet{mir88}.} 
\tablenotetext{d}{Parentheses indicate values of $E(B-V)$ 
estimated by other authors from intrinsic colors
at maximum/$t_2$-time proposed by \citet{van87}. See Section 
\ref{van_den_bergh} for details.} 
\tablenotetext{e}{Parentheses indicate values estimated 
by other authors using the method proposed by \citet{mir88}.
See Section \ref{miroshnichenko} for details.} 
\end{deluxetable}

%Fig.29
%\placefigure{color_color_diagram_templ_rs_oph_v446_her_v533_her}

%%\begin{figure}
%%\end{figure}

%Fig.30
%\placefigure{color_color_diagram_t_pyx_lv_vul_iv_cep_nq_vul}

%%\begin{figure}
%%\end{figure}

%Fig.31
%\placefigure{color_color_diagram_v1370_aql_gq_mus_qu_vul_os_and}

%%\end{figure}

%Fig.32
%\placefigure{color_color_diagram_qv_vul_v443_sct_v1419_aql_v705_cas}

%Fig.33
%\placefigure{color_color_diagram_v382_vel_v2274_cyg_v475_sct_v5114_sgr}

%%\begin{figure}
%%\end{figure}

%Fig.34
%\placefigure{distance_reddening_rs_oph_v446_her_lv_vul_iv_cep}

%%\begin{figure}
%%\end{figure}

%Fig.35
%\placefigure{distance_reddening_nq_vul_v1370_aql_gq_mus_pw_vul}

%%\end{figure}

%Fig.36
%\placefigure{distance_reddening_qu_vul_qv_vul_v443_sct_v1419_aql}

%%\begin{figure}
%%\end{figure}

%Fig.37
%\placefigure{distance_reddening_v382_vel_v2274_cyg_v475_sct_v5114_sgr}

%%\begin{figure}
%%\end{figure}

\subsection{RS~Oph 1958}
\label{rs_oph_color}
RS~Oph is a recurrent nova with six recorded outbursts, 
in 1898, 1933, 1958, 1967, 1985, and 2006.
We plot the $V$ light curve, $(B-V)_0$ and $(U-B)_0$ color 
evolution of RS~Oph outbursts in the Appendix (Figure 
\ref{rs_oph_v446_her_v1500_cyg_v1668_cyg_v382_vel_v_color_logscale})
and the color-color evolution of the 1958 outburst
in Figure \ref{color_color_diagram_templ_rs_oph_v446_her_v533_her}(b),
where the $UBV$ data are taken from \citet{con70}.
We obtained $E(B-V)=0.65\pm0.05$ by fitting.
This value is roughly consistent with those obtained by \citet{sni87}, i.e.,
$E(B-V)=0.73\pm 0.06$ from the \ion{He}{2} line ratio of 
1640\AA\  and 3203\AA, and $E(B-V)=0.73\pm 0.10$ from the 2715\AA\ 
interstellar dust absorption feature.

We examine the distance-reddening relation using the
data given by \citet{mar06} as shown in
Figure \ref{distance_reddening_rs_oph_v446_her_lv_vul_iv_cep}(a).
The galactic coordinates of RS~Oph are
$(l, b)=(19\fdg7995,+10\fdg3721)$.
In the figure, we plot two relations in the direction
close to RS~Oph.  These two different directions are
$(l, b)=(19\fdg75,10\fdg00)$ (red open squares) and 
$(20\fdg0,10\fdg00)$ (green filled squares), both with error bars.
Here we also plot the distance modulus of RS~Oph, $(m-M)_V = 12.8$ 
(blue solid line), which was calculated using
the time-stretching method in the Appendix (Figure
\ref{rs_oph_v446_her_v1500_cyg_v1668_cyg_v382_vel_v_color_logscale}).
We also obtained another distance-reddening relation by UV~1455\AA\  
fitting (magenta solid line) in Figure 
\ref{distance_reddening_rs_oph_v446_her_lv_vul_iv_cep}(a).

In Figure 
\ref{distance_reddening_rs_oph_v446_her_lv_vul_iv_cep}(a),
four trends, $E(B-V)=0.65$ (vertical black solid line),
$(m-M)_V = 12.8$ (blue solid line), UV~1455\AA\  fitting (magenta solid line),
and distance-reddening relation of \citet{mar06},
consistently cross at $E(B-V)\sim0.65$ and $d\sim1.4$~kpc.
If we adopt $d\sim1.4$~kpc, the galactic height 
of RS~Oph is $z\sim260$~pc, much above the scale height of 
the galactic matter distribution \citep[$\sim 125$~pc,][]{mar06}.
The NASA/IPAC galactic dust absorption map gives $E(B-V)=0.64 \pm 0.03$
in the direction toward RS~Oph, which is also consistent with our value
of $E(B-V)=0.65\pm0.05$ if circumstellar absorption is negligible.

\subsection{V446~Her 1960}
\label{v446_her_color}
V446~Her was discovered by Olaf Hassel at fifth magnitude
on UT 1960 March 7, a few days past maximum \citep{cra60, may63}.  
We plot the $V$ light curve, $(B-V)_0$ and $(U-B)_0$ color 
evolution of V446~Her in the Appendix (Figure 
\ref{rs_oph_v446_her_v1500_cyg_v1668_cyg_v382_vel_v_color_logscale}).
Its maximum magnitude was $m_V\sim2.75$ \citep{may63,coh85}.
Then it rapidly and smoothly declined with $t_2=5$~days \citep[e.g.,][]{coh85}
and $t_3=16$~days \citep[e.g.,][]{due81}.  We plot the color-color 
evolution of V446~Her in Figure 
\ref{color_color_diagram_templ_rs_oph_v446_her_v533_her}(c),
where the $UBV$ data are taken mainly from \citet{ross60} and \citet{bro61}.
These observations started after the nova had already decayed
to $m_V=5.5$, so we missed the early evolution of this nova.
We also add three data points (magenta open circles) from IAU Circular 
No. 1730, which start at about 2.5 mag below the maximum.
The observed color evolution started from point 4' (or point F)
and reached point 5', as shown in Figure 
\ref{color_color_diagram_templ_rs_oph_v446_her_v533_her}(c).
We obtained a reddening of $E(B-V)=0.40\pm0.05$ 
mainly from the data of \citet{ross60}, 
because their data follow the track from point 4' to 5'.
The color evolution is very similar to that of PW~Vul.

The reddening toward V446~Her was determined
to be $E(B-V)=0.45\pm0.1$ from the galactic absorption \citep{bro61},
$E(B-V)=A_V/3.1= (1.7\pm0.5)/3.1= 0.55\pm0.15$
from the galactic absorption \citep{due81}, 
$E(B-V)=A_V/3.1=(1.44\pm0.12)/3.1=0.46\pm0.04$
from the value at the stabilization stage \citep{mir88}, 
$E(B-V)=A_V/3.1=1.12/3.2=0.35$ from the 2200\AA\  feature \citep{gil94},
$E(B-V)=0.25$ from the depth of the interstellar
bump at 2175\AA\  \citep{sel04}.
These reddening estimates range from 0.25 to 0.55 and we
obtained a simple average of $E(B-V)=0.41\pm0.15$, which is
consistent with our value.

We further analyzed the distance-reddening relation based on the
data given by \citet{mar06}, as shown in Figure 
\ref{distance_reddening_rs_oph_v446_her_lv_vul_iv_cep}(b).
The galactic coordinates of V446~Her are $(l, b)= (45\fdg4092,+4\fdg7075)$.
In the figure, we plot four relations in directions
close to V446~Her: $(l, b)=(45\fdg25,4\fdg50)$ (red open squares), 
$(45\fdg50,4\fdg50)$ (green filled squares),
$(45\fdg25,4\fdg75)$  (blue asterisks), and $(45\fdg50,4\fdg75)$
(magenta open circles), all with error bars.
The last magenta open circles correspond to the closest direction.
We also plot the distance modulus of V446~Her,
$(m-M)_V = 11.7$ (blue solid line),
which was calculated using the time-stretching method in the Appendix
(Figure
\ref{rs_oph_v446_her_v1500_cyg_v1668_cyg_v382_vel_v_color_logscale}).
Three trends, $E(B-V)=0.40$ (vertical black solid line),
$(m-M)_V = 11.7$ (blue solid line),
and the distance-reddening relation of \citet{mar06},
consistently cross at $E(B-V)\sim0.40$ and $d\sim1.2$~kpc.

\subsection{V533~Her 1963}
\label{v533_her_color}
V533~Her was discovered by Leslie Peltier at fourth magnitude
on UT 1963 February 6 \citep{may63, gen63}.  We plot the $V$ light curve, 
and the $(B-V)_0$ and $(U-B)_0$ color evolution of V533~Her in the Appendix
(Figure
\ref{pw_vul_v1668_cyg_v1500_cyg_v1974_cyg_v_bv_ub_color_curve_logscale_no2}).
It reached third magnitude at maximum on UT January 31 \citep{may63, gen63}.   
Then it gradually declined with $t_2=22$~days and $t_3=46$~days
\citep[e.g.,][]{dow00}.  We plot the color-color evolution of V533~Her
in Figure 
\ref{color_color_diagram_templ_rs_oph_v446_her_v533_her}(d),
where the $UBV$ data are taken from \citet{gen63}, \citet{chi64} and
\citet{she64}.
These three observations started after the nova had already decayed
to $m_V\sim4.8$, so we missed its early evolution.
The observed color evolution started at point 4' 
(near point F) and reached point 5', as shown in Figure 
\ref{color_color_diagram_templ_rs_oph_v446_her_v533_her}(d).
We obtained a reddening value of $E(B-V)=0.05\pm0.05$,
mainly from the early phase data of
\citet{she64}, because these data (magenta open diamonds)
follow the track from point 4' to 5'.

The reddening toward V533~Her was determined to be $E(B-V)\sim0.0$
\citep{ver87}, $E(B-V)=A_V/3.1\sim0.0$ \citep{gil94}, 
both from the 2200\AA\  feature, and $E(B-V)=A_V/3.1=0.16/3.1\sim0.05$ 
from the value at the stabilization stage \citep{mir88}.  
The NASA/IPAC galactic dust absorption map
gives $E(B-V)=0.038 \pm 0.002$ in the direction toward V533~Her,
whose galactic coordinates are $(l, b)= (69\fdg1887, +24\fdg2733)$.
Our value of $E(B-V)=0.05\pm0.05$ is consistent with these estimates.

\subsection{T~Pyx 1966}
\label{t_pyx_color}
T~Pyx is a recurrent nova with known outbursts in 1890,
1902, 1920, 1944, 1966, and 2011.  We plot the color evolution
of the 1966 outburst in Figure 
\ref{color_color_diagram_t_pyx_lv_vul_iv_cep_nq_vul}(a),
the $UBV$ data of which are taken from \citet{lan70} and \citet{egg67}.
We obtained $E(B-V)=0.25\pm0.05$ by fitting.  \citet{gil07} also
obtained $E(B-V)=0.25\pm 0.02$ from the 2175\AA\  interstellar
absorption feature in the {\it IUE} spectra of T~Pyx, which is consistent
with our estimate.  

The galactic coordinates of T~Pyx are $(l, b)=(257\fdg2072,+9\fdg7067)$. 
If we assume a distance of $\sim 4.8$~kpc \citep{sok13},
its galactic height is $z\sim800$~pc, which is much above the galactic matter
distribution of $\sim125$~pc.  The NASA/IPAC galactic dust absorption map
gives $E(B-V)=0.24 \pm 0.01$ in the direction toward T~Pyx.
This value is consistent with our estimate of $E(B-V)=0.25\pm0.05$
but inconsistent with the value of $E(B-V)=0.5\pm 0.1$ recently estimated 
by \citet{sho11} from the diffuse interstellar band (DIB)
features of the 2011 outburst.  
We think that there is a large scatter in the $E(B-V)$ versus
$W_\lambda$ (5780\AA) relation near $W_\lambda =300$m\AA\  that Shore
et al. used, where $W_\lambda$(5780\AA) is the equivalent width
of the 5780\AA\  DIB in units of m\AA\   \citep[see Figure 4 of][]{fri11}.
Therefore, we consider that this empirical relation for the 5780\AA\  DIB  
does not accurately determine the individual extinctions
at least for a range of $E(B-V)=0.2$--$0.8$.

\subsection{LV~Vul 1968 No.1}
\label{lv_vul_color}
LV~Vul was discovered by \citet{alc68} on UT 1968 April 15.  We plot 
the $V$ light curve and the $(B-V)_0$ and $(U-B)_0$ colors of LV~Vul in 
the Appendix (Figure
\ref{v5114_sgr_v1500_cyg_v1494_aql_iv_cep_lv_vul_v_color_logscale}).
It reached its optical maximum at $m_V=4.7$ on April 17.
LV~Vul has $t_2=20.2$ days and $t_3=37$ days \citep{tem72}
so it belongs to the class of fast novae.
\citet{and86} took a spectrum at the pre-maximum phase that
showed an F-type spectrum.  Therefore,
we expect that the early color evolution of LV~Vul follows
the nova-giant sequence in the color-color diagram.
Figure \ref{color_color_diagram_t_pyx_lv_vul_iv_cep_nq_vul}(b)
shows the color-color evolution,
where the $UBV$ data are taken from \citet{abu69}, \citet{dor69},
and \citet{fer69}.  
The color-color evolution of LV~Vul resembles those of V1500~Cyg and
V1668~Cyg.  We obtained $E(B-V)=0.60\pm0.05$ by fitting.

\citet{fer69} determined the reddening toward LV~Vul to be 
$E(B-V)=0.6\pm 0.2$ from the color excesses of 14 B stars near 
the line of sight, which is consistent with our value.
\citet{tem72} obtained $E(B-V)=0.55$ from the color at optical maximum;
i.e., $E(B-V)=(B-V)_{\rm max} - (B-V)_{0, \rm max}= 0.9 - 0.35 = 0.55$.
He adopted $(B-V)_{0, \rm max}= + 0.35$ \citep{sch57} instead of 
$(B-V)_{0, \rm max}=+0.23$ \citep{van87}.

Using $E(B-V)=0.60$ together with the distance modulus 
$(m-M)_{V, \rm LV~Vul}=11.9$ calculated by the time-stretching
method in the Appendix (Equation 
(\ref{v5114_sgr_v1500_cyg_v1494_aql_iv_cep_lv_vul_distance})),
we obtain a distance of $d=1.0$~kpc.
This is consistent with the value of $d=0.92\pm 0.08$~kpc 
obtained by \citet{sla95} by the expansion parallax method.
We compare our results with the distance-reddening relation proposed 
by \citet{mar06} as shown in Figure
\ref{distance_reddening_rs_oph_v446_her_lv_vul_iv_cep}(c).
Four distance-reddening relations are plotted: 
toward $(l,b)=(63\fdg25,0\fdg75)$ (red open squares), 
toward $(l,b)=(63\fdg50,0\fdg75)$ (green filled squares), 
toward $(l,b)=(63\fdg25,1\fdg00)$ (blue asterisks),
toward $(l,b)=(63\fdg50,1\fdg00)$ (magenta open circles).
The closest direction is that denoted by red open squares because 
the galactic coordinates of LV~Vul are $(l,b)=(63\fdg3030,+0\fdg8464)$.
Marshall et al.'s relation gives $d\sim3.0$~kpc at $E(B-V)=0.60$.
This is not consistent with our estimate of $d=1.0$~kpc.
On the other hand, \citet{dow00} obtained 
$E(B-V)=0.56\pm0.14 = A_V/3.1=(1.75\pm0.42)/3.1$ using
the galactic extinction model of \citet{hak97}.  Thus, we 
plot the distance-reddening relation calculated using
the model of \citet{hak97}, as shown in Figure
\ref{distance_reddening_rs_oph_v446_her_lv_vul_iv_cep}(c)
(black open triangles).  This model is consistent with our estimate.

%%% 063.3030 +00.8464

\subsection{IV~Cep 1971}
\label{iv_cep_color}
IV~Cep was discovered by Kuwano on UT 1971 July 10 with $m_v = 8.0$
\citep{kuw71} a few days after its optical maximum of $m_{V,\rm max}=7.5$
\citep[e.g.,][]{sat73, ros75}.  The $V$ light curve and the $(B-V)_0$ 
and $(U-B)_0$ color evolution of IV~Cep are shown in the Appendix
(Figures 
\ref{v5114_sgr_v1500_cyg_v1494_aql_iv_cep_lv_vul_v_color_logscale},
\ref{v2491_cyg_v2468_cyg_v1500_cyg_v1668_cyg_iv_cep_v_color_logscale},
and 
\ref{v2467_cyg_v2468_cyg_v1668_cyg_iv_cep_v_color_logscale}),
and the color-color evolution is plotted in Figure 
\ref{color_color_diagram_t_pyx_lv_vul_iv_cep_nq_vul}(c), where
the $UBV$ data are taken from \citet{mac72} (red filled circles)
and \citet{koh73} (open blue circles).  We obtained $E(B-V)=0.70\pm0.05$
by fitting the cluster of red points with point F (or point 4'').
All the $UBV$ data were obtained after the optical maximum, 
so the nova had already reached point 4'' (near point F)
in Figure \ref{color_color_diagram_four_typical_example}(c).  
It then moved gradually leftward, like V1974~Cyg, 
owing to the effects of emission lines.
If we use the distance modulus of $(m-M)_V = 14.7$ calculated
by the time-stretching method in the Appendix (Equation 
(\ref{v2491_cyg_v2468_cyg_v1500_cyg_v1668_cyg_iv_cep_distance})),
we obtain a distance of $d=3.2$~kpc.

The reddening toward IV~Cep was estimated to be $E(B-V)= 0.8$ 
\citep{sat73} from the interstellar absorption in the Cepheus region,
and $E(B-V)=A_V/3.1 = 1.8/3.1 = 0.58$ \citep{tho73} and  $E(B-V)=A_V/3.1 = 
1.7/3.1 = 0.55$ \citep{koh73}, both from the absorption--distance
relation given by \citet{nec67}.  We reanalyzed the data using a new
distance-reddening relation given by \citet{mar06}, as shown in
Figure \ref{distance_reddening_rs_oph_v446_her_lv_vul_iv_cep}(d).
We plot four relations in directions
close to IV~Cep, $(l, b)=(99\fdg6137,-1\fdg6381)$:
$(l, b)=(99\fdg5,-1\fdg5)$ (red open squares), 
$(99\fdg75,-1\fdg5)$ (green filled squares),
$(99\fdg5,-1\fdg75)$ (blue asterisks),
and $(99\fdg75,-1\fdg75)$ (magenta open circles).
The closest direction is that denoted
by red open squares.  The two trends, i.e., the distance modulus
of $(m-M)_V = 14.7$ (blue thick solid line) and the distance-reddening
relation denoted by red open circles, cross consistently 
at or near $E(B-V)\approx0.70$ and $d\approx 3.2$~kpc.
This supports our estimate of $E(B-V)=0.70\pm0.05$
from the general tracks.

\subsection{NQ~Vul 1976}
\label{nq_vul_color}
NQ~Vul was discovered by G. E. D. Alcock \citep{mil76}
on UT 1976 October 21.7 near optical maximum \citep{mil76}.
It then rose to $m_V\approx6.0$ at maximum 
on UT November 2.  We plot the $V$ light curve and the $(B-V)_0$
and $(U-B)_0$ color evolution of NQ~Vul in the Appendix (Figures 
\ref{v475_sct_t_pyx_nq_vul_dq_her_v_bv_ub_color_logscale_no6} and
\ref{v496_sct_v2274_cyg_fh_ser_nq_vul_v_bv_ub_color_logscale}),
where the $UBV$ data are taken from \citet{yam77}, \citet{lan77},
\citet{cha77}, and \citet{due79}.  We also plot the $V$ and visual
magnitudes taken from \citet{dip78} and the archive of the
American Association of Variable Star Observers (AAVSO), respectively.
The $V$ magnitude of \citet{yam77} is systematically
brighter by 0.2 mag than the other data, so we shifted them down
by 0.2 mag.  In addition the $B-V$ and $U-B$ colors of \citet{yam77} are 
systematically bluer by 0.05 and 0.2 mag, respectively, than the other data. 
Therefore, we shift them down by 0.05 and 0.2 mag, respectively.
Figure \ref{color_color_diagram_t_pyx_lv_vul_iv_cep_nq_vul}(d) 
shows the color-color evolution of NQ~Vul.  
We obtained $E(B-V)=1.00\pm0.05$ by fitting, 
mainly from Chambliss' data because they have less scatter.

The overall distribution of the observational data points is in reasonable 
agreement with our general tracks in the color-color diagram.
\citet{yam77} wrote that 
the absorption spectrum on October 22 closely resembled those of 
F5Ia supergiants.  This suggests that the early data follow 
the nova-giant sequence, at least in the pre-maximum phase.  
The color-color data of NQ~Vul in Figure 
\ref{color_color_diagram_t_pyx_lv_vul_iv_cep_nq_vul}(d)
start from near point 2, i.e., $(B-V)_0 \approx +0.4$,
and then run along the nova-giant sequence up to 
near point D and remain there for a while.
The nova then darkened suddenly owing to dust formation,
and we stop following the color evolution.

Using $E(B-V)=1.00$ and the distance modulus of $(m-M)_V=13.6$
calculated in the Appendix (Equation
(\ref{v475_sct_t_pyx_nq_vul_dq_her_distance})), we obtain
a distance of $d=1.25$~kpc to NQ~Vul.
The distance to NQ~Vul was estimated by \citet{dow00}
to be $d=1.16$~kpc from an expansion parallax method, which is
consistent with our estimate. 

The following values for reddening toward NQ~Vul were obtained,
in increasing order of $E(B-V)$: $E(B-V)=0.7\pm 0.2$ \citep{you80}
from the equivalent width of the interstellar band at 6614\AA,
$E(B-V)=0.8$ \citep{yam77} from the color of spectral type F5Ia 
on October 22, $E(B-V)=0.9\pm 0.3$ \citep{mar77}
from the interstellar polarization, and $E(B-V)=1.2\pm 0.2$ 
\citep{car77} from the interstellar CH and CH$^+$ equivalent widths.
The simple arithmetic mean of these values is $E(B-V)=0.9\pm 0.2$,
which is consistent with our value of $E(B-V)=1.00\pm0.05$.

The galactic coordinates of NQ~Vul are $(l,b)=(55\fdg3549,+1\fdg2904)$.
We plot the distance-reddening relation toward NQ~Vul in Figure
\ref{distance_reddening_nq_vul_v1370_aql_gq_mus_pw_vul}(a), the data for
which are taken from \citet{mar06}.  The data show four directions:
$(55\fdg25,1\fdg25)$ (red open squares),
$(55\fdg50,1\fdg25)$ (green filled squares),
$(55\fdg25,1\fdg50)$ (blue asterisks),
$(55\fdg50,1\fdg50)$ (magenta open circles).
Our estimates of $E(B-V)=1.00\pm0.05$ and a distance of
$d\sim1.25$~kpc are roughly consistent with Marshall et al.'s 
distance-reddening relation.

%%% 055.3549 +01.2904 

\subsection{V1370~Aql 1982}
\label{v1370_aql_color}
V1370~Aql was discovered by Honda on UT 1982 January 27.85 
at about 6--7 mag \citep{kos82}. 
The $V$ light and $(B-V)_0$ and $(U-B)_0$ color curves of V1370~Aql
are plotted in the Appendix (Figure 
\ref{v1370_aql_v1668_cyg_os_and_v_bv_ub_color_logscale}).
It rapidly decayed to $m_V=11.2$ on UT April 2 and entered 
a shallow dust blackout phase \citep[e.g.,][]{ros83}.   
Then it came back to the usual decline of classical novae about 90 days
after the outburst.
We plot the color-color evolution of V1370~Aql in Figure 
\ref{color_color_diagram_v1370_aql_gq_mus_qu_vul_os_and}(a),
where the $UBV$ data are taken from \citet{oka86} and
IAU Circular No. 3689.  Only a few observational points are available,
but we were able to obtain a reddening of $E(B-V)=0.35\pm0.05$.
The novae stayed at or around point F (or point 4' or point 4''),
as shown in Figure
\ref{color_color_diagram_v1370_aql_gq_mus_qu_vul_os_and}(a).

The following values of the reddening toward V1370~Aql were obtained:
$E(B-V)=0.55\pm0.15$ \citep{sni82} from the 2175\AA\  feature, 
$E(B-V)=A_V/3.1\sim3.0/3.1\sim1.0$ \citep{wil84} from the assumed
spectral type of O8--9 and from the nearby cluster NGC 6755 
$(l=38\fdg6,b=-1\fdg7)$.  
These two values are much larger than our estimate of $E(B-V)=0.35\pm0.05$.

Figure \ref{distance_reddening_nq_vul_v1370_aql_gq_mus_pw_vul}(b) shows
the distance-reddening relation toward V1370~Aql, whose
galactic coordinates are $(l, b)= (38\fdg8126,-5\fdg9465)$.
Here we plot four nearby directions using data from \citet{mar06}:
$(l, b)= (38\fdg75,-5\fdg75)$ (red open squares),
$(39\fdg00,-5\fdg75)$ (green filled squares), 
$(38\fdg75,-6\fdg00)$ (blue asterisks), and
$(39\fdg00,-6\fdg00)$ (magenta open circles).
The closest one is that denoted by blue asterisks.
We also plot the distance-reddening relation based on
the distance modulus of $(m-M)_V=15.2$ (blue thick solid line),
which is calculated by the time-stretching method in the Appendix
(Equation(\ref{v1370_aql_v1668_cyg_os_and_distance})). 
Our estimate of $E(B-V)=0.35$ is consistent with the distance-reddening
relation, which gives a set of
$d\sim6.7$~kpc and $E(B-V)\approx0.35$ for V1370~Aql.
The NASA/IPAC galactic dust absorption map
gives $E(B-V)=0.41 \pm 0.02$ in the direction toward V1370~Aql,
which is also consistent with our estimate.

\subsection{GQ~Mus 1983}
\label{gq_mus_color}
GQ~Mus is a fast nova with $t_2\sim17$~days \citep{war95}.
Although its peak was missed, \citet{kra84} estimated
the peak brightness to be $m_{V,\rm max}\approx7.0$ 
(or $m_{V,\rm max}<7.3$).  In this paper, we adopt
$m_{V,\rm max}\approx7.2$ after \citet{hac08kc}.  
We plot the $V$ light curve and $(B-V)_0$ and $(U-B)_0$ 
color evolution in the Appendix (Figure
\ref{qu_vul_pw_vul_gq_mus_v_bv_ub_color_logscale}).
The $UBV$ data for GQ~Mus are taken from \citet{bud83} and \citet{whi84}
whereas the visual (or $V$) data are from 
the Fine Error Sensor monitor on board {\it IUE}
and the visual photometric data collected by the Royal
Astronomical Society of New Zealand and by the AAVSO
\citep[see][for more details]{hac08kc}.

We plot the color-color diagram of GQ~Mus  in Figure
\ref{color_color_diagram_v1370_aql_gq_mus_qu_vul_os_and}(b).
It seems that these data started a few days
after optical maximum \citep[e.g.,][]{hac08kc} and its color-color
evolution is consistent with the path from point 4 to point 5.
We adopted $E(B-V)=0.45\pm0.05$ by fitting
our general tracks to that of GQ~Mus.

The color excess of GQ~Mus was also determined to be $E(B-V)=0.43$ 
\citep{pac85} and $E(B-V)=0.50\pm0.05$ \citep{peq93} both from 
the hydrogen Balmer lines.  Similar values were reported by
\citet{kra84} and \citet{has90}, who found $E(B-V) = 0.45$ and 0.50,
respectively, on the basis of the 2175 \AA\  feature
in the early {\it IUE} spectra.  \citet{hac08kc} obtained 
$E(B-V) = 0.55\pm 0.05$ on the basis of the 2175\AA\  feature
and various line ratios.  These estimates are
all consistent with our estimate of $E(B-V)=0.45\pm0.05$.

Its distance was estimated to be $d\sim5$~kpc by \citet{kra84},
\citet{whi84}, and \citet{hac08kc}.  Using $E(B-V)=0.45$ and
the distance modulus of $(m-M)_V=15.7$ calculated from the
time-stretching method in the Appendix
(Equation(\ref{qu_vul_pw_vul_gq_mus_distance})),
we obtained a distance to GQ~Mus of 7.3~kpc.
The galactic coordinates of GQ~Mus are $(l, b)=(297\fdg2118,-4\fdg9959)$.
Thus, its height is $z\sim-630$~pc 
below the galactic plane, which is much larger than the scale hight 
$\sim 125$~pc of the galactic matter distribution.
The NASA/IPAC galactic dust absorption map
gives $E(B-V)=0.42 \pm 0.01$ in the direction toward GQ~Mus, which is
consistent with our estimated value.
We also obtained another distance-reddening relation by UV~1455\AA\  
fitting (magenta solid line) in Figure 
\ref{distance_reddening_nq_vul_v1370_aql_gq_mus_pw_vul}(c), which
is calculated in the Appendix (Figure
\ref{qu_vul_pw_vul_gq_mus_v_bv_ub_color_logscale}(a)).

We also plot four distance-reddening relations close to 
GQ~Mus, as shown in Figure 
\ref{distance_reddening_nq_vul_v1370_aql_gq_mus_pw_vul}(c), the data for
which are taken from \citet{mar06}: 
$(l, b)= (297\fdg00,-4\fdg75)$ (red open squares),
$(297\fdg25,-4\fdg75)$ (green filled squares), 
$(297\fdg00,-5\fdg00)$ (blue asterisks), and
$(297\fdg25,-5\fdg00)$ (magenta open circles).
The closest one is that denoted by magenta open circles.
Four trends, $E(B-V)=0.45$ (vertical black solid line),
$(m-M)_V = 15.7$ (blue solid line), UV~1455\AA\  fitting 
(magenta solid line), and distance-reddening relation of \citet{mar06},
consistently cross at $E(B-V)\sim0.45$ and $d\sim7.3$~kpc.

\subsection{QU~Vul 1984 No.2}
\label{qu_vul_color}
QU~Vul was discovered by Collins on UT 1984 December 22.13 
at about 6.8 mag \citep{col84}.  We plot the $V$ light curve and
$(B-V)_0$ and $(U-B)_0$ color evolution in the Appendix (Figure
\ref{qu_vul_pw_vul_gq_mus_v_bv_ub_color_logscale}).
The $UBV$ data for QU~Vul are taken from IAU Circular No. 4033,
\citet{kol88}, \citet{ber88}, \citet{ros92}, and the AAVSO archive. 
It rose to $m_V\approx5.5$ at maximum on UT December 27 
%%% 2446061.5
\citep[e.g.,][]{ros92}.   
Then it gradually declined with $t_2=22$~days and $t_3=49$~days
\citep[e.g.,][]{dow00}.  
The nova is a fast neon nova \citep{geh85}.

Figure \ref{color_color_diagram_v1370_aql_gq_mus_qu_vul_os_and}(c) 
shows the color-color evolution of QU~Vul.
The color evolution started near point F (or point 4'') and
then moved leftward to point 5''.  
We obtained $E(B-V)=0.55\pm0.05$ by fitting.

The following values were obtained for the reddening toward QU~Vul:
$E(B-V)=A_V/3.1\sim1.0/3.1=0.3$ \citep{geh86a} from the galactic
absorption toward $b=-6^\circ$;
$E(B-V)=0.5$ \citep{ros92} from the difference between the
F2--F5 spectral types of the nova at maximum, 
which correspond to $(B-V)_0=0.35$ and the observed one $B-V=0.8$; and
$E(B-V)=0.61\pm0.1$ \citep{sai92} from the \ion{He}{2} 1640/4686\AA\  ratio
in the nebular phase and from the 2200\AA\  UV spectral feature. 
\citet{del97} adopted $E(B-V)=A_V/3.1\sim1.7/3.1=0.55$
for dereddening the spectra.  \citet{sch02} used Saizar et al.'s  
$E(B-V)=0.61$ for obtaining UV fluxes that were consistent 
with the optical fluxes.  These values are roughly consistent
with our estimate of $E(B-V)=0.55\pm0.05$.

To further examine the reddening,
we plot the distance-reddening relation toward QU~Vul in Figure
\ref{distance_reddening_qu_vul_qv_vul_v443_sct_v1419_aql}(a).
The galactic coordinates of QU~Vul are $(l, b)= (68\fdg5108,-6\fdg0263)$.
Here we plot four nearby directions using data from \citet{mar06}:
$(l, b)= (68\fdg50,-6\fdg00)$ (red open squares),
$(68\fdg75,-6\fdg00)$ (green filled squares), 
$(68\fdg50,-6\fdg25)$ (blue asterisks), and
$(68\fdg75,-6\fdg25)$ (magenta open circles).
The closest one is that denoted by red open squares.

We also plot the distance-reddening relation of $(m-M)_V=13.5$ (blue thick
solid line in Figure 
\ref{distance_reddening_qu_vul_qv_vul_v443_sct_v1419_aql}(a))
calculated from the time-stretching method in the Appendix
(Figure \ref{qu_vul_pw_vul_gq_mus_v_bv_ub_color_logscale}).
We obtained another distance-reddening relation by UV~1455\AA\  
fitting (magenta solid line in Figure 
\ref{distance_reddening_qu_vul_qv_vul_v443_sct_v1419_aql}(a)), which is
calculated in the Appendix (Figure 
\ref{qu_vul_pw_vul_gq_mus_v_bv_ub_color_logscale}(a)).

Four trends, $E(B-V)=0.55$ (vertical black solid line),
$(m-M)_V = 13.5$ (blue solid line), UV~1455\AA\  fitting 
(magenta solid line), and distance-reddening relation of \citet{mar06},
consistently cross at $E(B-V)\sim0.55$ and $d\sim2.3$~kpc
in Figure \ref{distance_reddening_qu_vul_qv_vul_v443_sct_v1419_aql}(a).
The NASA/IPAC galactic dust absorption map
gives $E(B-V)=0.55 \pm 0.03$ in the direction toward QU~Vul,
which is also consistent with our estimate.

%Fig.38
%\placefigure{os_and_absorption}

\begin{figure}
%%\epsscale{0.75}
%%\epsscale{1.0}
\epsscale{1.15}
\plotone{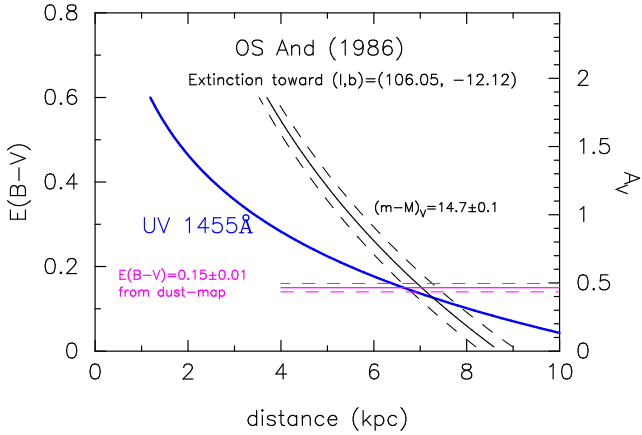}
%\plotone{os_and_absorption.epsi}
%\plotfiddle{evolution1.ps}{5.0cm}{270}{0.4}{0.4}{-170}{220}
\caption{
Same as Figure \ref{v1668_cyg_absorption}, but for OS~And.
Two distance-reddening relations are plotted: one is
calculated from the UV~1455\AA~  flux fit (blue solid line),
and the other is the optical $V$ fit with our free-free emission
model light curve of the $1.0~M_\sun$ WD in Figure
\ref{all_mass_os_and_v1668_cyg_x35z02c10o20_new}.
We also plot the extinction estimate from the dust-map calculated
on NASA/IPAC web site (magenta solid line flanked by dashed lines).
These three trends cross at $d\approx7.0$~kpc and $E(B-V)\approx0.15$.
\label{os_and_absorption}}
\end{figure}

\subsection{OS~And 1986}
\label{os_and_color}
OS~And was discovered on UT 1986 December 5.44 at about 8.0 mag 
\citep{bec87}.  The $V$ and UV~1455\AA\  light curves are plotted in Figure 
\ref{all_mass_os_and_v1668_cyg_x35z02c10o20_new} with those of V1668~Cyg. 
The $V$ light curve and the $(B-V)_0$ and $(U-B)_0$ color evolution
are plotted in the  Appendix
(Figure \ref{v1370_aql_v1668_cyg_os_and_v_bv_ub_color_logscale}).
The optical maximum of $m_{V,\rm max}=6.2$ 
was reached on UT December 7.5$\pm1.0$ \citep{kik88}.
Then it gradually decayed with $t_3\sim20$ days, followed by a sudden
drop by $\sim1.5$ mag about 30 days after discovery
because of dust formation \citep{kik88}.  
Here we adopted the $UBV$ data from \citet{kik88}; 
IAU Circular Nos. 4306, 4342, and 4452; and \citet{ohm87}.
Because \citet{ohm87} gave only the differences between
OS~And and the comparison star 7~And, we used the $UBV$ magnitudes of 
7~And given by \citet{har94} and obtained the $UBV$ magnitudes of OS~And.
Kikuchi et al.'s $UBV$ data were converted from the magnitudes of several
narrower bands, and they could differ greatly from the true
$UBV$ magnitudes when emission lines contribute strongly to the $UBV$ bands.
We compared Kikuchi et al.'s (red filled circles) and Ohmori \& Kaga's 
(red open triangles) data with the other $B-V$ and $U-B$ colors taken
from IAU Circulars Nos. 4282, 4293, 4306, 4342, and 4452 (red open diamonds),
and we shifted Kikuchi et al.'s data by $\Delta(B-V)=-0.1$
and $\Delta(U-B)=-0.15$, and Ohmori \& Kaga's data by $\Delta(B-V)=0.2$
and $\Delta(U-B)=-0.15$ in this study.

We plot the color-color diagram of OS~And in Figure
\ref{color_color_diagram_v1370_aql_gq_mus_qu_vul_os_and}(d).
The small number of data points prevents us from 
confidently estimating the color excess using our general tracks.
Therefore, we determined the color excess by other methods.

The reddening toward OS~And was estimated as
$E(B-V)=0.26\pm 0.04$ \citep{kik88} from $(B-V)_{0,t2}=-0.02\pm0.04$ 
together with $(B-V)_{t2}=0.24\pm0.02$, and
$E(B-V)= 0.25\pm0.05$ \citep{sch97} 
from an average of four values based on four empirical relations, i.e., 
$E(B-V)=0.27$ from $(B-V)_{0,t2}=-0.02\pm0.04$ \citep{van87},
$E(B-V)=0.21$ from $(B-V)_{0,\rm max}=0.2$
\citep{all73} together with $(B-V)_{\rm max}=+0.41$, 
$E(B-V)=0.34$ from $(B-V)_{0,\rm ss}=-0.11$
at the stabilization stage \citep{mir88},
and $E(B-V)=0.24$ from the extinction map toward OS~And. 

On the other hand, the recent NASA/IPAC galactic dust absorption map
gives $E(B-V)=0.15 \pm 0.01$ in the direction toward OS~And, whose
galactic coordinates are $(l, b)= (106\fdg0514, -12\fdg1173)$.
This value is not consistent with the above estimates but is somewhat
smaller.

It has been noted that OS~And is similar to V1668~Cyg except for the depth
of the dust blackout.  Thus, we plot light curves of these two novae
in Figure \ref{all_mass_os_and_v1668_cyg_x35z02c10o20_new}
with the free-free emission model light curves taken from \citet{hac10k}.
This figure also shows the UV~1455\AA\  light curves corresponding to
each optical light curve model. 
The $1.0~M_\sun$ WD (red solid line) model shows the best fit with 
both the optical and {\it IUE} UV~1455\AA\  fluxes of OS~And.
Here, we assumed a chemical composition of the WD envelope of $X=0.35$,
$Y=0.33$, $Z=0.02$, $X_{\rm C+O}=0.30$.  From the optical light curve
fitting, we obtained a distance modulus of
$(m-M)_V=m_{\rm w}-M_{\rm w}=16.1-(+1.4)=14.7$,
where $m_{\rm w}= 16.1$ was read directly from the end point
of the free-free emission model light curve (large open circle at
the bottom of the line) in
Figure \ref{all_mass_os_and_v1668_cyg_x35z02c10o20_new},
and $M_{\rm w}$ is taken from Table 2 of \citet{hac10k} as
$M_{\rm w}= +1.4$  for the $M_{\rm WD}=1.0~M_\sun$ model.
We also obtained a distance-reddening relation from the UV1455\AA\  
light curve fitting of Equation(\ref{v1668_cyg_uv1455_distance_modulus})
for OS~And.  We plot these two distance-reddening relations in Figure
\ref{os_and_absorption}.  We see that three trends,
the distance modulus $(m-M)_V=14.7$ together with Equation
(\ref{distance_modulus_extinction}), the UV~1455\AA\  fitting
in Figure \ref{all_mass_os_and_v1668_cyg_x35z02c10o20_new} together
with Equation (\ref{v1668_cyg_uv1455_distance_modulus}),
and the dust-map $E(B-V)=0.15$, cross consistently at
$d\approx7.0$~kpc and $E(B-V)\approx0.15$. 
Therefore, we adopted $E(B-V)=0.15$ for OS~And.  Using this value,
we plot the color-color diagram in Figure
\ref{color_color_diagram_v1370_aql_gq_mus_qu_vul_os_and}(d). 
The resultant distribution of the data seems to be reasonable,
although the number of data points is small.

\subsection{QV~Vul 1987}
\label{qv_vul_color}
QV~Vul was discovered on UT 1987 November 15 at about 7.0 mag
\citep{bec87}.  The $V$ light curve, $(B-V)_0$ and $(U-B)_0$ 
color evolution are plotted in the Appendix (Figure 
\ref{v2615_oph_v705_cas_qv_vul_fh_ser_v_bv_ub_color_logscale}),
where the $UBV$ data are taken from \citet{ohs88}.
%%% and IAU Circular Nos. 4493, 4511, and 4524.  
The optical maximum had already been reached at discovery \citep{ohs88}.
It gradually decayed and suddenly
dropped about 60 days after discovery because of dust formation.  
We also plot the color-color evolution of QV~Vul in Figure
\ref{color_color_diagram_qv_vul_v443_sct_v1419_aql_v705_cas}(a).
There are several data points just after maximum, so we set Ohshima's
data points (red filled circles) on the nova-giant sequence as shown
in the figure and obtained a reddening of $E(B-V)=0.60\pm0.05$.

The reddening toward QV~Vul was also determined to be $E(B-V)=0.40\pm 0.05$ 
\citep{sco94} from the H$\gamma$/H$\beta$ line ratio and
to be $E(B-V)=A_V/3.1 = 1.0/3.1 = 0.32$ \citep{geh92} from various loose  
constraints on the luminosities and spectra of QV~Vul.
To further examine the reddening,
we plot the distance-reddening relation toward QV~Vul in Figure
\ref{distance_reddening_qu_vul_qv_vul_v443_sct_v1419_aql}(b), whose
galactic coordinates are $(l, b)= (53\fdg8585,+6\fdg9741)$.
Here we plot four nearby directions from the data of \citet{mar06}:
$(l, b)= (53\fdg75,+6\fdg75)$ (red open squares),
$(54\fdg00,+6\fdg75)$ (green filled squares), 
$(53\fdg75,+7\fdg00)$ (blue asterisks), and
$(54\fdg00,+7\fdg00)$ (magenta open circles).
The closest one is that denoted by blue asterisks or magenta open circles.  
We also plot the distance-reddening relation of $(m-M)_V=14.0$
(blue thick solid line) calculated by the time-stretching method
in the Appendix 
(Equation(\ref{v2615_oph_v475_sct_v705_cas_qv_vul_fh_ser_distance})).  
These trends cross at $E(B-V)\approx0.60$ and $d\approx2.7$~kpc, which 
are consistent with our estimate above.
The NASA/IPAC galactic dust absorption map
gives $E(B-V)=0.64 \pm 0.02$ in the direction toward QV~Vul,
which is also consistent with our estimate.

\subsection{V443~Sct 1989}
\label{v443_sct_color}
V443~Sct was discovered on UT 1989 September 20 at about 10.5 mag
\citep{wil89}.  The optical maximum ($\sim 7.5$ mag) was already
reached a week before discovery \citep[see, e.g., Figure 1 of][]{ros91}.
We plot the $V$ light curve, and $(B-V)_0$ and $(U-B)_0$ color evolution
of V443~Sct in the Appendix (red symbols in Figure 
\ref{v458_vul_v443_sct_pw_vul_v_bv_ub_color_logscale}),
which shows a semi-periodic oscillation with a periodicity of $\sim 17$~days.
The light curve is similar to that of PW~Vul (denoted by blue symbols
in the same figure).
We plot the color-color evolution of V443~Sct in Figure
\ref{color_color_diagram_qv_vul_v443_sct_v1419_aql_v705_cas}(b),
where the $UBV$ data are taken from IAU Circular Nos. 4862, 4868,
and 4873.  Although there are only several data points after optical maximum,
we determine the reddening to be $E(B-V)=0.40\pm0.05$.

The reddening toward V443~Sct was estimated by \citet{anu92}
to be $E(B-V)=0.40$ from the Balmer/Paschen line ratios, which is consistent
with our estimate.   To further examine the reddening,
we plot the distance-reddening relation toward V443~Sct, whose
galactic coordinates are $(l, b)= (27\fdg2183,-2\fdg4211)$,  in Figure
\ref{distance_reddening_qu_vul_qv_vul_v443_sct_v1419_aql}(c).
Here we plot four nearby directions from the data of \citet{mar06}:  
$(l, b)= (27\fdg00,-2\fdg25)$ (red open squares),
$(27\fdg25,-2\fdg25)$ (green filled squares), 
$(27\fdg00,-2\fdg50)$ (blue asterisks), and
$(27\fdg25,-2\fdg50)$ (magenta open circles). 
The closest one is that denoted by magenta open circles.
We also plot the distance-reddening relation of $(m-M)_V=15.5$ 
calculated in the Appendix (Equation 
(\ref{v458_vul_v443_sct_pw_vul_distance})).
Our two trends intersect at $E(B-V)\approx0.40$
and the distance $d\approx7.1$~kpc.  Considering the small number of
data points in Figure
\ref{color_color_diagram_qv_vul_v443_sct_v1419_aql_v705_cas}(b),
our estimated extinction is barely consistent with
the distance-reddening relation given by \citet{mar06}.

\subsection{V1419~Aql 1993}
\label{v1419_aql_color}
V1419~Aql was discovered by Yamamoto on UT 1993 May 14 at about 7.6 mag
\citep{hir93}.  The $V$ light curve and $(B-V)_0$ and $(U-B)_0$ 
color evolution are plotted in the Appendix (Figure 
\ref{v496_sct_v1065_cen_v1419_aql_v1668_cyg_v_bv_ub_color_logscale}),
where the $UBV$ data are taken from \citet{mun94a} and IAU Circular Nos. 
5794, 5802, 5807, and 5829.  
The $V$ magnitude reached a maximum of $m_V=7.5$ on May 23.
%%% (JD~2449130.5).
Then it gradually decayed and sharply dropped about 35--40 days
after discovery owing to dust shell formation.
We also plot the color-color evolution of V1419~Aql in Figure 
\ref{color_color_diagram_qv_vul_v443_sct_v1419_aql_v705_cas}(c).
The color data in the IAU Circulars
are scattered, so we used mainly Munari et al.'s data
in our fitting procedure.
The first color data were obtained near maximum, on May 26, so 
we set this point on the nova-giant sequence, as shown in Figure
\ref{color_color_diagram_qv_vul_v443_sct_v1419_aql_v705_cas}(c),
and estimated the reddening to be $E(B-V)=0.50\pm0.05$.
The color evolution of V1419~Aql started near point D
and remained there for a while, as seen in Figures
\ref{v496_sct_v1065_cen_v1419_aql_v1668_cyg_v_bv_ub_color_logscale}(b)
and (c).
After that, the nova suddenly darkened owing to dust shell formation.
We do not follow the color evolution
after the dust blackout in the color-color diagram.

\citet{lyn95} obtained the following values for the reddening
toward V1419~Aql: $E(B-V)=1.02\pm 0.19$ from an average of
$E(B-V)=0.74\pm 0.46$ in June and
$E(B-V)=1.10\pm 0.32$ in September,
both from the Paschen series line ratios,
$E(B-V)= 1.22\pm 0.13$ in June and
$E(B-V)= 1.05\pm 0.12$ in September,
both from the \ion{O}{1} lines $\lambda 8446$, $\lambda 11287$, 
and $\lambda 13164$.  On the other hand, 
\citet{mun94b} estimated the reddening to be $E(B-V)= 0.55\pm 0.15$
from the star's color, i.e., the average of
$E(B-V)= 0.6$ from the intrinsic color at $t_2$ time \citep{van87}
and $E(B-V)= 0.5$ from the intrinsic color at maximum \citep{all73}.
Thus, these reddening estimates show a large difference between 
these two groups.

To further examine the reddening, we plot the distance-reddening
relation of \citet{mar06} in Figure
\ref{distance_reddening_qu_vul_qv_vul_v443_sct_v1419_aql}(d),
where the galactic coordinates of V1419~Aql are
$(l, b)= (36\fdg8110, -4\fdg1000)$.
We plot four nearby directions from the data from \citet{mar06}:
$(l, b)= (36\fdg75,-4\fdg00)$ (red open squares),
$(37\fdg00,-4\fdg00)$ (green filled squares), 
$(36\fdg75,-4\fdg25)$ (blue asterisks), and
$(37\fdg00,-4\fdg25)$ (magenta open circles). 
The closest one is that denoted by red open squares.  
We also plot the distance-reddening relation of $(m-M)_V=14.6$ 
calculated by the time-stretching method in the Appendix (Equation 
(\ref{v496_sct_v1065_cen_v1419_aql_v1668_cyg_lv_vul_distance})).
The two trends of $(m-M)_V=14.6$ and $E(B-V)=0.50$ cross at
$d\approx4.1$~kpc and $E(B-V)\approx0.50$, which are consistent with
the distance-reddening relation of \citet{mar06}.
The NASA/IPAC galactic dust absorption map
gives $E(B-V)=0.55 \pm 0.01$ in the direction toward V1419~Aql,
which is also consistent with our estimate.

\subsection{V705~Cas 1993}
\label{v705_cas_color}
V705~Cas was discovered by Kanatsu on UT 1993 December 7 at about 6.5 mag
\citep{nak93}.   The $V$ light curve and $(B-V)_0$ and $(U-B)_0$ 
color evolution are plotted in the Appendix (Figure
\ref{v2615_oph_v705_cas_qv_vul_fh_ser_v_bv_ub_color_logscale}),
where the $UBV$ data are taken from \citet{mun94b}, \citet{hri98},
and IAU Circular Nos. 5920 and 5929.  It rose to $m_V=5.5$ 10 days 
after discovery, i.e., on UT December 17.  
A deep minimum appeared about 60 days after discovery
because of dust shell formation.
Therefore, we only plot the data for $t < 60$ days
for the color-color evolution of V705~Cas in Figure 
\ref{color_color_diagram_qv_vul_v443_sct_v1419_aql_v705_cas}(d).
The color-color data of Munari et al. are scattered, so 
we determine the reddening to be $E(B-V)=0.45\pm0.05$,
mainly from the data of Hric et al. and the IAU Circulars.
Combining the distance modulus of $(m-M)_V=13.6$ calculated in the Appendix
(Equation (\ref{v2615_oph_v475_sct_v705_cas_qv_vul_fh_ser_distance}))
and $E(B-V)=0.45$, we obtained a distance of $d=2.8$~kpc.
The color-color evolution of V705~Cas started near point 3 (or point C).
Then, the nova ascended along the nova-giant sequence to near point 4''
(near point F) and stayed there for a while.
After that, the nova suddenly darkened owing to dust shell formation.

The reddening toward V705~Cas was estimated by \citet{hri98}
to be $E(B-V)=0.38$ from an intercomparison of the color indices
of stars surrounding the nova selected from the SAO catalog.  They
also obtained $E(B-V)= (B-V)_{\rm ss}-(B-V)_{0, \rm ss}=0.32-(-0.11)=0.43$
from the intrinsic color at the stabilization stage \citep{mir88}.
\citet{hau95} obtained $E(B-V)=0.5$ assuming that
the total (optical + UV) luminosity in the early phase is constant
\citep[see also][]{sho94}.  The simple arithmetic mean of these values 
is $E(B-V)=0.44\pm 0.05$, which is consistent with our estimate
of $E(B-V)=0.45\pm0.05$.
If we use the relations given by \citet{van87},
we obtain $E(B-V)= (B-V)_{\rm max}
- (B-V)_{0, \rm max} = 0.56 - (+0.23) = 0.33$ and 
$E(B-V)= (B-V)_{t_2} - (B-V)_{0, t_2} = 0.34 - (-0.02) = 0.36$,
the color data of which are taken from \citet{hri98}.  
These values are slightly smaller than our value of $E(B-V)=0.45\pm0.05$.
The galactic coordinates of V705~Cas are $(l,b)=(113\fdg6595,-4\fdg0959)$.
Unfortunately, there are no data on the dust extinction map calculated
by \citet{mar06} because their data are given 
for $-100\fdg0 \le l \le 100\fdg0$ and $-10\fdg0 \le b \le +10\fdg0$. 
The NASA/IPAC galactic dust absorption map gives 
$E(B-V)=0.48 \pm 0.02$ in the direction toward V705~Cas, 
which is consistent with our value.

\subsection{V382~Vel 1999}
\label{v382_vel_color}
V382~Vel is a very fast nova identified as a neon nova \citep{woo99}.
The $V$ light curve and $(B-V)_0$ and $(U-B)_0$ color evolution 
are plotted in the Appendix (Figures 
\ref{rs_oph_v446_her_v1500_cyg_v1668_cyg_v382_vel_v_color_logscale}).
The nova reached $m_V = 2.7$ at maximum on UT 1999 May 23.
The $UBV$ color-color evolution is plotted in Figure 
\ref{color_color_diagram_v382_vel_v2274_cyg_v475_sct_v5114_sgr}(a), where
the data are taken from IAU Circular Nos. 7176, 7179, 7196, 7209, 7216, 
7226, and 7232 (observed mainly by A. C. Gilmore and P. M. Kilmartin).  
We obtained $E(B-V)=0.15\pm0.05$ by fitting.
\citet{muk01} obtained a hydrogen column density toward V382~Vel
of $N_{\rm H}= (1.01 \pm 0.05)\times 10^{21}$ from early hard
X-ray spectrum fittings.  This value can be converted to
$E(B-V)= N_{\rm H}/ 5.8 \times 10^{21} \approx 0.2$ \citep{boh78}
or $E(B-V)= N_{\rm H}/ 8.3 \times 10^{21} \approx 0.12$ \citep{lis14},
which are consistent with our estimated value. 
\citet{sho03} obtained $E(B-V)= 0.20$ and \citet{del02} estimated
$E(B-V)=0.05$--$0.099$ from various line ratios
and the \ion{Na}{1}~D interstellar absorption features. 
These values are all consistent with our estimated 
value of $E(B-V)=0.15\pm0.05$.

The galactic coordinates of V382~Vel are $(l, b)=(284\fdg1674, +5\fdg7715)$.
We plot the distance-reddening relation calculated by \citet{mar06}
in Figure 
\ref{distance_reddening_v382_vel_v2274_cyg_v475_sct_v5114_sgr}(a).
Here we plot four nearby directions:
$(l, b)= (284\fdg00,+5\fdg75)$ (red open squares),
$(284\fdg25,+5\fdg75)$ (green filled squares), 
$(284\fdg00,+6\fdg00)$ (blue asterisks), and
$(284\fdg25,+6\fdg00)$ (magenta open circles).
The closest one is that denoted by green filled squares.
We also plot the distance-reddening relation $(m-M)_V=11.4$ calculated
in the Appendix (Equation 
(\ref{rs_oph_v446_her_v1500_cyg_v1668_cyg_v382_vel_distance})). 
It seems that our estimate of $E(B-V)=0.15$ is slightly smaller than 
Marshall et al.'s distance-reddening relation.

\subsection{V2274~Cyg 2001\#1}
\label{v2274_cyg_color}
V2274~Cyg was discovered by Nakamura on UT 2001 July 13.65 
at about 11.9 mag \citep{sat01}.  We plot the $V$ light curve 
and $(B-V)_0$ and $(U-B)_0$ color evolution in the Appendix (Figure 
\ref{v496_sct_v2274_cyg_fh_ser_nq_vul_v_bv_ub_color_logscale}),
where the $UBV$ data are taken from 
%%% IAU Circular Nos. 7687, 7691, and 7712.
\citet{vol02b}.
It rose to $m_V=11.8$ on UT August 20.09 \citep{waa01}.   
Then it declined with large oscillatory variations until dust
blackout began about 50 days after discovery.
We also plot the color-color evolution of V2274~Cyg in Figure 
\ref{color_color_diagram_v382_vel_v2274_cyg_v475_sct_v5114_sgr}(b);
we obtained $E(B-V)=1.35\pm0.1$ by fitting.  There is large scatter in the 
color-color data, so we have a relatively large error of determination.

\citet{rud03} estimated the reddening to be $E(B-V)=1.30\pm0.2$ from
three \ion{O}{1} line fluxes based on the reddening law given by
\citet{dra89}, which is consistent with our estimate of $E(B-V)=1.35\pm0.1$.
\citet{rud03} also obtained a distance of $d=10.8^{+4.2}_{-3.1}$~kpc
to the nova from della Valle \& Livio's (1995) MMRD relation
together with $t_2=33\pm4$.  Then the apparent distance modulus
in $V$ becomes $(m-M)_V=19.16\pm0.7$,
including $3\sigma$ scatter in the MMRD relation.  

The galactic coordinates of V2274~Cyg are $(l, b)=(73\fdg0415,+1\fdg9910)$,
and we plot the distance-reddening relation calculated by \citet{mar06}
in Figure 
\ref{distance_reddening_v382_vel_v2274_cyg_v475_sct_v5114_sgr}(b).
Here we plot four nearby directions:
$(l, b)= (73\fdg00,2\fdg00)$ (red open squares),
$(73\fdg25,2\fdg00)$ (green filled squares), 
$(73\fdg00,1\fdg75)$ (blue asterisks), and
$(73\fdg25,1\fdg75)$ (magenta open circles).
The closest one is that denoted by red open squares.
Our two trends, $(m-M)_V=18.7$ (thick blue solid line) calculated in 
the Appendix (Equation
(\ref{v496_sct_v2274_cyg_fh_ser_nq_vul_distance})) 
and $E(B-V)=1.35$ (vertical black solid line), cross at $d=8.0$~kpc.
This point is consistent with Marshall et al.'s distance-reddening 
relation.

\subsection{V475~Sct 2003}
\label{v475_sct_color}
V475~Sct was discovered by Nishimura on UT 2003 August 28.58 
at about 8.5 mag \citep{nak03}.
We plot the $V$ light curve and $(B-V)_0$ and $(U-B)_0$ color
evolution of V475~Sct in the Appendix (Figure
\ref{v475_sct_t_pyx_nq_vul_dq_her_v_bv_ub_color_logscale_no6})
together with those of DQ~Her, NQ~Vul, and T~Pyx.
It rose to $m_V=8.4$ on UT September 2 \citep[e.g.,][]{str06}.
Then it gradually declined with $t_2=48$~days and $t_3=53$~days
\citep[e.g.,][]{cho05}.  The nova started a dust blackout about 60 days
after discovery.  We plot the color-color evolution of V457~Sct
in Figure \ref{color_color_diagram_v382_vel_v2274_cyg_v475_sct_v5114_sgr}(c),
where the $UBV$ data are taken from \citet{cho05}.
The color evolution followed the nova-giant sequence in the early phase,
as shown in Figure
\ref{color_color_diagram_v382_vel_v2274_cyg_v475_sct_v5114_sgr}(c),
so we determine the reddening to be $E(B-V)=0.55\pm0.10$.

\citet{cho05} obtained a value of $E(B-V)=0.69\pm0.05$ for
the reddening toward V475~Sct averaging various estimates:
$E(B-V)= (B-V)_{\rm max}- (B-V)_{0, \rm max} = 0.91 - 0.2 = 0.71$,
$E(B-V)= (B-V)_{t2} - (B-V)_{0, t2} = 0.45 - (-0.02) = 0.47$,
$E(B-V)= (B-V)_{\rm ss} - (B-V)_{0, \rm ss} = 0.61 - (-0.11) = 0.72$,
$E(B-V)= (B-V)_{\rm F2} - (B-V)_{0, \rm F2} = 1.08 - (0.23) = 0.85$,
where the spectral type of V475~Sct observed on UT August 31.97 is
F2 supergiant and $(B-V)_{0, \rm F2}$ is the intrinsic color of
F2 supergiants \citep{cox00}, and
$E(B-V)=0.70$ from the interstellar \ion{K}{1} line.
\citet{cho05} derived values of $(m-M)_V=15.57\pm0.15$ and $d=4.8\pm0.9$~kpc
for the distance modulus in the $V$ band and the distance, respectively.
The distance was also estimated by \citet{str06} to be
$(m-M)_0=13.7$ ($d\sim5.5$~kpc) from the MMRD relations with
$t_2=46$, and $t_3=53$ days together with 
$E(B-V)=(B-V)_{\rm F2} - (B-V)_{0,\rm F2}=0.78-0.23 =0.55$.
The latter value of $E(B-V)=0.55$ is consistent with our estimated value.

To examine the distance-reddening relation toward V475~Sct, we compared
our results with that of \citet{mar06}.
The galactic coordinates of V475~Sct are $(l, b)= (24\fdg2015,-3\fdg9466)$.
We plot the distance-reddening relation calculated by \citet{mar06}
in Figure 
\ref{distance_reddening_v382_vel_v2274_cyg_v475_sct_v5114_sgr}(c).
Here we plot four nearby directions: 
$(l, b)= (24\fdg00,-4\fdg00)$ (red open squares),
$(24\fdg25,-4\fdg00)$ (green filled squares), 
$(24\fdg00,-3\fdg75)$ (blue asterisks), and
$(24\fdg25,-3\fdg75)$ (magenta open circles).
The closest one is that denoted by green filled squares.
We also plot the distance-reddening relation
of $(m-M)_V=15.6$ calculated in the Appendix (Equation 
(\ref{v475_sct_t_pyx_nq_vul_dq_her_distance})).
The two trends of $(m-M)_V=15.6$ 
(thick blue solid line) and $E(B-V)=0.55$ (black solid line)
cross at $d=6.0$~kpc, which is consistent with Marshall et al.'s trend.
The NASA/IPAC galactic dust absorption map gives $E(B-V)=0.59 \pm 0.05$
in the direction toward V475~Sct, which is also consistent with our estimate.

\subsection{V5114~Sgr 2004}
\label{v5114_sgr_color}
V5114~Sgr was discovered independently by Nishimura on UT 2004 March 15.82 
at about 9.4 mag and by Liller on 2004 March 17.34 at about 8.2 mag 
\citep{nis04}.  We plot the $V$ light curve and $(B-V)_0$ and $(U-B)_0$ 
color evolution in the Appendix (Figure
\ref{v5114_sgr_v1500_cyg_v1494_aql_iv_cep_lv_vul_v_color_logscale}).  
It rose to $m_V=8.0$ on UT March 17.17 \citep{ede06}.   
Then it gradually declined with $t_2=11$~days and $t_3=21$~days.
We plot the color-color evolution of V5114~Sgr
in Figure \ref{color_color_diagram_v382_vel_v2274_cyg_v475_sct_v5114_sgr}(d),
where the $UBV$ data are taken from \citet{ede06}.
The color evolution followed the nova-giant sequence in the very early phase
from point D to point 4''
and then followed the path from point 4'' to 5'', 
so we obtained a reddening of $E(B-V)=0.45\pm0.05$.
This color-evolution path is very similar to that of V1974~Cyg.
Here we omitted some data that differ greatly from the others to reduce
the scatter in the color data.  

The reddening toward V5114~Sgr was determined by \citet{ede06}
to be $E(B-V)=0.45$ from the equivalent width of the interstellar line 
\ion{K}{1}$~\lambda$7699, $E(B-V)= (B-V)_{\rm max}
- (B-V)_{0, \rm max} = 0.66 - (0.23 \pm 0.06) = 0.43\pm 0.06$,
$E(B-V)= (B-V)_{t2} - (B-V)_{0, t2} = 0.38 - (-0.02\pm 0.04) = 0.40\pm 0.04$,
$E(B-V)=0.58$ from the galactic dust map \citep{sch98},
$E(B-V)=0.57$ from the line ratio of \ion{He}{2}~$\lambda$4686
and $\lambda$10124, 
$E(B-V)=0.65$ from the line ratio of \ion{He}{1}~$\lambda$4861
and $\lambda$10049, and so on.  We checked the NASA/IPAC galactic
dust absorption map and obtained $E(B-V)=0.49 \pm 0.02$
in the direction toward V5114~Sgr, whose galactic coordinates are
$(l, b)= (3\fdg9429,-6\fdg3121)$.  The simple arithmetic mean of the
first six values above, given by \citet{ede06}, is $E(B-V)=0.51\pm 0.09$,
which is consistent with our value of $E(B-V)=0.45\pm0.05$.

To further check the reddening-distance relation toward V5114~Sgr,
we compared our results with that of \citet{mar06}.  In Figure 
\ref{distance_reddening_v382_vel_v2274_cyg_v475_sct_v5114_sgr}(d),
We plot four nearby directions:
$(l, b)= (3\fdg75,-6\fdg50)$ (red open squares),
$(4\fdg00,-6\fdg50)$ (green filled squares), 
$(3\fdg75,-6\fdg25)$ (blue asterisks), and
$(4\fdg00,-6\fdg25)$ (magenta open circles).
The closest one is that denoted by magenta open circles.
These three trends, Marshall et al.'s
magenta open circles, $(m-M)_V=16.5$ (thick blue solid line)
calculated in the Appendix (Equation 
(\ref{v5114_sgr_v1500_cyg_v1494_aql_iv_cep_lv_vul_distance})),
and $E(B-V)=0.45$ (vertical black solid line), are all consistent with
$d\sim10.5$~kpc and $E(B-V)\sim0.45$.

%Fig.39
%\placefigure{hr_diagram_5types_novae_one}

\begin{figure}
%%\epsscale{0.85}
%%\epsscale{1.0}
\epsscale{1.15}
\plotone{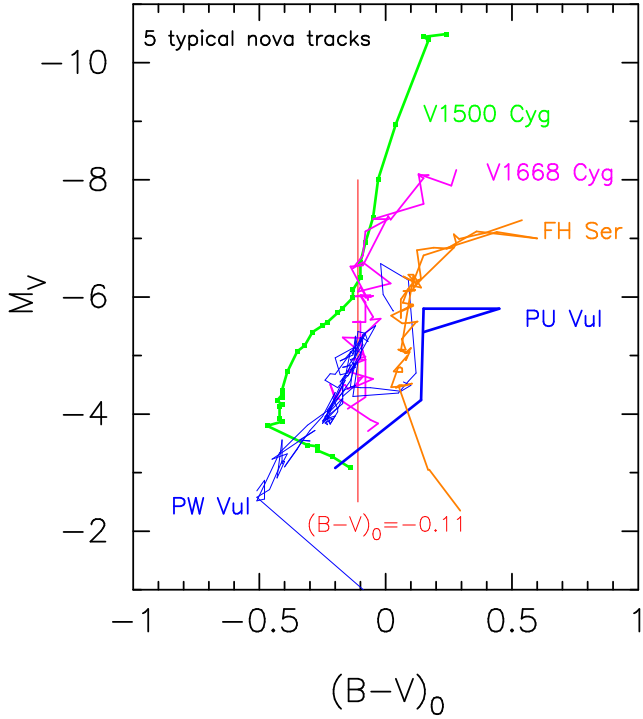}
%\plotone{hr_diagram_5types_novae_one.epsi}
%\plotfiddle{evolution1.ps}{5.0cm}{270}{0.4}{0.4}{-170}{220}
\caption{
Five typical nova tracks in the color-magnitude diagram:
from left to right,
V1500~Cyg (green solid), V1668~Cyg (magenta solid), PW~Vul (blue thin solid), 
FH~Ser (ocher solid), and PU~Vul (blue thick solid line).
Vertical red solid line indicates stabilization track of
$(B-V)_0=-0.11$ \citep{mir88}.
\label{hr_diagram_5types_novae_one}}
\end{figure}

\section{Discussion}
\label{discussion}

\subsection{Comparison with Miroshnichenko's Relation}
\label{miroshnichenko}
\citet{mir88} identified a trend in which the $B-V$ and $U-B$ colors
remain constant for a while soon after optical maximum, which usually has
the reddest color, and further that 
this stage showed a general trend of $(B-V)_{0,\rm ss} = -0.11\pm 0.02$.
He called this stage ``the stabilization stage'' and
determined the $V$ band absorptions ($A_V$) of 23 novae assuming 
that they all have the same intrinsic $(B-V)_0$ color
at this stabilization stage, i.e., $E(B-V)=(B-V)_{\rm ss} - (B-V)_{0,\rm ss}
= (B-V)_{\rm ss} + 0.11$ and 
$A_V= R_V E(B-V)= 3.13 E(B-V)$, where $(B-V)_{\rm ss}$ is
the observed $B-V$ color at the stabilization stage.
This method looks fascinating but sometimes results in a large
difference from our value (see Table 
\ref{extinction_from_two_color_diagram}). 

Figure \ref{hr_diagram_5types_novae_one} summarizes 
the color-magnitude diagram for our 28 novae.  There are five typical
nova tracks, each of which represents a similar speed class as well as
other characteristic properties.
We found that the two types of V1668~Cyg and FH~Ser
clearly show a stabilization stage, in which the $B-V$ color is almost
constant during a decay of a few magnitudes in $V$.  In the V1668~Cyg
type of novae (fast novae with optically-thin dust formation) 
the stabilization stage clearly appears as a straight vertical track
of $(B-V)_0\approx-0.05$ (point 4''), which is close to $(B-V)_0=-0.11$
(the red vertical line).  On the other hand, the FH~Ser type 
(moderately fast/slow novae with optically
thick dust formation) lies between $(B-V)_0=0.0$ and 0.1.
In the V1500~Cyg and PW~Vul types of novae, the $(B-V)_0$ colors
are close to $(B-V)_0\sim-0.03$ at $t_2$ because they remain for a while
near point F (see Section \ref{van_den_bergh}).  Because
their colors gradually become blue around $t_2$, their color could be
consistent with $(B-V)_{0,\rm ss}=-0.11$.  In the PU~Vul type of novae,
including V723~Cas and HR~Del, the color path is much redder because
they stay near point 0, whose color is $(B-V)_0=+0.13$ (see Figure
\ref{color_color_diagram_pu_vul_points_postmax} for PU~Vul and Figure
\ref{color_color_diagram_hr_del_v723_cas_postmax} for V723~Cas/HR~Del).
Thus, three types, the V1668~Cyg, V1500~Cyg, and PW~Vul types,
could have a stabilization stage with $(B-V)_{0,\rm ss}\approx-0.11$.
For the other two types, the FH~Ser and PU~Vul types, Miroshnichenko's 
method may result in a larger extinction than the true value,
as seen in Table \ref{extinction_from_two_color_diagram}.

\subsection{Comparison with van den Bergh and Younger's Relations}
\label{van_den_bergh}
\citet{van87} derived the general trends of color evolution
in nova light curves, i.e., $(B-V)_{0,\rm max} = 0.23\pm 0.06$ at maximum
and $(B-V)_{0,t2} = -0.02\pm 0.04$ at $t_2$.  
Using these relations, one can obtain $E(B-V)$ for individual novae.
These two methods, however, often give very different values
from those obtained by other methods.  Table 
\ref{extinction_from_two_color_diagram} summarizes these values of 
$E(B-V)$, including our results from the previous sections.
Here we examine van den Bergh and Younger's empirical relations
for our 28 novae.

In this paper, we showed that novae generally follow the nova-giant
sequence when the photospheric emission dominates the spectrum
in the optical region.
The optical maximum corresponds to the reddest point of the journey
along the nova-giant sequence.  This reddest point varies among novae.
Thus, the color at maximum, $(B-V)_{0,\rm max}$,
is not the same for all the novae.  To understand 
the difference among various types of novae, 
we plot five typical nova tracks in the color-magnitude diagram
(Figure \ref{hr_diagram_5types_novae_one}).
A nova outburst evolves from the peak magnitude (top of a line)
and then moves down.  V1500~Cyg and V1668~Cyg showed colors consistent
with $(B-V)_{0, \rm max} = 0.23\pm 0.06$ at maximum \citep{van87}.
In this way, many fast and very fast novae are consistent with
van den Bergh \& Younger's $(B-V)_{0, \rm max} = 0.23\pm 0.06$
except for some novae like V1974~Cyg.  V1974~Cyg is not consistent 
with this because of its too-short journey along the
nova-giant sequence from point D, where $(B-V)_{0, \rm D}=+0.04$
(see Figure \ref{color_color_diagram_pw_vul_v1500_cyg_v1668_cyg_1974_cyg}(d)).
For the other types of novae,
e.g., FH~Ser and PU~Vul are also not consistent with this,
because their long journeys along the nova-giant sequence reach
$(B-V)_0\sim0.6$ far beyond $(B-V)_0=0.23$. 
Thus, van den Bergh and Younger's law of 
$(B-V)_{0,\rm max} = 0.23\pm 0.06$ is usually not applicable to 
slow/very slow novae and sometimes fast novae like V1974~Cyg.  

The second law of $(B-V)_{0,t2} = -0.02\pm0.04$
at $t_2$, on the other hand, shows rough agreement with our estimated values.
This is because $(B-V)_{0,t2} = -0.02\pm0.04$ is very close to 
the $(B-V)_0$ color at point F ($-0.03$), point 4'' ($-0.05$), 
and point 4' ($-0.08$), where fast novae usually remain for a while
when free-free emission dominates the spectrum after optical maximum
(see Figure \ref{hr_diagram_5types_novae_one}).
However, some of the very slow novae, e.g., PU~Vul, V5558~Sgr, V723~Cas, 
and HR~Del, remain near point 0 (not point F) for a while as
mentioned in Sections \ref{pu_vul} and \ref{each_slow_novae},
and its color is $(B-V)_0=+0.13$, which is not consistent with
$(B-V)_{0,t2} = -0.02\pm0.04$.  Thus,
the PU~Vul and FH~Ser type tracks show much redder colors at $t_2$.


%% Included in this acknowledgments section are examples of the
%% AASTeX hypertext markup commands. Use \url without the optional [HREF]
%% argument when you want to print the url directly in the text. Otherwise,
%% use either \url or \anchor, with the HREF as the first argument and the
%% text to be printed in the second.

\section{Conclusions}
\label{conclusions}
We extensively examined the color-color evolutions
of nova outbursts and found several important properties of
nova color evolution.  Our main results are summarized as follows.\\

\noindent
1. We compiled or obtained the distance and extinction 
for eight well-observed novae including all the speed classes.
Based on the revised distances and extinctions, we plotted the dereddened
$(B-V)_0$ versus $(U-B)_0$ color-color diagrams of these novae 
and found a general course for the color evolution of nova outbursts.
We further found that a number of novae follow this general course
in the color-color diagram.\\

\noindent
2. The general tracks of nova outbursts consist mainly of three 
branches, i.e., the nova-giant sequence phase and free-free emission 
(point F) phase, followed by development of strong emission lines
(horizontal blueward excursion):
(1) In the early phase of a nova outburst, i.e., in the pre-maximum
phase, a nova evolves from near the blackbody sequence and 
follows a new sequence redward.  After optical maximum,
the nova quickly evolves back blueward along this new sequence.
This new sequence is located parallel to, but $\Delta (U-B)\approx -0.2$
mag bluer than, the supergiant sequence.  We call the new sequence 
``the nova-giant sequence'' after the supergiant sequence.  The
spectra of novae on the nova-giant sequence resemble those of A--F type
supergiants.  
(2) Subsequently, the spectra of novae become that of free-free emission
in the optical and near IR regions.
Therefore, novae stay at the point of free-free emission 
($B-V=-0.03$, $U-B=-0.97$) for a while.
Thus, we call this point ``point F'' in the color-color diagram
after free-free emission.  (3) After this stage, novae evolve leftward
(blueward in $B-V$ but almost constant in $U-B$) mainly because of
the development of strong emission lines.  In this work, 
we stopped following the color evolution when the $V$ magnitude drops by 
3 -- 4 mag from the maximum because strong emission lines make an
increasingly large contribution to the $(U-B)_0$ and $(B-V)_0$ colors,
and their effects cloud the overall evolution of colors.\\

\noindent
3. We can determine the color excess $E(B-V)$ of a target nova
by fitting its color evolution track 
with our general course in the color-color diagram.
This is a new and convenient method for obtaining accurate
color excesses for classical novae.  In this paper, we redetermined
the color excesses of 19 novae by fitting with our general track.
They are shown in Figures
\ref{color_color_diagram_templ_rs_oph_v446_her_v533_her} --
\ref{color_color_diagram_v382_vel_v2274_cyg_v475_sct_v5114_sgr}, 
and the obtained values are tabulated in Table
\ref{extinction_from_two_color_diagram}. \\

\noindent
4. Using a time-stretching method for nova light curves \citep{hac10k},
we can estimate the apparent distance modulus $(m-M)_V$ of a target nova.
We have confirmed that this time-stretching method is applicable  
for several novae with known distances.
Then, we determined the distance moduli of other novae,
which are shown in Figures 
\ref{v603_aql_gk_per_v1500_cyg_x55z02o10ne03_revised} --
\ref{v458_vul_v443_sct_pw_vul_v_bv_ub_color_logscale}, 
and the obtained values of $(m-M)_V$ are tabulated in Table
\ref{extinction_from_two_color_diagram}. 
Comparing some other available distance-reddening relations with our 
apparent distance modulus $(m-M)_V$ for a target nova, 
we examined and confirmed our estimated color excess for each nova
in Figures \ref{distance_reddening_rs_oph_v446_her_lv_vul_iv_cep} ---
\ref{distance_reddening_v382_vel_v2274_cyg_v475_sct_v5114_sgr}
and showed that our estimated values of $E(B-V)$ are
in good agreement with these relations, thus supporting the validity
of our new method. \\

\acknowledgments
     We are grateful to S. Shugarov and D. Chochol for permitting us to
use digital data for PU~Vul's $UBV$ magnitudes, and also to
R. Gonz\'alez-Riestra for providing us with her
new reddening estimates of V723 Cas 1995, and A. Cassatella
for providing us with their machine readable UV 1455 \AA\  data
for various novae.  We also thank
the American Association of Variable Star Observers
(AAVSO) and the Variable Star Observers League of Japan (VSOLJ)
for archival nova data.  We are also grateful
to the anonymous referee for useful comments to improve the manuscript.
This research used the NASA/IPAC Infrared Science Archive,
which is operated by the Jet Propulsion Laboratory, 
California Institute of Technology,
under contract with the National Aeronautics and Space Administration.
This research was supported in part by Grants-in-Aid for
Scientific Research (22540254, 24540227) 
from the Japan Society for the Promotion of Science.



%\appendix

\appendix

\section{Time-Stretching Method for Nova Light Curves}
\label{time-stretching_nova_light_curve}
\citet{hac06kb} found that nova light curves follow a universal 
decline law when free-free emission dominates the optical and IR
flux.  Using the universal decline law, \citet{hac10k} found that,
if two nova light curves overlap each other after one of 
the two is squeezed/stretched by a factor
of $f_s$ ($t'=t/f_s$) in the time direction, the brightnesses of the
two novae obey the relation of $m'_V = m_V - 2.5 \log f_s$.  
Using this result and calibrated nova light curves, we can estimate
the absolute magnitude of a target nova.
In the following, we determine the distance moduli of 28 novae.
The results are tabulated in Table \ref{extinction_from_two_color_diagram}.

%Fig.40
%\placefigure{v603_aql_gk_per_v1500_cyg_x55z02o10ne03_revised}

%%\begin{figure}
\begin{figure*}
\epsscale{1.0}
%\epsscale{1.15}
\plotone{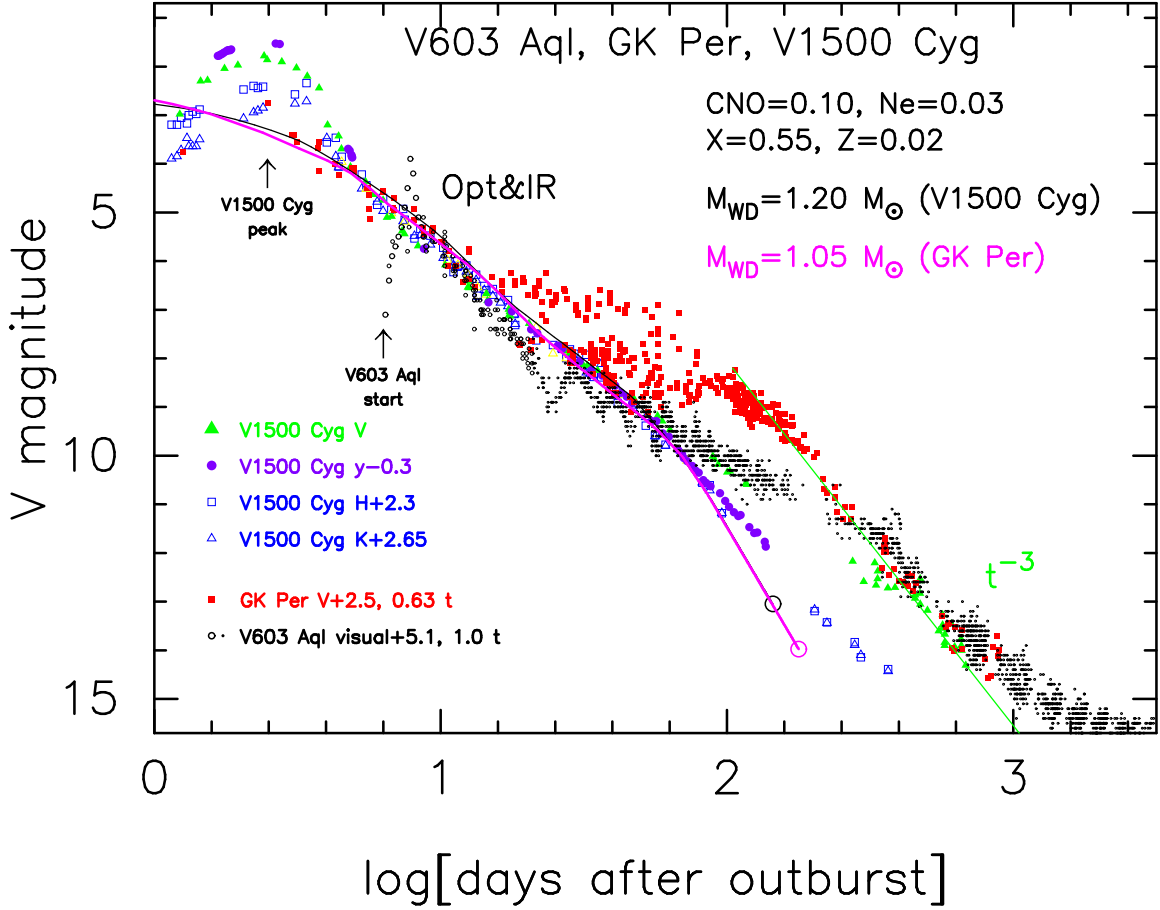}
%\plotone{v603_aql_gk_per_v1500_cyg_x55z02o10ne03_revised.epsi}
%\plotfiddle{evolution1.ps}{5.0cm}{270}{0.4}{0.4}{-170}{220}
\caption{
Light curves of V1500 Cyg, GK~Per, and V603~Aql.  Light curves of 
free-free emission are almost independent of wavelength, 
so the $V$, $y$, $H$, and $K$ light curves of V1500 Cyg 
almost overlap from several days to $\sim 65$ days 
after the outburst.  The $V$ light curve of GK Per is shifted
downward by $\Delta V=2.5$ mag and leftward by $\Delta \log t=-0.2$,
whereas the visual light curve of V603~Aql is shifted downward
by $\Delta V=5.1$ mag but not shifted in the time direction.
The smooth light curve of V1500 Cyg is 
fitted with the bottom line of the GK Per light curve 
during the transition oscillations of GK~Per, 
whereas it is fitted with the top line of the V603~Aql light curve 
during the transition oscillations of V603~Aql.
We added two model light curves of free-free emission for WDs with
masses of $1.2~M_\sun$ (black solid line) and $1.05~M_\sun$ 
(magenta solid line).  The light curve of the $1.05~M_\sun$ WD is
shifted against that of the $1.2~M_\sun$ WD 
by $\Delta \log t= \log f_s =-0.2$
and by $\Delta V=-2.5 \log f_s = 0.50$ mag to make them overlap,
as shown in the figure.
}
\label{v603_aql_gk_per_v1500_cyg_x55z02o10ne03_revised}
\end{figure*}
%%\end{figure}

First, we calibrate the distance modulus of V1500~Cyg with those of
GK~Per and V603~Aql, because the distances to GK~Per and V603~Aql
were recently obtained by \citet{har13} using trigonometric 
parallax, i.e., $d=477^{+28}_{-25}$~pc for GK~Per and 
$d=249^{+9}_{-8}$~pc for V603~Aql.  The value of GK~Per is consistent 
within a $1\sigma$ error with the previous estimate of $d=455$~pc
by \citet{sla95} using the nebular expansion parallax method.
The distance modulus of GK~Per is 
$(m-M)_V=5\log 477^{+28}_{-25}/10 + 3.1\times0.3 = 9.3\pm0.1$ for
$E(B-V)=0.3$ \citep{wu89}.
The distance modulus of V603~Aql is 
$(m-M)_V=5\log 249^{+9}_{-8}/10 + 3.1\times0.07 = 7.2\pm0.07$ for
$E(B-V)=0.07$ \citep{gal74}.

Figure \ref{v603_aql_gk_per_v1500_cyg_x55z02o10ne03_revised} shows
the light curves of V1500~Cyg in the $V$ and near-infrared (NIR) bands.
The NIR data are taken from \citet{enn77}, \citet{gal76}, and \citet{kaw76}
and the $y$ and $V$ data are from \citet{loc76} and \citet{tem79}.
When we shift each light curve of V1500~Cyg in the vertical direction, 
all the light curves merge into one line in the mid-time region.
This property was explained by \citet{hac06kb} as a characteristic
property of free-free emission.  
Figure \ref{v603_aql_gk_per_v1500_cyg_x55z02o10ne03_revised} also 
shows the light curves of GK~Per and V603~Aql.
The optical $V$ data for GK~Per are those in Figure 2
of \citet{hac07k}.
The data for V603~Aql are taken from \citet{cam23} and the AAVSO.
These light curves are shifted in the vertical direction so that they
overlap with the free-free emission model light curves
(black and magenta solid lines explained later).  
In addition, the light curve of 
GK~Per is shifted by $\Delta \log t=-0.2$
in the horizontal direction so that it overlaps 
with the final decline of the $t^{-3}$ law for V1500~Cyg shown
in Figure \ref{v603_aql_gk_per_v1500_cyg_x55z02o10ne03_revised}.
We do not shift the light curve of V603~Aql in the time direction
because its $t^{-3}$ slope accidentally overlaps with that of V1500~Cyg.
This means that the timescale of V603~Aql is almost the same as that of
V1500~Cyg. 
Thus, we can determine the time-scaling factors as 0.63 for GK~Per
and 1.0 for V603~Aql, and the vertical shifts as
$\Delta V=2.5$ mag for GK~Per and $\Delta V=5.1$ mag for V603~Aql.
The start time of V603~Aql is located in the middle
because of its smaller envelope mass.

GK~Per shows transition oscillations in the middle 
part of the light curve, and the bottom line of these oscillations
almost coincides with our free-free emission model light curve.  
V603~Aql also shows transition oscillations, and
its top line almost overlaps our model light curve.
Both of these light curves agree with that of V1500~Cyg
in the early phase and in the later phase of the $t^{-3}$ law. 
The time-stretching method proposed by \citet{hac10k}
is based on the relation 
$m'_V = m_V - 2.5 \log f_s$ between two light curves that overlap
each other.  In Figure \ref{v603_aql_gk_per_v1500_cyg_x55z02o10ne03_revised},
we obtain the following relation among the apparent distance moduli
of the three novae:
\begin{eqnarray}
(m-M)_{V, \rm V1500~Cyg} &=& 12.3\cr
&=&(m-M)_{V,\rm GK~Per} + \Delta V - 2.5 \log 0.63/1.0 \cr
&\approx& 9.3\pm0.1 + 2.5 + 0.50 = 12.3\pm0.1\cr
&=&(m-M)_{V,\rm V603~Aql} + \Delta V + 2.5 \log 1.0/1.0 \cr
&=& 7.2\pm0.07 + 5.1 - 0.0 = 12.3\pm0.07,
\label{v1500_cyg_distance_mod}
\end{eqnarray}
where we use distance moduli of 
$(m-M)_V=9.3\pm0.1$ for GK~Per and
$(m-M)_V=7.2\pm0.07$ for V603~Aql
from Harrison et al.'s results obtained using the trigonometric parallax
method.  The latter two values in the third and fifth lines of Equation 
(\ref{v1500_cyg_distance_mod}) agree very well with our distance modulus
in the first line of Equation (\ref{v1500_cyg_distance_mod})
presented in Section \ref{v1500_cyg}.  
Moreover, the mutual agreement in the two novae 
with their known trigonometric distances indicates 
that the time-stretching method is reliable.

Now we determine the WD masses of V1500~Cyg, GK~Per, and V603~Aql
using model light curve analysis of free-free emission.
The model light curves are taken from \citet{hac10k}.
The best-fit model is $M_{\rm WD}=1.20~M_\sun$ (black solid line)
for V1500~Cyg.  We select the $1.20~M_\sun$ WD from the position at
which a bend or break appears on the $y$, $H$, and $K$ light curves
at $\sim65$ days ($\log t_{\rm break}\sim1.8$) after
the outburst \citep[see][for more details]{hac06kb}.
Here we search for the best-fit model by eye by changing
the WD mass in $0.05~M_\sun$ steps.  This figure also shows the model 
light curve for a 1.05$~M_\sun$ WD shifted in the time direction by 
a scaling factor of 0.63 to overlap that of 1.2$~M_\sun$ WD. 
The resultant curve fits that of GK~Per shifted by the same
scaling factor of 0.63.
Therefore, we adopted $M_{\rm WD}=1.05~M_\sun$ for GK~Per.
For V603~Aql, the timescale is almost the same as that of V1500~Cyg,
The WD mass of V603~Aql is almost the same, i.e., $M_{\rm WD}=1.20~M_\sun$.
This is very consistent with the dynamical mass
of $M_{\rm WD}=1.2\pm0.2~M_\sun$ obtained by \citet{are00}.

\citet{hac10k} calibrated the absolute magnitudes of their free-free
model light curves.  Using their absolute magnitudes,
we obtained the distance modulus of GK~Per as
\begin{equation}
\left[ (m-M)_{V,\rm GK~Per} \right]_{\rm FF}
= m_{\rm w} - M_{\rm w}= 10.9 - 1.6 = 9.3
\label{model_fit_gk_per}
\end{equation}
from $m_{\rm w}= 10.9$ and $M_{\rm w}= 1.6$ ($M_{\rm WD}=1.05~M_\sun$)
in Table 3 of \citet{hac10k}.  We obtained the distance modulus of
V1500~Cyg as 
\begin{equation}
\left[ (m-M)_{V,\rm V1500~Cyg} \right]_{\rm FF}
= m_{\rm w} - M_{\rm w}= 13.0 - 0.7 = 12.3
\label{model_fit_v1500_cyg}
\end{equation}
from $m_{\rm w}= 13.0$ and $M_{\rm w}= 0.7$ ($M_{\rm WD}=1.20~M_\sun$)
in Table 3 of \citet{hac10k}.  Here, $m_{\rm w}$/$M_{\rm w}$ 
is the apparent/absolute magnitude at the end point of
optically-thick winds (denoted by a large open circle at the end
of each black or magenta free-free model light curve). 
The value of $m_{\rm w}$ is read directly from Figure 
\ref{v603_aql_gk_per_v1500_cyg_x55z02o10ne03_revised}
and $M_{\rm w}$ was calibrated and given in Table 3 of \citet{hac10k}. 
These results from Equations (\ref{model_fit_gk_per}) and
(\ref{model_fit_v1500_cyg}) are consistent with the results 
of Equation (\ref{v1500_cyg_distance_mod}).  Thus, we confirm
the validity of our time-stretching method; at the same time, we
show that the absolute magnitudes of our model light curves, which are
based on the free-free emission light curves, are consistently
calibrated with the absolute magnitudes of GK~Per and V603~Aql,
which are based on trigonometric parallax.

\begin{figure}
%%\epsscale{0.75}
\epsscale{1.0}
%%\epsscale{1.15}
\plotone{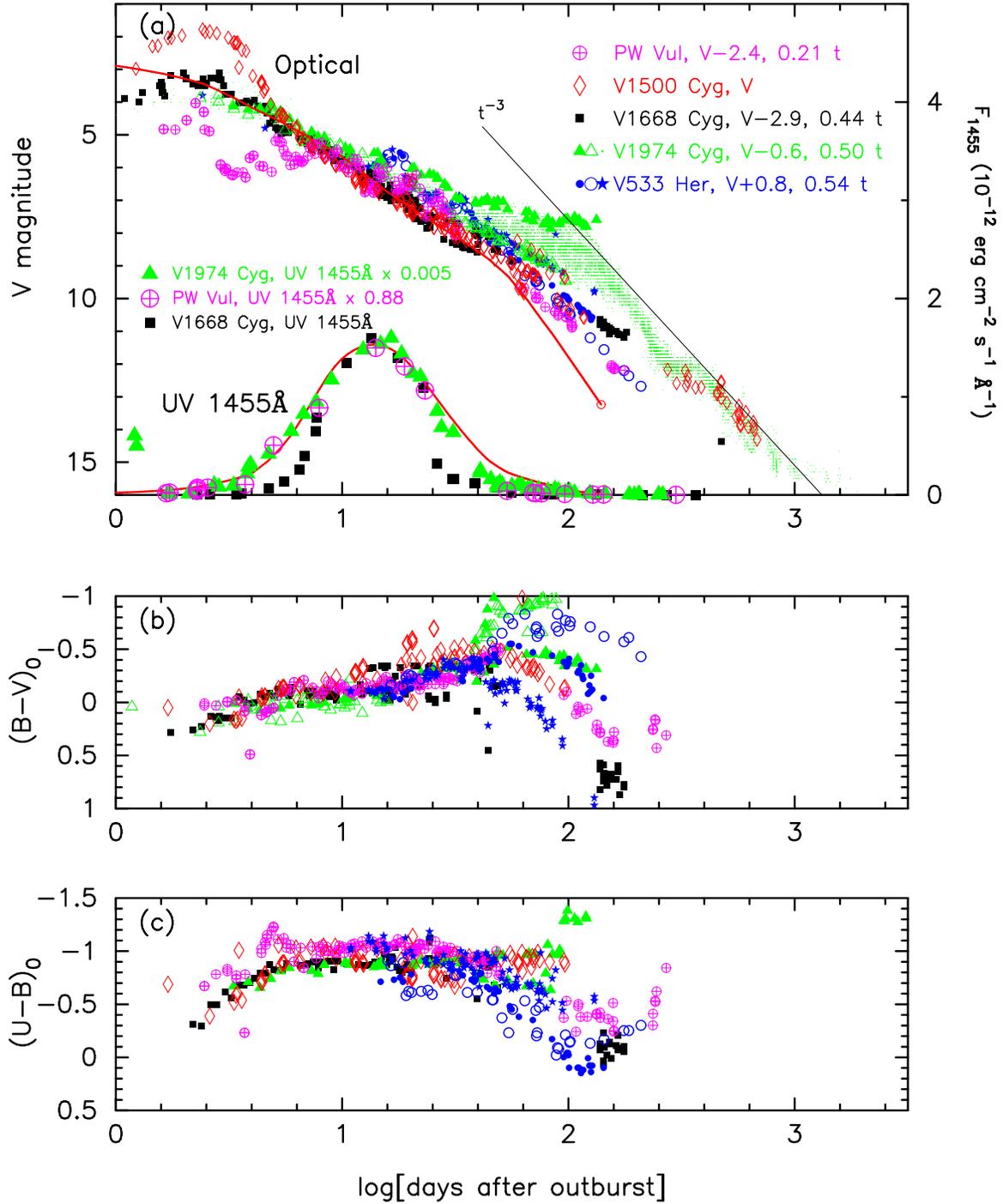}
%\plotone{pw_vul_v1668_cyg_v1500_cyg_v1974_cyg_v_bv_ub_color_curve_logscale_no2.epsi}
%\plotfiddle{evolution1.ps}{5.0cm}{270}{0.4}{0.4}{-170}{220}
\caption{
(a) $V$ band and UV 1455\AA\ narrow band light curves, 
(b) $(B-V)_0$, and (c) $(U-B)_0$ color curves
for PW~Vul (magenta open circles with plus sign), 
V1500~Cyg (red open diamonds), V1668~Cyg (black filled squares), 
and V1974~Cyg (green open and filled triangles).
Those of V533~Her are also shown for comparison.
Blue filled circles are taken from \citet{gen63}, 
blue open circles are from \citet{chi64}, and blue star symbols 
are from \citet{she64}.
To make them overlap in the early decline phase,
we stretched the light curves of PW~Vul, V1668~Cyg, 
V1974~Cyg, and V533~Her by 0.21, 0.44, 0.50, and 0.54, 
and shifted their magnitudes by $-2.4$, $-2.9$, $-0.6$, 
and $+0.8$ mag, respectively, as indicated in the figure.
UV~1455\AA\  fluxes of each nova are also rescaled as indicated
in the figure. 
\label{pw_vul_v1668_cyg_v1500_cyg_v1974_cyg_v_bv_ub_color_curve_logscale_no2}}
\end{figure}

Figure 
\ref{pw_vul_v1668_cyg_v1500_cyg_v1974_cyg_v_bv_ub_color_curve_logscale_no2}
shows the $V$ light curves and $(B-V)_0$ and $(U-B)_0$ color evolution
of PW~Vul, V1500~Cyg, V1668~Cyg, and V1974~Cyg as well
as those of V533~Her.
We squeezed the light curves of PW~Vul, V1668~Cyg, V1974~Cyg, and V533~Her
in the time direction by factors of 0.21, 0.44, 0.50, and 0.54
and shifted them by $\Delta V=-1.6$, $\Delta V=-2.6$, $\Delta V=-0.4$, 
and $\Delta V=+0.8$~mag, respectively, against that of V1500~Cyg.  
Then, we obtained the apparent distance modulus in the $V$ band as
\begin{eqnarray}
(m-M)_{V, \rm V1500~Cyg} &=& 12.3\cr
&=&(m-M)_{V,\rm PW~Vul} + \Delta V - 2.5 \log 0.21/1.0 \cr
&\approx& 13.0 - 2.4 + 1.7 = 12.3\cr
&=&(m-M)_{V,\rm V1668~Cyg} + \Delta V - 2.5 \log 0.44/1.0 \cr
&\approx& 14.25 - 2.9 + 0.90 = 12.25\cr
&=& (m-M)_{V,\rm V1974~Cyg} + \Delta V - 2.5 \log 0.50/1.0 \cr
&\approx& 12.2 - 0.6 + 0.75 = 12.35\cr
&=& (m-M)_{V,\rm V533~Her} + \Delta V - 2.5 \log 0.54/1.0 \cr
&\approx& 10.8 + 0.8 + 0.67 = 12.27,
\end{eqnarray}
where the apparent distance moduli of PW~Vul, V1500~Cyg, V1668~Cyg, and
V1974~Cyg were already obtained as
$(m-M)_{V,\rm PW~Vul} = 13.0$ in Section \ref{pw_vul},
$(m-M)_{V,\rm V1500~Cyg} = 12.3$ in Section \ref{v1500_cyg},
$(m-M)_{V,\rm V1668~Cyg} = 14.25$ in Section \ref{v1668_cyg}, and
$(m-M)_{V,\rm V1974~Cyg} = 12.2$ in Section \ref{v1974_cyg}.
These are all consistent with each other.
Therefore, we can use this method to obtain the distance modulus
of a target nova; the distance modulus of V533~Her is found 
to be $(m-M)_{V,\rm V533~Her} = 10.8$.

In Figure 
\ref{pw_vul_v1668_cyg_v1500_cyg_v1974_cyg_v_bv_ub_color_curve_logscale_no2},
we also plotted UV~1455\AA\  fluxes for PW~Vul (magenta open circles
with plus sign), V1668~Cyg (black filled squares), and 
V1974~Cyg (green filled triangles).
UV~1455\AA\  fluxes of V1974~Cyg and PW~Vul are scaled down by 0.005
and 0.88 to fit its shape with those of V1668~Cyg.
In this figure we adopted $M_{\rm WD}=0.83~M_\sun$ for PW~Vul
as a best-fit model for our assumed chemical composition of 
$X=0.55$, $Y=0.23$, $Z=0.02$, and $X_{\rm CNO}=0.20$
(magenta solid lines). 
The distance-reddening relation (magenta solid line in Figure 
\ref{distance_reddening_nq_vul_v1370_aql_gq_mus_pw_vul}(d)) of 
UV~1455\AA\  flux fitting is calculated from 
Equation (\ref{v1668_cyg_uv1455_distance_modulus}) for PW~Vul.
We adopted $F_{\lambda}^{\rm obs}= 4.5\times10^{-12}$ 
erg~s$^{-1}$~cm$^{-2}$~\AA$^{-1}$, 
which is the observed flux at $\lambda=1455$\AA\  corresponding to the
upper bound of Figure
\ref{pw_vul_v1668_cyg_v1500_cyg_v1974_cyg_v_bv_ub_color_curve_logscale_no2}(a),
and $F_{\lambda}^{\rm mod}= 1.1\times10^{-11}$ 
erg~s$^{-1}$~cm$^{-2}$~\AA$^{-1}$, which is the model flux of the
$0.83~M_\sun$ WD corresponding to the upper bound of Figure
\ref{pw_vul_v1668_cyg_v1500_cyg_v1974_cyg_v_bv_ub_color_curve_logscale_no2}(a).

\begin{figure}
%%\epsscale{0.75}
\epsscale{1.0}
%%\epsscale{1.15}
\plotone{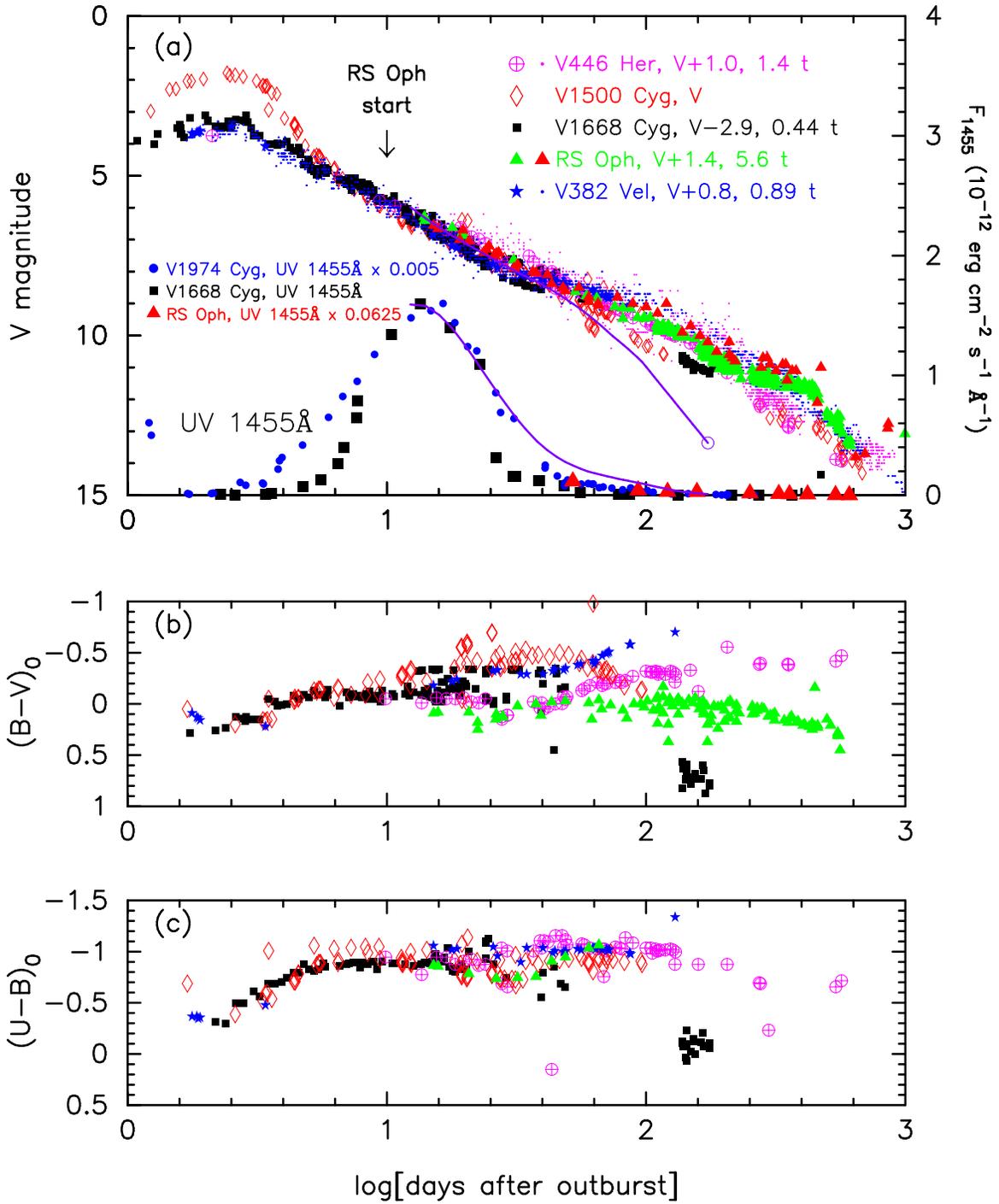}
%\plotone{rs_oph_v446_her_v1500_cyg_v1668_cyg_v382_vel_v_color_logscale.epsi}
%\plotfiddle{evolution1.ps}{5.0cm}{270}{0.4}{0.4}{-170}{220}
\caption{
Same as Figure 
\ref{pw_vul_v1668_cyg_v1500_cyg_v1974_cyg_v_bv_ub_color_curve_logscale_no2}, 
but for V446~Her, V1500~Cyg, V1668~Cyg, RS~Oph, and V382~Vel.
Time-stretching factors and vertical shifts of these novae are shown in
the figure.  We arranged the origin of the time of RS~Oph according to
that of V1668~Cyg.  The UV~1455\AA\ fluxes of V1668~Cyg, V1974~Cyg,
and RS~Oph are added.  Model light curve is also added for RS~Oph 
($1.37~M_\sun$ WD, purple solid lines).  See text for more details. 
\label{rs_oph_v446_her_v1500_cyg_v1668_cyg_v382_vel_v_color_logscale}}
\end{figure}

In this way, we obtained the distance moduli for various novae studied
in Section \ref{extinction_novae}, using the time-stretching method.
From Figure 
\ref{rs_oph_v446_her_v1500_cyg_v1668_cyg_v382_vel_v_color_logscale}, 
we obtain the apparent distance moduli in the $V$ band as
\begin{eqnarray}
(m-M)_{V, \rm RS~Oph} &=& 12.8 \cr
&=& (m-M)_{V,\rm V1500~Cyg} + \Delta V - 2.5 \log 1.0/5.6\cr
&=& 12.3 + (+0.0 - 1.4) + 1.88 = 12.78\cr
&=& (m-M)_{V,\rm V1668~Cyg} + \Delta V - 2.5 \log 0.44/5.6\cr
&\approx& 14.25 + (-2.9 - 1.4) + 2.76 = 12.71\cr
&=& (m-M)_{V,\rm V446~Her} + \Delta V - 2.5 \log 1.41/5.6\cr
&\approx& 11.7 + (+1.0 - 1.4) + 1.50 = 12.80\cr
&=& (m-M)_{V,\rm V382~Vel} + \Delta V - 2.5 \log 0.89/5.6\cr
&\approx& 11.4 + (+0.8 - 1.4) + 2.0 = 12.80.
\label{rs_oph_v446_her_v1500_cyg_v1668_cyg_v382_vel_distance}
\end{eqnarray}
In Figure \ref{rs_oph_v446_her_v1500_cyg_v1668_cyg_v382_vel_v_color_logscale}, 
we also plotted the UV~1455\AA\  fluxes of the 1985 RS~Oph outburst (red 
filled triangles) together with those of V1668~Cyg (black filled squares)
and V1974~Cyg (blue filled circles).
The UV~1455\AA\  flux of RS~Oph is scaled down by 0.0625$(=1/16)$ to fit its
shape with those of V1974~Cyg/V1668~Cyg.
In this figure we adopted $M_{\rm WD}=1.37~M_\sun$ for RS~Oph
as a best-fit model for our assumed chemical composition of 
$X=0.55$, $Y=0.23$, $Z=0.02$, and $X_{\rm CNO}=0.20$
(magenta solid lines). 
The distance-reddening relation (magenta solid line in Figure 
\ref{distance_reddening_rs_oph_v446_her_lv_vul_iv_cep}(a)) of 
UV~1455\AA\  flux fitting is calculated from 
Equation (\ref{v1668_cyg_uv1455_distance_modulus}) for RS~Oph.
We adopted $F_{\lambda}^{\rm obs}= 5.0\times10^{-11}$ 
erg~s$^{-1}$~cm$^{-2}$~\AA$^{-1}$, 
which is the observed flux at $\lambda=1455$\AA\  corresponding to the
upper bound of Figure
\ref{rs_oph_v446_her_v1500_cyg_v1668_cyg_v382_vel_v_color_logscale}(a),
and $F_{\lambda}^{\rm mod}= 1.7\times10^{-11}$ 
erg~s$^{-1}$~cm$^{-2}$~\AA$^{-1}$, which is the model flux of the
$1.37~M_\sun$ WD corresponding to the upper bound of Figure
\ref{rs_oph_v446_her_v1500_cyg_v1668_cyg_v382_vel_v_color_logscale}(a).

\begin{figure}
%%\epsscale{0.75}
\epsscale{1.0}
%%\epsscale{1.15}
\plotone{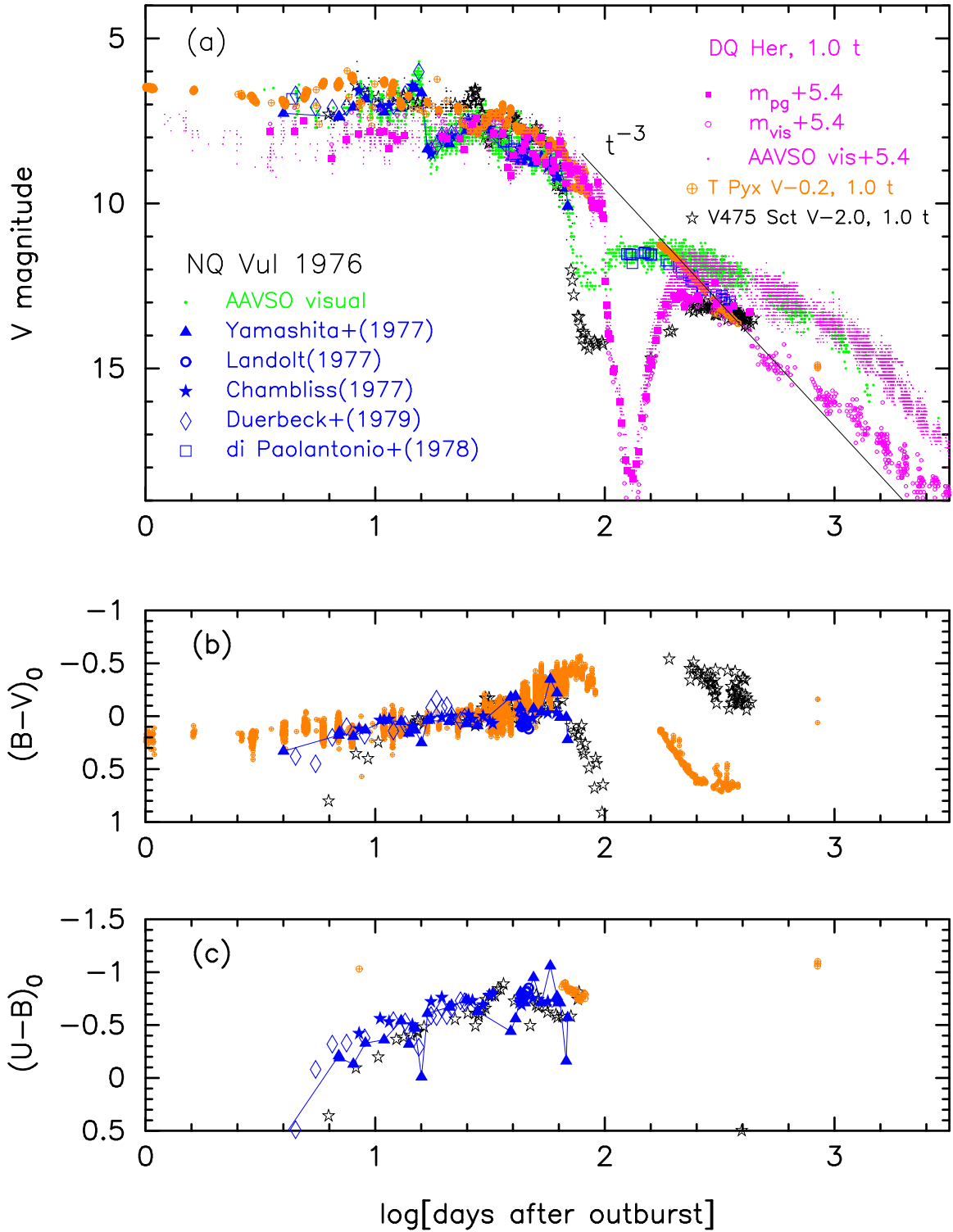}
%\plotone{v475_sct_t_pyx_nq_vul_dq_her_v_bv_ub_color_logscale_no6.epsi}
%\plotfiddle{evolution1.ps}{5.0cm}{270}{0.4}{0.4}{-170}{220}
\caption{
Same as Figure 
\ref{pw_vul_v1668_cyg_v1500_cyg_v1974_cyg_v_bv_ub_color_curve_logscale_no2}, 
but for DQ~Her, NQ~Vul, V475~Sct, and T~Pyx.  For T~Pyx, 
we take the start of the day about 20 days after the outburst. 
\label{v475_sct_t_pyx_nq_vul_dq_her_v_bv_ub_color_logscale_no6}}
\end{figure}

In Figure
\ref{v475_sct_t_pyx_nq_vul_dq_her_v_bv_ub_color_logscale_no6},
the $V$ light curve of T~Pyx almost overlaps that of DQ~Her
from the maximum to the final $t^{-3}$ decline law (thin solid black line)
except for the dust blackout period of DQ~Her.
The final $t^{-3}$ decline law can be explained as free-free emission
from a homologously expanding nebula \citep[see, e.g.,][]{hac06kb}
after the optically-thick winds ended.
In this figure, we took a start of day about 20 days after
the outburst of T~Pyx in order to make the final declines of each
nova overlap.  Because the overall timescales of these four novae
(DQ~Her, NQ~Vul, V475~Sct, and T~Pyx) are almost the same,
we consider that their brightnesses are almost the same,
i.e., 
\begin{eqnarray}
(m-M)_{V, \rm T~Pyx} &=& 13.8 \cr
&=& (m-M)_{V,\rm DQ~Her} + \Delta V \cr
&=& 8.2 + (+5.4 + 0.2) =13.8 \cr
&=& (m-M)_{V,\rm NQ~Vul} + \Delta V \cr
&=& 13.6 + (+0.0 + 0.2) =13.8 \cr
&=& (m-M)_{V,\rm V475~Sct} + \Delta V \cr
&=& 15.6 + (-2.0 + 0.2) =13.8.
\label{v475_sct_t_pyx_nq_vul_dq_her_distance}
\end{eqnarray}

\begin{figure}
%%\epsscale{0.75}
\epsscale{1.0}
%%\epsscale{1.15}
\plotone{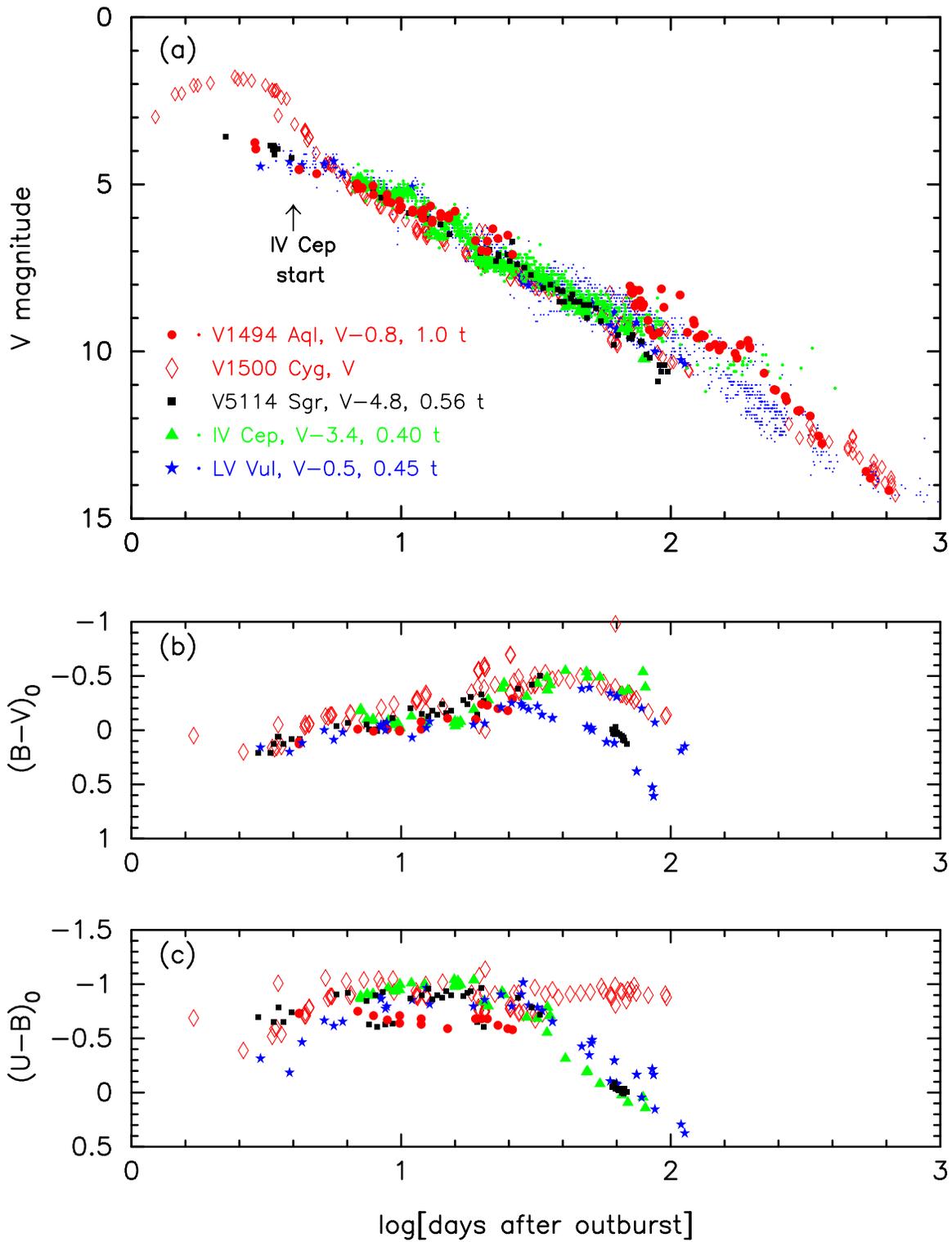}
%\plotone{v5114_sgr_v1500_cyg_v1494_aql_iv_cep_lv_vul_v_color_logscale.epsi}
%\plotfiddle{evolution1.ps}{5.0cm}{270}{0.4}{0.4}{-170}{220}
\caption{
Same as Figure 
\ref{pw_vul_v1668_cyg_v1500_cyg_v1974_cyg_v_bv_ub_color_curve_logscale_no2}, 
but for LV~Vul (blue filled star symbols and dots), 
IV~Cep (green filled triangles and dots), 
V1494~Aql (red filled circles), V5114~Sgr (black filled squares),
and V1500~Cyg (red open diamonds).
The visual data (small dots) of each nova are taken from 
the AAVSO archive.  To make them overlap in the early decline phase,
we stretched the light curves of LV~Vul, IV~Cep, 
V1494~Aql, and V5114~Sgr by 0.45, 0.40, 1.0, and 0.56,
and shifted their magnitudes up by $0.5$,  $3.4$, 
$0.8$, and $4.8$~mag, respectively, against V1500~Cyg, 
as indicated in the figure.
\label{v5114_sgr_v1500_cyg_v1494_aql_iv_cep_lv_vul_v_color_logscale}}
\end{figure}

From Figure 
\ref{v5114_sgr_v1500_cyg_v1494_aql_iv_cep_lv_vul_v_color_logscale},
we derive
\begin{eqnarray}
(m-M)_{V, \rm LV~Vul} &=&11.9\cr
&=& (m-M)_{V,\rm V1500~Cyg} + \Delta V - 2.5 \log 1.0/0.45 \cr
&\approx& 12.3 + (+0.0 + 0.5) - 0.87 = 11.93\cr
&=& (m-M)_{V,\rm IV~Cep} + \Delta V - 2.5 \log 0.40/0.45 \cr
&\approx& 14.7 + (-3.4 + 0.5) + 0.13 = 11.93\cr
&=& (m-M)_{V,\rm V1494~Aql} + \Delta V - 2.5 \log 1.0/0.45 \cr
&\approx& 13.1 + (-0.8 + 0.5) - 0.87 = 11.93\cr
&=& (m-M)_{V,\rm V5114~Sgr} + \Delta V - 2.5 \log 0.56/0.45 \cr
&\approx& 16.5 + (-4.8 + 0.5) - 0.24 = 11.96.
\label{v5114_sgr_v1500_cyg_v1494_aql_iv_cep_lv_vul_distance}
\end{eqnarray}

\begin{figure}
%%\epsscale{0.75}
\epsscale{1.0}
%%\epsscale{1.15}
\plotone{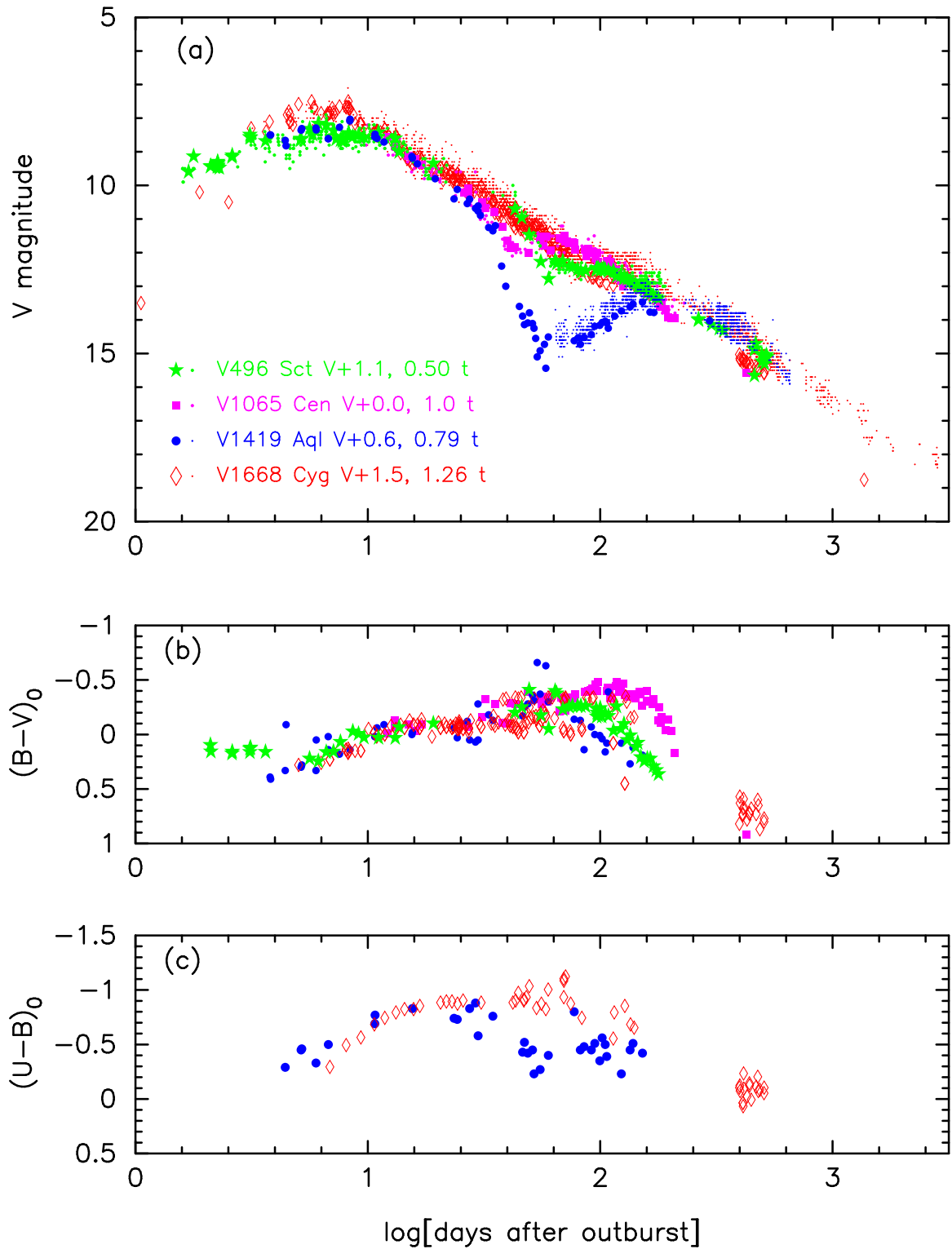}
%\plotone{v496_sct_v1065_cen_v1419_aql_v1668_cyg_v_bv_ub_color_logscale.epsi}
%\plotfiddle{evolution1.ps}{5.0cm}{270}{0.4}{0.4}{-170}{220}
\caption{
Same as Figure 
\ref{pw_vul_v1668_cyg_v1500_cyg_v1974_cyg_v_bv_ub_color_curve_logscale_no2}, 
but for V496~Sct (green star symbols and dots),
V1065~Cen (magenta filled squares and dots), 
V1419~Aql (blue filled circles and dots), 
and V1668~Cyg (red open diamonds and dots).
The time of V1668~Cyg, V1419~Aql, V1065~Cen, and V496~Sct are 
stretched by a factor of 1.26, 0.79, 1.0, and 0.50, respectively.
We shifted the visual and $V$ light curves of V1668~Cyg, V1419~Aql,
V1065~Cen, and V496~Sct, down by 1.5, 0.6, 0.0, and 1.1 mag,
respectively.  These four $V$ light curves almost overlap 
in the early decline phase.
\label{v496_sct_v1065_cen_v1419_aql_v1668_cyg_v_bv_ub_color_logscale}}
\end{figure}

From Figure 
\ref{v496_sct_v1065_cen_v1419_aql_v1668_cyg_v_bv_ub_color_logscale},
\begin{eqnarray}
(m-M)_{V, \rm V1419~Aql} &=&14.6\cr
&=& (m-M)_{V,\rm V1668~Cyg} + \Delta V - 2.5 \log 1.26/0.79 \cr
&\approx& 14.25 + (+1.5 - 0.6) - 0.51 = 14.64\cr
&=& (m-M)_{V,\rm V1065~Cen} + \Delta V - 2.5 \log 1.0/0.79 \cr
&=& 15.5 + (+0.0 - 0.6) - 0.26 = 14.64\cr
&=& (m-M)_{V,\rm V496~Sct} + \Delta V - 2.5 \log 0.50/0.79 \cr
&\approx& 13.6 + (+1.1 - 0.6) + 0.50 = 14.59.
\label{v496_sct_v1065_cen_v1419_aql_v1668_cyg_lv_vul_distance}
\end{eqnarray}

\begin{figure}
%%\epsscale{0.75}
\epsscale{1.0}
%%\epsscale{1.15}
\plotone{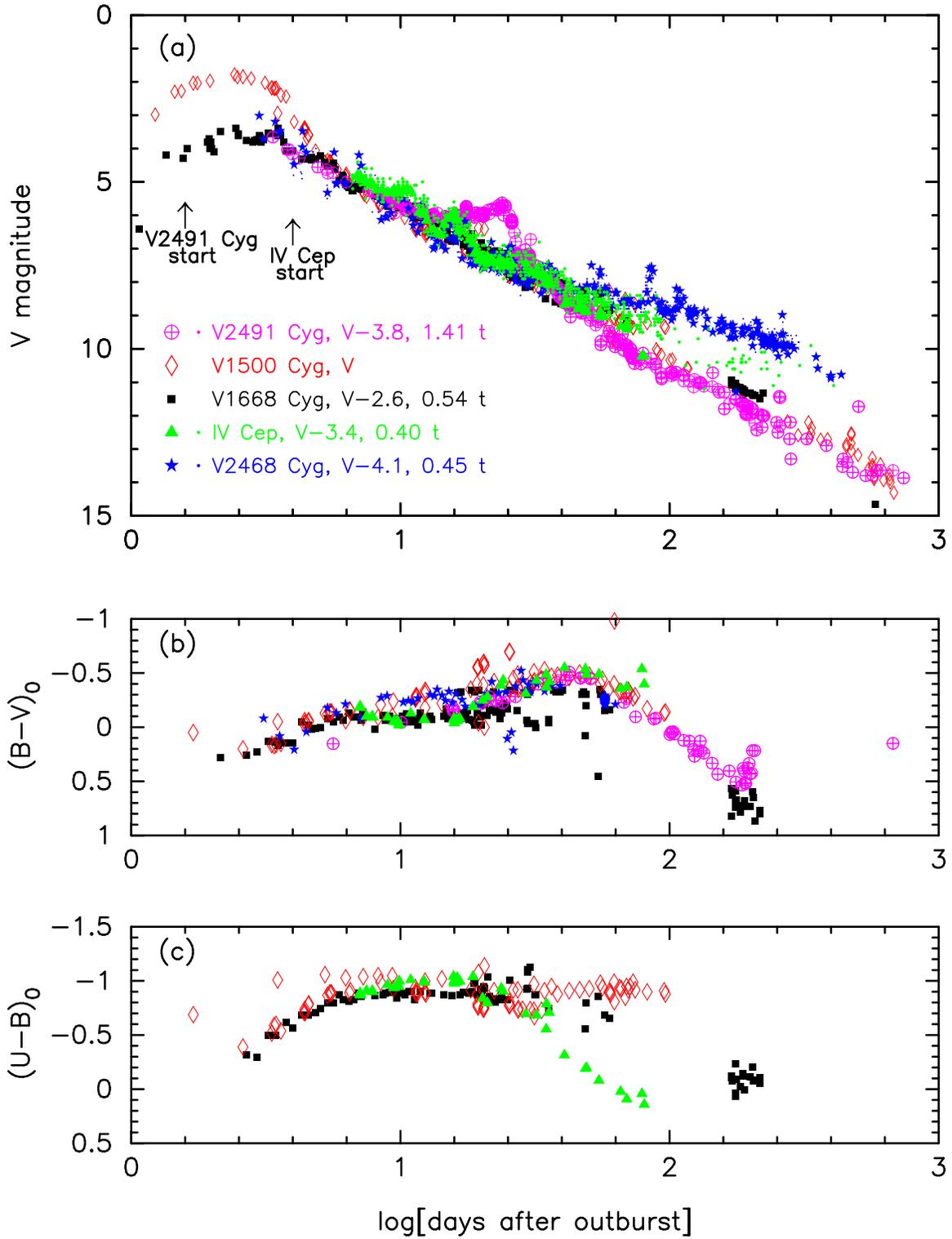}
%\plotone{v2491_cyg_v2468_cyg_v1500_cyg_v1668_cyg_iv_cep_v_color_logscale.epsi}
%\plotfiddle{evolution1.ps}{5.0cm}{270}{0.4}{0.4}{-170}{220}
\caption{
Same as Figure 
\ref{pw_vul_v1668_cyg_v1500_cyg_v1974_cyg_v_bv_ub_color_curve_logscale_no2}, 
but for V2491~Cyg, V1500~Cyg, V1668~Cyg, IV~Cep, and V2468~Cyg.
The start of the outburst is indicated by an arrow for IV~Cep and V2491~Cyg.
\label{v2491_cyg_v2468_cyg_v1500_cyg_v1668_cyg_iv_cep_v_color_logscale}}
\end{figure}

From Figure
\ref{v2491_cyg_v2468_cyg_v1500_cyg_v1668_cyg_iv_cep_v_color_logscale},
\begin{eqnarray}
(m-M)_{V, \rm IV~Cep} &=& 14.7\cr
&=& (m-M)_{V,\rm V1500~Cyg} + \Delta V - 2.5 \log 1.0/0.40 \cr
&\approx& 12.3 + (+0.0 + 3.4) - 0.99 = 14.71 \cr
&=& (m-M)_{V,\rm V1668~Cyg} + \Delta V - 2.5 \log 0.54/0.40 \cr
&\approx& 14.25 + (-2.6 + 3.4)- 0.33 = 14.72 \cr
&=& (m-M)_{V,\rm V2468~Cyg} + \Delta V - 2.5 \log 0.45/0.40 \cr
&\approx& 15.6 + (-4.1 + 3.4)- 0.13 = 14.77 \cr
&=& (m-M)_{V,\rm V2491~Cyg} + \Delta V - 2.5 \log 1.41/0.40 \cr
&\approx& 16.5 + (-3.8 + 3.4) - 1.34 = 14.73,
\label{v2491_cyg_v2468_cyg_v1500_cyg_v1668_cyg_iv_cep_distance}
\end{eqnarray}

\begin{figure}
%%\epsscale{0.75}
\epsscale{1.0}
%%\epsscale{1.15}
\plotone{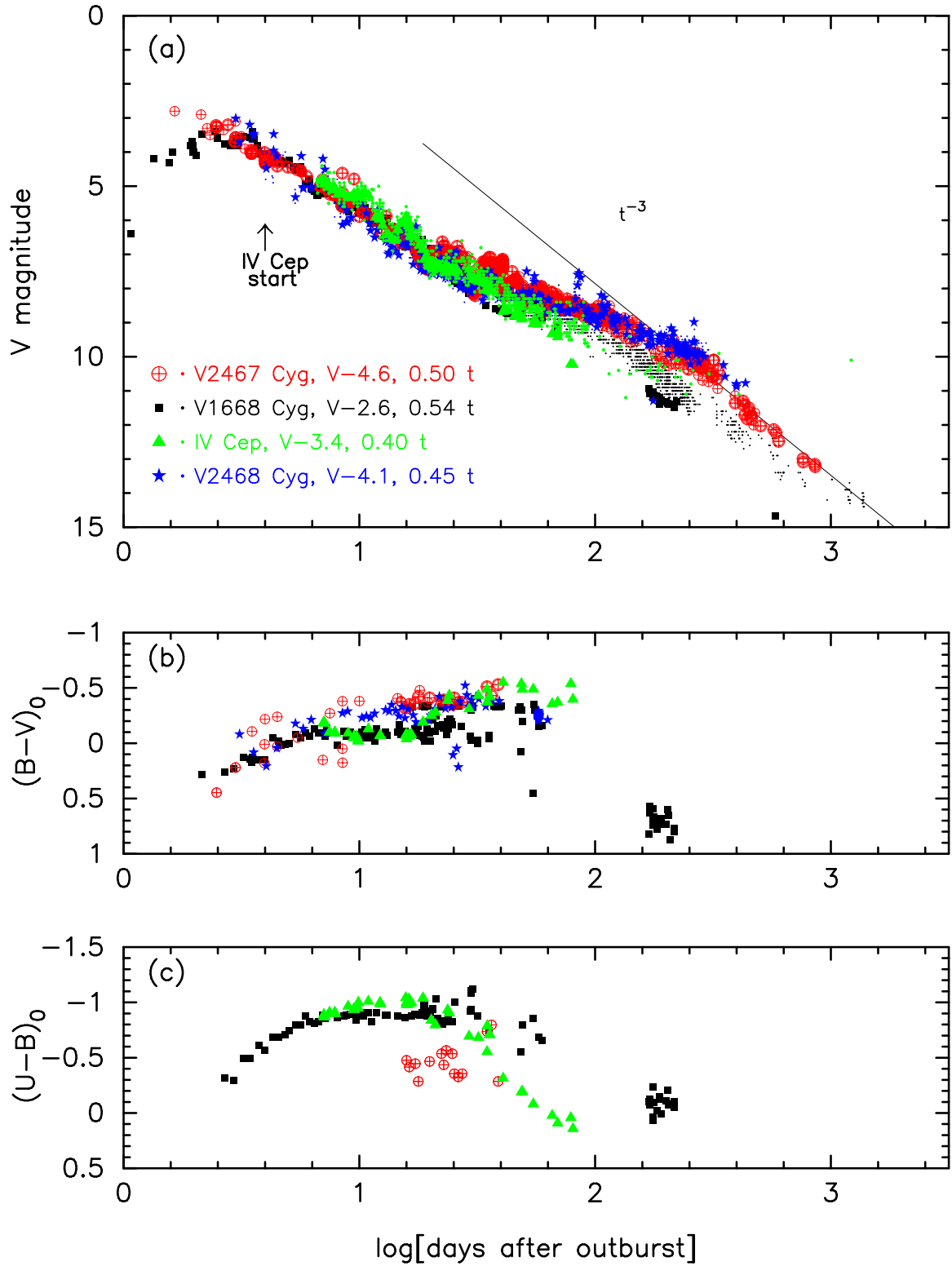}
%\plotone{v2467_cyg_v2468_cyg_v1668_cyg_iv_cep_v_color_logscale.epsi}
%\plotfiddle{evolution1.ps}{5.0cm}{270}{0.4}{0.4}{-170}{220}
\caption{
Same as Figure 
\ref{pw_vul_v1668_cyg_v1500_cyg_v1974_cyg_v_bv_ub_color_curve_logscale_no2},
but for V2467~Cyg, V1668~Cyg, IV~Cep, and V2468~Cyg.
\label{v2467_cyg_v2468_cyg_v1668_cyg_iv_cep_v_color_logscale}}
\end{figure}

From Figure
\ref{v2467_cyg_v2468_cyg_v1668_cyg_iv_cep_v_color_logscale},
\begin{eqnarray}
(m-M)_{V, \rm IV~Cep} &=& 14.7\cr
&=& (m-M)_{V,\rm V1668~Cyg} + \Delta V - 2.5 \log 0.54/0.40 \cr
&\approx& 14.25 + (-2.6 + 3.4)- 0.33 = 14.72 \cr
&=& (m-M)_{V,\rm V2467~Cyg} + \Delta V - 2.5 \log 0.50/0.40 \cr
&\approx& 16.2 + (-4.6 + 3.4)- 0.24 = 14.76 \cr
&=& (m-M)_{V,\rm V2468~Cyg} + \Delta V - 2.5 \log 0.45/0.40 \cr
&\approx& 15.6 + (-4.1 + 3.4)- 0.13 = 14.77.
\end{eqnarray}

\begin{figure}
%%\epsscale{0.75}
\epsscale{1.0}
%%\epsscale{1.15}
\plotone{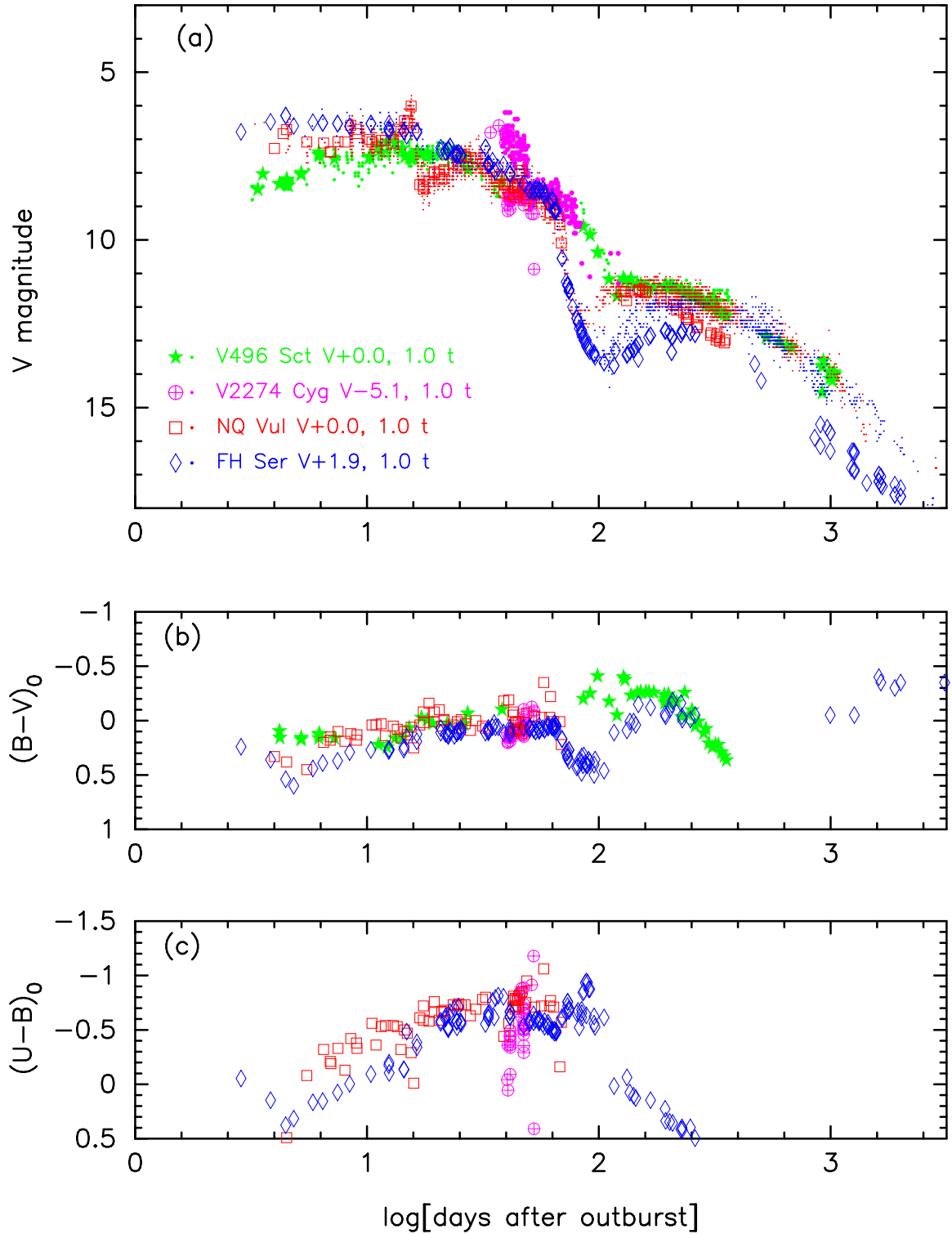}
%\plotone{v496_sct_v2274_cyg_fh_ser_nq_vul_v_bv_ub_color_logscale.epsi}
%\plotfiddle{evolution1.ps}{5.0cm}{270}{0.4}{0.4}{-170}{220}
\caption{
Same as Figures 
\ref{v475_sct_t_pyx_nq_vul_dq_her_v_bv_ub_color_logscale_no6},
but for V496~Sct (green symbols), V2274~Cyg (magenta symbols), 
NQ~Vul (red symbols, including the data from all the authors in Figure
\ref{v475_sct_t_pyx_nq_vul_dq_her_v_bv_ub_color_logscale_no6}), and
FH~Ser (blue symbols, including the data from all the authors in Figure
\ref{v475_sct_t_pyx_nq_vul_dq_her_v_bv_ub_color_logscale_no6}).
The data for V2274~Cyg are taken from \citet{vol02a},
\citet{vol02b}, and IAU Circular Nos. 7666 and 7668
(open circles with a plus sign inside), and the AAVSO archive (small dots).  
\label{v496_sct_v2274_cyg_fh_ser_nq_vul_v_bv_ub_color_logscale}}
\end{figure}

From Figure
\ref{v496_sct_v2274_cyg_fh_ser_nq_vul_v_bv_ub_color_logscale},
\begin{eqnarray}
(m-M)_{V, \rm V496~Sct}&=&13.6 \cr 
&=& (m-M)_{V,\rm FH~Ser} + \Delta V \cr
&=& 11.7 + (+1.9 - 0.0) = 13.6 \cr
&=& (m-M)_{V,\rm NQ~Vul} + \Delta V \cr
&=& 13.6 + (+0.0 - 0.0) = 13.6 \cr
&=& (m-M)_{V,\rm V2274~Cyg} + \Delta V \cr
&=& 18.7 + (-5.1 - 0.0) = 13.6.
\label{v496_sct_v2274_cyg_fh_ser_nq_vul_distance}
\end{eqnarray}

\begin{figure}
%%\epsscale{0.75}
\epsscale{1.0}
%%\epsscale{1.15}
\plotone{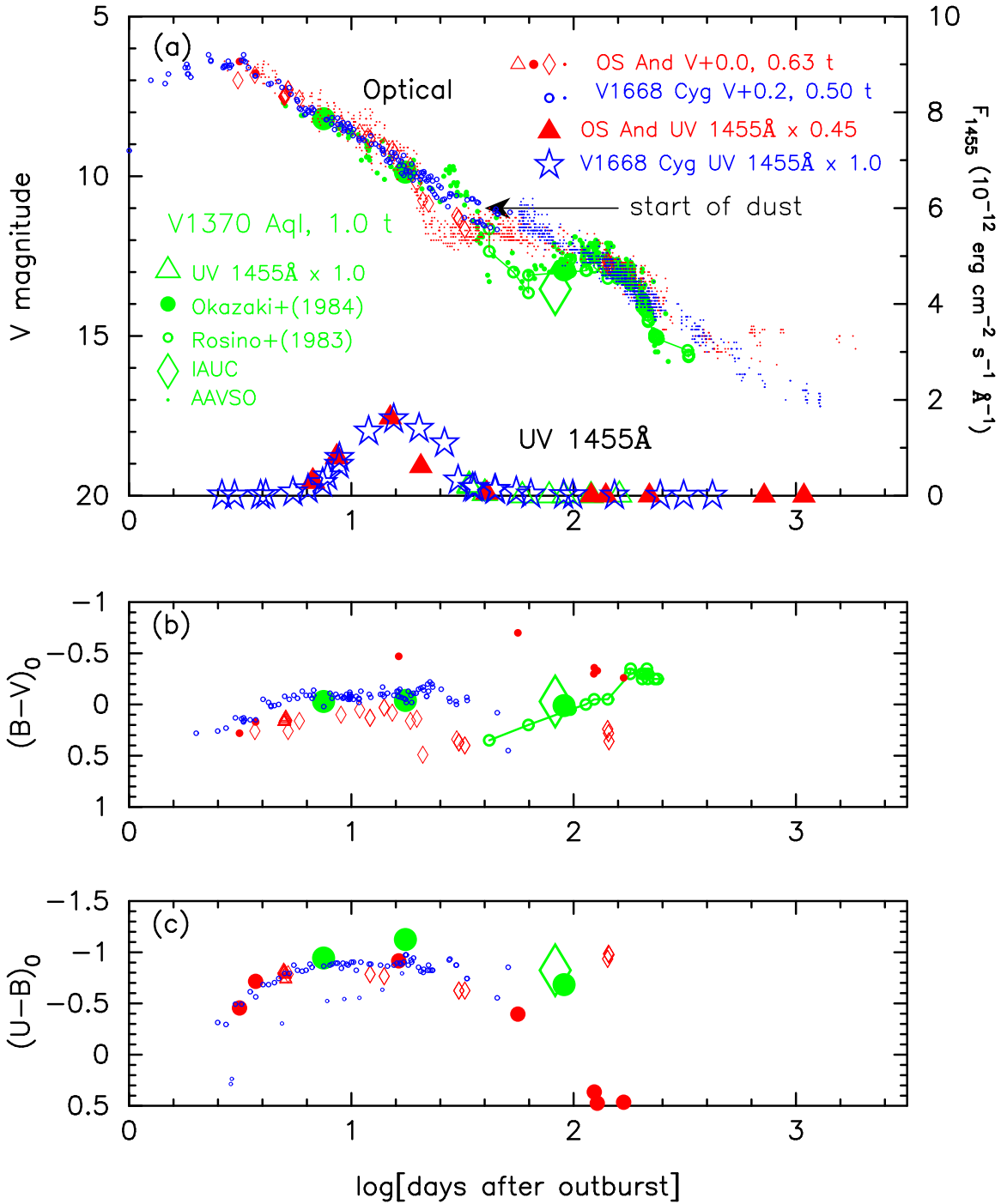}
%\plotone{v1370_aql_v1668_cyg_os_and_v_bv_ub_color_logscale.epsi}
%\plotfiddle{evolution1.ps}{5.0cm}{270}{0.4}{0.4}{-170}{220}
\caption{
Same as Figure 
\ref{pw_vul_v1668_cyg_v1500_cyg_v1974_cyg_v_bv_ub_color_curve_logscale_no2},
but for V1370~Aql,
V1668~Cyg, and OS~And.  The $V$ light curves of V1668~Cyg and OS~And
are squeezed by factors of 0.50 and 0.63 and shifted down by 0.2 
and 0.0 mag, respectively, against that of V1370~Aql.  
The UV~1455\AA\  data for OS~And and V1668~Cyg are also plotted.
The flux scale is linear between $0.0$ and 
$1.0\times 10^{-11}$ erg~s$^{-1}$~cm$^{-11}$~\AA$^{-1}$
for V1668~Cyg (from the bottom to the top of the frame)
but scaled down by 2.2 times for OS~And to match them with 
those of V1668~Cyg.
\label{v1370_aql_v1668_cyg_os_and_v_bv_ub_color_logscale}}
\end{figure}

From Figure
\ref{v1370_aql_v1668_cyg_os_and_v_bv_ub_color_logscale},
\begin{eqnarray}
(m-M)_{V, \rm V1370~Aql} &=& 15.2 \cr
&=&(m-M)_{V,\rm V1668~Cyg} + \Delta V - 2.5 \log 0.50/1.0 \cr
&\approx& 14.25 + 0.2 + 0.75 = 15.2\cr
&=& (m-M)_{V,\rm OS~And} + \Delta V - 2.5 \log 0.63/1.0 \cr
&\approx& 14.7 + 0.0 + 0.50 = 15.2.
\label{v1370_aql_v1668_cyg_os_and_distance}
\end{eqnarray}

\begin{figure}
%%\epsscale{0.75}
\epsscale{1.0}
%%\epsscale{1.15}
\plotone{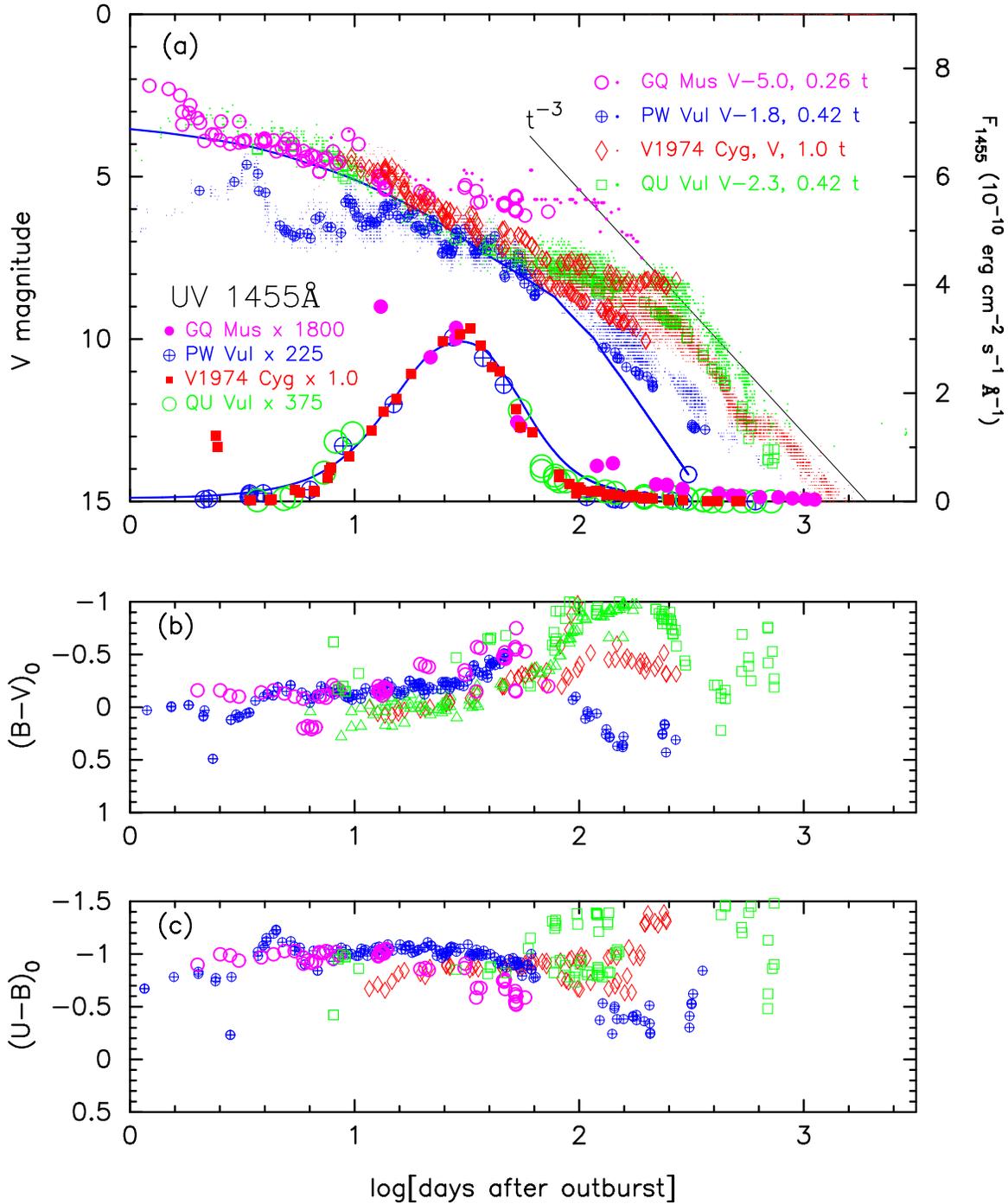}
%\plotone{qu_vul_pw_vul_gq_mus_v_bv_ub_color_logscale.epsi}
%\plotfiddle{evolution1.ps}{5.0cm}{270}{0.4}{0.4}{-170}{220}
\caption{
Same as Figure 
\ref{pw_vul_v1668_cyg_v1500_cyg_v1974_cyg_v_bv_ub_color_curve_logscale_no2},
but for GQ~Mus (magenta open circles for optical/magenta filled 
circles for UV~1455\AA), PW~Vul (blue circles with a plus sign for both
optical and UV~1455\AA), 
V1974~Cyg (red open diamonds/red filled squares), 
and QU~Vul (green open squares/open circles).
Model light curves are also added for PW~Vul
($0.83~M_\sun$ WD, blue solid lines).  See text for more details.
\label{qu_vul_pw_vul_gq_mus_v_bv_ub_color_logscale}}
\end{figure}

From Figure
\ref{qu_vul_pw_vul_gq_mus_v_bv_ub_color_logscale},
\begin{eqnarray}
(m-M)_{V, \rm GQ~Mus} &=& 15.7\cr
&=& (m-M)_{V,\rm V1974~Cyg} + \Delta V - 2.5 \log 1.0/0.26 \cr
&=& 12.2 + (-0.0 + 5.0) - 1.46 = 15.74 \cr
&=& (m-M)_{V,\rm PW~Vul} + \Delta V - 2.5 \log 0.42/0.26 \cr
&=& 13.0 + (-1.8 + 5.0) - 0.52 = 15.68 \cr
&=& (m-M)_{V,\rm QU~Vul} + \Delta V - 2.5 \log 0.42/0.26 \cr
&=& 13.5 + (-2.3 + 5.0) - 0.52 = 15.68.
\label{qu_vul_pw_vul_gq_mus_distance}
\end{eqnarray}

In Figure 
\ref{qu_vul_pw_vul_gq_mus_v_bv_ub_color_logscale},
we plotted UV~1455\AA\  fluxes for PW~Vul (blue open circles
with plus sign), GQ~Mus (magenta filled circles), V1974~Cyg 
(red filled squares), and QU~Vul (green open circles).
In this figure we adopted $M_{\rm WD}=0.65~M_\sun$ for GQ~Mus
as a best-fit model for our assumed chemical composition of 
$X=0.35$, $Y=0.33$, $Z=0.02$, and $X_{\rm CNO}=0.30$
(blue solid lines). 
The distance-reddening relation (magenta solid line in Figure 
\ref{distance_reddening_nq_vul_v1370_aql_gq_mus_pw_vul}(c)) of 
UV~1455\AA\  flux fitting is calculated from 
Equation (\ref{v1668_cyg_uv1455_distance_modulus}) for GQ~Mus.
We adopted $F_{\lambda}^{\rm obs}= 5.0\times10^{-13}$ 
erg~s$^{-1}$~cm$^{-2}$~\AA$^{-1}$, 
which is the observed flux at $\lambda=1455$\AA\  corresponding to the
upper bound of Figure
\ref{qu_vul_pw_vul_gq_mus_v_bv_ub_color_logscale}(a),
and $F_{\lambda}^{\rm mod}= 8.5\times10^{-12}$ 
erg~s$^{-1}$~cm$^{-2}$~\AA$^{-1}$, which is the model flux of the
$0.65~M_\sun$ WD corresponding to the upper bound of Figure
\ref{qu_vul_pw_vul_gq_mus_v_bv_ub_color_logscale}(a).
We also adopted $M_{\rm WD}=0.95~M_\sun$ for QU~Vul
as a best-fit model for our assumed chemical composition of 
$X=0.65$, $Y=0.27$, $Z=0.02$, $X_{\rm CNO}=0.03$, and $X_{\rm Ne}=0.03$.
The distance-reddening relation (magenta solid line in Figure 
\ref{distance_reddening_qu_vul_qv_vul_v443_sct_v1419_aql}(a)) of 
UV~1455\AA\  flux fitting is calculated from 
Equation (\ref{v1668_cyg_uv1455_distance_modulus}) for QU~Vul.
We adopted $F_{\lambda}^{\rm obs}= 3.0\times10^{-12}$ 
erg~s$^{-1}$~cm$^{-2}$~\AA$^{-1}$, 
which is the observed flux at $\lambda=1455$\AA\  corresponding to the
upper bound of Figure
\ref{qu_vul_pw_vul_gq_mus_v_bv_ub_color_logscale}(a),
and $F_{\lambda}^{\rm mod}= 1.1\times10^{-11}$ 
erg~s$^{-1}$~cm$^{-2}$~\AA$^{-1}$, which is the model flux of the
$0.95~M_\sun$ WD corresponding to the upper bound of Figure
\ref{qu_vul_pw_vul_gq_mus_v_bv_ub_color_logscale}(a).

\begin{figure}
%%\epsscale{0.75}
\epsscale{1.0}
%%\epsscale{1.15}
\plotone{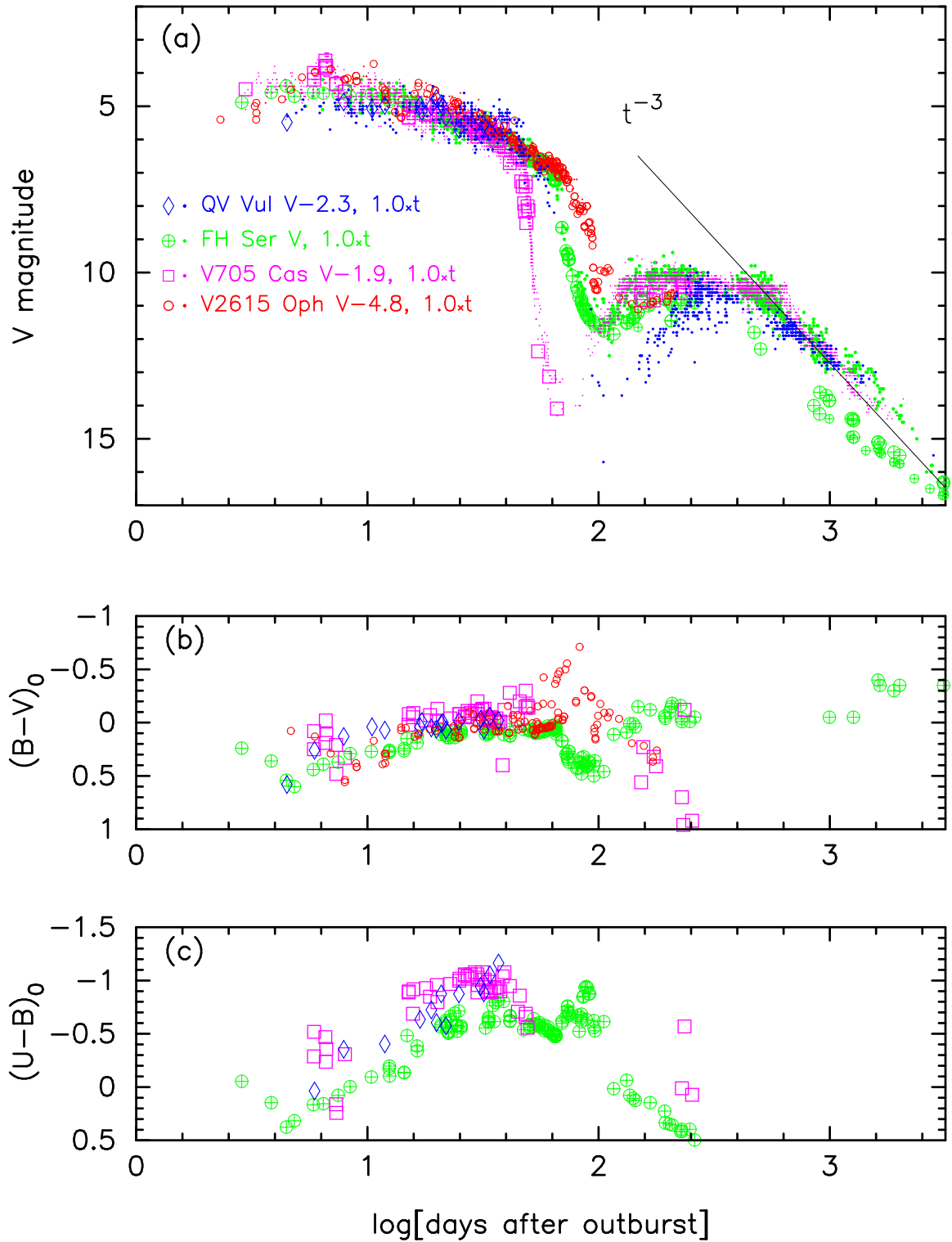}
%\plotone{v2615_oph_v705_cas_qv_vul_fh_ser_v_bv_ub_color_logscale.epsi}
%\plotfiddle{evolution1.ps}{5.0cm}{270}{0.4}{0.4}{-170}{220}
\caption{
Same as Figures 
\ref{v475_sct_t_pyx_nq_vul_dq_her_v_bv_ub_color_logscale_no6},
but for QV~Vul (blue symbols), FH~Ser (green symbols),
V705~Cas (magenta symbols), and V2615~Oph (red symbols).
\label{v2615_oph_v705_cas_qv_vul_fh_ser_v_bv_ub_color_logscale}}
\end{figure}

From Figure 
\ref{v2615_oph_v705_cas_qv_vul_fh_ser_v_bv_ub_color_logscale},
\begin{eqnarray}
(m-M)_{V,\rm QV~Vul} &=& 14.0 \cr
&=& (m-M)_{V,\rm FH~Ser} + \Delta V \cr
&=& 11.7 +(-0.0 + 2.3) = 14.0\cr
&=& (m-M)_{V,\rm V705~Cas} + \Delta V \cr
&=& 13.4 +(-1.7 + 2.3) = 14.0\cr
&=& (m-M)_{V,\rm V2615~Oph} + \Delta V \cr
&=& 16.5 +(-4.8 + 2.3) = 14.0.
\label{v2615_oph_v475_sct_v705_cas_qv_vul_fh_ser_distance}
\end{eqnarray}

\begin{figure}
%%\epsscale{0.75}
\epsscale{1.0}
%%\epsscale{1.15}
\plotone{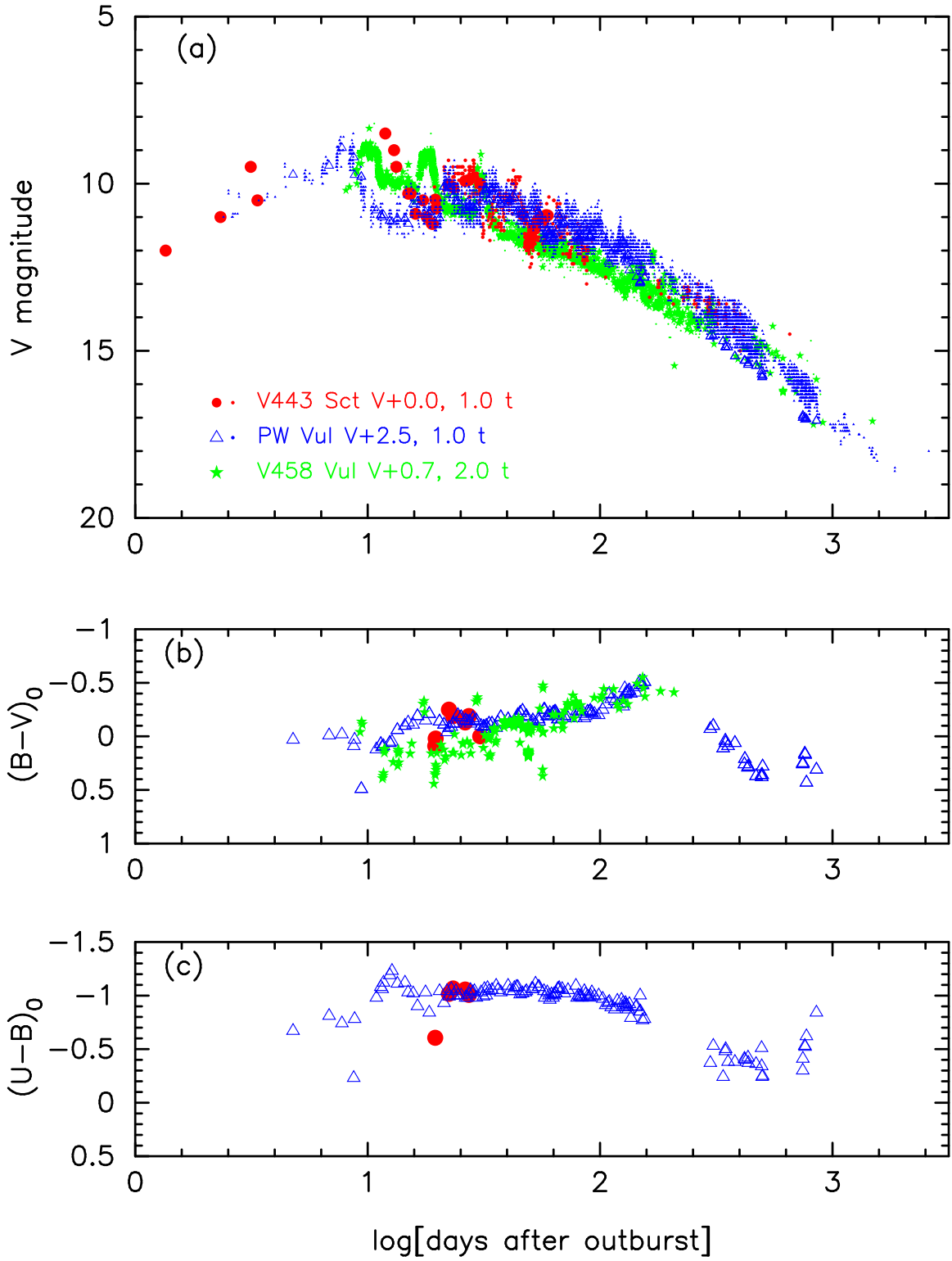}
%\plotone{v458_vul_v443_sct_pw_vul_v_bv_ub_color_logscale.epsi}
%\plotfiddle{evolution1.ps}{5.0cm}{270}{0.4}{0.4}{-170}{220}
\caption{
Same as Figure 
\ref{pw_vul_v1668_cyg_v1500_cyg_v1974_cyg_v_bv_ub_color_curve_logscale_no2},
but for V443~Sct and V458~Vul together with PW~Vul.
\label{v458_vul_v443_sct_pw_vul_v_bv_ub_color_logscale}}
\end{figure}

From Figure
\ref{v458_vul_v443_sct_pw_vul_v_bv_ub_color_logscale},
\begin{eqnarray}
(m-M)_{V,\rm V443~Sct} &=& 15.5 \cr
&=& (m-M)_{V,\rm PW~Vul} + \Delta V - 2.5 \log 1.0\cr 
&=& 13.0 + 2.5 + 0.0 = 15.5 \cr
&=& (m-M)_{V,\rm V458~Vul} + \Delta V - 2.5 \log 2.0\cr 
&\approx& 15.5 + 0.7 - 0.75 = 15.45.
\label{v458_vul_v443_sct_pw_vul_distance}
\end{eqnarray}

The obtained values of $(m-M)_V$ are summarized in Table  
\ref{extinction_from_two_color_diagram}.

%Fig.41
%\placefigure{pw_vul_v1668_cyg_v1500_cyg_v1974_cyg_v_bv_ub_color_curve_logscale_no2}

%Fig.42
%\placefigure{rs_oph_v446_her_v1500_cyg_v1668_cyg_v382_vel_v_color_logscale}

%Fig.43
%\placefigure{v475_sct_t_pyx_nq_vul_dq_her_v_bv_ub_color_logscale_no6}

%Fig.44
%\placefigure{v5114_sgr_v1500_cyg_v1494_aql_iv_cep_lv_vul_v_color_logscale}

%Fig.45
%\placefigure{v496_sct_v1065_cen_v1419_aql_v1668_cyg_v_bv_ub_color_logscale}

%Fig.46
%\placefigure{v2491_cyg_v2468_cyg_v1500_cyg_v1668_cyg_iv_cep_v_color_logscale}

%Fig.47
%\placefigure{v2467_cyg_v2468_cyg_v1668_cyg_iv_cep_v_color_logscale}

%Fig.48
%\placefigure{v5114_sgr_v1500_cyg_v1494_aql_iv_cep_lv_vul_v_color_logscale}

%Fig.49
%\placefigure{v496_sct_v1065_cen_v1419_aql_v1668_cyg_v_bv_ub_color_logscale}

%Fig.50
%\placefigure{qu_vul_pw_vul_gq_mus_v_bv_ub_color_logscale}

%Fig.51
%\placefigure{v2615_oph_v705_cas_qv_vul_fh_ser_v_bv_ub_color_logscale}

%Fig.52
%\placefigure{v458_vul_v443_sct_pw_vul_v_bv_ub_color_logscale}

\end{document}